\documentclass[aps,pre,groupedaddress]{revtex4-2}

\usepackage{xcolor}
\usepackage{graphicx}
\usepackage{amsmath}
\usepackage{amssymb}
\usepackage{bm}
\usepackage{comment}
\usepackage{float}
\usepackage[ruled,lined]{algorithm2e}
\usepackage{url}
\usepackage{hyperref}
\usepackage{lineno}

\def\x{{\bf x}}
\def\nc{n_{\rm c}}
\def\Nte{N_{\rm te}}
\def\Xte{X_{\rm te}}

\def\<{\left<}
\def\>{\right>}

\def\d{{\rm d}}
\def\u{{\bf u}}
\def\z{{\bf z}}

\def\y{{\bf y}}
\def\h{{\bf h}}
\def\pl{{\rm pl}}

\begin{document}

\title{Inverse generalized spin models of answers to questionnaires}

\author{Arianna Armanetti}
\affiliation{NETWORKS research unit, IMT School for Advanced Studies, P.zza San Francesco 19, 55100 Lucca (Italy)}
\author{Luca Cecchetti}
\affiliation{MOMILAB research unit, IMT School for Advanced Studies, P.zza San Francesco 19, 55100 Lucca (Italy)}
\author{Paolo Sarti}
\affiliation{NETWORKS research unit, IMT School for Advanced Studies, P.zza San Francesco 19, 55100 Lucca (Italy)}
\affiliation{Department of Computer Science, University of Pisa, Largo B. Pontecorvo, 3, 56127 Pisa PI (Italy)}
\author{Diego Garlaschelli}
\affiliation{NETWORKS research unit, IMT School for Advanced Studies, P.zza San Francesco 19, 55100 Lucca (Italy)}
\affiliation{INdAM-GNAMPA Istituto Nazionale di Alta Matematica `Francesco Severi', P.le Aldo Moro 5, 00185 Rome (Italy)}
\affiliation{Lorentz Institute for Theoretical Physics, University of Leiden, Einsteinweg 55, 2333 NL Leiden (The Netherlands)}
\author{Miguel Ib{\'a}{\~n}ez-Berganza}
\affiliation{NETWORKS research unit, IMT School for Advanced Studies, P.zza San Francesco 19, 55100 Lucca (Italy)}
\affiliation{INdAM-GNAMPA Istituto Nazionale di Alta Matematica `Francesco Severi', P.le Aldo Moro 5, 00185 Rome (Italy)}

\date{\today}

\begin{abstract}
Network psychometrics conceptualises psychological constructs as emergent properties of systems of interacting items. Energy-based probabilistic models have gained popularity as models of these interactions, but their psychometric application has so far been limited to binary responses, bilinear interactions, and approximated inference methods. To fill these gaps, we here infer and analyse three generalized-spin models of ordinal questionnaire data: the Ising, Blume-Capel (BC), and Blume-Emery-Griffiths (BEG) models. These are maximum-entropy models that accommodate ordinal responses on Likert-type scales with an arbitrary number of options, allowing for single-site anisotropy (BC, BEG) and bi-quadratic item interactions (BEG). We prove the concavity of the maximum likelihood estimation of their parameters, as well as the gauge invariance of the Ising and BC models. We introduce a stochastic gradient ascent algorithm for maximum likelihood inference, and apply this procedure to eleven psychometric and sociological questionnaire datasets.
Evaluating the predictive ability of the inferred models reveals that the BEG model systematically outperforms Factor Analysis and the other spin models in capturing the distributions of factors and distances of subject answers to the mean, across all datasets. By leveraging a competitive interplay between the quadratic and bi-quadratic energy terms, the BEG model uniquely captures individual average-extremist response styles, alongside standard latent factor positioning. Moreover, only the spin models can account for the non-concavity and multi-modality of factor histograms in the most polarizing questionnaires.  
Finally, the analysis reveals other highly non-linear traits of ordinal data---such as the fat-tailed distribution of Mahalanobis distances to the mean---that escape satisfactory description by both factor and spin models.
\end{abstract}

\maketitle

\section{Introduction\label{sec:intro}}
In conventional psychometric frameworks, observed indicators (e.g., questionnaire items such as “I work hard to accomplish my goals” and “I strive for excellence in everything I do”) are treated as effects of an underlying latent variable (e.g., a personality trait such as conscientiousness), consistent with a cause-and-effect interpretation \cite{cronbach1955}. Although this approach has dominated psychometrics for much of the past century, alternative perspectives have emerged in recent years. In particular, approaches rooted in network theory conceptualize psychological phenomena as emergent properties of complex systems \cite{van2006,borsboom2008,cramer2010,cramer2012}. From this perspective, a construct is not assumed to be the common cause of its indicators but is instead understood as arising from interactions among them, reflecting a mereological (i.e., part–whole) relation \cite{borsboom2013}.
Within this framework, a range of psychological constructs can be conceptualized as systems of interacting components. For example, personality may be viewed as a network of affective, cognitive, and behavioral processes \cite{costantini2019,christensen2020}; intelligence as emerging from mutualistic interactions among cognitive abilities during development \cite{van2017,savi2019}; psychopathology as arising from interdependencies among symptoms \cite{cramer2016,robinaugh2019,robinaugh2024}; and attitudes as reflecting interactions among beliefs, feelings, and behaviors toward an object \cite{dalege2016,dalege2018}.
This perspective has motivated the development of formal models that differ from those traditionally used in latent variable approaches, with network models providing one such class of tools \cite{borsboom2021}. Importantly, conceptual differences between network and latent variable frameworks do not imply a strict separation at the statistical level. A growing body of work has shown that several network models are equivalent to latent variable models—such as item response theory \cite{marsman2017} and factor analysis \cite{epskamp2017,vanbork2019}—under specific conditions. Moreover, network models can recover meaningful structure even when the underlying data-generating process follows a common-cause model \cite{marsman2017,fried2020}. Thus, although network theory entails specific assumptions about the nature of psychological constructs, the associated statistical models are not tied to a single data-generating mechanism.
To represent interactions in network models, researchers commonly employ Pairwise Markov Random Fields, in which nodes represent variables (e.g., questionnaire items) and edges represent conditional dependencies \cite{borsboom2021,isvoranu2022}. Estimating such networks is a central and nontrivial task \cite{forbes2017,neal2023}, as it involves determining both the presence and the strength of conditional relationships between variables. This step is critical because commonly used network indices—such as measures of global structure, node centrality, and community organization—depend directly on the estimated edge strengths \cite{borsboom2021}. Accordingly, a range of methods has been developed to estimate network structures from psychometric data (see, for instance, \cite{van2014,golino2017,epskamp2018,waldorp2018,golino2020, haslbeck2020,marsman2025,van2026}, with ongoing refinements driven by critical evaluations in the literature \cite{brusco2019,brusco2021,brusco2022}).
Among these methods, the Ising model has received increasing attention in the psychometric literature. Originally developed in statistical physics to describe ferromagnetism, it represents a system of discrete elements that can take one of two states, whose configuration is determined by local interactions and external influences \cite{brush1967}. Within this framework, the probability of a given system state is defined by an energy function, and its behavior can, under certain conditions, be approximated using methods such as mean-field models. In psychometrics, the Ising model has been adopted as a general framework for modeling interactions among dichotomous variables \cite{van2014}. For example, it can represent binary responses---such as endorsement versus rejection of an item (e.g., pro vs.\ contra attitudes, coded as $+1$ and $-1$)---and examine how these responses interact with one another and with external variables, such as involvement \cite{dalege2018} or attention \cite{dalege2025}. This approach has been applied to the study of attitudes \cite{dalege2016}, political beliefs \cite{brandt2021}, polarization \cite{van2020}, and choice behavior \cite{kruis2020}. Similarly, in the network approach to psychopathology \cite{robinaugh2019}, the Ising framework is used to model interactions among symptoms, which can be either absent (coded as 0) or present (coded as 1), as well as perturbations induced by external variables such as stress \cite{cramer2016}. This framework has been used to account for phenomena such as recovery, chronicity, and resilience \cite{lunansky2025}.
Despite its usefulness, the standard Ising model is limited by its assumption of binary variables. Most psychological data, however, are collected on ordinal scales, and practices such as dichotomization can result in information loss and potential bias \cite{hoffman2017,marsman2025}. Furthermore, intermediate response options—such as neutral or undecided categories (e.g., “neither agree nor disagree”)—often carry substantive meaning that is not adequately captured by binary representations \cite{van2026}. One way to address this limitation is to consider extensions of the Ising model developed in the physics literature (see \cite{finnemann2026} for an accessible overview), such as general-spin formulations that allow for more than two discrete states \cite{rabe1994}. These models have only recently been introduced into psychometrics as a framework for modeling responses on visual analogue scales, with accompanying work examining their mean-field behavior and dynamical properties, including pitchfork bifurcations and hysteresis \cite{waldorp2025}. 
Another extension is the ternary spin Blume–Capel model, originally developed to describe magnetic systems with vacancies \cite{blume1966,capel1966}. In addition to the interaction and external field terms of the standard Ising model, the Blume–Capel formulation includes a single-site anisotropy term that modulates the tendency of spins to occupy the zero state, thereby regulating the balance between active ($\pm 1$) and inactive (0) states. In a psychometric context, this property directly addresses a key limitation of the standard Ising model by allowing nodes to occupy a neutral or inactive state, providing a natural representation of neutrality or uncertainty in item responses \cite{van2026}. It has also been suggested that such models can offer a unified framework for studying different forms of psychopathology, including bipolar and major depressive disorders \cite{van2026}. In parallel, dedicated software for analyzing psychometric data within the Blume–Capel framework is currently under development \cite{waldorp2026}.
A relatively underexplored extension of the Ising model in network psychometrics is the Blume–Emery–Griffiths (BEG) model \cite{blume1971}. This model builds on the Blume–Capel formulation by retaining the single-site term and introducing an additional biquadratic coupling term that captures intensity associations—that is, the tendency of two nodes to exhibit strong activation simultaneously, irrespective of their sign. Consequently, when applied to psychometric data, the BEG model can account for neutrality and uncertainty while also enabling the modeling of tendencies toward extreme responding \cite{ferri2022}.
A final aspect concerns the notions of alignment and energy minimization in Ising models. In general, the energy of the system is minimized when nodes (i.e., questionnaire items) adopt the same state (e.g., +1 or -1) and are aligned both with one another (i.e., responses tend to be consistent across items) and with external influences acting on the system \cite{finnemann2026}. 
This reasoning entails the constraint that interactions between nodes, when present, are predominantly positive, a consideration particularly relevant for network structure estimation, a core aspect of network psychometrics. In practice, this implies that items can be coded consistently with respect to the emergent property of the system. 
For example, if the item “Artificial Intelligence (AI) will make this world a better place” is coded positively, then “I prefer technologies that do not feature AI” should be reverse-coded to maintain consistency and reduce attitudinal energy \cite{dalege2018}, thereby facilitating the emergence of a coherent positive attitude toward AI \cite{stein2024}. 
Although these assumptions may be reasonable in some domains, such as attitude research, they represent a substantial simplification for more complex systems and are likely to be violated in multi-construct settings, where variables may differ in meaning and direction. 
Allowing for both positive and negative interactions introduces structural frustration, in which not all pairwise relations can be simultaneously satisfied, resulting in more complex system behavior but also greater modeling flexibility \cite{moessner2006,finnemann2026}.

To summarize, the application of inverse spin models to psychometrics has so far been mostly restricted to binary responses, bilinear interactions, and approximated inference methods. Crucially, these restrictions can preclude the characterization of the very non-linear collective features---such as polarization and complex or segregated response patterns---that, beyond the reach of linear approaches such as Factor Analysis, network psychometrics aims to uncover.

In light of these considerations, we consider the {\it generalized Ising, BC and BEG models} as a framework for describing cross-sectional questionnaire data. 
This approach is designed to accommodate graded responses on an ordinal scale of arbitrary length, as well as tendencies toward neutral and extreme response categories. By relaxing the constraint of predominantly positive pairwise associations between items, these models are also suited to multi-construct settings, such as personality research. 
We infer the parameters of the spin models from eleven empirical cross-sectional questionnaire datasets spanning diverse psychological constructs, including both unipolar and bipolar scales. 
Ultimately, we compare the inferred generative models against simpler probabilistic benchmarks and examine their predictive properties, including the statistics of distances to the mean and the emergence of multi-stability. 

In particular, we first define the models, demonstrate their identifiability and the concavity of the maximum likelihood inference of their parameters, and prove the gauge invariance for the generalized Ising and BC models. We then implement a protocol of full likelihood maximization, based on the estimation of the stochastic gradients of the log-likelihood via Markov Chain Monte Carlo Gibbs sampling, combined with Adaptive Moment Estimation (ADAM) and Persistent Contrastive Divergence (PCD). 
The description of the models, the numerical methods, the data observables that we use to evaluate the predictive power of the models (such as histograms of latent factors and of distances to the mean), and the benchmark probabilistic models, are detailed in Sec. \ref{sec:methods}. 
Sec. \ref{sec:results} presents an assessment of the out-of-sample accuracy of the inferred models, an analysis of their adherence to the empirical data, and an interpretation of the role plaid by the BEG quadratic and bi-quadratic parameters. We draw our conclusions and discuss future research perspectives in Sec. \ref{sec:conclusions}.

\section{Methods\label{sec:methods}}
\subsection{Notation}
Let a given questionnaire $X\in {\mathbb X} = {\mathbb S}^{ N\times M}$ be a matrix of $N$ subjects by $M$ item answers, whose elements $X_{sj}$ (the answer of the $s$-th subject to the $j$-th question) lie in a set $X_{sj}\in{\mathbb S} =\{v_1,\ldots,v_R\}$ of discrete ordinal answers with cardinality $R$, where $v_q=-(R-1)/2+(q-1)$ is the $q$-th value of the item, $q=1,\ldots,R$. In this way, the values $v_q$ are chosen to be symmetric around zero, i.e., $v_q=-v_{R+1-q}$, so that $\sum_{q=1}^Rv_q=0$. As an example, ${\mathbb S}=\{-1,0,1\}$ if $R=3$ and ${\mathbb S}=\{-3/2,-1/2,1/2,3/2\}$ if $R=4$. We refer to $\x_s = X_{s*}$ as the $s$-th row vector of $X$, representing the vector of answers of the $s$-th data subject across all $M$ questions.

\subsection{Probabilistic generative models with which we infer the data \label{sec:models}}

Given a dataset, we divide the subject indices into two sets of length $N$ and $\Nte$, with which we define the training- and the test-set matrices, $X$ and $\Xte$, of dimension $N\times M$ and $\Nte\times M$, respectively. We perform unsupervised inference of the training-set data $X$. This requires a likelihood probability distribution $P_{\cal M}(\cdot|\theta_{\cal M}):{\mathbb S}^{ M}\to [0,1]$, corresponding to model $\cal M$, whose parameters $\theta_{\cal M}$ are fixed according to certain criteria defined in Sec. \ref{sec:inference}. We will consider two classes of probabilistic models. First, {\it spin models}, which are energy-based models with discrete support in $\mathbb{S}^{ M}$, corresponding to three different Hamiltonian functions. Second, two {\it simple} probabilistic models: {\it Factor Analysis} and the {\it categorical-independent model}. We now describe both the spin and the simple models.

\paragraph{Spin models. \label{sec:spinmodels}} First, we consider, as likelihoods, three energy-based probability distributions whose support is the space of vectors $\x\in \mathbb S^{ M}$ of $M$ discrete answers. In the jargon of statistical physics, such likelihood distributions correspond to the Boltzmann probability measure over the space of spin configurations $\x$ of three different {\it spin models} in the canonical ensemble. The considered spin models are the generalized Ising, Blume-Capel (BC) and Blume-Emery-Griffiths (BEG) models \cite{ising1925,onsager1944,blume1966,capel1966,blume1971,baxter1985,mussardo2010,pathria2011}, where {\it generalized spin models} (also referred to as higher-spin models) denotes that the variables $x_i\in {\mathbb S}$ (the item answers) belong to a domain taking $R$ possible values (with $R\ge 2$ for Ising and $R\ge 3$ for BC,BEG) \cite{krinsky1975,camp1976,berker1976blume,lawrie1984theory}. The probability mass and Hamiltonian of the BEG model are: 

\begin{subequations}\label{eq:boltzmann}
\begin{align}
P(\x|\theta) &= \frac{1}{Z_\theta}\exp \left(-H(\x|\theta) \right) \\
H(\x|\theta) &= - \left( \h^\dag\cdot\x + \frac{1}{2}\x^\dag\cdot J\cdot\x +\frac{1}{2}\sum_{i,j=1}^M x_i^2x_j^2 K_{ij} \right), \label{eq:hamiltonian} \\
& \qquad J_{ij}=J_{ji}\, \forall i\ne j, K_{ij}=K_{ji}, K_{ii}=0\, \forall i,j
,
\end{align}
\end{subequations}
where the parameters $\theta=(\h,J,K)$, with $\h \in {\mathbb R}^M$, and $J,K$ being real symmetric matrices, with $K$ having null diagonal. 

The three coupling terms in the Hamiltonian play distinct physical roles. The first, linear term $\h^\dag\cdot\x = \sum_i h_i x_i$ encodes the local bias acting on each item: $h_i>0$ ($h_i<0$) favors positive (negative) responses to item $i$, independently of the other items. The second, bilinear term $\frac{1}{2}\x^\dag\cdot J\cdot\x = \frac{1}{2}\sum_{i,j}J_{ij}x_ix_j$ encodes pairwise linear interactions between pairs of items: $J_{ij}>0$ ($J_{ij}<0$) favors aligned (anti-aligned) responses for items $i$ and $j$. The diagonal elements $J_{ii}$ act as single-site quadratic anisotropy terms, penalizing or favoring (for $J_{ii}<0$ and $>0$ respectively) responses far from zero at item $i$, regardless of the other items. In the particular case of the BC model, the quadratic couplings $J_{ii}$ penalize or favor responses close to any reference value $a$ (not necessarily to $a=0$), since the maximum likelihood value of the parameter $J_{ii}$ is invariant (see Appendix \ref{sec:gauge}) under shifting the training data by item-dependent constants $x_i\to x_i+a_i$ (meaning that centering the variables $x_j$ at the empirical mean of the original scale does not alter the maximum likelihood quadratic couplings of the BC model). The bi-quadratic interaction term, exclusive to the BEG model, $\frac{1}{2}\sum_{i,j}K_{ij}x_i^2 x_j^2$, encodes intensity--intensity couplings: $K_{ij}>0$ favors items $i$ and $j$ simultaneously taking extreme values, irrespective of their sign. 

The BC model results from taking $K=0_{M}$ in the BEG Hamiltonian in Eq. (\ref{eq:hamiltonian}) (i.e., switching off the biquadratic couplings), while the Ising model results from taking $K=0_{M}$ and $J_{ii}=0,\, \forall i \in\{1,\ldots,M\}$, so that the only interactions are linear and bilinear. Consequently, the number of parameters of the Ising, BC and BEG models are $M(M+1)/2,M(M+3)/2$ and $M(M+1)$, respectively.

The spin models are {\it maximum entropy models} in the sense that they are the maximum entropy probability distributions with support on $\mathbb{S}^{ M}$, subject to fixed expectation values for some specific operators, $\<o_\mu(\cdot)\>_{P(\cdot|\theta)}$ \cite{jaynes1957,jaynes1957-2,nguyen2017}. 
These operators $o_\mu$ are: $\{x_i\}_i$ and $\{x_i x_j\}_{i,j\ne i}$ for the Ising model; $\{x_i\}_i$, $\{x_i x_j\}_{i,j\ne i}$ and $\{x_i^2\}_{i}$ for the BC model; $\{x_i\}_i$, $\{x_i x_j\}_{i,j\ne i}$, $\{x_i^2\}_{i}$ and $\{x_i^2x_j^2\}_{i,j\ne i}$ for the BEG model. The parameters $\theta$ of each model will be fitted to data according to two different criteria: the maximum likelihood and maximum pseudo-likelihood, as detailed in Sec. \ref{sec:inference}.

Importantly, most results presented in this study are independent of the choice of the item values $v_q$, that we here choose as $v_q=-(R-1)/2+q-1$. These are the observables (see Sec. \ref{sec:quantities}) that we sample from the learned Ising and BC models. 
In Appendix \ref{sec:gauge}, we prove that the Ising and BC models are {\it gauge invariant}, in the sense that, for any translation of the spin values $\x\to\x+{\bf a}$ with ${\bf a}\in\mathbb{R}^M$, there exists a unique transformation of the (same) model parameters $\theta\to\tilde \theta_{\bf a}$ such that $P_{\cal M}(\x+{\bf a}|\tilde\theta_{\bf a})=P_{\cal M}(\x|\theta)$ for every $\x\in {\mathbb S}$. As a consequence of gauge invariance, central moments or histograms of centered quantities (as the Euclidean and Mahalanobis distances, or the latent factors) sampled from the maximum likelihood distributions cannot depend on $\bf a$. The BEG model, however, is not gauge-invariant (see Appendix \ref{sec:gauge}); as a consequence, sampling results may, in principle, depend on the particular map of the original Likert scale variables into the spin variables $x_j\in{\mathbb S}$ considered here --only for the BEG model. \\

\paragraph{Simple models.\label{sec:simplemodels}} As a reference, we will fit as well two further probabilistic models, here called {\it simple} as opposed to the above defined spin models (and where simple merely refers to the fact that the fitting procedure of these models is simpler). These are: the {\it Factor Analysis} (FA) and the {\it categorical-independent} (cat-ind) models.

\begin{enumerate}
    \item The {\bf Factor Analysis} (FA) is simply the multivariate Gaussian ${\cal N}({\bm\mu},\Sigma)$ over the vectors of $M$ real numbers, whose vector of means and covariance matrix ${\bm \mu},\Sigma$ are, respectively: the sample estimator of the mean the training set, ${\bm\mu}=\<{\x}\>_{\rho_X}$ (where $\<o(\x)\>_{Q(\x)}$ denotes the expected value of the observable $o$ according to the distribution $Q$, and $\rho(\x)=(1/N)\sum_{s=1}^N\delta_{\x,\x_s}$ is the empirical distribution of the data), and the maximum likelihood covariance matrix with a low-rank plus diagonal noise structure, $\Sigma=\Lambda^\dag\cdot\Lambda+\Psi$, where $\Lambda \in {\mathbb R}^{F\times M}$ and $\Psi$ is a diagonal matrix with positive entries. Matrix $\Sigma$ is fitted by Singular Value Decomposition without rotation, using the python module {\tt sklearn-factor-analysis} in \cite{sklearn_factor_analysis,Armanetti2026_qspin_pkg} with $F=5$ factors (see below for a $F=M$ benchmark). The number of fitted parameters of the FA model is $M (F+2) - F(F-1)/2$. 
    \item The {\bf cat-ind model} (cat-ind) consists of item-independent categorical distributions, such that the probability of each item $j$ assuming the $q$-th value, $x_j = v_q$, coincides with that of the empirical distribution, $p_{j,q}:=\<\delta_{x_j,v_q}\>_{\rho_X}$. In other words, the probability mass of answering a vector $\x$ in the cat-ind model is $\text{prob}(\x)=\prod_{j=1}^M \kappa(x_j;{\bf p}_{j,*})$, where $\kappa(v;{\bf p})=\sum_{q=1}^R\delta_{v,v_q}p_{q}$ is the density of the categorical distribution with vector of probabilities $\bf p$. The number of fitted parameters of the cat-ind model is $RM$.
\end{enumerate}

We define a further simple, reference model for consistency tests, it is:

\begin{itemize}
    \item The {\bf Gaussian model} is simply the multivariate Gaussian ${\cal N}({\bm\mu},C)$ as in FA but using the sample estimator of the covariance matrix $C=\<\x\x^\dag\>_{\rho_X}-\<\x\>_{\rho_X}\<\x\>_{\rho_X}^\dag$. The number of fitted parameters of the Gaussian model is $M+M(M-1)/2$. This number of parameter coincides with that of FA when $F=M$. Although the Gaussian and FA with $F=M$ models are not strictly equal \cite{ledermann1937}, their maximum likelihood estimates of $\Sigma=C$ coincide for $F=M$ and, therefore, the maximum likelihood densities of both models are identical.  
\end{itemize}

Indeed, all the prediction of the FA model that we describe in this article are qualitatively identical to those of the Gaussian model. For this reason, we do not include the latter in our comparisons. The equivalence of FA and the Gauss model also serves as a robustness test of the choice of $F$, since {\it the maximum likelihood} Gaussian model coincides with that of FA for $F=M$. The possible models that we analyze will therefore be referred to as ${\cal M}=\{$Ising, BC, BEG, FA, cat-ind$\}$.

\subsection{Inference strategies for the spin models \label{sec:inference}}

We infer the parameters $\theta_{\cal M}$ via a two-step procedure whose full algorithmic details are given in Appendices \ref{sec:PseudoLik} and \ref{sec:PCD}.
\bigskip
\paragraph*{{\bf Step 1 --- Pseudo-likelihood maximization.}\cite{besag1975,aurell2012,ekeberg2013,ekeberg2014,decelle2014,nguyen2017,cocco2018}}
The normalization constant $Z_\theta$ requires summing $R^M$ terms, making direct evaluation of the log-likelihood intractable for large $M$. As a first step, we therefore maximize the \emph{pseudo-likelihood}, which replaces the joint distribution with a product of the $M$ single-spin conditional probabilities $P_{{\cal M},i}(x_i|\x_{\setminus i},\theta)$, each requiring only $R$ terms to normalize:
\begin{align}
\theta_{\pl}^* = \arg\max_\theta \sum_{s=1}^N\ln P_{\pl,{\cal M}}(\x_s|\theta), \qquad
\ln P_{\pl,{\cal M}}(\x|\theta) := \sum_{i=1}^M \ln P_{{\cal M},i}(x_i|\x_{\setminus i},\theta). \label{eq:maxpselik}
\end{align}
This is a concave optimization problem \cite{ravikumar2010} with analytically computable gradients, that can be efficiently solved with deterministic gradient descent. We employ the L-BFGS-B algorithm \cite{liu1989,nocedal2006,L-BFGS-B}.
The explicit conditional log-probabilities and the gradient formulas are given in Appendix~\ref{sec:PseudoLik} (Algorithm~\ref{alg:pselik}).

As we will explain in Sec.~\ref{sec:datasets}, we achieve convergence of the parameters for most of the considered datasets. However, the pseudo-likelihood estimator is \emph{non-consistent}: a zero pseudo-likelihood gradient does not imply that the inferred model reproduces the sufficient statistics, i.e., $\<o_\mu(\x)\>_{P(\x|\theta_\pl^*)} = \<o_\mu(\x)\>_{\rho_X}$ is not guaranteed (see Appendix~\ref{sec:nonconsistency}). The estimate $\theta_\pl^*$ can therefore be used as a standalone approximation when computational resources are limited, but it does not in general yield a moment-matching distribution.
\bigskip
\paragraph*{{\bf Step 2 --- Full likelihood maximization via Persistent Contrastive Divergence (PCD).}\cite{hinton2002,tieleman2008,hinton2012,decelle2016}} 

The maximization of the full likelihood of discrete spin models is generally non-trivial \cite{nguyen2017}. The generalized Ising, BC and BEG models are {\it identifiable} in the sense that their probability distributions are uniquely determined given their parameters. This property, demonstrated in Appendix \ref{sec:gauge}, is related to the concavity of the maximum likelihood estimation of their parameters, also formally demonstrated in Appendix \ref{sec:gauge} (see \cite{wainwright2008,mackay2003} as well). Likelihood concavity justifies the learning algorithm employed here. We approximately maximize the log-likelihood $\max_\theta\sum_s \ln P(\x_s|\theta)$ using a stochastic gradient descent algorithm ---specifically, Persistent Contrastive Divergence (PCD) \cite{hinton2002,tieleman2008,hinton2012}), combined with the ADAM optimizer \cite{kingma2017}. As an initial condition for the optimization algorithm, we use the maximum-pseudo-likelihood parameters $\theta_\pl^*$, inferred as explained in the previous paragraph. The full likelihood gradients are at the $\tt t$-th gradient iteration are:
\begin{subequations}
\label{eq:CDgradients}
\begin{align}
\label{eq:gradh}
\left. \frac{\partial}{\partial h_j}\right|_{\theta({\tt t})} \ln P(X|\theta)/N & =  \<x_j\>_{\rho_X} - \<x_j\>_{P(\x|\theta({\tt t}))}  \\
\label{eq:gradJ}
\left. \frac{\partial}{\partial J_{ij}}\right|_{\theta({\tt t})} \ln P(X|\theta)/N & = \<x_i x_j\>_{\rho_X} -  \<x_ix_j\>_{P(\x|\theta({\tt t}))} \qquad i\ne j \\
\label{eq:gradJdiag}
\left. \frac{\partial}{\partial J_{ii}}\right|_{\theta({\tt t})} \ln P(X|\theta)/N & =  \frac{1}{2} \Big(\<x_i^2\>_{\rho_X} -  \<x_i^2\>_{P(\x|\theta({\tt t}))} \Big) \\
\label{eq:gradK}
\left. \frac{\partial}{\partial K_{ij}}\right|_{\theta({\tt t})} \ln P(X|\theta)/N & =  \<x^2_i x^2_j\>_{\rho_X} -  \<x_i^2 x^2_j\>_{P(\x|\theta({\tt t}))} \qquad i\ne j
\end{align}
\end{subequations}
Each gradient is the difference between an empirical average and the corresponding model average (the so called {\it sleep term}), a difference that vanishes when the moment-matching conditions are satisfied. The model averages $\<\cdot\>_{P(\x|\theta)}$ cannot be computed analytically, since the partition function $Z_\theta$ involves a sum over an exponentially large space of $R^M$ configurations --preventing the explicit sum already for $M \gtrsim 12$. 
We therefore approximate the gradients via Markov Chain Monte Carlo (MCMC) Gibbs sampling \cite{pelissetto1993,sokal1997,binder1997}: at each gradient step $\tt t$, we run $\nc$ independent copies of a Markov chain of $\tau_{\rm PCD}$ sweeps each. We therefore approximate the theoretical (sleep) term of the gradients $\<o_\mu\>_{P(\cdot|\theta)}$ as an average over the $\nc\times\tau_{\rm PCD}$ sampled configurations, and update the couplings $\theta({\tt t}+1)$ with an Euler step $=\theta({\tt t})+\eta \nabla_\theta|_{\theta({\tt t})}\ln P(X|\theta)$, optionally augmented with the ADAM momentum terms.

The Markov Chains are \emph{persistent}: the final configuration of the $n$-th copy of the Markov Chain at step $\tt t$ is used as the initial condition at step $\tt t+1$, avoiding re-thermalization at every step. At the beginning of each gradient-descent block (every ${\tt t}_{\rm r}$ gradient steps), all chains are independently re-initialized to randomly drawn empirical configurations $\x_s$.
The full rationale for the $\nc$ parallel copies, the persistence scheme, and the ${\tt t}_{\rm r}$ reset are discussed in Appendix~\ref{sec:PCD} (Algorithm~\ref{alg:PCD}). 
Since the full likelihood is unfeasible to compute explicitly, convergence is monitored through three loss functions, $L_{\bf h},L_J,L_K$ equal to the squared norms of the gradients in Eqs.~(\ref{eq:CDgradients}), in such a way that $L_{\bf h} = \sum_j [ \<x_j\>_{\rho_X} - \<x_j\>_{P(\cdot|\theta({\tt t}))}]^2$, and analogously for $L_J,L_K$ (see Eq. \ref{eq:losses} in Appendix \ref{sec:PCD}). The losses tend to diminish, up to stochastic fluctuations, as the algorithm converges toward the moment-matching conditions, as will be explained below. The PCD protocol, ADAM hyper-parameters, and block structure are detailed in Appendix~\ref{sec:PCD}.

In Sec. \ref{sec:results}, we present the comparison between the empirical, training-set histograms ${\sf h}_{o,\rho_X}$ for a set of observables $o$ (defined in Sec. \ref{sec:quantities}), and their equivalent ${\sf h}_{o,Q_{\cal M}}$, where $Q_{\cal M}$ are the likelihoods $P_{\cal M}(\cdot|\theta^*)$ of the spin models defined above in Sec. \ref{sec:models}, whose parameters $\theta^*$ {\it are fitted with the PCD algorithm of approximated maximization of the full likelihood}. The pseudo-likelihood maximization algorithm is used, in Sec. \ref{sec:results}, only as a guess initial condition $\theta({\tt t}=0)$ for the maximization of the parameters with the PCD algorithm. 

As we will say below, we obtain qualitatively similar, although non identical results using the maximum pseudo-likelihood parameters, instead of the  maximum likelihood ones (see Sec. \ref{sec:validity} and Appendix \ref{sec:nonconsistency}). This partial consistency suggests that the relative advantage of the BEG model with respect to the Ising and BC models, and that of the spin models with respect to the simple models, do not depend much on the details of the maximization protocol.


\subsection{Quantities that we sample from the inferred generative models \label{sec:quantities}}

Once learned, we compare the probabilistic generative models with the data, through some {\it quantities} that we estimate both from the empirical histogram $\rho_X$, and from the likelihood probability distribution of the learned models $P_{\cal M}(\cdot|\theta_{\cal M})$, through MCMC sampling. We can therefore assess to what extent the spin models, that are constructed in such a way that they reproduce by construction some empirical quantities (the sufficient statistics $\<o_\mu\>_{P(\cdot|\theta_{\cal M})}=\<o_\mu\>_{\rho_X}$), also reproduce different empirical quantities that are not required to reproduce. This distinction is the central diagnostic of our study: a model that is consistent with its training constraints (the sufficient statistics) may nonetheless fail to capture other structural features of the data, and such failures reveal which physical ingredients are missing from the model. While in the next subsection we describe the sampling protocol, we now describe which are these quantities:

\begin{enumerate}
    \item The histogram of Euclidean (squared) distance to the mean. We estimate, according to the data and to all the models the histogram of Euclidean (squared) distances to the sample mean $\bm\mu=\<\x\>_{\rho_X}$:
    \begin{equation}
    d_{\x} := {||\x-{\bm\mu}||_2}^2 = \sum_{j=1}^M (x_j-\mu_j)^2
    \label{eq:Ed2}
    \end{equation}
    We therefore compare the histogram of Euclidean distances according to the fitted model ${\sf h}_{d_{\x},P_{\cal M}}$, with that of the empirical data, ${\sf h}_{d_{\x},\rho_{X}}$ where ${\sf h}_{o,Q}(\cdot)$ refers to the histogram of the quantity $o:{\mathbb S}^{ M}\to{\mathbb R}$ over the probability density $Q$:
    \begin{align}
    {\sf h}_{o,Q}(z) = \sum_{\x} Q(\x) \delta(o({\x})-z).
    \end{align}
    We notice that one disposes of the analytical expression for ${\sf h}_{d_{\x},\nu(\x)}$, i.e., for the particular case of the Gaussian model. Such distribution corresponds to that of the sum of $M$ $\chi^2$ variables, weighted by the eigenvalues $\bm\lambda$ of the sample covariance matrix $C$ --a distribution of which we know its characteristic function (see Appendix \ref{sec:gaussian}).
    \item We consider as well the histogram of the Mahalanobis distance:
    \begin{align}
    \label{eq:maha}
    d^{({\rm M})}_{\x} &:= \frac{1}{2}(\x-{\bm\mu})^\dag \cdot C^{-1} \cdot (\x-{\bm\mu})
    \end{align}
    where $C$ is the sample estimator of the covariance of the data $X$. In the language of statistical models, the Mahalanobis distance is the energy (minus the exponent) of the probability density of the multivariate Gaussian exhibiting the same mean and covariance of the data $X$. Unlike the Euclidean distance, the Mahalanobis distance accounts for the correlation structure of the data: it rescales each direction in response space by the corresponding variance, so that directions of high variance count less. A response vector $\x$ that deviates from the mean along a high-variance principal component will have a smaller Mahalanobis distance than one deviating by the same Euclidean amount along a low-variance direction. The two distances are therefore complementary: the Euclidean distance is sensitive to the overall spread, while the Mahalanobis distance is sensitive to the shape of the distribution relative to its covariance structure. Again, one disposes of an analytical expression for ${\sf h}_{d^{({\rm M})}_{\x},\nu(\x)}$ in the particular case of the Gaussian model. This is the Gamma distribution with shape parameter $M/2$ (see Appendix \ref{sec:gaussian}).
%
%
    \item The histogram of the first three latent factors, ${\sf h}_{f_j}$ with $j=1,2,3$. In factor analysis, latent factors $\mathbf{f}$ are unobserved variables that account for the shared covariance structure of the manifest variables: the model assumes $\x = {\bm\mu} + \Lambda \mathbf{f} + {\bm\epsilon}$, where $\Lambda$ is the $M\times F$ loading matrix, $\mathbf{f}\sim\mathcal{N}(0,1_F)$ are the $F$ latent factors, and ${\bm\epsilon}\sim\mathcal{N}(0,\Psi)$ being $\Psi$ a diagonal matrix of positive entries. This implies $C = \Lambda\Lambda^T + \Psi$, where $\Psi$ captures item-specific variances. Factor scores of each subject $\x$ are estimated as the posterior mean, $\hat{f}_j = [\Lambda^T C^{-1}(\x-{\bm\mu})]_j$. The factor histograms explore the full shape of the marginal distributions along the latent dimensions. Different clusters in the data should reflect in the onset of convexity or even multi-modality of factor distributions.
\end{enumerate}

As a benchmark reference, we have considered a further observable:

\begin{itemize}
\item The histogram of the first tree principal components ${\sf h}_{x'_j}$ with $j=1,2,3$, where $x'_j = {\bf u_j}^\dag\cdot (\x-{\bm \mu})$ is the $j$-th principal component of vector $\x$, and where ${\bf u}_j$ is the $j$-th eigenvector (corresponding to the $j$-th eigenvalue $\lambda_j$ in decreasing order $\lambda_1\ge\lambda_2\ge\cdots$) of the sample covariance matrix $C$.
\end{itemize}

Remarkably, all the results of the article regarding histograms of factors are qualitatively identical for histograms of principal components. For brevity, we therefore omit to report the latter.

\subsection{Sampling strategy from the inferred generative models}

Once the parameters of each model $\theta_{\cal M}$ have been learned, we sample the quantities from the corresponding probability distributions $P_{\cal M}(\cdot|\theta_{\cal M})$. For this scope, 
we employ MCMC Gibbs sampling again. 
The MCMC sampling protocol (different from the one employed to estimate the gradient terms in Eqs. (\ref{eq:CDgradients})) consists in $\tau_{\rm s}=10^7$ Monte Carlo Gibbs sampling sweeps, of which we collect all the consecutive configurations ${\bm \sigma}(\tau)$. Importantly, every $\tau_{\rm c}$ we reset the configuration ${\bm\sigma}(\tau):=\x_i$ to the vector of answers of a random empirical subject $i$, with $\tau_{\rm c}=10^4$. In this way, there are essentially ${\rm int}(\tau_{\rm s}/\tau_{\rm c})$ independent Markov chains, each one starting from a random empirical subject. Finally, we estimate the theoretical expectation values $\<\cdot\>_{P_{\cal M}}$ and histograms of the observables ${\sf h}_{o,P_{\cal M}}$ as the sample mean and histograms of the sampled configurations $\{{\bm\sigma}(\tau)\}_{\tau=1}^{\tau_{\rm s}}$.

This protocol will also serve as a consistency test of the PCD algorithm, when comparing the consistency between theoretical and empirical sufficient statistics after learning (see Appendix \ref{sec:consistency}). We observe (see Appendix \ref{sec:consistency}) that the agreement between empirical and theoretical sufficient statistics $\<o_\mu\>_{P_{\cal M}}$ is consistent with the one found during the training.

\subsection{Considered datasets \label{sec:datasets}}

We briefly describe the considered datasets in Table \ref{table:questionnaires}. Initially, we considered $14$ datasets, spanning a heterogeneous range of psychometric instruments: clinical and diagnostic scales ({\sf dass}), measures of opinions and beliefs ({\sf gcbs}, {\sf pwe}, {\sf rwas}, {\sf cfcs}), and personality-trait inventories---including both broad multi-scale instruments ({\sf big5}, {\sf hexaco}, {\sf msscq}) and narrower constructs such as narcissism ({\sf hsns}), Machiavellianism ({\sf mach}), the Dark Triad ({\sf sd3}), and three empathy-related scales ({\sf iri}, {\sf ei}, {\sf acme}). The datasets also differ markedly in size: the number of items per questionnaire $M$ ranges from $12$ ({\sf cfcs}) to $240$ ({\sf hexaco}), and the number of subjects $N$ from a few hundreds to $10^5$. Most datasets are taken from the {\it Open-Source Psychometrics Project} website~\cite{openpsy}; three empathy inventories ({\sf iri}~\cite{albiero2006iri}, {\sf ei}~\cite{jordan2002ei}, {\sf acme}~\cite{vachon2016acme}) were collected in-house and are described in the following. 
We therefore excluded those datasets for which the number of parameters of the BEG model $n_{\rm BEG}=M(M+1)$ exceeds the number of sample points, $n_{\rm BEG}>N$; these are also the questionnaires ({\sf acme}, {\sf hexaco}, {\sf msscq}) for which the pseudo-likelihood maximization algorithm (that we use as an initial condition for the PCD algorithm) turns out not to converge. We consequently consider, for the analysis of the present article, the remaining $11$ questionnaires.

The inversion of the BEG model analyzed here can be, in principle, also performed for such under-sampled datasets, for which the number of parameters to infer is of the same order of the number of data samples. First, one could split the questionnaire into two or more sets of questions exhibiting particularly low inter-set correlation, learn the parameter of such sets independently, and use the resulting inferred set of couplings (with zero inter-set couplings) as an initial condition. Second, and more simply, one could apply a regularization (like L2 or Lasso), to reduce the effective number of inferred parameters. Our publicly available repository \cite{Armanetti2026_qspin_pkg} actually allows one to apply an L2 regularization, which, however, in the present study is set to zero since, as we explain in Sec. \ref{sec:results}, a systematic accuracy/complexity trade-off analysis of the inference problem at hand is out of our present scope, and left for future studies.

\begin{table}
\caption{Analyzed questionnaires and the corresponding number of subjects and items $N, M$; the subjects per item $N/M$; the number of BEG parameters per subject $n_{\rm BEG}/N$; the number of possible answers $R$; the reference and description of each dataset. \label{table:questionnaires}}
\begin{ruledtabular}
\begin{tabular}{c|c|c|c|c|c|c|c}
name 		& $N$ 		& $M$ 		& $N/M$ 	& $n_{\rm BEG}/N$ & $R$ & source 	& description	\\
\hline
{\sf acme} 	& 1009		& 36		& $28.0$	& 1.320		& 5	& in-house	& Affective and Cognitive Measure of Empathy \\
{\sf big5} 	& 8000		& 50		& $160.0$	& 0.318		& 5	&\cite{openpsy}	& Big-5 personality test	\\
{\sf cfcs} 	& 8000		& 12		& $666.0$	& 0.019		& 5	&\cite{openpsy}	& Consideration of Future Consequences Scale	\\
{\sf dass} 	& 2840		& 42		& $67.6$	& 0.635		& 4	& in-house	& Depression Anxiety Stress Scale	\\
{\sf ei} 	& 1009		& 14		& $72.1$	& 0.635		& 5	& in-house	& Empathy Index \\
{\sf hexaco} 	& 8000		& 240		& $33.3$	& 7.23		& 7	&\cite{openpsy}	& HEXACO personality test \\
{\sf gcbs} 	& 1026		& 15		& $68.4$	& 0.233		& 5	&\cite{openpsy}	& Generic Conspiracist Beliefs Scale	\\
{\sf hsns} 	& 8000		& 22		& $363.6$	& 0.063		& 5	&\cite{openpsy}	& Hypersensitive Narcissism Scale	\\
{\sf iri} 	& 1009		& 28		& $36.0$	& 0.804		& 5	&\cite{openpsy}	& Interpersonal Reactivity Index	\\
{\sf mach} 	& 8000		& 20		& $400.0$	& 0.052		& 5	&\cite{openpsy}	& Machiavellianism Test	\\
{\sf msscq} 	& 8000		& 100		& $80.0$	& 1.262		& 5	&\cite{openpsy}	& Multidimensional Sexual Self-Concept Questionnaire	\\
{\sf pwe} 	& 713		& 19		& $37.5$	& 0.532		& 5	&\cite{openpsy}	& Protestant Work Ethic Scale	\\
{\sf rwas} 	& 7744		& 22		& $352.0$	& 0.065		& 9	&\cite{openpsy}	& Right-wing Authoritarianism Scale	\\
{\sf sd3} 	& 8000		& 27		& $296.3$	& 0.094		& 5	&\cite{openpsy}	& Short Dark Triad (2011)	\\
\end{tabular}
\end{ruledtabular}
\end{table}

\section{Results\label{sec:results}}
\noindent The results are organized in three groups. Sec. \ref{sec:validity} concerns the accuracy, significance and out-of-sample error of the inferred spin models. Section~\ref{sec:predictiveitem} analyses the emergence of non-Gaussian structure at the level of item statistics and multi-stability in the inferred Gibbs measure. Sec. \ref{sec:predictivesubject} examines how well the models reproduce the distributions of distances of subject responses to the mean response. We finally provide an interpretation of the BEG model non-linear parameters in Sec. \ref{sec:interpretation}.

\subsection{Significance and out-of-sample accuracy of the inferred models. \label{sec:validity}}

\paragraph{Moment matching.} The maximum-likelihood spin models fitted with PCD approximately reproduce the corresponding sufficient statistics, up to the final values achieved by the losses $L_{\bf h},L_{J},L_{K}$ at the end of training. These are shown in Figs. \ref{fig:losses_mach} and \ref{fig:momentmatching_sampling_mach} of Appendix \ref{sec:consistency}, for the {\sf mach} data. The equivalent figures for the rest of the datasets are Figs. \ref{fig:losses_all},\ref{fig:momentmatching_sampling_all} of Appendix \ref{sec:other}. 

As we show in Appendix \ref{sec:consistency}, the employed learning protocol, common to all datasets, is such that the attained value of the training losses $L_{o_\mu}$ is of the same order of magnitude of the variance error $V_{o_\mu}$ of the empirical sufficient statistics $\<o_\mu\>_{\rho_X}$, estimated through bootstrapping (see Appendix \ref{sec:consistency}). Given the employed learning protocol, common for all the questionnaires, whether $L_{o_\mu}$ is larger or smaller than $V_{o_\mu}$ depends primarily on the number of subjects $N$ in each dataset. For under-sampled questionnaires presenting low number of samples per feature $N/M$, the variance error of the empirical sufficient statistics is larger than the training loss. This indicates that incorporating explicit regularization beyond unregularized maximum likelihood inference  will enhance the out-of-sample accuracy of the inferred model, more than increasing the number of gradient ascent iterations. Conversely, when the losses are larger than the variance error of the empirical sufficient statistics in well sampled questionnaires, one can afford the inference of all the parameters, , making the accuracy of the PCD algorithm the primary bottleneck for the model's generalization ability. 

We now discuss, from the point of view of the accuracy-complexity trade-off, both the significance of the inferred couplings, and the generalization ability of the inferred models. 

\paragraph{Significance of the inferred couplings.} Both sources of error discussed above ---bias error induced by the stochasticity of the PCD algorithm, and variance error stemming from the finite-sample sufficient statistics---contribute to the relative statistical significance of the couplings $\theta_\mu$. In Appendix \ref{sec:significance}, we present a bare statistical significance assessment of the inferred couplings. Such an analysis consists, first, in sampling a synthetic sample $Y$ of $N$ vectors from the maximum likelihood distribution ${\bf y}_s\sim P(\cdot|\theta^*)$, and inferring the maximum likelihood couplings $(\theta^{**})$ of the synthetic data $Y$; afterwards, we compute the element-wise relative error $|\theta_\mu^{**}-\theta_\mu^*|/|\theta_\mu^*|$ between empirical vs re-sampled fitted couplings, and consider barely significant those coupling elements $\theta^*_\mu$ whose error is lower than the $10\%$ of their value. Such a simple (and probably conservative, of high specificity) significance criterion does not aim to be exhaustive but to demonstrate that our interpretation of the inferred BEG model couplings in Sec. \ref{sec:interpretation} is not an artifact of non-significant or over-fitted couplings (see Sec. \ref{sec:interpretation} and Appendix \ref{sec:significance}). 

\paragraph{The BEG and BC models present higher out-of-sample pseudo-likelihood than Ising.} Both bias and variance errors limit the generalization ability of the spin models inferred by maximum likelihood. Yet, these models present better out-of-sample scores (higher test-pseudo-likelihood and lower test-{\it completion-error}) than the simple models and, moreover, the BEG model presents better out-of-sample scores than the Ising and BC models (see Appendix \ref{sec:out}). This suggests that the BEG model is the best model also accounting, in the sense of the accuracy-complexity trade-off, for its higher complexity. The model selection assessment of Appendix \ref{sec:out} is limited since, as we discuss in Appendix \ref{sec:out}, the exact value of the likelihood function is not computationally accessible in this case. The main scope of this article is, however, not to perform an exhaustive model comparison between spin and simple models but, rather, to examine which essential ingredients are necessary in the Hamiltonian of the spin models, to reproduce the observables defined in Sec. \ref{sec:quantities} and their non-linear traits. 

Beyond their sufficient statistics, the maximum-likelihood, maximum-entropy models approximately reproduce as well other quantities that they are not required to reproduce by construction. At the item-level, these are the histograms of item responses and factors (Sec. \ref{sec:predictiveitem}). At the subject level, these are the histograms of Euclidean and Mahalanobis distances to the mean response (Sec. \ref{sec:predictivesubject}). The predictive power assessment of the inferred models that we present below in Secs. \ref{sec:predictiveitem},\ref{sec:predictivesubject} is performed on training-set quantities. We therefore assess the statistical significance of the empirical training-set histograms reporting, as error-bars in the histogram figures, the Wilson score confidence interval with a significance $\alpha=0.05$ (see Appendix \ref{sec:out} for further details and for an alternative assessment of statistical significance of training-set histograms).

\paragraph{Alternative training protocols.} For those datasets for which $V_{o_\mu}$ is not significantly lower than $L_{o_\mu}$, some regularization strategy, as Lasso or L2 (available in the repository {\tt psyspin}), will help enhancing the significance of the inferred couplings (Appendix \ref{sec:significance}) and the out-of-sample scores of the models (Appendix \ref{sec:out}). Such alternative training protocols are left for future studies. The lack of significance of some of the inferred couplings has no impact on the results of the current study: all of them stay qualitatively identical when sampling from the likelihood $P(\cdot|\theta^{**})$ with the re-fitted couplings (see the significance analysis in Appendix \ref{sec:significance}), instead of from the maximum likelihood distributions with parameters $\theta^*$. The model predictions are actually qualitatively similar even sampling from the maximum pseudo-likelihood distribution $P(\cdot|\theta_{\rm pl})$ (see Appendix \ref{sec:nonconsistency}), even though these couplings lead to losses that are two orders of magnitude larger (see the initial iteration ${\tt t}=0$ losses in Fig. \ref{fig:losses_mach} in Appendix \ref{sec:consistency}). This suggests that our conclusions on the relative descriptive power of the spin models do not rely on fine-tuned nor particularly accurate estimations of the model parameters.

    Further considerations regarding the inference with regularization and the influence of the ADAM protocol can be found in Appendix \ref{sec:adam}.

\subsection{Predictive power of the spin models on item quantities.\label{sec:predictiveitem}}

\paragraph{Histograms of item responses and factors.} First, the spin models approximately reproduce the histograms of item responses ${\sf h}_{x_i}$ and of item factors ${\sf h}_{f_i}$ (see, respectively, Figs. \ref{fig:itemhist_all},\ref{fig:factorhist_all} in Appendix \ref{sec:other}). Generally speaking, the BEG model tends to reproduce better than the Ising and BC models the histogram of item values, as can be seen by inspection from the histograms of item responses (Fig. \ref{fig:itemhist_all} in Appendix \ref{sec:other}). This is consistent with the relatively lower completion error of the BEG model shown above. More quantitatively, the spin models reproduce better than FA the empirical item histograms, as results from the analysis of Fig. \ref{fig:gof_item} in appendix \ref{sec:gof_hist} in terms of inter-histogram Jensen–Shannon Divergence and Wasserstein Distance. 

\begin{figure}
\includegraphics[width=0.7\columnwidth]{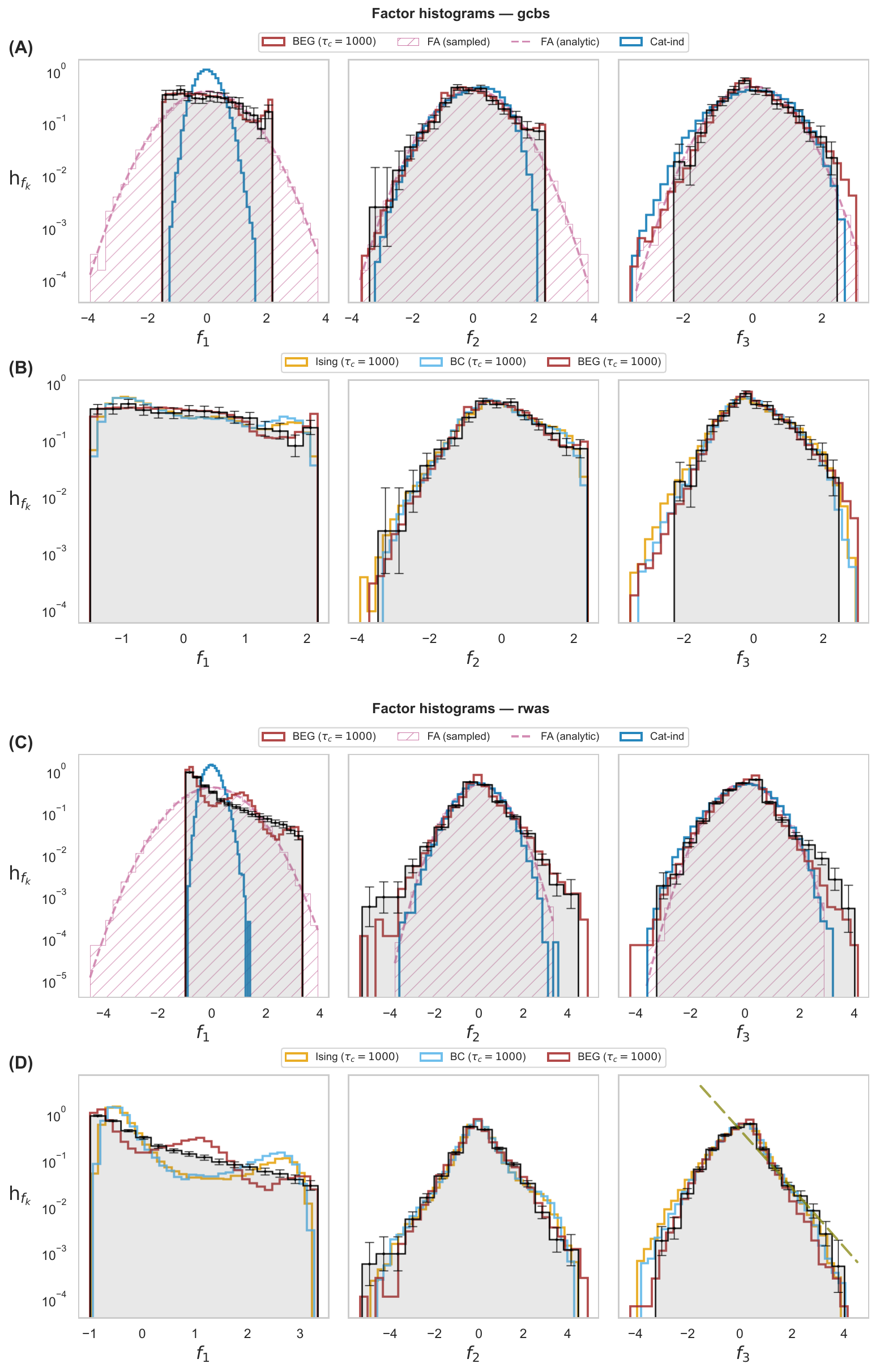}
\caption{Histograms of the first three factors ${\sf h}_{f_k}$ ($k=1,2,3$) for the {\sf gcbs} (A,B) and {\sf rwas} (C,D) questionnaires. (A,C) are the comparison between empirical data and the three spin models (Ising, BC, BEG). (B,D) are the comparison between empirical data, the BEG model, and the simple models (FA, cat-ind). The BEG model reproduces the bi-modal empirical structure of the $f_1,f_2$ for {\sf gcbs}, not so the simple (A) nor the Ising and BC models (B). For {\sf rwas} data (panel D), the Ising and BC models each exhibit two local maxima of ${\sf h}_{f_1}$ and the BEG model exhibits three, while the empirical distribution does not show multi-modal structure. The dashed green line in the $f_3$ histogram in panel D evidences the convexity of ${\sf h}_{f_3}$ in the {\sf rwas} data: it crosses the confidence intervals of six points, but leaves {\it above the line} several other empirical points, and their confidence intervals. Such a convexity is captured at least by the Ising and BC models, not so the simple models in (C). Error bars are Wilson score confidence intervals at $\alpha=0.05$.}\label{fig:fahist_rwas}
\end{figure}

\paragraph{Non-Gaussian structure and metastability in the inferred models.} For some datasets, the histograms of factors ${\sf h}_{f_k}$ and principal components ${\sf h}_{x'_k}$ exhibit a change in the sign of the curvature $\d^2{\sf h}_{f_k}(z)/\d z^2$ --from concave to convex and again concave-- (see Fig. \ref{fig:factorhist_all} in Appendix \ref{sec:other}, where these cases are itemized). In some cases, such a non-Gaussian trait is qualitatively well reproduced by one or more spin models, as it is the case of the $f_3$ of {\sf rwas} in Fig. \ref{fig:fahist_rwas}-D. The emergence of convexity in the factor histograms is a non-linear trait, incompatible with the FA model density ${\sf h}_{f_j}(f)=\nu(f;0,\lambda_j)$ and with the cat-ind model for moderate or large values of $M$, since the histogram of a linear combination of independent variables quickly converges in $M$ to a concave function. 

For some questionnaires, the non-linearity in the histograms of factors and PCs is so strong that they do not only develop convexity but also multi-modality --as for the {\sf gcbs} ($f_1,f_2$) in Fig. \ref{fig:fahist_rwas}-A,B. 
The multi-modal distributions of factors are also developed by the fitted spin models in some cases. For them, they correspond, at least in the language of the naif mean-field picture of Appendix \ref{sec:meanfield}, to the emergence of second (or third) metastable phases, characterized by higher local minima of the mean-field variational free energy. We observe multi-modal histograms of $f_1$ in the Ising-BC models inferred from the datasets {\sf gcbs}, {\sf rwas}, and in the BEG model inferred from the {\sf cfcs}, {\sf dass}, {\sf hsns}, {\sf mach} and {\sf rwas} data (see Fig. \ref{fig:factorhist_all} in Appendix \ref{sec:other}). Two qualitatively different situations arise. 

In the first, the multi-modality in the model mirrors a genuine multi-modal structure present in the empirical data, as in the case of the {\sf gcbs} questionnaire (see Fig. \ref{fig:fahist_rwas}-A,B), where the BEG model captures the empirical multi-modality of $f_1,f_2$, that is absent in the FA and cat-ind models (Fig. \ref{fig:fahist_rwas}-A), and not well captured by the Ising and BC models (Fig. \ref{fig:fahist_rwas}-B). Notably, the metastable phase developed by the BEG model inferred from the {\sf gcbs} questionnaire is reflected in the presence of a second and a third peak in the histogram of Euclidean distances ${\sf h_{d_\x}}$ (see Fig. \ref{fig:E2dhist_gcbs}) that, also in this case, are present as well in the data and absent in the Ising and BC models.

In the second scenario, the multi-modality appears in the inferred model but not in the empirical data --as observed in {\sf rwas}, {\sf cfcs}, {\sf hsns}, and {\sf mach}. For the {\sf rwas} dataset, see Fig. \ref{fig:fahist_rwas}-C,D, the Ising and BC models exhibit two maxima of ${\sf h}_{f_1}$, while the BEG model exhibits three local maxima. Here, the inferred model develops a metastable state that is not reflected in the empirical histogram, indicating that there are essential features of the data eluding an accurate description in terms of the spin models.

In any case, and as a matter of fact, the spin models capture better the factor histograms ${\sf h}_{f_j}$ in questionnaires developing strong non-linearities, as quantitatively reported in Figs. \ref{fig:gof_f1},\ref{fig:gof_fmean} of Appendix \ref{sec:gof_hist}, where we compare, in terms of Jensen–Shannon Divergence and of Wasserstein Distance, the empirical and theoretical histograms, according to all the models. The spin models actually outperform the FA model in terms of both metrics in the more non-linear questionnaires: {\sf cfcs}, {\sf gcbs}, {\sf mach}, {\sf pwe} and {\sf rwas}. Interestingly, these are as well those questionnaires that result significantly clustered according to recently developed clustering methods, applied to the same data \cite{armanetti2026}.

As said in the previous section, these results are completely equivalent in terms of principal components $x'_j$ instead of factors $f_j$.  

\paragraph{Mean-field interpretation of the metastable states.} 

The multi-stability of the Ising model described above is qualitatively and partially quantitatively predicted by the naif mean-field approximation, as discussed in Appendix \ref{sec:meanfield}. Under the hypotheses of the mean-field approximation and of {\it weak enough disorder}, the histogram of the first PC $x'_1$ {\it can develop} (for sufficiently low fields $|{\bf h}|$ and sufficiently high interaction strength) two local maxima, corresponding to the coexistence between a thermodynamically stable and a metastable phases. In this theoretical picture, the first principal component plays the role of the {\it order parameter}, or the {\it generalized magnetization} $\phi=\sum_{i=1}^M x_i U_{1i}$, generalizing the standard magnetization $\phi=\sum_{i=1}^M x_i/M$ in the presence of a weakly disordered coupling matrix. Overall, the analysis in Appendix \ref{sec:meanfield} reveals that, despite the inaccuracy of the mean-field approximation to describe such finite-$M$ systems with inhomogeneous inferred couplings $J$ and fields $\bf h$, the weak-disorder hypotheses leading to the bi-modality of the histogram of $f_1$ are at least approximately satisfied in those datasets for which the inferred Ising model actually exhibits a double peak in ${\sf h}_{f_1}$. 

In conclusion, the double peaks in the histograms of factors and PC's in polarizing questionnaires are strongly non-linear features, escaping a description in terms of linear models. When captured by spin models, such bimodal histograms can actually be interpreted, in the light of mean field theory, as the coexistence between stable and metastable states.

The simple mean field arguments in Appendix \ref{sec:meanfield} are in terms of principal components. However, the results apply for factors as well, since the largest-variance factors are so aligned to principal components in empirical data, to behave in a qualitatively identical manner.

%

\subsection{Predictive power of the spin models on subject quantities: distributions of distances from the mean.\label{sec:predictivesubject}}

\begin{figure}
\includegraphics[width=0.9\columnwidth]{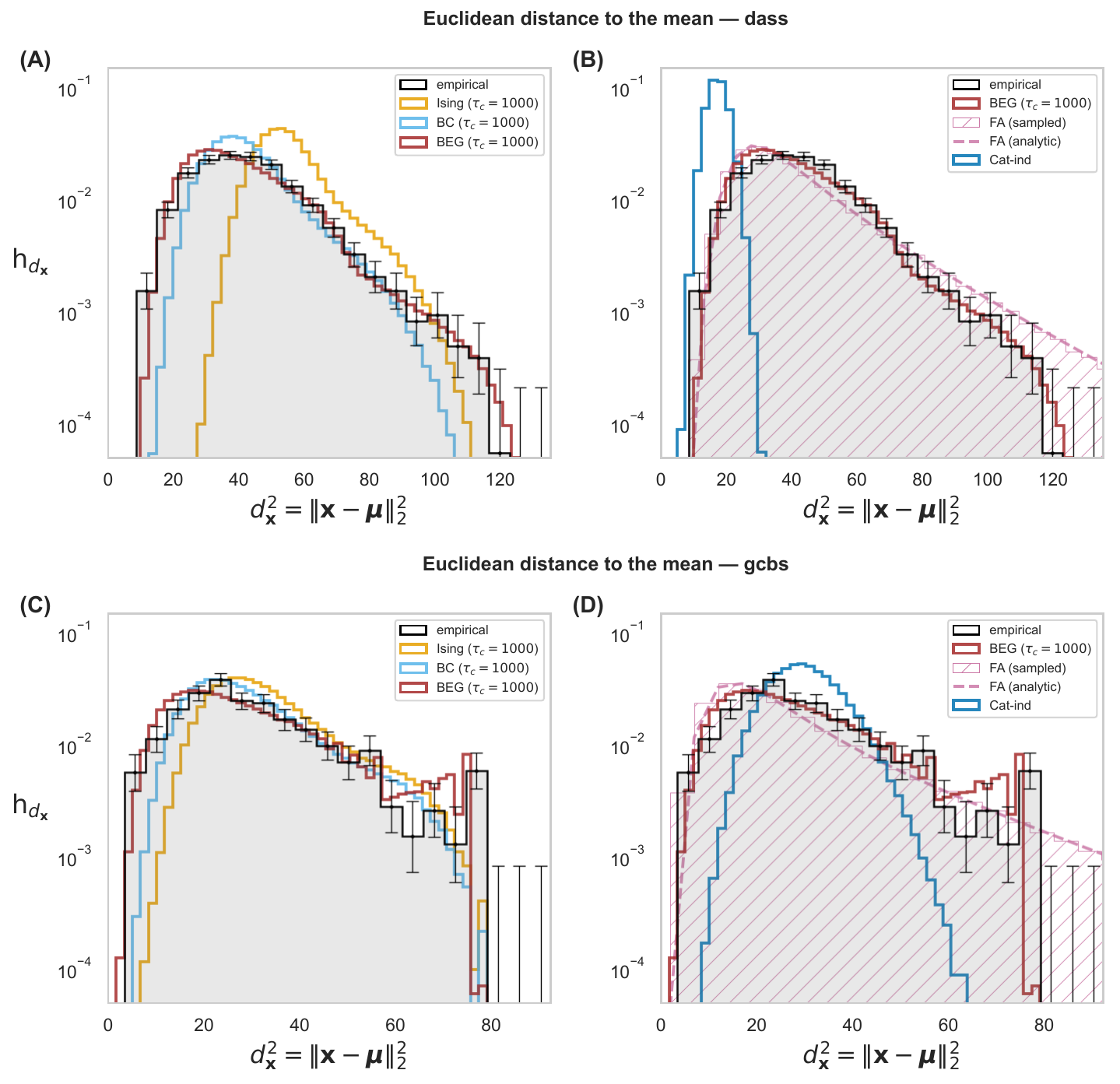}
\caption{Histograms of Euclidean distances to the mean, ${\sf h}_{d_\x}$. (AB) and (CD) correspond to {\sf dass} and {\sf gcbs} questionnaires, respectively. (AC): comparison between empirical data and the three spin models (Ising, BC, BEG); (BD): comparison between empirical data, the BEG model and the simple models. The BEG model reproduces at least the two more pronounced maxima of the {\sf gcbs} empirical histograms, which are absent in the Ising, BC (C) and simple (D) models. Error bars are Wilson score confidence intervals at $\alpha=0.05$.}
\label{fig:E2dhist_gcbs}
\end{figure}

\paragraph{The BEG model captures better than Ising, BC, and simple models the Euclidean distance to the mean.}
The BEG model captures better than the Ising and BC models the histograms of Euclidean distances to the mean, ${\sf h}_{d_\x}$. 
In particular, the BEG model accounts better for the high probability density of subjects close to and far from the mean (and particularly better than Ising for the high probability of subjects close to the mean). This effect is illustrated in Fig. \ref{fig:E2dhist_gcbs} for the {\sf dass} and {\sf gcbs} questionnaires, although it is not a peculiarity of such datasets --see the same quantity for all the datasets in Fig. \ref{fig:E2dhist_all} in Appendix \ref{sec:other}. The BEG model reproduces the histogram of Euclidean distances also better than the simple models. This is shown  in Fig. \ref{fig:E2dhist_gcbs}-B for the {\sf dass} data, and in Fig. \ref{fig:E2dhist_simple_all} in Appendix \ref{sec:other} for all the datasets. A quantitative comparison in terms of Jensen–Shannon Divergence and of Wasserstein Distance can be found in Fig. \ref{fig:gof_e2d} of Appendix \ref{sec:gof_hist}. 

\begin{figure}
\includegraphics[width=0.9\columnwidth]{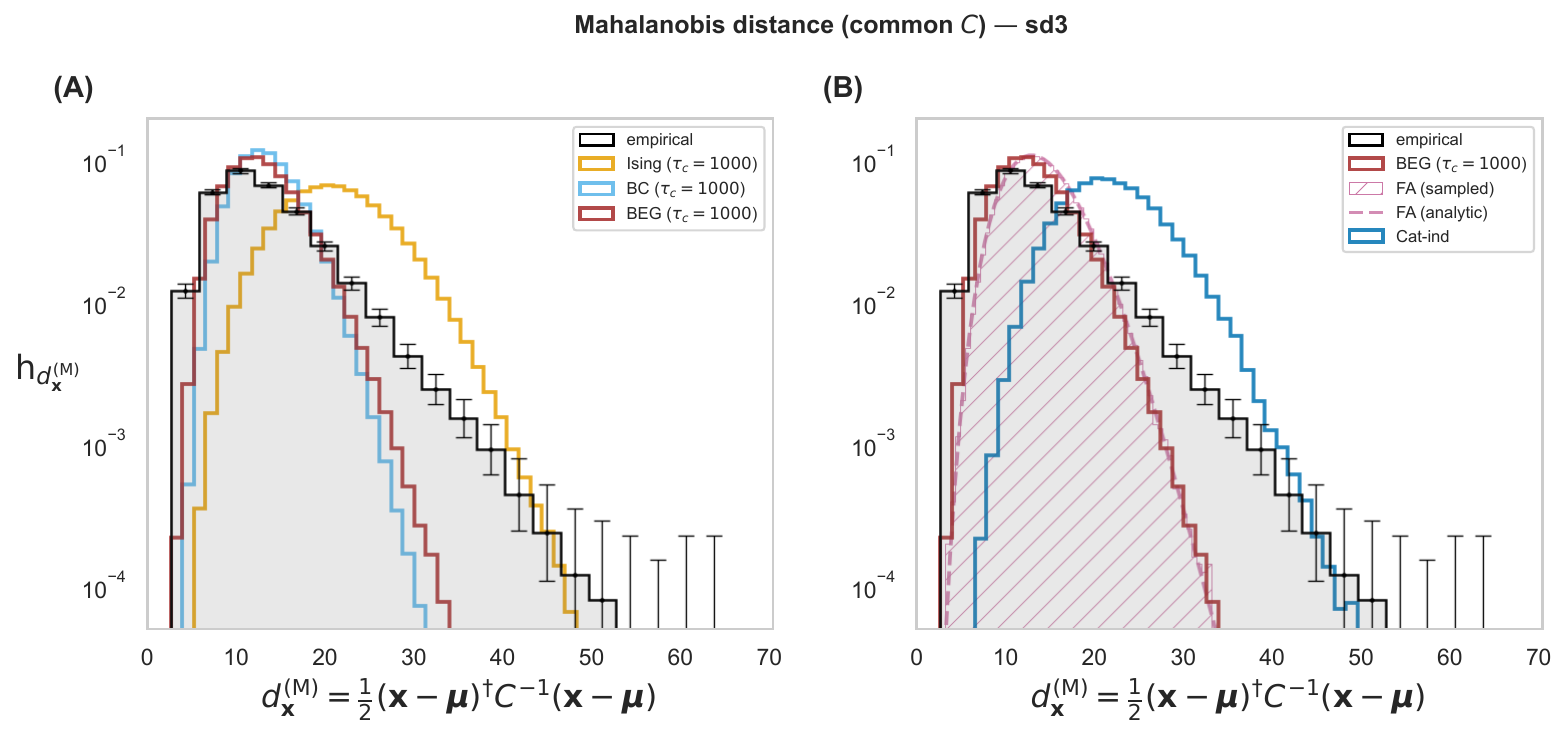}
\caption{Histogram of the Mahalanobis distance to the mean ${\sf h}_{d_\x^{({\rm M})}}$ for the {\sf sd3} questionnaire. (A): comparison between empirical data and the three spin models. (B): comparison between empirical data, the BEG model and the simple models. All models underestimate the frequency of subjects close to and far from the mean. Error bars are Wilson score confidence intervals at $\alpha=0.05$.}
\label{fig:energy_begcopulas_sd3}
\end{figure}

\paragraph{The Mahalanobis distance to the mean is generally not well reproduced by any model, the BEG model being the one that reproduces it better.} All the models underestimate the frequency of empirical subjects that are close to and especially far away from the mean in the sense of the Mahalanobis distance. We show this effect in Fig. \ref{fig:energy_begcopulas_sd3} for the {\sf sd3} data. Figs. \ref{fig:energy_all},\ref{fig:energy_simple_all} in Appendix \ref{sec:other} show that this occurs for many, if not for all the considered questionnaires. In Appendix \ref{sec:mahalanobis} we show that the differences between theoretical and empirical Mahalanobis distances cannot be simply attributed to the error in the model estimation of the sample covariance matrix $C$. A quantitative demonstration of the fact that the BEG model systematically reproduces better than other models the empirical histograms of Mahalanobis distances to the mean, is provided in Fig. \ref{fig:gof_mahala} in Appendix \ref{sec:gof_hist}. 

Unlike the Euclidean distance, the Mahalanobis distance takes into account the correlation or factor structure of the data which, by construction, is present in psychometric answers. A small deviation from the mean along a principal axis with low associated variance would lead to a small Euclidean distance to the mean, but to a large Mahalanobis distance. As a consequence, this quantity is more subtle and difficult to fit than the Euclidean distance. The systematic failure of all models to reproduce its heavy tails indicates that the empirical data exhibit essential features that eludes a satisfactory description not only in terms of Gaussian or factor models, but even in terms of spin models. 

Finally, in contrast to the above limitations of the spin models to describe the Mahalanobis distance, (only) the BEG model turns out to successfully reproduce the empirical anomalous proliferation of subjects close to the mean in the {\sf rwas} data, as we show in Appendix \ref{sec:rwasenergy}.

%
%

%

\subsection{Interpretation of the quadratic and biquadratic coupling constants. \label{sec:interpretation}}

\begin{figure}
\includegraphics[width=0.9\columnwidth]{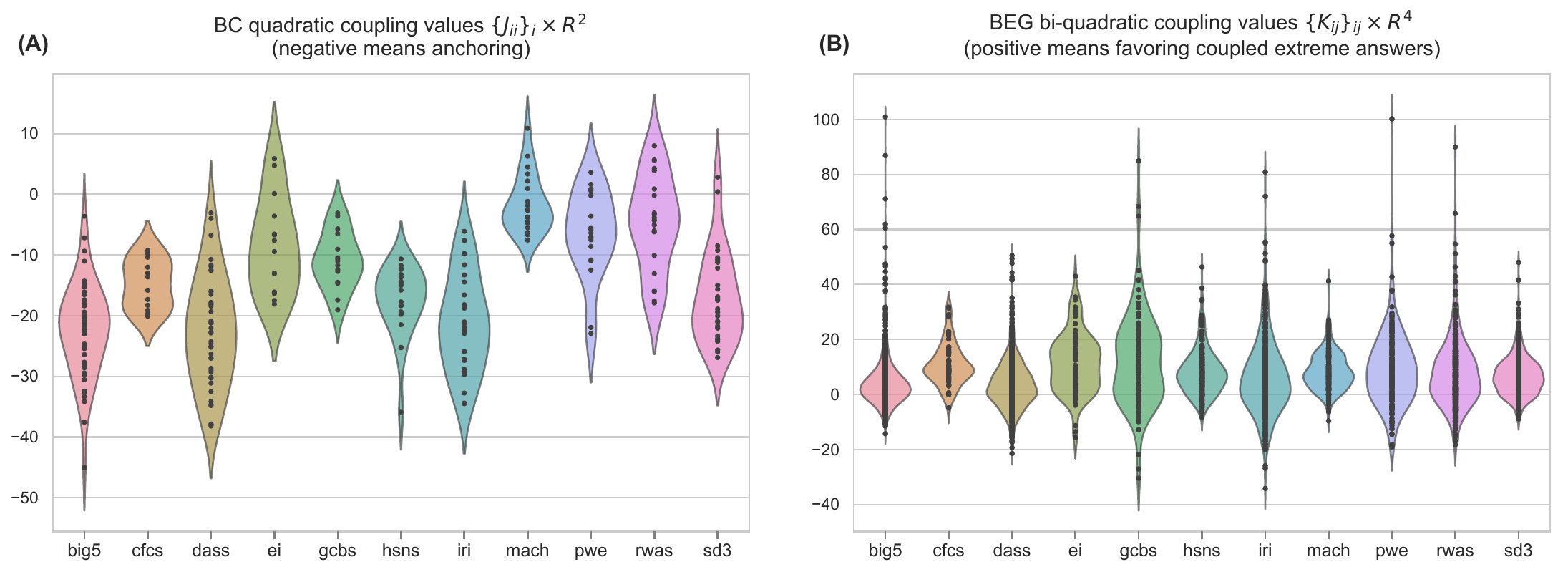}
\caption{Histograms of BEG model inferred quadratic $(J_{ii})_i$ (A) and bi-quadratic $(K_{ij})_{ij}$ (B) couplings for all the questionnaires. The quadratic and bi-quadratic couplings are multiplied by $R^2$ and $R^4$ respectively, to allow for an equal-footing comparison between questionnaires with different values of $R$.}
\label{fig:couplings}
\end{figure}

We now examine the BEG model quadratic $(J_{ii})_i$ and bi-quadratic $K$ coupling constants inferred from the data. In the sign convention of Eqs. (\ref{eq:boltzmann}), a positive quadratic coupling constant $J_{ii}>0$ favors responses with extreme values of $x_i$, being $x_j=\pm r$ the most probable ones, while a negative $J_{ii}$ favors low-spin values and $x_j=0$ in particular. Analogously, a positive bi-quadratic constant $K_{ij}$ favors coupled extreme values of $(x_i,x_j)$, with $(\pm r,\pm r)$ and $(\pm r,\mp r)$ most probable values, while a negative $K_{ij}$ favors coupled moderate values, in particular $(0,v), (v,0)$, where $v$ is any value in $\mathbb{S}$. An inspection of these coupling constants reveals that, for all the questionnaires, most of the quadratic couplings are negative (see Fig. \ref{fig:couplings}-A). Interestingly, the socio-political surveys are those exhibiting the larger fraction of $J_{ii}$ positive elements, that may represent polarizing items, triggering extreme responses. Similarly, Fig. \ref{fig:couplings}-B reveals that, for all the questionnaires, most of the bi-quadratic couplings $K_{ij}$ are positive and favors couples of extreme responses. 

\begin{figure}
\includegraphics[width=0.65\columnwidth]{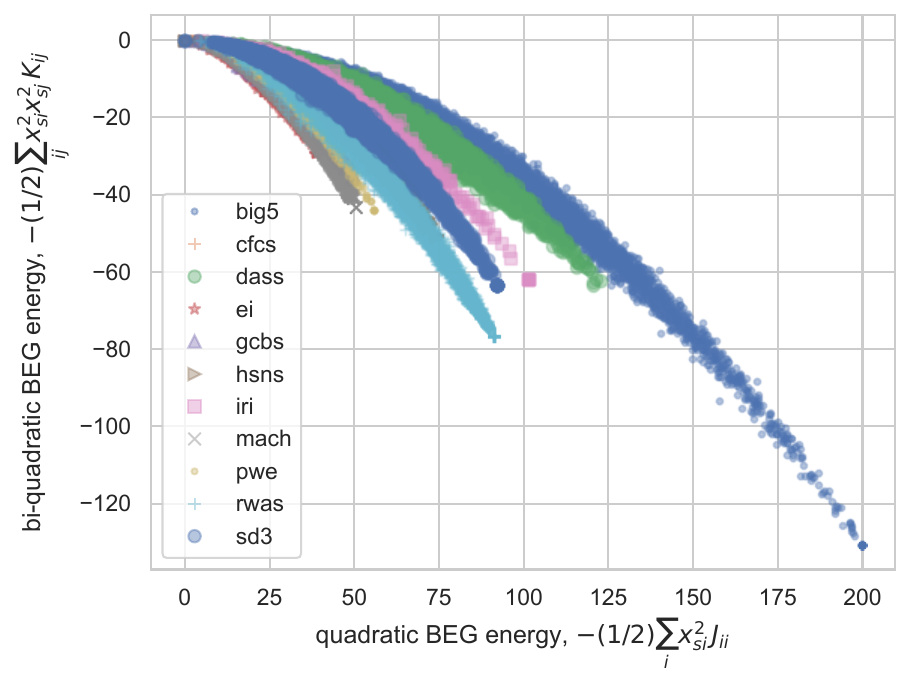}
\caption{Scatter plot of bi-quadratic vs quadratic energy terms, $H_{22}(\x_s)$ vs $H_{2}(\x_s)$, according to the inferred BEG model, for all the subjects in the training set (each point corresponds to a subject $s$), for all the considered datasets (different colors corresponding to different datasets).}
\label{fig:energy_scatter}
\end{figure}

The prevalence of negative quadratic couplings and of positive bi-quadratic couplings suggests a possible interplay mechanism between quadratic and bi-quadratic energies such that the maximum likelihood probability distribution is able to accommodate both ``conformist'' subjects (with prevalence of low intensity responses) and ``extremist'' subjects (exhibiting prevalence of high-absolute value coupled responses). According to this mechanism, the conformists would exhibit a low quadratic energy and a high bi-quadratic energy, and vice-versa, in such a way that none of both antithetical types are penalized too much in probability. This hypothesis is consistent with the scatter plot in Fig. \ref{fig:energy_scatter}, in which we confront the quadratic ($H_2(\x)=-(1/2)\sum_j x_{sj}^2 J_{jj}$, abscissa) and bi-quadratic ($H_{22}(\x)=-(1/2)\sum_{ij} x_{si}^2x_{sj}^2 K_{ij}$, ordinates) energies of all the subjects $s$, for all the considered questionnaires. The figure shows an evident anti-correlation with small dispersion. The bi-quadratic energy $H_{22}(\x_s)$ of a subject $s$ is consequently a meaningful estimation of the subject's relative tendency to answer (coupled) extreme values according to the inferred BEG model. 

In Appendix \ref{sec:significance} we repeat the above analysis in Figs. \ref{fig:couplings},\ref{fig:energy_scatter}, but only for those coupling elements that result to be statistically significant according to the bare variance error assessment analysis explained in Sec. \ref{sec:validity}. This demonstrates that the quadratic/bi-quadratic interplay described above is not an artifact of the smaller, non-significant inferred couplings.

Importantly, in the BEG model, a shift of the scale $x_j\to x_j+a_j$ by an item-dependent integer number $a_j$ can lead to different maximum likelihood quadratic and bi-quadratic couplings (see Appendix \ref{sec:gauge}). The quadratic and bi-quadratic energies, $H_{2}(\x_s)$ and $H_{22}(\x_s)$, consequently depend, in the case of the BEG model, on the choice of how the original Likert scale variables $\bf y$ map into the spin variables $\x$ (see Sec. \ref{sec:spinmodels}). For these quantities, different modeling choices can be more appropriate, depending on the case of study, than simply centering the original variable $x_i=y_j-a$ so that $-r,r$ are the highest and lowest values. One of these alternative choices is centering the variable with respect to the mean value $x_j={\sf int}(y_j-\bar y_j)$, where $\bar y_j$ is the original variable sample mean, ${\sf int}(\cdot)$ is the integer part, and where one lets $x_j$ take values below $-r$ and above $+r$. Three further casting choices $\bf y\to \x$ are presented and analyzed in terms of inference of higher-order spin models in reference \cite{declercq2026}.\\

\section{Conclusions and perspectives\label{sec:conclusions}}
We have studied the generalized Ising, Blume-Capel (BC), and Blume-Emery-Griffiths (BEG) spin models as inverse probabilistic models of questionnaire data, fitting them to eleven psychometric and sociological questionnaires and comparing their ability to reproduce a range of observables beyond their training constraints. The main findings of this work can be summarized as follows.

First, we have analytically established the concavity of the joint data likelihood in terms of their parameters, and the gauge invariance of the Ising and BC models. Leveraging these properties, we have implemented an algorithm of full likelihood maximization, based on the Markov Chain Monte Carlo estimation of the gradients, which is publicly accessible \cite{Armanetti2026_qspin_pkg}. 
Full likelihood maximization is necessary in order to achieve moment-matching (Appendix~\ref{sec:consistency}), as pseudo-likelihood maximization yields parameters whose theoretical moments do not match the empirical ones (Appendix~\ref{sec:nonconsistency}). Methodologically, and to the best of our knowledge, these developments represent an advancement in the literature of inverse spin models of real questionnaire data, from the points of view of the algorithm of full likelihood maximization \cite{waldorp2026,finnemann2021,marsman2025}, of the inference of generalized spin models of general value of the spin \cite{marsman2025}, and of the study of the BEG model, which has received less attention in the literature \cite{finnemann2026}.

These improvements translate directly into enhanced empirical accuracy. First, the BEG model consistently outperforms Factor Analysis (FA), as well as the Ising and BC models, in out-of-sample predictive accuracy (tables~\ref{table:testpselik}--\ref{table:testcomperr} and Fig. \ref{fig:compl-error} in Appendix~\ref{sec:out}). 
Second, generalized spin models capture non-linear traits of polarization in some questionnaires, especially those of socio-political character, as the convexity or bi-modality of the distribution of factors, that remain fundamentally inaccessible to linear methods as FA (Fig. \ref{fig:fahist_rwas}, and Fig. \ref{fig:gof_f1} in Appendix \ref{sec:gof_hist}). 
Third, the BEG model faithfully reproduces the distribution of distances of subject vectors of answers $\x_s$ to the empirical mean ${\bm\mu}=\<\x_s\>_s$, better than the Ising, BC, and FA models, across all datasets (Figs. \ref{fig:gof_e2d},\ref{fig:gof_mahala} in Appendix \ref{sec:gof_hist}). In particular, the BEG model accounts better than the others for the high density of respondents who either tightly align with or strongly deviate from the average response $\bm\mu$.

Interestingly, this predictive superiority admits an interpretation in terms of the specific architecture of the BEG Hamiltonian. Our analysis reveals a competitive interplay between the prevalently negative quadratic couplings (which anchor responses toward moderate values), and positive bi-quadratic couplings (which favor extreme response pairs). Through this energetic competition that the BEG model likelihood can assign high probability to average respondents, without penalizing too much the less common extremist profiles. The two corresponding terms in the energy are, indeed, strongly anti-correlated across subjects (Fig. \ref{fig:energy_scatter}), suggesting that the bi-quadratic energy $H_{22}(\x)$ of a subject answer $\x$ may actually be considered as a meaningful estimator of the subject's relative propensity to extreme responses. Remarkably, we observe this interplay across all eleven questionnaires (Fig. \ref{fig:energy_scatter}), suggesting that the trade-off between conformist and extremist response styles represents a universal statistical signature of multi-state survey data that the BEG model captures within a single, domain-agnostic statistical framework.

While the psychological interpretation of outlier subjects --whose response patterns deviate significantly from the mean— is well established across various psychological domains (such as attitudes, personality, and emotions), the interpretation of subjects whose responses closely align with the mean remains elusive. Specifically, interpreting such average responses as ``conformity'' or ``social desirability'' may be valid in attitudinal research \cite{furnham1986}, but it does not necessarily generalize to other constructs, such as personality. By simultaneously capturing both average and extreme response profiles across highly heterogeneous questionnaires, the BEG model demonstrates, again, its potential to provide a unified descriptive framework. 

Our findings open several prominent avenues for future inquiry. Crucially, our analysis uncovers several features of the empirical data, like the heavy tails in the subject-to-mean Mahalanobis distance, that remain elusive not only to the FA, Gaussian and categorical-independent models, but also to spin models with bi-quadratic interaction. Additionally, because spin models can occasionally introduce artifactual peaks in latent factor distributions, future implementations could suppress the discontinuous onset of metastable states—for instance, by fitting the Ising model to de-meaned data with zero external fields. Addressing the elusive distribution of factors in some questionnaires, as well as the systematic underestimation of heavy-tailed Mahalanobis distances (likely driven by population heterogeneity), remains an open structural challenge. This motivates the exploration of different probabilistic models, such as energy-based models with spin interactions of arbitrary order through advanced model-selection schemes \cite{demulatier2025,declercq2026}, mixtures of energy-based models, or latent variable models with discrete support \cite{kim2023,bauer2004,muthen1984}. 

Another natural extension is the generalization of the mean-field theory in Appendix~\ref{sec:meanfield} to the BC and BEG models, and a systematic assessment of its validity against MCMC sampling. More broadly, the multi-stability identified in Sec. \ref{sec:results} raises the question of the nature of the low-temperature phases of the inferred models. An in-depth analysis of overlap histograms and of the overlap matrix spectra in high-dimensional questionnaires could, in principle, reveal a (finite-size) Gibbs measure $P(\cdot|\theta)$ featuring a multiplicity of well-separated local maxima. Such a finding would open the door to a new, interaction-based descriptive paradigm for psychometrics beyond both the factor paradigm and the metastability picture of Appendix \ref{sec:meanfield}. In this new paradigm, the spin models would describe a multiplicity of clusters of response profiles emerging as those energetically compatible with the constraints imposed by the couplings, in a picture similar to replica-symmetry-breaking phenomena in disordered spin systems \cite{mezard1988}. Conceptually, this regime naturally emerges when the weak disorder approximation in Appendix \ref{sec:meanfield} breaks down. The variational free energy $F$ in Eq.~\ref{eq:Fmuprime} of Appendix~\ref{sec:meanfield} presents, in this case, non-negligible fourth-order interaction terms with random, disordered coefficients, which can induce a proliferation of local minima of $F$ whose number grows rapidly with $M$ \cite{bray1980,fyodorov2004,mezard1988}.

Finally, further developments of the learning algorithms---including investigating the influence of regularization terms, sensitivity to the ADAM algorithm, and benchmarking against alternative inference methods such as naive mean-field or message-passing \cite{nguyen2017}-- will guide the application of the present framework to larger behavioral streams.


\section*{Data availability}
Most of the questionnaire datasets analysed in this study are publicly available
from the \emph{Open-Source Psychometrics Project} (\url{https://openpsychometrics.org})~\cite{openpsy}.
The remaining datasets, collected in-house, are available from the corresponding author upon reasonable request.

\section*{Code availability}
The code used to infer the generalized-spin models (Ising, Blume--Capel and Blume--Emery--Griffiths)
and to reproduce all the analyses reported in this article is publicly available as the
open-source Python package \textsf{psyspin}~\cite{Armanetti2026_qspin_pkg}
(\url{https://github.com/armanetti/psyspin}, DOI \url{https://doi.org/10.5281/zenodo.21679124}).

\section*{Acknowledgements}
We acknowledge Eric Cator, Han van der Maas, Matteo Marsili, Cl\'elia de Mulatier and Lourens Waldorp
for useful discussions.


\section*{Competing interests}
The authors declare no competing interests.

\section*{Funding}
This publication is part of the projects ``Network renormalization: from theoretical physics to the resilience of societies’’ with file number NWA.1418.24.029 of the research programme NWA L3 - Innovative projects within routes 2024 and ``Redefining renormalization for complex networks’’ with file number OCENW.M.24.039 of the research programme Open Competition Domain Science Package 24-1, which are (partly) financed by the Dutch Research Council (NWO) under the grants \url{https://doi.org/10.61686/AOIJP05368} and \url{https://doi.org/10.61686/PBSEC42210}.

\appendix

\section{Identifiability, gauge invariance and strict concavity of the log-likelihood}
\label{sec:gauge}

This appendix establishes three interrelated mathematical properties of the spin models used in this paper: \textit{gauge invariance, identifiability and strict concavity} of the log-likelihood.

We first introduce gauge transformations (induced by translations of the spin values) and their effect on the parameters (\S\ref{subsec:GaugeTransf}), we define gauge invariance as invariance under these transformations, then recall the notion of model identifiability and its connection to the strict concavity of the log-likelihood (\S\ref{subsec:IdModel}--\ref{subsec:LearnGuarantees}). Finally, we verify explicitly that these three properties - gauge invariance, identifiability and strict concavity of the log-likelihood- hold for the Ising, BC and BEG models (\S\ref{subsec:GaugeComp}--\ref{subsec:IdComp}). 
More precisely, gauge invariance for the BEG model requires including symmetric mixed cubic terms, while identifiability holds for all three models whenever $R\geq 3$ (and already for $R=2$ in the Ising case).

\noindent\textit{Notation.} A statistical model $\mathcal{M}$ is specified by a parameter space $\Theta$ and a likelihood function $P_{\mathcal{M}}(\cdot|\theta),\,\theta\in\Theta,$ on a finite state space $\mathbb{X}$, which we suppose to be contained in $\mathbb{R}^{N\times M}$.
 For the spin models considered here, $\Theta = \mathbb{R}^k$ where $k$ is the number of maximum entropy constraints, and $\mathbb{X} = \mathbb{S}^M$ if $N=1$, while $\mathbb{X} = \mathbb{S}^{N\times M}$ for a sample of $N$ subjects, where $\mathbb{S} = \{v_1, \ldots, v_R\}$ is the common set of $R$ spin values for each item. We also write $\x = (x_1,\ldots,x_M) \in \mathbb{S}^M$ for the response vector of a single subject, while $X = (\x_1,\ldots,\x_N)$ will be the matrix corresponding to a sample of $N$ subjects, with $\x_i$ denoting response vectors of subjects in the sample. There will be cases when we want to talk about the case of one or more subjects simultaneously; we will then just write $\x$ for a state in $\mathbb{X}$, whether it be the response vector of a single subject or the matrix corresponding to a sample of subjects.

\subsection{Gauge transformations and gauge invariance}\label{subsec:GaugeTransf}

Given a shift vector $a\in\mathbb{R}^M$, let $\mathbb{X}' = \mathbb{X}+a$ denote a  translated state space. We say that the pair $\xi = (a, f)$ defines a \emph{gauge transformation} from $\mathcal{M}'$ (on $\mathbb{X}'$) to $\mathcal{M}$ (on $\mathbb{X}$) if $f:\Theta'\to\Theta$ is invertible and
\begin{align}\label{eq:GaugeDef0}
    P_{\mathbb{X}+a}(\x+a\,|\,\theta') = P_{\mathbb{X}}(\x\,|\,f(\theta')) \qquad \forall\, \x \in \mathbb{X},\; \theta' \in \Theta'.
\end{align}
Since both $f$ and the translation $\x\mapsto\x+a$ are invertible, if $\xi=(a,f)$ is a gauge transformation and $\theta^*$ maximizes $\log P_{\mathcal{M}}(\x^*|\cdot)$ for some fixed state $\x^*$, then $\tilde\theta = f^{-1}(\theta^*)$ maximizes $\log P_{\mathcal{M}'}(\x^*+a|\cdot)$. Maximum-likelihood estimators in $\mathcal{M}$ and $\mathcal{M}'$ are thus related by $\xi$, and one can freely work in whichever gauge is most convenient.

Taking logarithms, $\xi$ defines a gauge transformation if and only if 
\begin{align}\label{eq:GaugeDef}
H_{\mathbb{X}+a}(\x+a\,|\,\theta) = H_{\mathbb{X}}(\x\,|\,f_a(\theta)) + c(\theta,a),
\end{align}
where $c$ is state-independent and $f_a(\theta)=\theta+g(\theta,a)$ is invertible.

In the rest of this section, we are interested in the following three classes $\mathcal{C}=(\mathcal{M})_{\mathcal{M}}$ of models: given $\mathbb{X}=\mathbb{S}^{N\times M}$, the classes of generalized Ising models, BC models and BEG models defined on the shifted states spaces $\mathbb{X}+a$.

We say that \emph{gauge invariance} or \emph{invariance under gauge transformations} holds for a certain class $\mathcal{C}$ of models and for vectors $a$ in a certain vector subspace $V$ of $\mathbb{R}^{N\times M}$ if, any time we have a gauge transformation $\xi=(a,f)$ between models $\mathcal{M}'$ and $\mathcal{M}$ (i.e. Eq. \ref{eq:GaugeDef0} or equivalently Eq. \ref{eq:GaugeDef} holds for $\mathcal{M}'$ and $\mathcal{M}$) with $a\in V$ and $\mathcal{M}$ in $\mathcal{C}$, then also $\mathcal{M}'$ is a model in the class $\mathcal{C}$. The specification of $V$ will be important when we will assess gauge invariance and identifiability in the case of the BEG model. 
If $\mathcal{C}$ is gauge invariant with respect to $V=\mathbb{R}^{N\times M}$, we simply say that it is gauge invariant.

Similarly, we say that \emph{gauge transitivity} holds for $\mathcal{C}$ (and $V$) if for any two models $\mathcal{M}$ and $\mathcal{M}'$ in $\mathcal{C}$ there is a gauge transformation $\xi=(a,f)$ from $\mathcal{M}'$ to $\mathcal{M}$. In our cases, this means that for any is equivalent to saying that $\mathbb{X}'=\mathbb{X}+a$ and that Eq. \ref{eq:GaugeDef} holds for some invertible $f$, where Hamiltonians in both sides have the same functional form (e.g., that of the Ising model or the BC model).

 We say that $\mathcal{C}$ is \emph{gauge-rigid} (with respect to $V$) if for any two models $\mathcal{M}'$ and $\mathcal{M}$ in $\mathcal{C}$, at most one gauge transformation between them exists.

\subsection{Identifiability}\label{subsec:IdModel}

A statistical model is \textit{identifiable} if $\theta \neq \theta'$ implies $P_{\mathcal{M}}(\cdot|\theta) \neq P_{\mathcal{M}}(\cdot|\theta')$. 
For a maximum-entropy model with Hamiltonian $H_{\mathcal{M}}(\x|\theta) = \sum_{\mu}\theta_\mu C_\mu(\x)$, two parameter vectors give the same distribution if and only if $H(\x|\theta) - H(\x|\theta')=c$ is constant over all states $\x\in\mathbb{X}$. Since in this case the Hamiltonian is linear in $\theta$, 
this is equal to $H(\x|\theta - \theta') = c$, so identifiability is equivalent to: \emph{$H(\x|\theta) = c$ for all $\x\in\mathbb{X}$ implies $\theta = 0$ - and, thus, also $c=0$}.
We observe that \emph{identifiability is invariant under gauge transformations} in the following sense: if a class of models is gauge-invariant, then the model is identifiable when considered on a state space $\mathbb{X}$ if and only if it is identifiable when considered on the state space $\mathbb{X}+a$, due to the invertibility of the gauge transformations.

Finally, for ease of notation, in this section we will follow a different kind of convention than the one used in the rest of this article: here, the BEG model (and the Ising and BC models, too, once we set $K_{ij}=0$ or $K_{ij}=0,\qquad J_{ii}=0$) will have Hamiltonian $H(\x|K,J,h)=\sum_{i<j}K_{ij}x_i^2x_j^2 + \sum_i J_{ii}x_i^2+\sum_{i<j}J_{ij}x_ix_j+\sum_i h_ix_i=0$.
\paragraph{Reduction to the single-subject case.}
The full-data log-likelihood factorizes as
\begin{align}
    \log P_{\mathcal{M}}(X|\theta)
    = -\sum_{s=1}^N H_{\mathcal{M}}(\x_s|\theta) - N\log Z_{\mathcal{M}}
    = -\sum_\mu \theta_\mu \Bigl(\sum_{s=1}^N C_\mu(\x_s)\Bigr) - N\log Z_{\mathcal{M}},
\end{align}
so the $N$-subject model is itself a maximum-entropy model with constraints $\sum_s C_\mu(\x_s)$.

Let us see that in order to verify both identifiability and gauge invariance for a \emph{maximum entropy model} with the full-data likelihood \emph{on $N>1$ subjects, it suffices to check them for $N=1$}.  

For what concerns gauge invariance, a transformation $\xi=(a,f)$ on single-subject states extends to $N$ subjects by applying the same translation to each subject: if $X=(\x_1,\ldots,\x_N)$, we have $X+a = (\x_1+a,\ldots,\x_N+a)$, while $f$ is the same as in the $N=1$ case.  

For what concerns identifiability, if gauge invariance holds for our (class of) model(s), then, due to invariance of identifiability (both for $N$ and for just one subject) under gauge transformations, we can suppose to have already shifted the spin values so that $0\in\mathbb{S}$. Suppose the single-subject model is identifiable, and suppose $P_{\mathcal{M}}(X|\theta)=c$ for all samples $X$ of $N$ subjects. In particular, this equality holds when the sample $X$ is such that $\x_s = 0$ for all $s>1$. Then $H_{\mathcal{M}}(X|\theta) = H_{\mathcal{M}}(\x_1|\theta)$, and the condition $H_{\mathcal{M}}(X|\theta) = c$ reduces to $H_{\mathcal{M}}(\x_1|\theta) = c$ for all $\x_1$, and this, since the model on one subject is identifiable by hypothesis, implies $\theta = 0$ and $c=0$.

Finally, we note that for our models, if gauge invariance holds then, since we can shift the spins so that $0\in\mathbb{S}$, it follows that also gauge transitivity holds; moreover, for maximum entropy models, if gauge invariance holds, then gauge-rigidity is equivalent to identifiability of a (hence, any) model $\mathcal{M}$. 
Indeed, if the model is identifiable, then gauge-rigidity is obvious; viceversa, if $H(X|\theta')=c$ for all $X$ and $\theta'\neq0$, then since $H$ is linear in the parameters, we have that $H(X|\theta+\theta')=H(X|\theta)+c$ for all $X$, so both $(x,\theta)\mapsto(x,\theta)$ and $(x,\theta)\mapsto(x,\theta+\theta')$ define a gauge transformation from $\mathcal{M}$ in itself.

\subsection{Strict concavity of the log-likelihood}\label{subsec:LearnGuarantees}

As shown in \cite[pp.~62--63]{wainwright2008}, a maximum-entropy model is identifiable in the sense of the definition in subsection \ref{subsec:IdModel} if and only if the Hessian of the log-likelihood is negative definite for all states and parameters (the term that is used in \cite[pp.~40,62--63]{wainwright2008} for an identifiable model is a \emph{model having a minimal representation}). Identifiability therefore implies that the log-likelihood is \emph{strictly concave} with at most one local (hence global) maximum. As a consequence, any gradient-ascent algorithm that maintains an ascent direction which is a sufficiently good approximation of the gradient of the log-likelihood is guaranteed to converge to the unique maximum-likelihood estimator, and to always strictly increase the log-likelihood at each step, for sufficiently small learning rates. We note that the discussion in the previous two subsections will allow us to establish for our maximum entropy models identifiability - and so, strict concavity of the log-likelihood - both in the case of $N=1$ and of $N>1$ subjects.

\subsection{Assessment of gauge invariance for the Ising, the BC and the BEG models}\label{subsec:GaugeComp}

Keeping in mind the reduction to the case of $N=1$ subjects in the previous subsections, in this and in the next subsection we will just consider the case of one subject. 

The key ingredient is the binomial identity: $(z+c)^l = z^l + (\text{lower-degree terms in }z)$, so the leading monomial in each spin variable is preserved under translation while only the lower-order parameters are shifted.

Let $a\in\mathbb{R}^M$ and $\x\in\mathbb{S}^M$.

\subsubsection{Ising model}
\begin{align*}
    H_{\mathbb{X}+a}(\x+a\,|\,J,h) &= \sum_{i<j}J_{ij}x_ix_j + \sum_i\Bigl(h_i+\sum_{j\neq i}J_{ij}a_j\Bigr)x_i + \mathrm{const}(J,h,a) \\
    &= H_{\mathbb{X}}(\x\,|\,J,\; h+g(J,a)) + c(J,h,a),
\end{align*}
where $g_i(J,a) = \sum_{j\neq i}J_{ij}a_j$. Since $J$ is preserved, $f_a$ is invertible with $f_a^{-1}(J,h) = (J,\; h-g(J,a))$. So, the Ising model is invariant under gauge transformations.

\subsubsection{BC model}
\begin{align*}
    H_{\mathbb{X}+a}(\x+a\,|\,J,h) &= \sum_i J_{ii}x_i^2 + \sum_{i<j}J_{ij}x_ix_j + \sum_i\Bigl(h_i + \sum_{j\neq i}J_{ij}a_j + 2J_{ii}a_i\Bigr)x_i + \mathrm{const}(J,h,a) \\
    &= H_{\mathbb{X}}(\x\,|\,J,\; h+g(J,a)) + c(J,h,a),
\end{align*}
with $g_i(J,a) = \sum_{j\neq i}J_{ij}a_j + 2J_{ii}a_i$, and $f_a^{-1}(J,h) = (J,\; h-g(J,a))$ as before. Therefore, also the BC model is invariant under gauge transformations.

\subsubsection{BEG model}
For the BEG model, a uniform translation $a_i = \alpha$ generates cross-terms of the form $x_i^2 x_j$ and $x_j^2 x_i$ ($i\neq j$), which are absent from the BEG Hamiltonian. Gauge invariance therefore does not hold within the plain BEG model; it does hold - if we restrict just to translations of the form just discussed - for the \emph{extended BEG model} obtained by adding symmetric mixed cubic couplings $S_{ij}$:
\begin{align*}
H_{\mathbb{X}}(\x\,|\,K,S,J,h) = \sum_{i<j}K_{ij}x_i^2x_j^2 + \sum_{i<j}S_{ij}(x_i^2x_j + x_j^2x_i) + \sum_i J_{ii}x_i^2 + \sum_{i<j}J_{ij}x_ix_j + \sum_i h_ix_i.
\end{align*}

The plain BEG model is the special case $S_{ij}=0$. For a uniform shift $a_i=\alpha$, the translated Hamiltonian satisfies $H_{\mathbb{X}+\alpha}(\x+\alpha\,|\,K,S,J,h) = H_{\mathbb{X}}(\x\,|\,K,\,S+g^{(S)},\,J+g^{(J)},\,h+g^{(h)}) + c(K,S,J,h,\alpha)$, where $K$ is preserved and the shifts $g^{(S)},g^{(J)},g^{(h)}$ are functions of $(K,S,J,h,\alpha)$. We omit the expression of the transformed Hamiltonian, which is analogous to the Ising and BC cases, though more cumbersome.
 Since $K$ is preserved, the map $(K,S,J,h)\mapsto(K,S,J,h)+g(K,S,J,h,\alpha)$ is invertible, establishing gauge invariance (under a uniform shift) for the extended BEG model.

Note that since we must admit the mixed, symmetric cubic terms, indeed gauge invariance does not hold for the plain BEG model, and in particular the couplings $K_{i,j}$ in the plain BEG model {\it are not} invariant under shifting, differently from what happens for the couplings $J_{i,j}$ in the BC model.

Note also that here we chose to restrict to uniform $a_i=\alpha$ in order to obtain \emph{symmetric} couplings $S_{i,j}$ and to obtain a minimal model invariant under gauge containing the original, plain BEG model.

This result has a practical consequence: for the Ising and BC models one can freely re-centre the spin values (useful, e.g., to ensure $0\in\mathbb{S}$ for the identifiability proofs in the following subsection), while for the BEG model the spin values should be chosen symmetric around zero from the start, as done throughout this paper; in fact, the choice to use the BEG model is \emph{explicit, and  not natural} - in the sense that the plain BEG model is not gauge invariant, so the version of it with centred spins is not retrievable via gauge transformations by the version of it with spins taking values in $\{1,\dots,R\}$, and viceversa - but it is the only choice which can ensure we can avoid considering $\binom{M}{2}$ additional parameters $S_{i,j}$, which would increase the complexity of the model and increase, with many of the questionnaires under consideration, the risk of overfitting the data or, worse, of having the number of parameters $n_{\mathcal{M}}>N$.

\subsection{Assessment of identifiability for the Ising, the BC and the BEG models}\label{subsec:IdComp}

The strategy is the same in all three cases. By gauge invariance (\S\ref{subsec:GaugeTransf}) and invariance of identifiability under gauge transformations (where we use the restricted type of gauge invariance with a uniform $a$ in the case of the extended BEG model, as will be seen) we may assume $0\in\mathbb{S}$, so that setting $\x=0$ in the condition $H(\x|\theta)=c$ immediately forces $c=0$. We then isolate each parameter in turn by \emph{activating} a certain set of spins, i.e. setting them to a (possibly nonzero) value and leaving all others at zero.

\subsubsection{Ising model}

We must show that $H(\x|J,h)=\sum_{i<j}J_{ij}x_ix_j+\sum_i h_ix_i=0$ for all $\x\in\mathbb{X}$ implies $(J,h)=0$.

\begin{itemize}
\item Activate \emph{one} spin: set $x_i=v\neq0$, $x_j=0$ for $j\neq i$. The Hamiltonian reduces to $h_iv=0$, giving $h_i=0$ for every $i$.
\item Activate \emph{two} spins: set $x_i=x_j=v\neq0$, $x_l=0$ for $l\neq i,j$. The Hamiltonian reduces to $J_{ij}v^2=0$, giving $J_{ij}=0$ for every $i<j$.
\end{itemize}
The Ising model is therefore identifiable for all $R\geq 2$.

\subsubsection{BC model}

We must show that $H(\x|J,h)=\sum_i J_{ii}x_i^2+\sum_{i<j}J_{ij}x_ix_j+\sum_i h_ix_i=0$ for all $\x$ implies $(J,h)=0$.

\begin{itemize}
\item Activate \emph{one} spin ($R\geq 3$): set $x_i=v\neq0$, $x_j=0$ for $j\neq i$. The Hamiltonian gives $J_{ii}v^2+h_iv=0$, i.e., $J_{ii}v+h_i=0$. Since $R\geq 3$ there exist two distinct nonzero values $v\neq w$ in $\mathbb{S}$; applying the same equation to $w$ and subtracting gives $J_{ii}(v-w)=0$, so $J_{ii}=0$ and hence $h_i=0$ for every $i$.

\item Once $J_{ii}=h_i=0$, the Hamiltonian reduces to the Ising form $\sum_{i<j}J_{ij}x_ix_j=0$, which by the Ising argument forces $J_{ij}=0$.
\end{itemize}
BC is therefore identifiable for $R\geq 3$.

\smallskip
For $R=2$, BC is \emph{not} identifiable: by gauge invariance one can centre the values to $\mathbb{S}=\{-q,q\}$, but then $x_i^2=q^2$ is constant and the diagonal terms $J_{ii}x_i^2$ become a state-independent additive constant, so the model reduces to an overparametrization of the Ising model.

\subsubsection{Extended BEG model and plain BEG model}

Let $R\geq 3$. We prove identifiability for the \emph{extended} BEG model (which includes the cubic couplings $S_{ij}$); since the plain BEG model is nested within it, identifiability of the plain model follows.

We must show that $H(\x|K,S,J,h)=0$ for all $\x$ implies $(K,S,J,h)=0$. As in the BC case, activating one spin at a time gives $J_{ii}=h_i=0$ for all $i$.

\smallskip
\noindent\textit{Case $R\geq 4$.} Activate two spins: set $x_i=x_j=v\neq0$, $x_l=0$ for $l\neq i,j$. The Hamiltonian gives
\begin{align*}
K_{ij}v^4 + 2S_{ij}v^3 + J_{ij}v^2 = 0
\quad\Longrightarrow\quad
K_{ij}v^2 + 2S_{ij}v + J_{ij} = 0.
\end{align*}
This is a degree-2 polynomial in $v$ that must vanish for at least three distinct nonzero values (since $R\geq 4$). A polynomial of degree $\leq 2$ with more than two roots is identically zero, so $K_{ij}=S_{ij}=J_{ij}=0$ for every $i<j$.

\smallskip
\noindent\textit{Case $R=3$.} Write $\mathbb{S}=\{-q_1,0,q_2\}$ with $q_1,q_2>0$. Activating pairs $(x_i,x_j)$ at the three possible nonzero combinations yields:
\begin{align*}
\mathrm{(1)}&\quad K_{ij}q_2^2+2S_{ij}q_2+J_{ij}=0 && (x_i=x_j=q_2)\\
\mathrm{(2)}&\quad K_{ij}q_1^2-2S_{ij}q_1+J_{ij}=0 && (x_i=x_j=-q_1)\\
\mathrm{(3)}&\quad -K_{ij}q_1q_2+S_{ij}(q_2-q_1)+J_{ij}=0 && (x_i=-q_1,\,x_j=q_2).
\end{align*}
Three unknowns, three equations. The combination $(1)+(2)-2\times(3)$ cancels $S_{ij}$ and $J_{ij}$, giving $(q_1+q_2)^2 K_{ij}=0$, hence $\mathbf{K_{ij}=0}$. With $K_{ij}=0$, the combination $(1)-(2)$ gives $2(q_1+q_2)S_{ij}=0$, hence $\mathbf{S_{ij}=0}$. Substituting back into (1) gives $\mathbf{J_{ij}=0}$.

\smallskip
The extended BEG model (and, thus, the plain BEG model) is therefore identifiable for both $R=3$ and $R\geq 4$.

\smallskip
For $R=2$, an argument analogous to the BC case shows that both the extended and the plain BEG models reduce to an overparametrization of the Ising model and are not identifiable.

\section{Pseudo-likelihood maximization: explicit form \label{sec:PseudoLik}}
The pseudo-likelihood is given by the product over the items of the conditional log-probability of each item $i$ given all other items. 
Fixing $\x_{\setminus i}$, the Hamiltonian of a generalized spin model reduces to a function of $x_i$ alone, so the conditional partition function requires only $R$ terms. One obtains:
\begin{subequations}
\label{eq:pselik}
\begin{align}
    \ln P_{\pl}(\x_s|\theta) &= \sum_{i=1}^M \ln P_i(x_{si}|\x_{s,\setminus i},\theta) \label{eq:pselik_sum} \\
    \ln P_i(z|\x_{\setminus i},\theta) &= z\,\phi_i(\x_{\setminus i},\theta) + z^2\,\varphi_i(\x_{\setminus i},\theta) - \ln\sum_{q=1}^R e^{v_q\,\phi_i(\x_{\setminus i},\theta) + v_q^2\,\varphi_i(\x_{\setminus i},\theta)} \label{eq:condlik} \\
    \phi_i(\x_{\setminus i},\theta) &:= h_i + \sum_{j\ne i} J_{ij}\,x_j \label{eq:phifield} \\
    \varphi_i(\x_{\setminus i},\theta) &:= \frac{1}{2}J_{ii} + \sum_{j\ne i} K_{ij}\,x^2_j \label{eq:varphifield}
\end{align}
\end{subequations}
where $z\in\mathbb{S}$ is a dummy variable. The quantity $\phi_i$ is the effective linear (cavity) field acting on item $i$, collecting the contributions of all other items through the pairwise couplings $J_{ij}$ and the external bias $h_i$ (this is the direct generalization of the Ising cavity field). 
The quantity $\varphi_i$ is the effective quadratic field: it modulates the energy cost for item $i$ to take any non-zero value, and has no analogue in the Ising model. 
For the BC model ($K=0$), $\varphi_i = \frac{1}{2}J_{ii}$ is a configuration-independent constant; for the Ising model ($K=0$, $J_{ii}=0$), $\varphi_i=0$ and only $\phi_i$ survives.

\noindent The gradients of the joint pseudo-likelihood with respect to the BEG parameters are:
\begin{subequations}\label{eq:PLgradients}
\begin{align}
\frac{\partial}{\partial h_i}\ln P_{\pl}(X|\theta) & = N \<x_i\>_{\rho_X} - \sum_{s=1}^N \<x_i\>_{P(x_i|\x_{s,\setminus i},\theta)} \label{eq:hgradpselik}\\
\frac{\partial}{\partial J_{ij}}\ln P_{\pl}(X|\theta) & = N \<x_i x_j\>_{\rho_X} - \sum_{s=1}^N  x_{sj}\<x_i\>_{P(x_i|\x_{s,\setminus i},\theta)}, \qquad i\ne j \label{eq:Jgradpselik} \\
\frac{\partial}{\partial J_{ii}}\ln P_{\pl}(X|\theta) & = \frac{N}{2} \Big(\<x_i^2\>_{\rho_X} - \sum_{s=1}^N  \<x_i^2\>_{P(x_i|\x_{s,\setminus i},\theta)} \Big) \\
\frac{\partial}{\partial K_{ij}}\ln P_{\pl}(X|\theta) & = N \<x^2_i x^2_j\>_{\rho_X} - \sum_{s=1}^N  x^2_{sj}\<x^2_i\>_{P(x_i|\x_{s,\setminus i},\theta)}, \qquad i\ne j \label{eq:Kgradpselik}
\end{align}
\end{subequations}
Each gradient is an empirical average minus a sum of model conditional averages over training samples; gradient ascent therefore drives the conditional expectations of the model toward those of the data. For the BC model it suffices to set $K=0$ and drop Eq.~(\ref{eq:Kgradpselik}); for the Ising model one additionally sets $J_{ii}=0$. Because the off-diagonal gradients $\nabla_{J_{ij}}$ and $\nabla_{K_{ij}}$ computed item-by-item are not automatically symmetric, we symmetrize them at each step before the L-BFGS-B update (Algorithm~\ref{alg:pselik}).

\begin{algorithm}
\caption{\textbf{Pselik Algorithm} (Pseudo-likelihood maximization) \label{alg:pselik}}
Requires: $T$, maximum number of L-BFGS-B steps, {\tt tol} is the tolerance\\
Initialize the couplings $\theta({\tt t=0})$ to some value\;
\For{${\tt t}= 0,\ldots,{\tt T} - 1$}{
    compute the {\tt loss}, $=-\ln P_{\pl}(X|\theta({\tt t}))$ in Eq.~(\ref{eq:pselik})\;
    \If{$\mathtt{loss} < \mathtt{tol}$}
        {\Return $\theta({\tt t})$}
    \For{$i=1,\ldots,M-1$}{
        compute $\nabla_{h_i} = \left.\partial_{h_i}\right|_{\theta({\tt t})}\ln P_{\pl}(X|\theta)$ in Eq.~(\ref{eq:hgradpselik})\;
        compute $\nabla_{J_{i*}} = \left.\partial_{J_{i*}}\right|_{\theta({\tt t})}\ln P_{\pl}(X|\theta)$ in Eq.~(\ref{eq:Jgradpselik})\;
        compute $\nabla_{K_{i*}} = \left.\partial_{K_{i*}}\right|_{\theta({\tt t})}\ln P_{\pl}(X|\theta)$ in Eq.~(\ref{eq:Kgradpselik})\;
    }
    symmetrize $\nabla_{J_{ij}}:=(\nabla_{J_{ij}}+\nabla_{J_{ji}})/2$, $\forall\,i<j$\;
    symmetrize $\nabla_{K_{ij}}:=(\nabla_{K_{ij}}+\nabla_{K_{ji}})/2$, $\forall\,i<j$\;
    L-BFGS-B update of $\theta({\tt t}+1)$ using gradients and loss at time $\tt t$\;
}
\KwRet $\theta({\tt T})$
\end{algorithm}

\section{Protocol of Contrastive Divergence learning \label{sec:PCD}}

 The PCD algorithm approximates the model averages $\<\cdot\>_{P(\x|\theta_{\tt t})}$ in Eqs.~(\ref{eq:CDgradients}) at each gradient step $\tt t$ by running $\nc$ independent MCMC Gibbs sampling chains \cite{pelissetto1993,sokal1997,binder1997} of length $\tau_{\rm PCD}$ sweeps. In each Gibbs sweep, all $M$ spins are updated sequentially: spin $i$ is sampled from its conditional distribution $P_{{\cal M},i}(x_i|\x_{\setminus i},\theta({\tt t}))$ defined in Eq.~(\ref{eq:condlik}), which requires summing only $R$ terms and is therefore computationally cheap even when the full partition function $Z_\theta \sim R^M$ is intractable for large $M$. The ensemble average is then estimated as the empirical mean over the $\nc\times\tau_{\rm PCD}$ sampled configurations:
\begin{align}
\<o(\x)\>_{P(\x|\theta_{\tt t})} \approx \frac{1}{\nc\,\tau_{\rm PCD}} \sum_{n=1}^{\nc}\sum_{\tau=1}^{\tau_{\rm PCD}} o\!\left({\bm\sigma}({\tt t},n,\tau)\right).
\end{align}

 The algorithm is \emph{persistent} (PCD \cite{tieleman2008}): rather than re-thermalizing $\nc$ fresh chains from scratch at each step --- which would require long burn-in periods and greatly increase the computational cost --- the final configuration of chain $n$ at step $\tt t$ is used as the initial configuration at step $\tt t+1$:
\begin{align}
{\bm\sigma}({\tt t},\tau=0,n) := {\bm\sigma}({\tt t}-1,\tau_{\rm PCD},n).
\end{align}

 Simulating $\nc$ parallel copies rather than a single chain of $\nc\tau_{\rm PCD}$ sweeps is advantageous for two reasons. First, it is statistically more efficient: the $\nc$ copies provide approximately independent gradient estimates, whereas the configurations of a single long chain are correlated on the timescale $\tau_o$ (see Appendix~\ref{sec:correlation}). 
Second, when the inferred model exhibits a multi-basin free-energy landscape — as we show occurs for several questionnaires — the parallel copies can simultaneously explore different basins, providing a more representative estimate of the model averages and preventing the gradient from being dominated by a single energy minimum.

 Every ${\tt t}_{\rm r}$ gradient steps, all $\nc$ chains are reset to independently drawn empirical configurations $\x_s$. In Appendix~\ref{sec:correlation} we show that the inter-reset interval ${\tt t}_{\rm i}$$\times\tau_{\rm PCD}$ is at least two orders of magnitude larger than the longest autocorrelation time $\tau_{x'_1}$ of the slowest linear combination of the item values, so that the reset will have no influence in the MCMC path within a block --unless, probably, in determining which of the eventual multiple local minima in variational free energy the chain is sampling. In fact, the rationale for this empirical resetting is that, in the case of onset of two or more metastable phases representing clusters of outliers in the data (see Appendix \ref{sec:meanfield}), we incentivate the algorithm to reproduce the empirical {\it relative frequency} of subjects in the different clusters (i.e., {\it their relative log-probability or free energy heights}). An assessment of the impact of this procedure in the learning protocol is left for future studies. 

 The PCD learning protocol (Algorithm \ref{alg:PCD}) is organized in sequential \emph{blocks} of length ${\tt t}_{\rm r}$, i.e. consisting of ${\tt t}_{\rm r}$ gradient-ascent iterations. Within each block the hyperparameters (${\tt t}_{\rm r}$, $\eta$, $\nc$, $\tau_{\rm PCD}$) are kept fixed; across blocks, the learning rate $\eta$ and the number of MCMC copies $\nc$ are adjusted to achieve progressive refinement. At the beginning of each block, the $\nc$ persistent Markov chains are re-initialized to independent, randomly drawn empirical configurations $\x_s$. Within a block, the chains evolve persistently: the final state of chain $n$ at iteration $\tt t$ is used as the initial state for iteration $\tt t+1$. No within-block reset is performed, so that the total number of MCMC sweeps between resets equals ${\tt t}_{\rm r}$$\times\tau_{\rm PCD} = 1.2\times10^4$ for the main blocks; see Appendix \ref{sec:correlation}).

 The ADAM optimizer \cite{kingma2017} is used throughout, with fixed hyperparameters $\beta_1=0.9$, $\beta_2=0.999$, $\varepsilon=10^{-8}$. The learning rate $\eta$ is reset at the beginning of each block (i.e., the first- and second-moment estimates of ADAM are re-initialized). The initial condition $\theta({\tt t}=0)$ for the first block is the maximum pseudo-likelihood estimate (Algorithm \ref{alg:pselik}).

 The rationale behind the block structure is as follows. In the early blocks, a small number of copies $\nc$ and a larger learning rate $\eta$ provide rapid, coarse progress toward the maximum likelihood manifold (in reality, this reduces just to a point in the cases considered, as shown in Appendix \ref{sec:gauge}). In the later blocks, $\nc$ is increased to reduce the variance of the stochastic gradient estimates and $\eta$ is decreased to allow the optimizer to settle into the minimum. The BEG model requires a more conservative initial block (small $\nc$ and ${\tt t}_{\rm r}$) because of its larger parameter space and higher correlation times (see Appendix \ref{sec:correlation}).

 The protocol parameters for each model are given in Tables \ref{table:pcd-protocol}.

 Convergence is monitored through the three pseudo-loss functions $L_{\bf h}$, $L_J$, $L_K$ defined as:

\begin{subequations}
\label{eq:losses}
\begin{align}
L_{\bf h} &= \sum_j \Bigl(\<x_j\>_{\rho_X} - \<x_j\>_{P(\x|\theta({\tt t}))}\Bigr)^2, \\
L_J &= \sum_{i < j}\Bigl(\<x_ix_j\>_{\rho_X} - \<x_ix_j\>_{P(\x|\theta({\tt t}))}\Bigr)^2 + \sum_{i}\frac{1}{4}\Bigl(\<x_i^2\>_{\rho_X} - \<x_i^2\>_{P(\x|\theta({\tt t}))}\Bigr)^2, \\
L_K &= \sum_{i<j}\Bigl(\<x_i^2x_j^2\>_{\rho_X} - \<x_i^2x_j^2\>_{P(\x|\theta({\tt t}))}\Bigr)^2.
\end{align}
\end{subequations}


Training is stopped after the total number of gradient iterations (equal to the sum of the parameter ${\tt t}_{\rm r}$ of each block, see Table \ref{table:pcd-protocol}). Although the full log-likelihood is a concave function of $\theta$ \cite{mackay2003,wainwright2008}, convergence of PCD is not guaranteed in finite time: the gradient estimates are stochastic (mean over $\nc\times\tau_{\rm PCD}$ correlated MCMC samples), and the estimation error decreases only as $(\nc\,\tau_{\rm PCD})^{-1/2}$. 

The sufficiency of such number of iterations, and the evolution of the losses are described in Appendix \ref{sec:consistency}.


\begin{algorithm}
\caption{\textbf{PCD Algorithm} (Likelihood maximization with PCD) \label{alg:PCD}}
Requires: ${\tt B}$, maximum number of blocks; {\tt tol}, tolerance\\
load and update hyperparameters $(\nc,\tau_{\rm PCD},\eta,{\tt t}_{\rm r})$ of block 1\;
initialize $\theta(b = 1, {\tt t}=0)$ to $\theta_{\pl}^*$ from Algorithm~\ref{alg:pselik}\;
\For{${b}=1,\ldots, {\tt B}$}{
    \If{$b>1$}{
        $\theta(b,{\tt t} = 0) = \theta(b-1, {\tt t}_r)$\;
        load and update hyperparameters $(\nc,\tau_{\rm PCD},\eta,{\tt t}_{\rm r})$ of block b\;
    }
    \For{${\tt t} = 0,\ldots,{\tt t}_r-1$}{
        \For{$n=1,\ldots,\nc$}{
            \eIf{${\tt t} = 0$}{
                (re)-initialize ${\bm\sigma}(b,{\tt t} = 0, n,\tau=0)$ to a random empirical vector $\x_{s(n)}$\;
            }{
                persistence: ${\bm\sigma}(b,{\tt t},\tau=0,n) := {\bm\sigma}({b,\tt t}-1,\tau_{\rm PCD},n)$\;
            }
            \For{$\tau=1,\ldots,\tau_{\rm PCD}$}{
                sample ${\bm\sigma}(b,{\tt t}, n)$ via a Gibbs sweep: for $j=1,\ldots,M$ sample $\sigma_j\sim P_{{\cal M},j}(\cdot|{\bm\sigma}_{\setminus j},\theta({\tt t}))$ using Eq.~(\ref{eq:condlik})\;
            }
        }
        estimate model averages $\<o\>_{P}\simeq \<o({\bm\sigma}(b,{\tt t},\tau,n))\>_{\tau,n}$, for $o=x_i,\,x_ix_j,\,x_i^2,\,x_i^2x_j^2$\;
        compute gradients $\nabla_{\bf h},\nabla_J,\nabla_K$ via Eqs.~(\ref{eq:CDgradients})\;
        ADAM update: $\theta(b, {\tt t}+1) = {\rm Adam}\!\left(\theta(b, {\tt t}),\,\nabla_\theta,\,\eta,\,\beta_1,\,\beta_2,\,\varepsilon\right)$\;
        save pseudo-losses $L_{\bf h}({\tt t}),\,L_J({\tt t}),\,L_K({\tt t})$ via Eqs.~(\ref{eq:losses})\;
        \If{$\mathrm{pseudo-losses}\, L_{\bf h}({\tt t}),\,L_J({\tt t}),\,L_K({\tt t}) < \mathtt{tol}$}{
            \KwRet $\theta(b, {\tt t}+1)$\;
        }
    }
}
\KwRet $\theta({\tt B}, {\tt t}_r)$
\end{algorithm}

\begin{table}[H]
\caption{PCD learning protocol for the Ising and BC models (a) and for BEG (b). 
Each row corresponds to one block; the initial condition for the first block is the pseudo-likelihood estimate. ${\tt t}_{\rm r}$: number of PCD iterations per block; $\eta$: ADAM learning rate; $n_c$: number of persistent MCMC copies; $\tau_{\rm PCD}$: MCMC sweeps per PCD iteration.
The learning protocol for the BEG model has an extra short exploratory stage with few copies, used to move the parameters close to the likelihood maximum before the main convergence phase.}
\label{table:pcd-protocol}
\begin{ruledtabular}
\flushleft{Inverse Ising and inverse BC}
\vspace{1mm}
\begin{tabular}{c|c|c|c|c}
Block & ${\tt t}_{\rm r}$ & $\eta$ & $\nc$ & $\tau_{\rm PCD}$ \\
\hline
1 & 120 & $10^{-3}$ & 100  & 100 \\
2 & 120 & $10^{-3}$ & 1000 & 100 \\
3 & 120 & $10^{-4}$ & 1000 & 100 \\
\end{tabular}
\end{ruledtabular}
\vspace{1mm}
\begin{ruledtabular}
\flushleft{Inverse BEG}
\vspace{1mm}
\begin{tabular}{c|c|c|c|c}
Block & ${\tt t}_{\rm r}$ & $\eta$ & $\nc$ & $\tau_{\rm PCD}$ \\
\hline
1 & 20  & $10^{-4}$ & 10   & 100 \\
2 & 120 & $10^{-4}$ & 100  & 100 \\
3 & 120 & $10^{-4}$ & 1000 & 100 \\
4 & 120 & $10^{-5}$ & 1000 & 100 \\
\end{tabular}
\end{ruledtabular}
\end{table}

\section{Lack of consistency of maximum pseudo-likelihood learning \label{sec:nonconsistency}}

\begin{figure}[H]
\center
    \includegraphics[width=0.8\columnwidth]{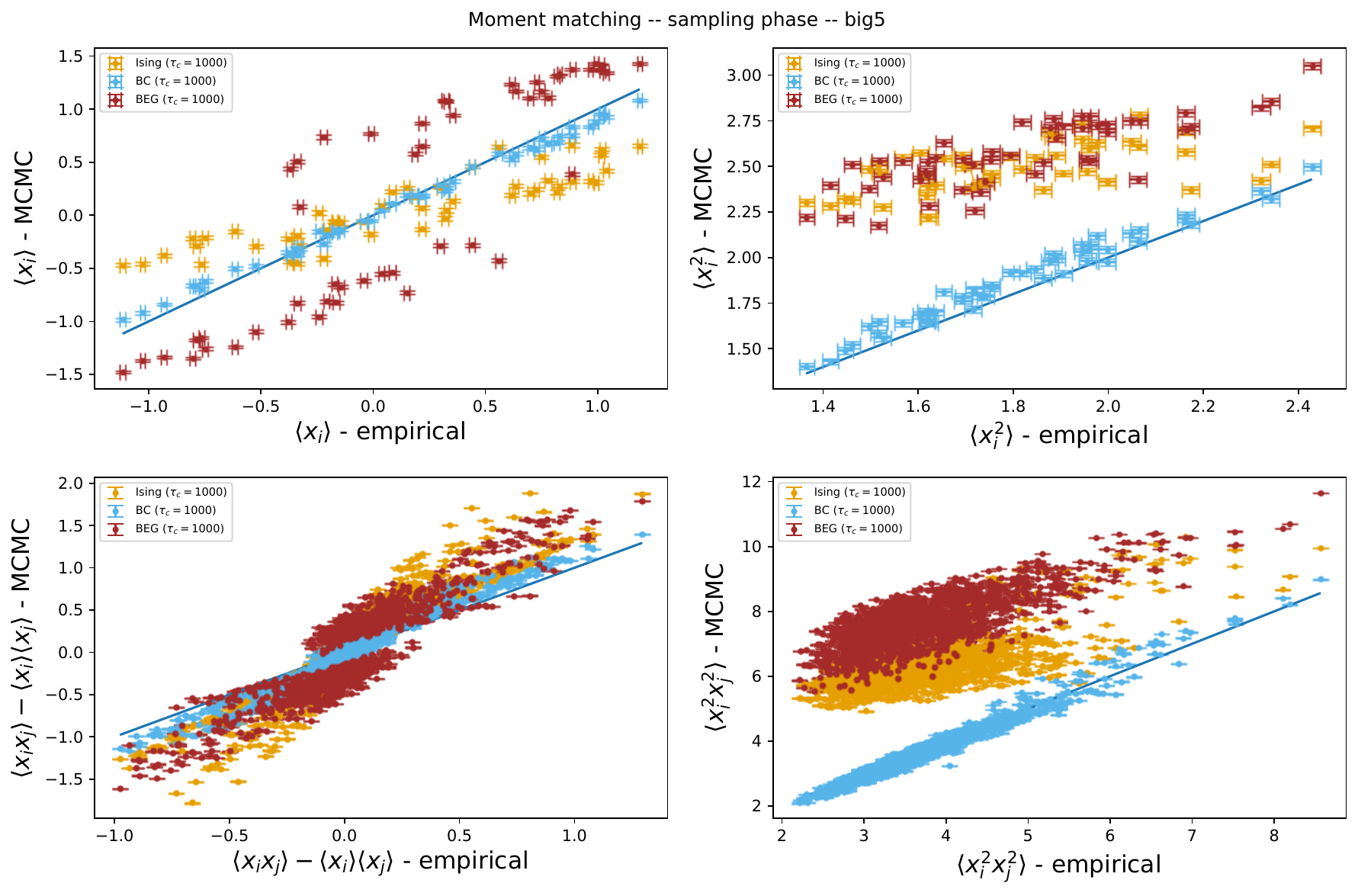}%
    \caption{Illustration of the lack of consistency of pseudo-likelihood maximization, for the {\sf big5} questionnaire. The stationary conditions Eqs.~(\ref{eq:maxpselik}) are satisfied after convergence of the L-BFGS-B algorithm, but the moment-matching conditions $\langle o(\x)\rangle_{P(\x|\theta_{\rm pl})}=\langle o(\x)\rangle_{\rho_X}$ are not: the theoretical moments of the spin model at the pseudo-likelihood optimum do not match the empirical moments. Grey crosses represent, for each abscissa, the sleep part of the pseudo-likelihood gradient, Eq. (\ref{eq:PLgradients}) in Appendix \ref{sec:PseudoLik}, that the pseudo-likelihood is required to match, when converged. \label{fig:momentmatching_pselik_sampling_big5}}
\end{figure}

Since the L-BFGS-B algorithm for the pseudo-likelihood maximization converges for all the datasets not leading to an underdetermined problem, the conditions in Eqs. (\ref{eq:maxpselik}) are satisfied after training with the pseudo-likelihood maximization algorithm, up to an arbitrarily low tollerance. This is illustrated, for the {\sf big5} data, in Fig. \ref{fig:momentmatching_pselik_sampling_big5}. The same is not true for the moment-matching conditions $\<o(\x)\>_{P(\x|\theta_{\rm pl})}=\<o(\x)\>_{\rho_X}$, that are not a consequence of the pseudo-likelihood maximization. In fact, when learning the data with the pseudo-likelihood maximization algorithm, Algorithm~\ref{alg:pselik}, we do not find consistency in general: the theoretical moments of the spin model with the learned parameters do not match the empirical moments. Such a lack of consistency is shown in Fig.~\ref{fig:momentmatching_pselik_sampling_big5} for the {\sf big5} data. The inconsistency persists when sampling from $\<o(\x)\>_{P(\x|\theta_{\rm pl})}$ after training, as visible in Figure~\ref{fig:momentmatching_pselik_sampling_big5}.
A systematic study of the validity of the pseudo-likelihood approximation is left for future studies. 

\begin{figure}[H]
\center
\includegraphics[width=0.9\columnwidth]{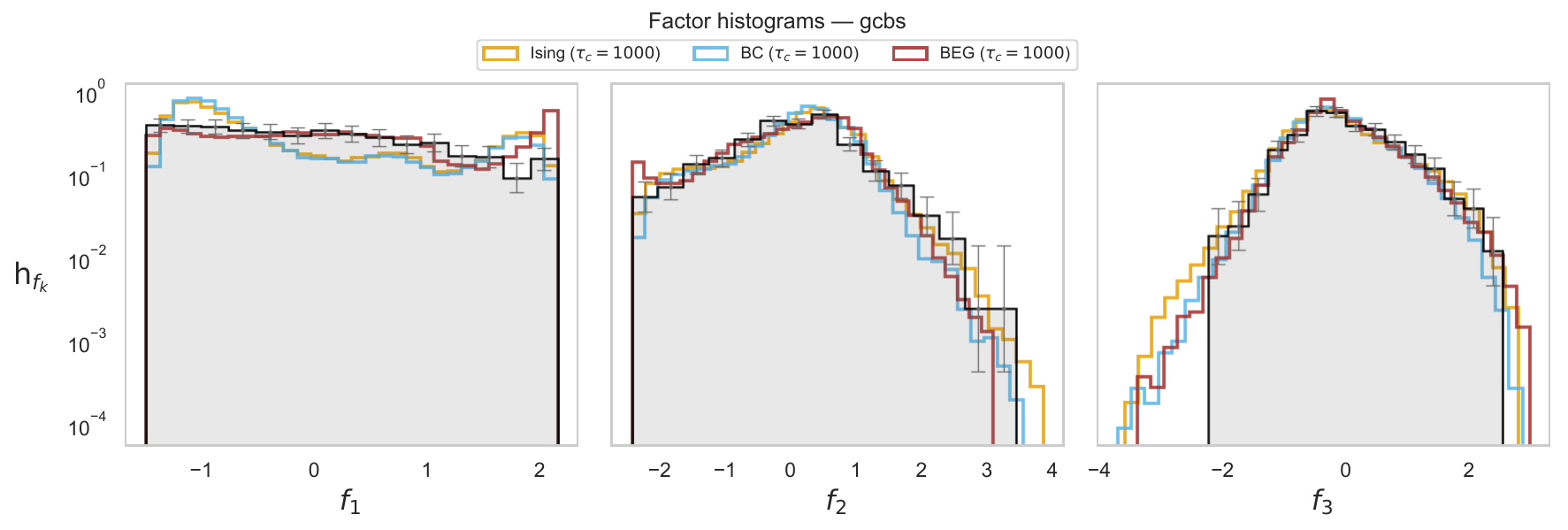}
\caption{Histograms of the first three factors ${\sf h}_{f_k}$ ($k=1,2,3$) for the {\sf gcbs} questionnaires, as in \ref{fig:factorhist_all} of the main text, but sampling from the maximum pseudo-likelihood distribution.}\label{fig:fahist_big5_pselik}
\end{figure}

\begin{figure}[H]
\center
\includegraphics[width=0.49\linewidth]{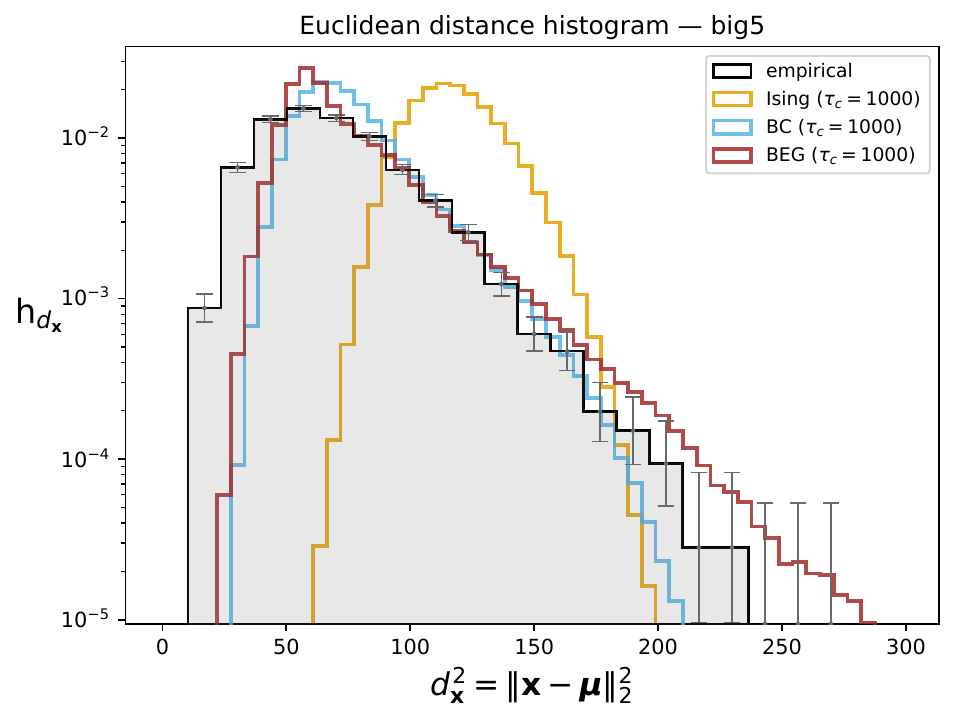}%
\caption{Histogram of the Euclidean distance to the mean, ${\sf h}_{d_\x}$, for the {\sf big5} questionnaire, as that of \ref{fig:E2dhist_all} in Appendix \ref{sec:other}, but sampling from the maximum pseudo-likelihood distribution.}
\label{fig:E2dhist_big5_pselik}
\end{figure}

\begin{figure}[H]
\center
\includegraphics[width=0.49\columnwidth]{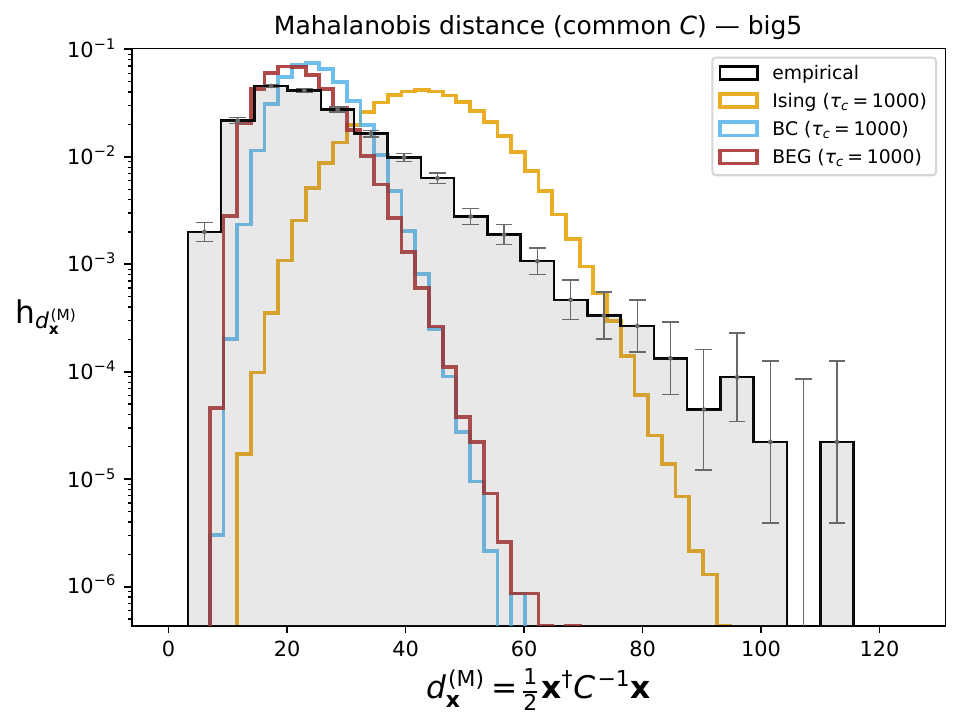}%
\caption{Histogram of the Mahalanobis distance to the mean ${\sf h}_{d_\x^{({\rm M})}}$ for the {\sf big5} questionnaire, as that of \ref{fig:energy_all}  in Appendix \ref{sec:other}, but sampling from the maximum pseudo-likelihood distribution. }
\label{fig:energy_big5_pselik}
\end{figure}

It is worth noticing that, despite such a lack of consistency, most of the results of this study, in particular the fact that the spin models capture non-linear traits of the factor histograms, that BEG model captures better than the other spin models the abundance of outliers and mean responders, and that the Mahalanobis distance to the mean is not reproduced by the spin models, remain qualitatively valid also for the maximum pseudo-likelihood distributions ${P(\x|\theta_{\rm pl})}$. This is illustrated in Figs. \ref{fig:fahist_big5_pselik},\ref{fig:E2dhist_big5_pselik},\ref{fig:energy_big5_pselik}.

%
%

\section{Analytical distributions of distances for the Gaussian model \label{sec:gaussian}}

\noindent We derive here the analytical expressions for the histograms of the Euclidean and Mahalanobis distances to the mean, ${\sf h}_{d_\x,\nu}$ and ${\sf h}_{d^{({\rm M})}_\x,\nu}$, in the particular case of the Gaussian model $\nu(\cdot)={\cal N}({\bm\mu},C)$, referenced in Sec.~\ref{sec:quantities}.

\noindent Let $\y = \x - {\bm\mu}$ and let $C = U^\dag \Lambda U$ be the eigendecomposition of the sample covariance matrix, with $\Lambda = {\rm diag}(\lambda_1,\ldots,\lambda_M)$ and $U$ orthogonal. The principal components $x'_j = (U\y)_j$ are independent and satisfy $x'_j \sim {\cal N}(0,\lambda_j)$. Since $U$ is orthogonal, $\|\y\|_2^2 = \|U\y\|_2^2 = \sum_j x_j'^2$, so that the Euclidean distance reads:
\begin{align}
d_\x = \|\y\|_2^2 = \sum_{j=1}^M x_j'^2 = \sum_{j=1}^M \lambda_j z_j^2,
\end{align}
where $z_j = x'_j/\sqrt{\lambda_j}\sim{\cal N}(0,1)$ are independent standard Gaussian variables. The variable $d_\x$ is therefore a \emph{weighted sum of independent $\chi^2_1$ variables}. For such distributions, closed-form expressions of the probability density generally do not exist, but the characteristic function is known exactly:
\begin{align}
\label{eq:charfun_Ed}
\varphi_{d_\x}(t) = \mathbb{E}_\nu\!\left[e^{it\,d_\x}\right] = \prod_{j=1}^M \left(1 - 2i\lambda_j t\right)^{-1/2}.
\end{align}
The probability density ${\sf h}_{d_\x,\nu}$ is recovered from Eq.~(\ref{eq:charfun_Ed}) via the Gil-Pelaez inversion formula \cite{gilpelaez1951}:
\begin{align}
{\sf h}_{d_\x,\nu}(u) = \frac{1}{\pi}\int_0^\infty {\rm Re}\!\left[e^{-itu}\,\varphi_{d_\x}(t)\right]{\rm d}t, \qquad u > 0.
\end{align}
In the isotropic case $\lambda_j = \bar\lambda$ for all $j$, Eq.~(\ref{eq:charfun_Ed}) reduces to $(1-2i\bar\lambda t)^{-M/2}$, the characteristic function of $\bar\lambda\chi^2_M$, and ${\sf h}_{d_\x,\nu}$ is the Gamma density with shape $M/2$ and scale $2\bar\lambda$.

\noindent For the Mahalanobis distance, let $z_j = x'_j/\sqrt{\lambda_j}\sim{\cal N}(0,1)$ i.i.d., so that:
\begin{align}
d^{({\rm M})}_\x = \frac{1}{2}\y^\dag C^{-1}\y = \frac{1}{2}\sum_{j=1}^M z_j^2 \equiv \frac{1}{2}Q, \qquad Q\sim\chi^2_M.
\end{align}
The probability density of $W = Q/2$ follows from the $\chi^2_M$ density by change of variable:
\begin{align}
{\sf h}_{d^{({\rm M})}_\x,\nu}(w) = e^{-w}\frac{w^{\frac{M}{2}-1}\,}{\Gamma(\frac{M}{2})}, \qquad w>0.
\end{align}
This is the ${\rm Gamma}(M/2,\,1)$ distribution: shape parameter $\alpha = M/2$, scale parameter $\theta = 1$ (equivalently, rate $\beta = 1$).

\section{Consistency of maximum likelihood learning \label{sec:consistency}}

\begin{figure}[H]
\begin{center}
\includegraphics[width=0.99\columnwidth]{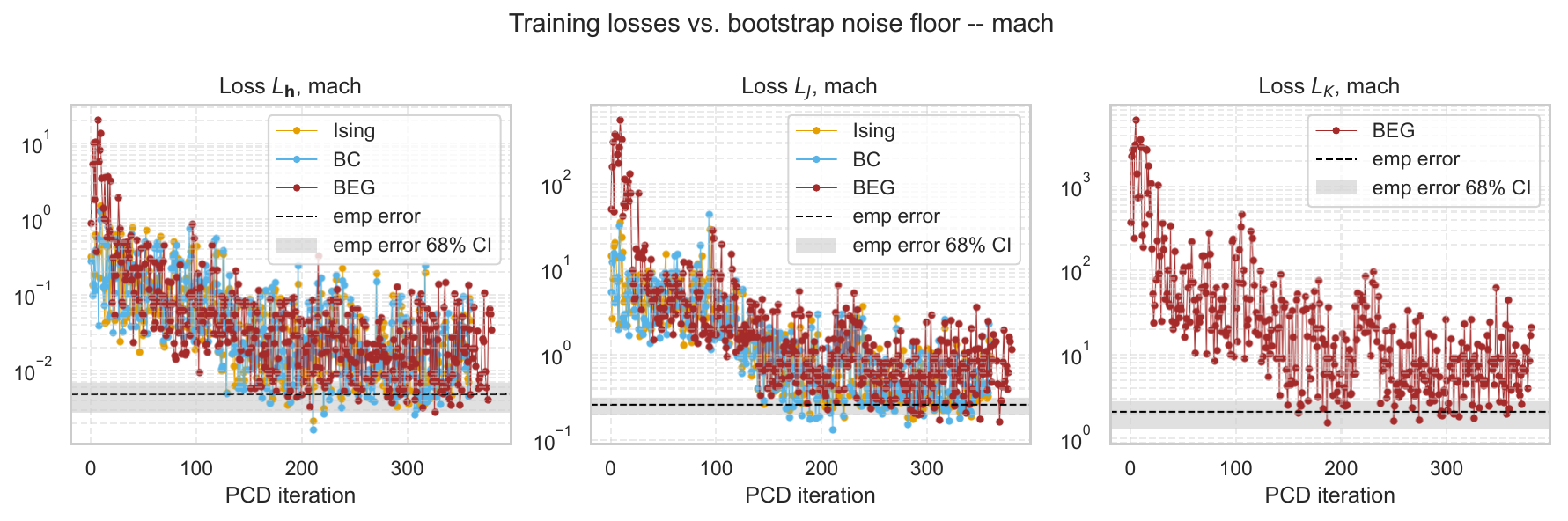}%
\caption{Loss functions $L_{\bf h}$, $L_J$ (and $L_K$ for the BEG model) as a function of PCD iteration number, for the three spin models (Ising, BC, BEG) trained on the {\sf mach} questionnaire. Convergence of the losses to small residual values confirms that the moment-matching conditions on the sufficient statistics are approximately satisfied at the end of training. \label{fig:losses_mach}}
\end{center}
\end{figure}

\begin{figure}[H]
\begin{center}
    \includegraphics[width=0.9\columnwidth]{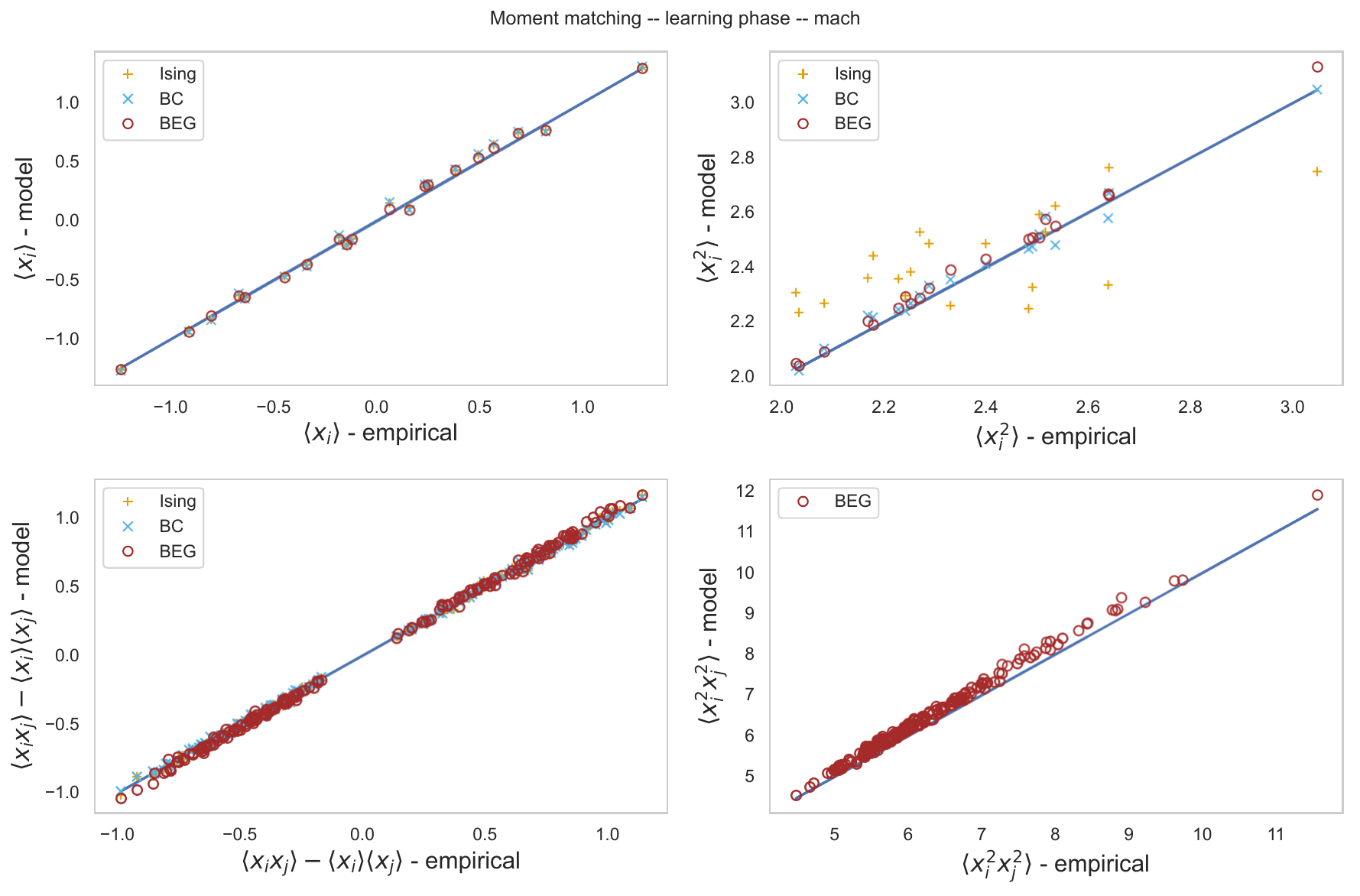}%
    \caption{Comparison between theoretical and empirical sufficient statistics after convergence of the PCD algorithm, for the {\sf mach} questionnaire. All three spin models approximately match the empirical means $\langle x_i\rangle$ and covariances $\langle x_i x_j\rangle$; the BC and BEG models additionally match $\langle x_i^2\rangle$; the BEG model additionally matches $\langle x_i^2 x_j^2\rangle$. The residual deviations are within the tolerance set by the final values of the losses shown in Fig.~\ref{fig:losses_mach}; see Sec.~\ref{sec:consistency}. \label{fig:momentmatching_learning_mach}}
\end{center}
\end{figure}

We illustrate the evolution of the $\bf h$-, $J$- and $K$-losses as a function of the number of PCD iterations in Fig. \ref{fig:losses_mach}, for the {\sf mach} dataset. After convergence, the match between theoretical and empirical sufficient statistics is as in Fig. \ref{fig:momentmatching_learning_mach}: all the models (Ising, BC, BEG) match, within the tolerance given by the last value of Fig. \ref{fig:losses_mach}, the means and the covariances, while BC also matches $\<x_i^2\>$, and BEG also matches $\<x_i^2\>$ and $\<x_i^2 x_j^2\>$, as they should. 

\begin{figure}[H]
\begin{center}
    \includegraphics[width=0.9\columnwidth]{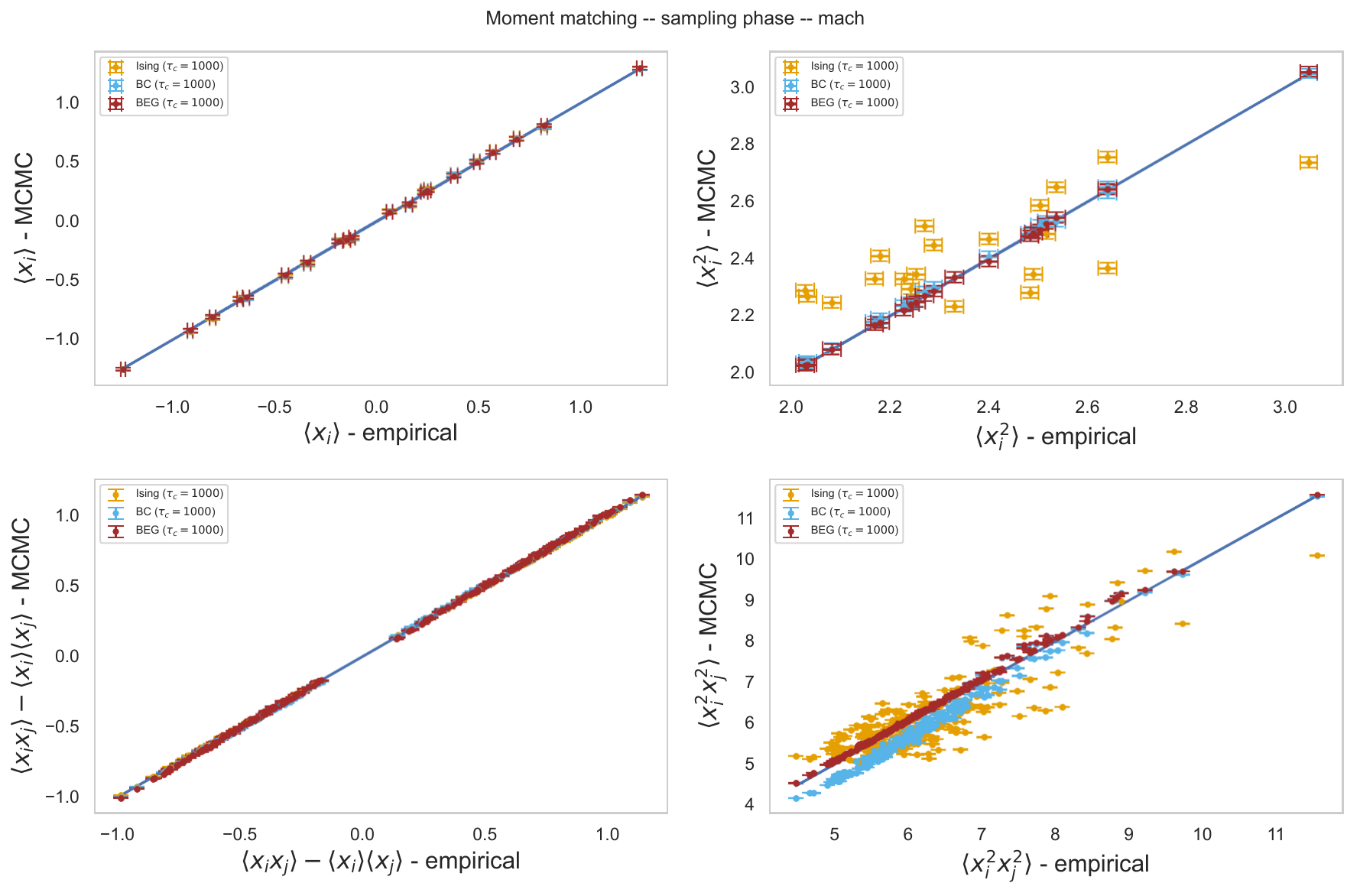}%
    \caption{Moment-matching consistency check in the sampling phase, for the {\sf mach} questionnaire. Comparison between empirical and theoretically sampled sufficient statistics ($\langle x_i\rangle$, $\langle x_i^2\rangle$, $\langle x_i x_j\rangle - \langle x_i \rangle \langle x_j\rangle$ and $\langle x_i^2 x_j^2\rangle$ as appropriate for each model). The matching is approximately preserved even though the MCMC sampling protocol used here differs from the one employed during gradient estimation in training. \label{fig:momentmatching_sampling_mach}}
\end{center}
\end{figure}

As a consistency check, we see that the matching is approximately satisfied also in the sampling phase cf.~Fig.~\ref{fig:momentmatching_sampling_mach}), even though the MCMC sampling protocol in this phase differs from the sampling protocol used in the estimation of the gradient during learning. 

In Fig. \ref{fig:losses_mach} we compare the value of the losses during training with the variance error $V_{o_\mu}$ of the sample estimation of the sufficient statistics $\<o_\mu\>_{\rho_X}$, with $o_\mu = x_i,x_ix_j,x_i^2,x_i^2x_j^2$ (dashed horizontal lines), estimated by bootstrapping as the average of the element-wise sum of squared errors (as in the definition of the losses $L_{o_\mu}$ in Eq. \ref{eq:losses} in Appendix \ref{sec:PCD}), over $2\cdot 10^4$ bootstrap samplings of the subject indices. The figure shows that the training losses are, for all the observables, of the same order of the sample variance error. 

In the perspective of the accuracy-complexity trade-off, this fact suggests the following considerations. On the one hand, a longer training protocol (with larger number of gradient iterations, number of copies of the Markov Chain, or contrastive divergence iteration steps per gradient ascent step) could improve the out-of-sample accuracy of the models. On the other hand, it would be pointless to make the training losses orders of magnitude lower than $V_{o_\mu}$, since the bottleneck of the models' generalization ability would become the variance error of the sufficient statistics, induced by the finiteness of $N$. Specially for those datasets in which the training losses are lower than the empirical errors, some sort of regularization (or, in the Bayesian language, an inference protocol beyond maximum likelihood) would increase the generalization accuracy of the model, more than increasing the accuracy of the approximated maximum likelihood algorithm.

We show the equivalent of Figs. \ref{fig:losses_mach},\ref{fig:momentmatching_sampling_mach} for all the questionnaires in Figs. \ref{fig:losses_all},\ref{fig:momentmatching_sampling_all} in Appendix \ref{sec:other}. For those questionnaires presenting low $N$ or low $N/M$ ratio, the training losses are lower than the empirical variance error. 

\section{Variant of the figures for all the considered questionnaires \label{sec:other}}

\begin{figure}[H]
\begin{center}
\includegraphics[width=0.45\columnwidth]{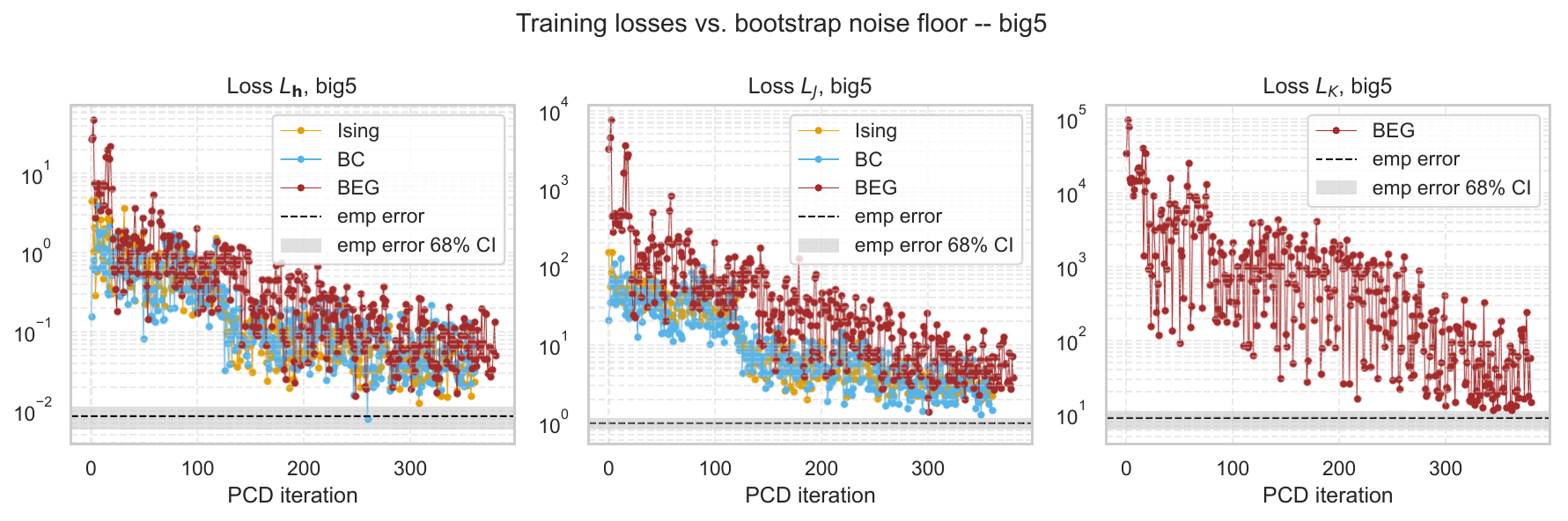}
\includegraphics[width=0.45\columnwidth]{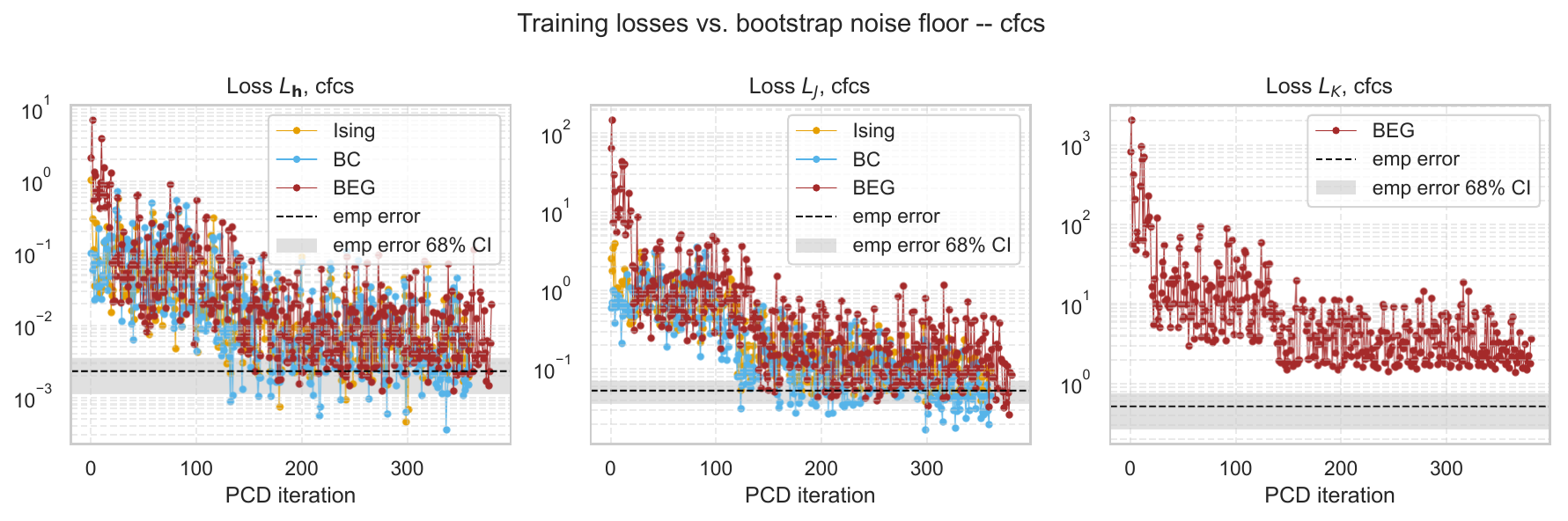}\\
\includegraphics[width=0.45\columnwidth]{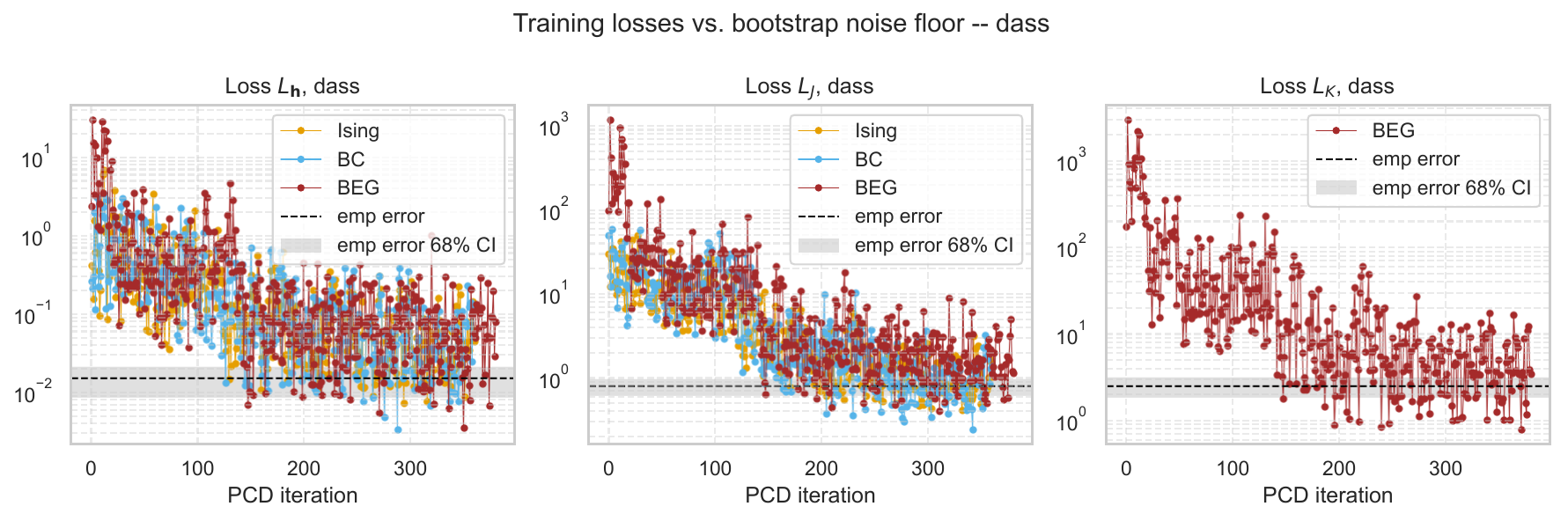} 
\includegraphics[width=0.45\columnwidth]{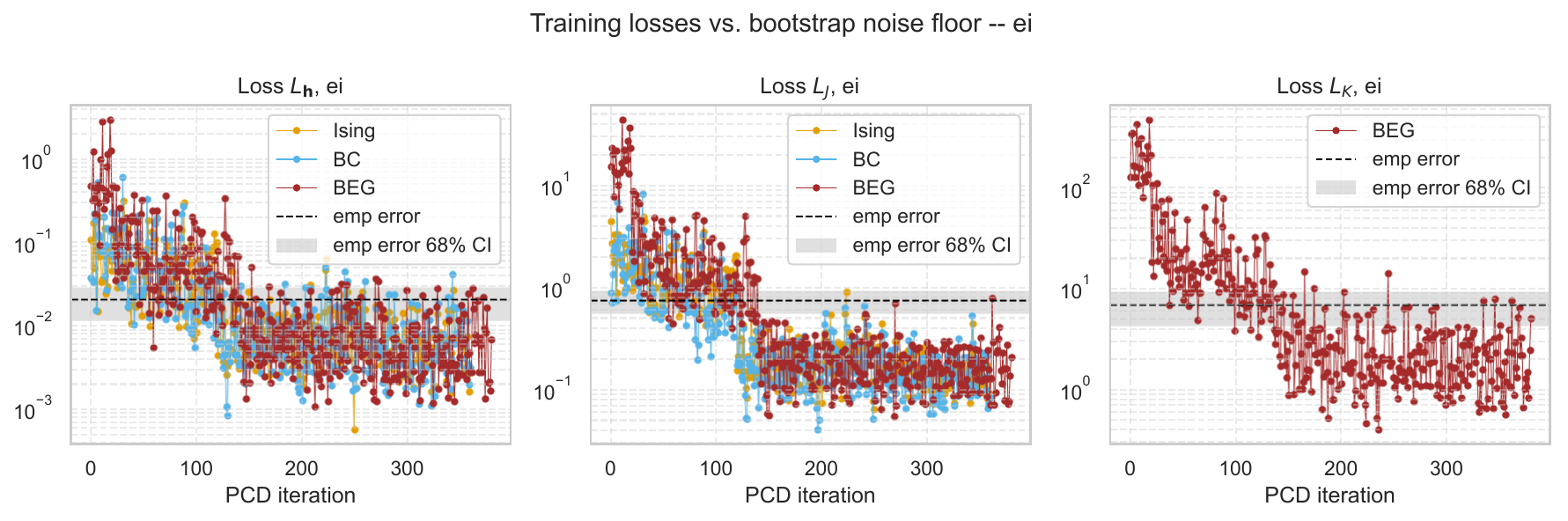}\\
\includegraphics[width=0.45\columnwidth]{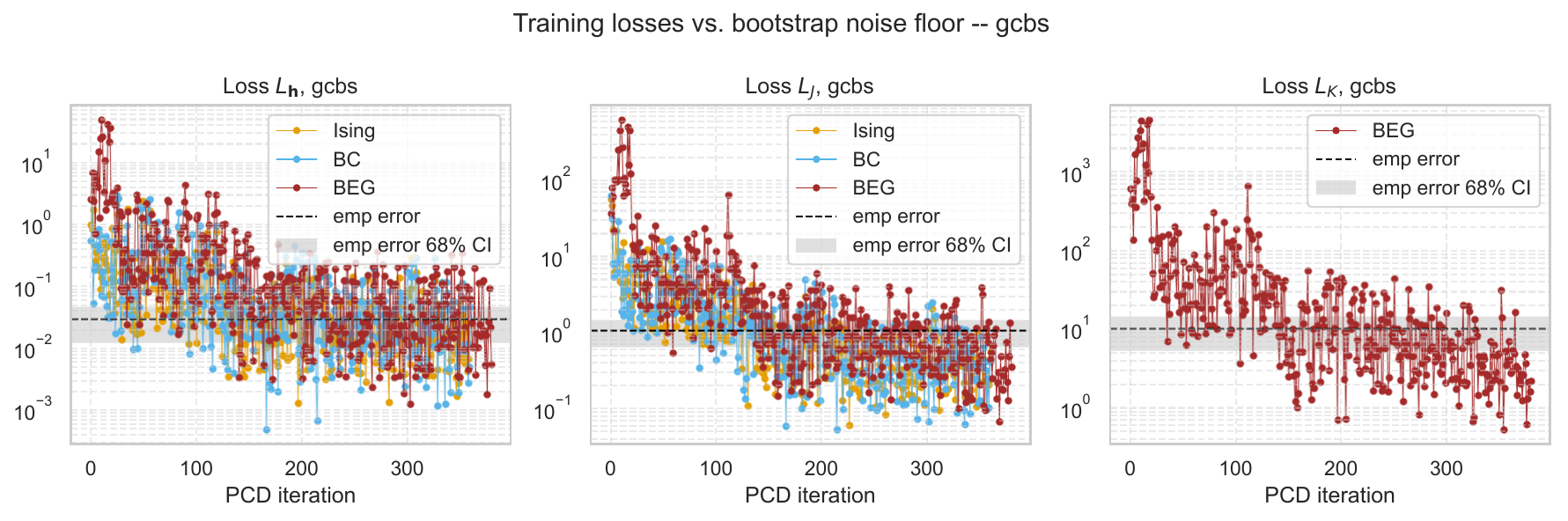}
\includegraphics[width=0.45\columnwidth]{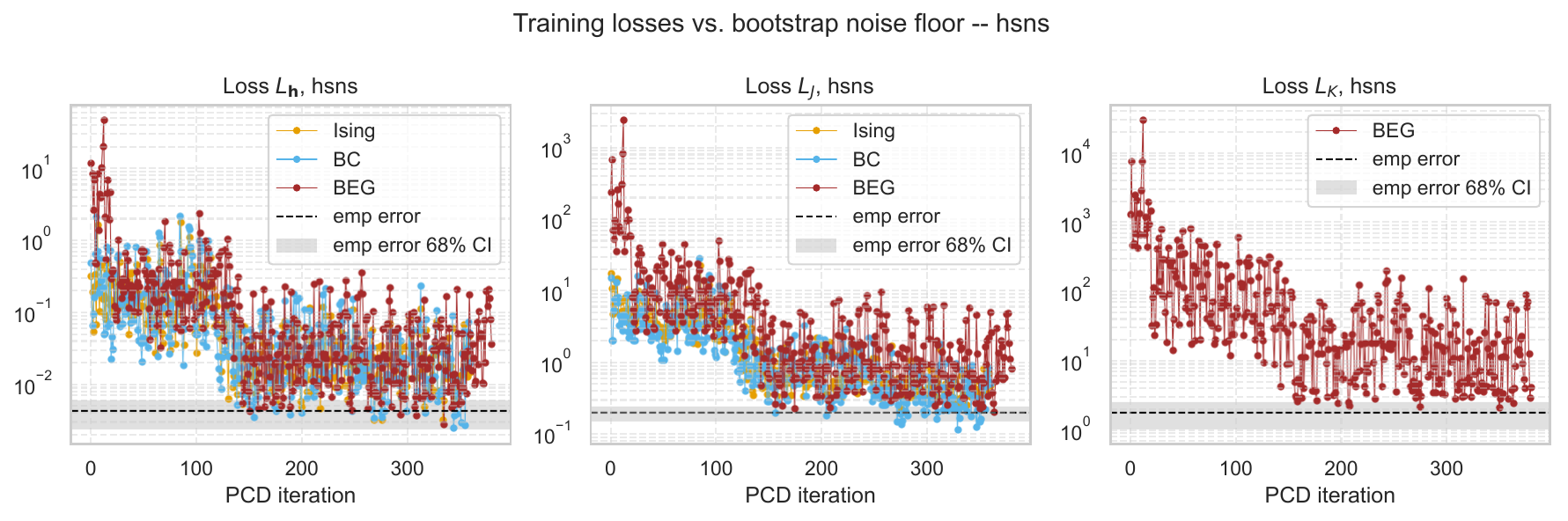} \\
\includegraphics[width=0.45\columnwidth]{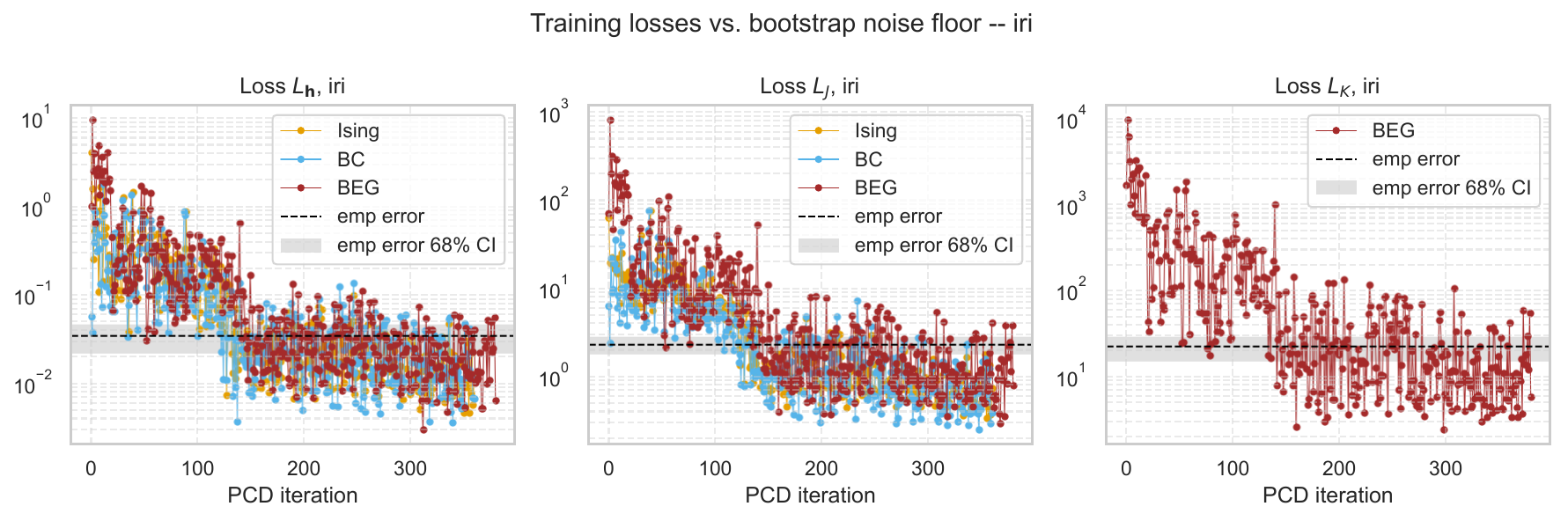}
\includegraphics[width=0.45\columnwidth]{figures/losses/losses_data_empfloormach}\\
\includegraphics[width=0.45\columnwidth]{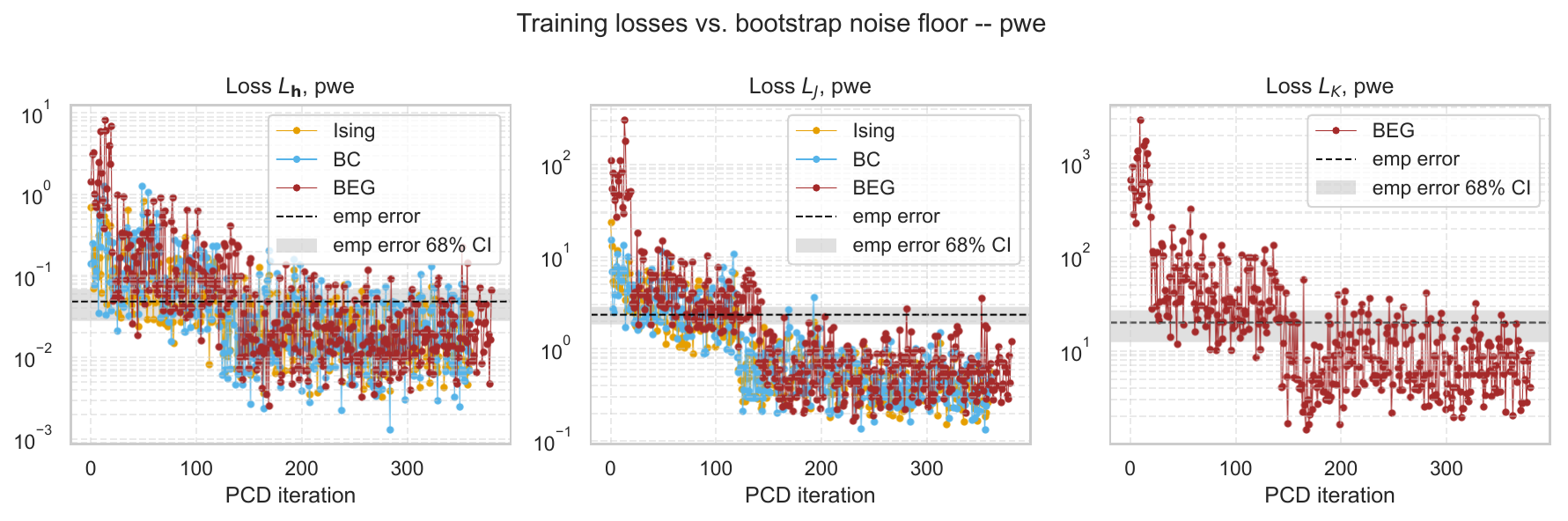} 
\includegraphics[width=0.45\columnwidth]{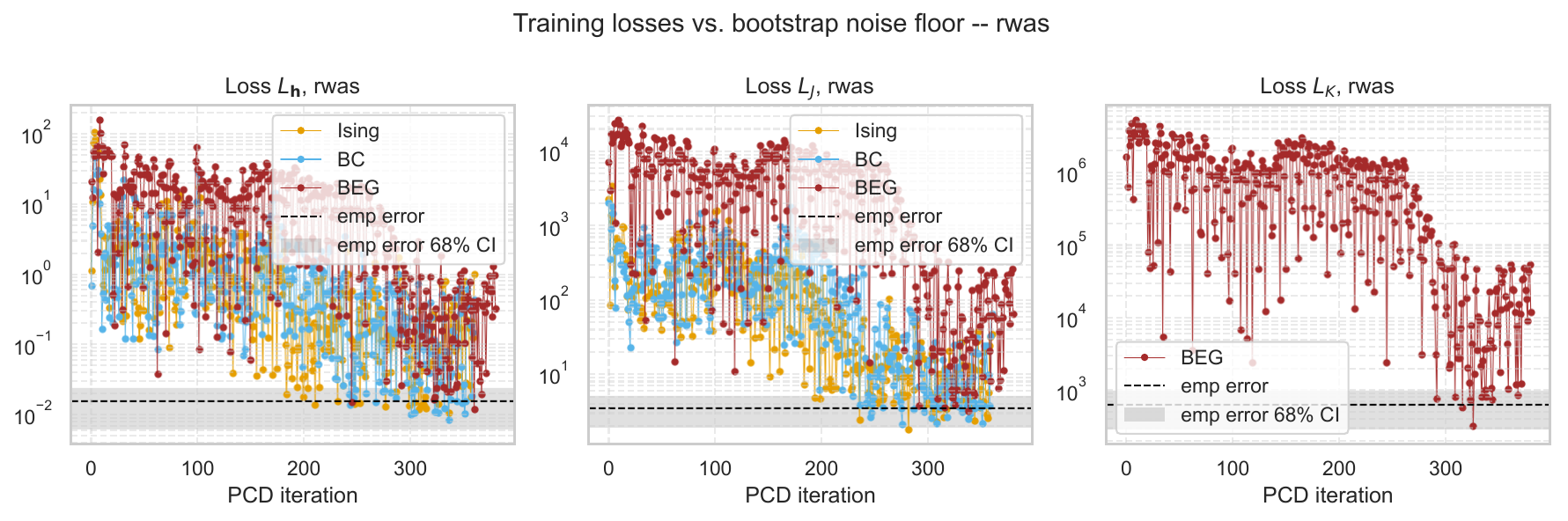}\\
\includegraphics[width=0.45\columnwidth]{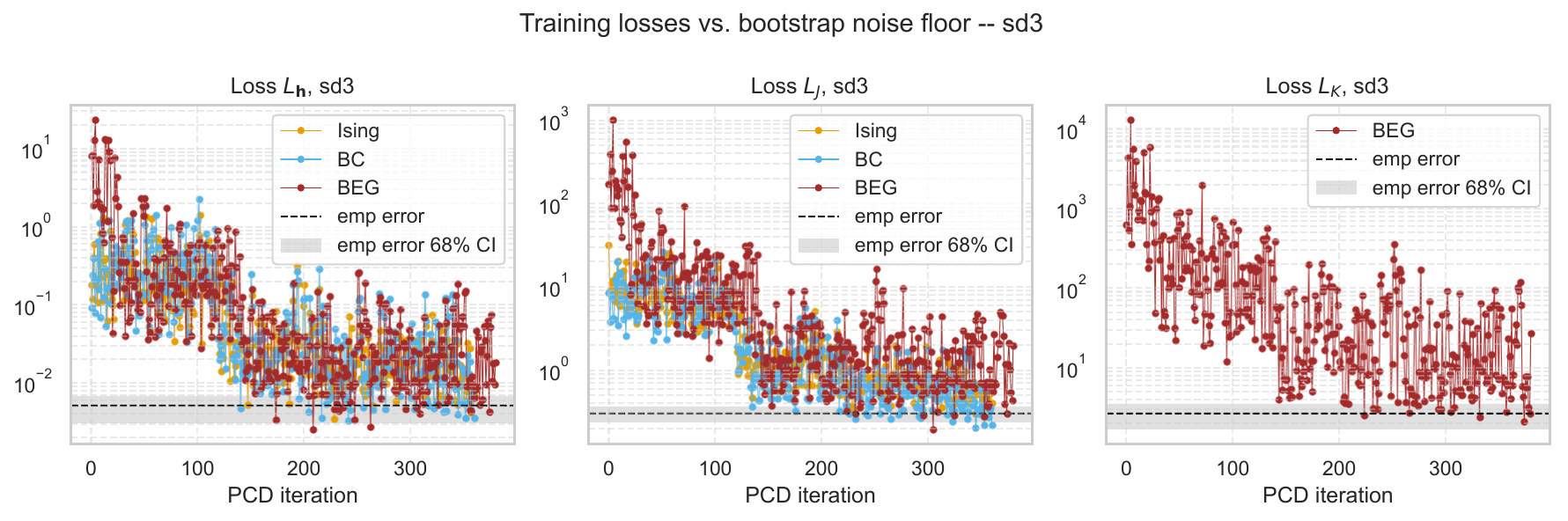}
\caption{Loss functions $L_{\bf h}$, $L_J$ (and $L_K$ for BEG) as a function of PCD iteration number, for all the analyzed questionnaires (each panel labeled by questionnaire name). Equivalent of Fig.~\ref{fig:losses_mach} for all datasets; see Sec.~\ref{sec:consistency}.}
\label{fig:losses_all}
\end{center}
\end{figure}

\begin{figure}[H]
\begin{center}
\includegraphics[width=0.45\columnwidth]{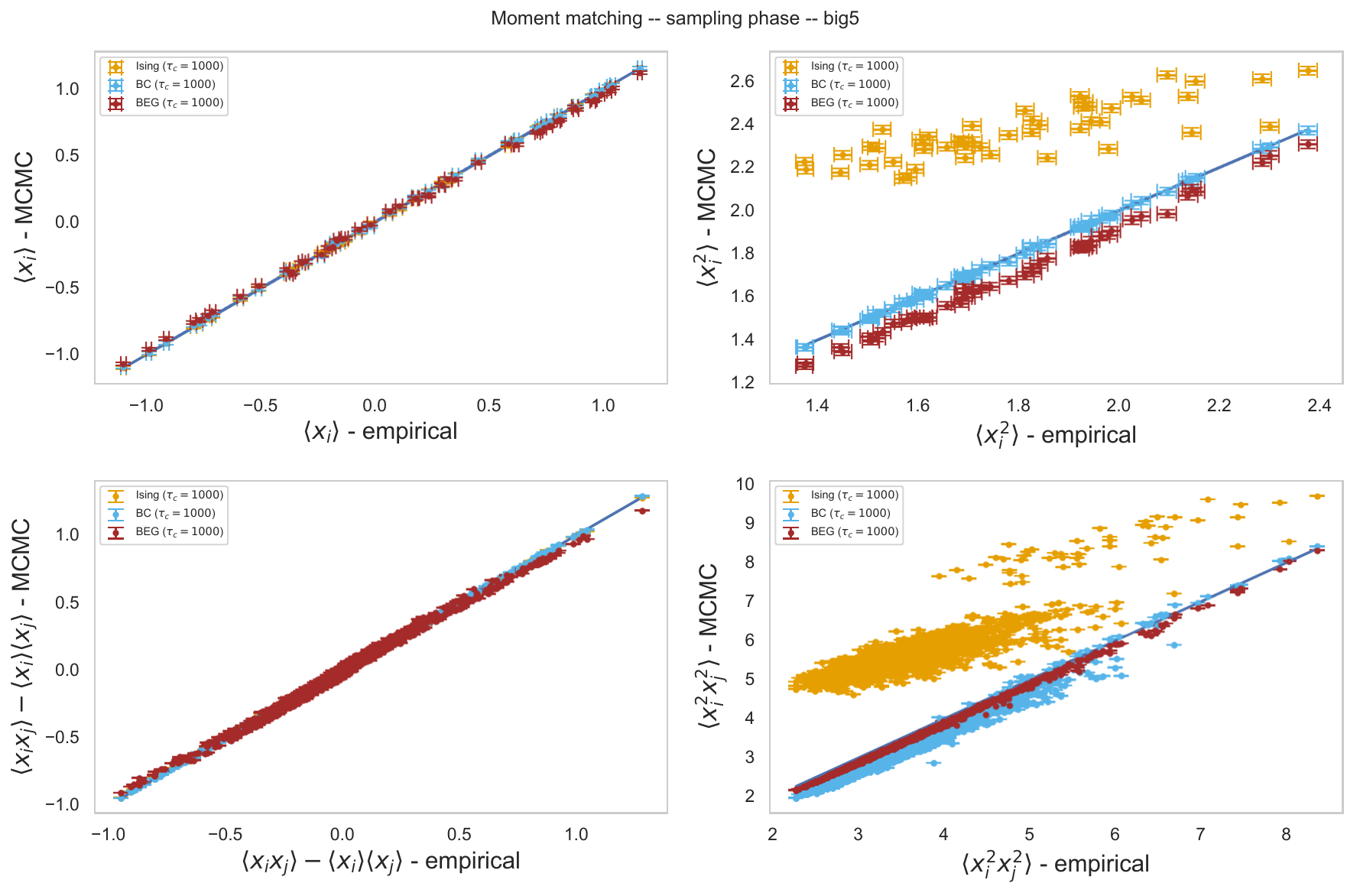}
\includegraphics[width=0.45\columnwidth]{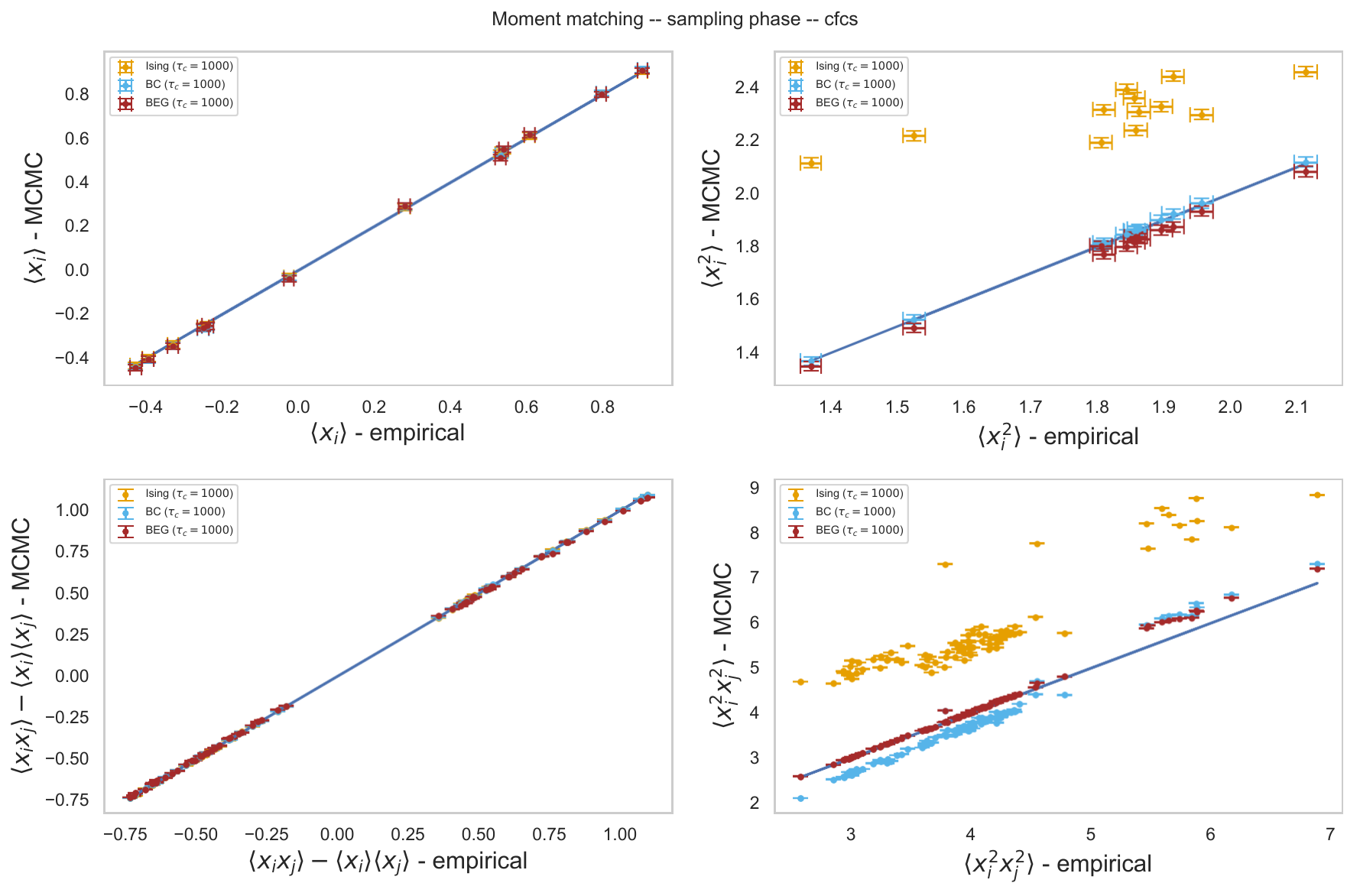}\\
\includegraphics[width=0.45\columnwidth]{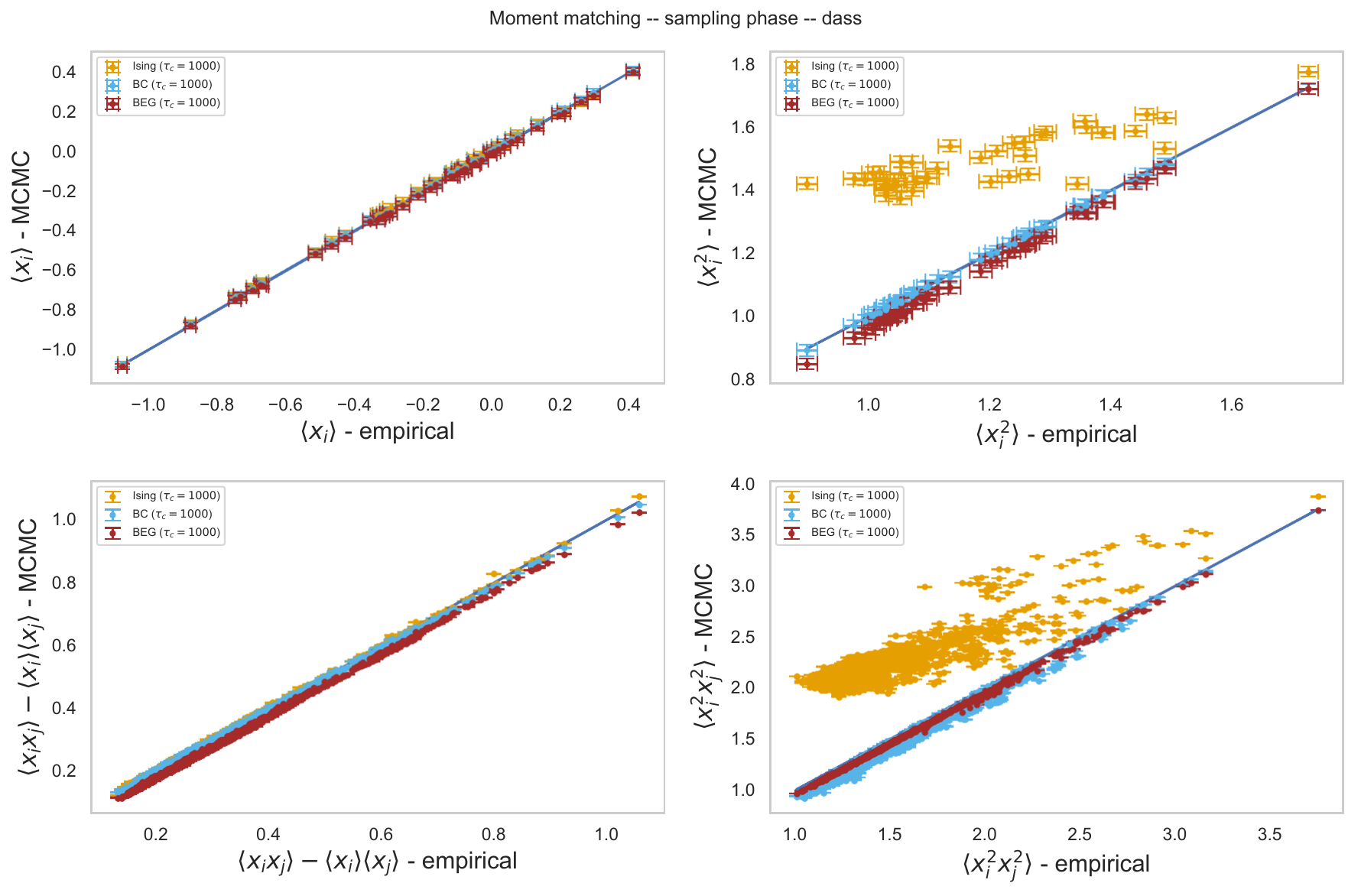} 
\includegraphics[width=0.45\columnwidth]{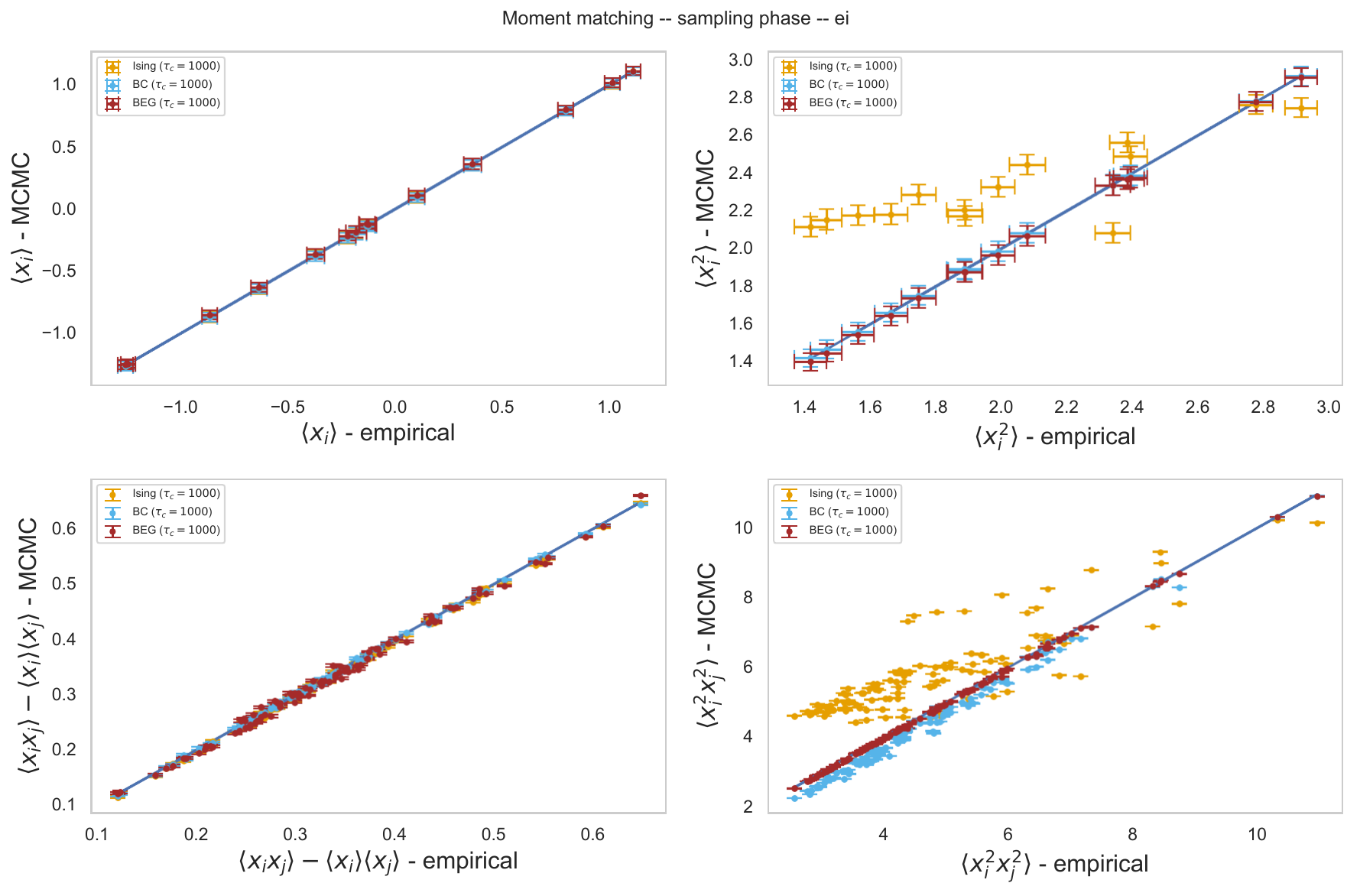}\\
\includegraphics[width=0.45\columnwidth]{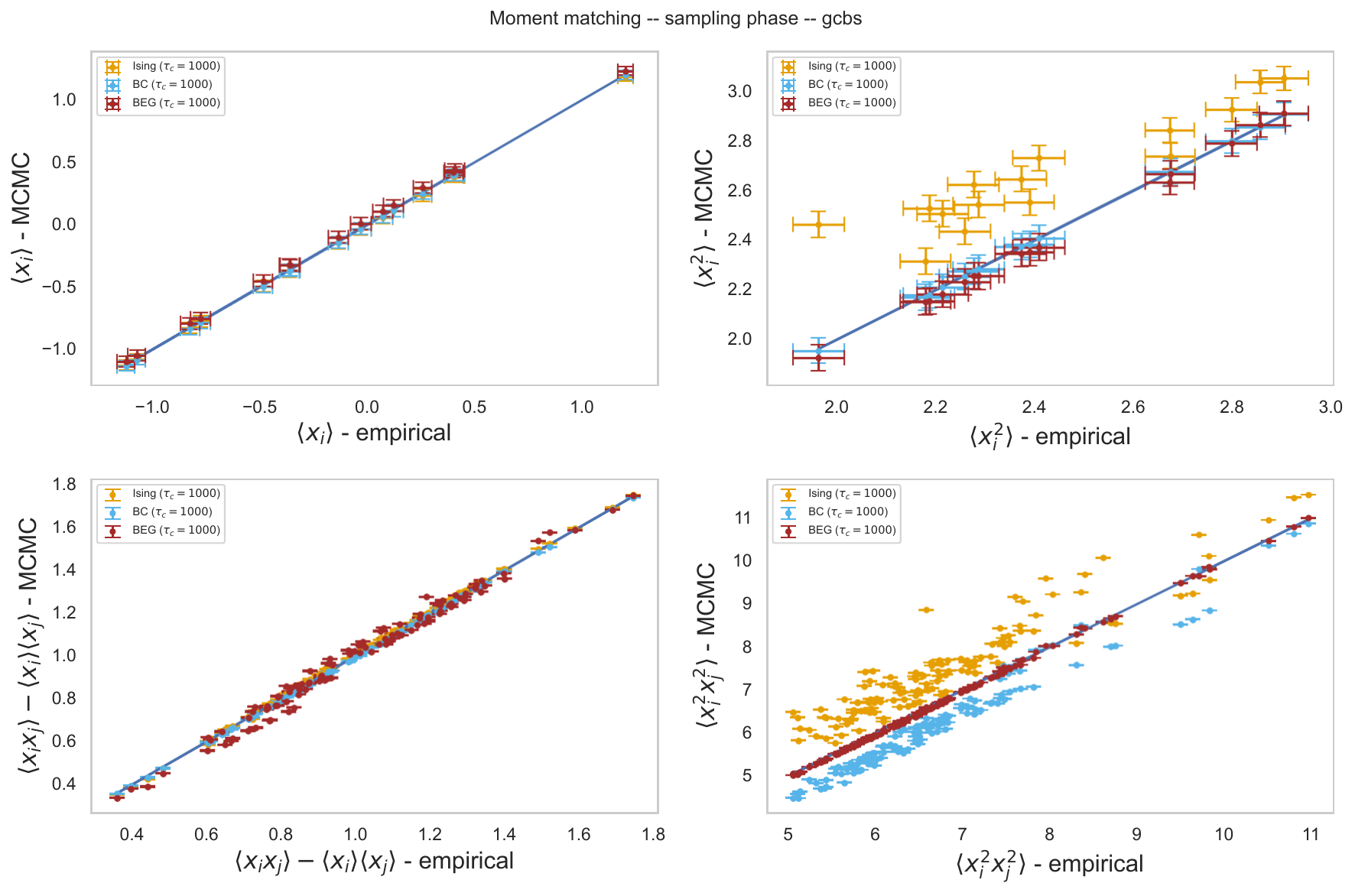}
\includegraphics[width=0.45\columnwidth]{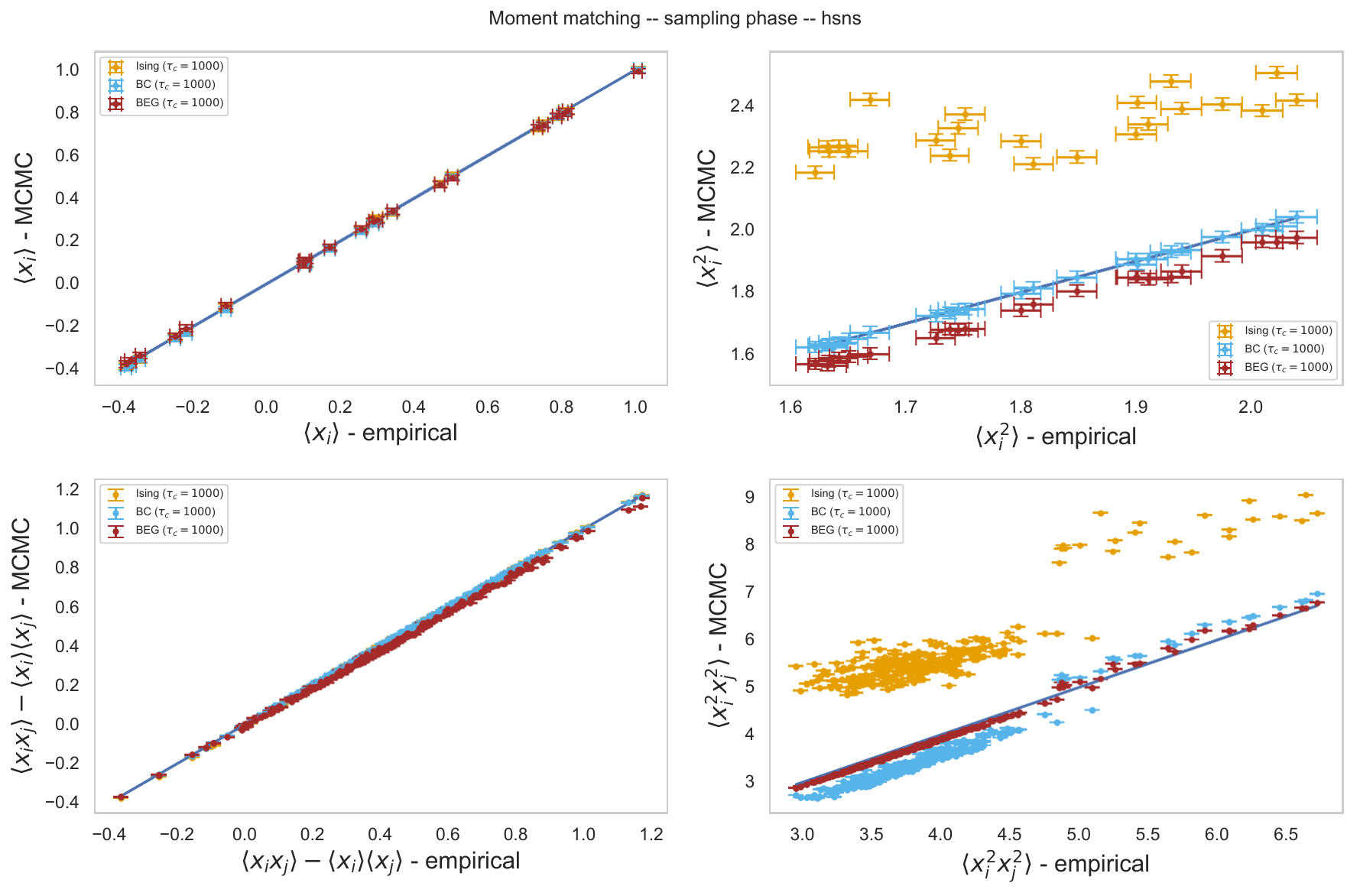} 

\caption{Moment-matching consistency check in the sampling phase for all the analyzed questionnaires: comparison between empirical and theoretically sampled sufficient statistics. Equivalent of Fig.~\ref{fig:momentmatching_sampling_mach} for all datasets; see Sec.~\ref{sec:consistency}.}
\label{fig:momentmatching_sampling_all}
\end{center}
\end{figure}

\begin{figure}[H]
\begin{center}
\includegraphics[width=0.45\columnwidth]{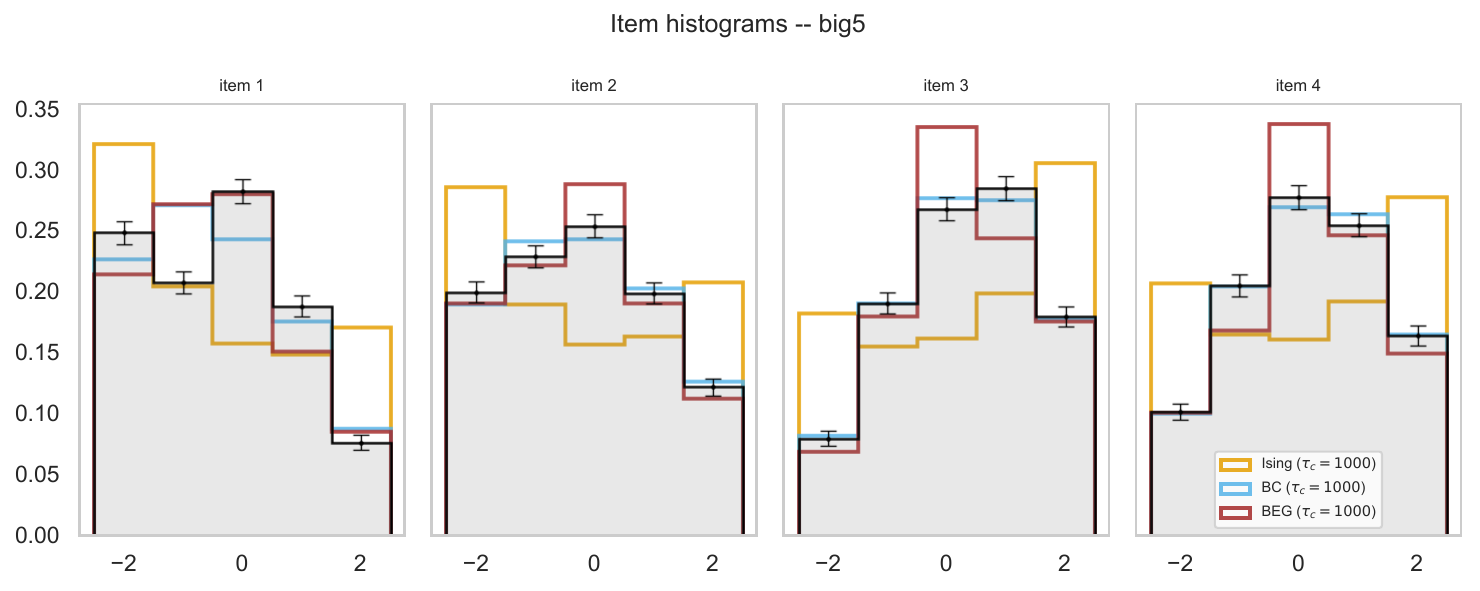}
\includegraphics[width=0.45\columnwidth]{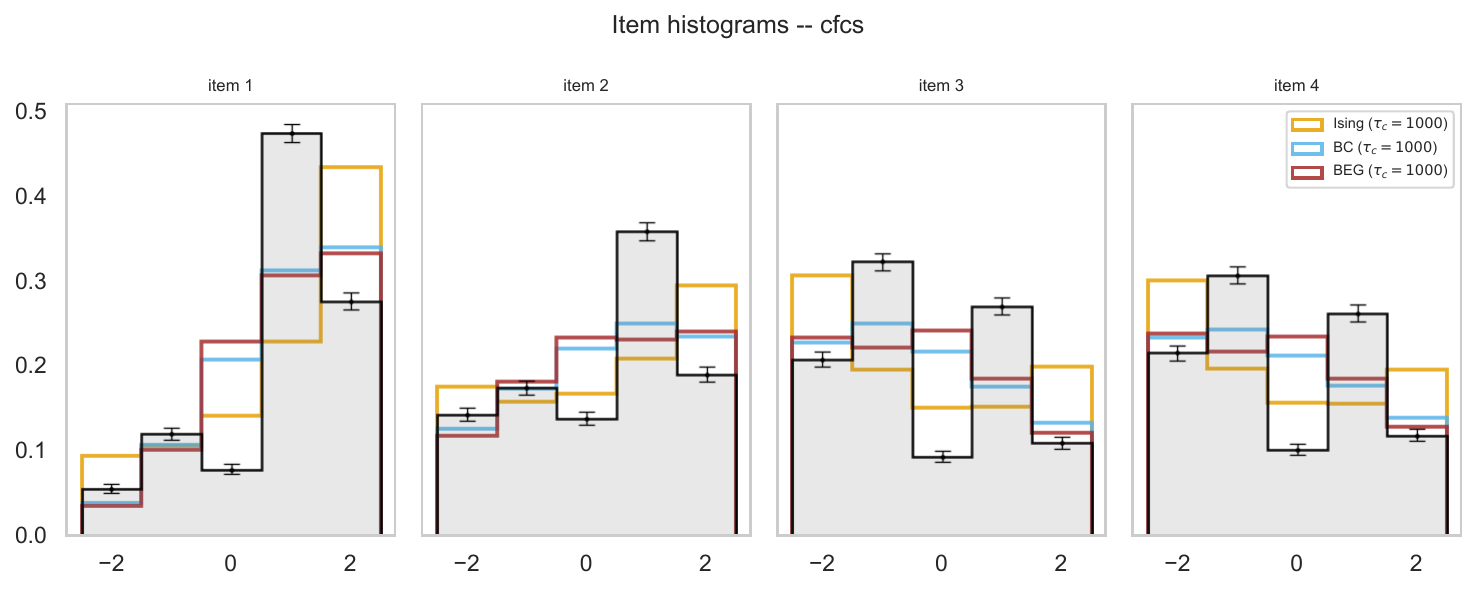} \\
\includegraphics[width=0.45\columnwidth]{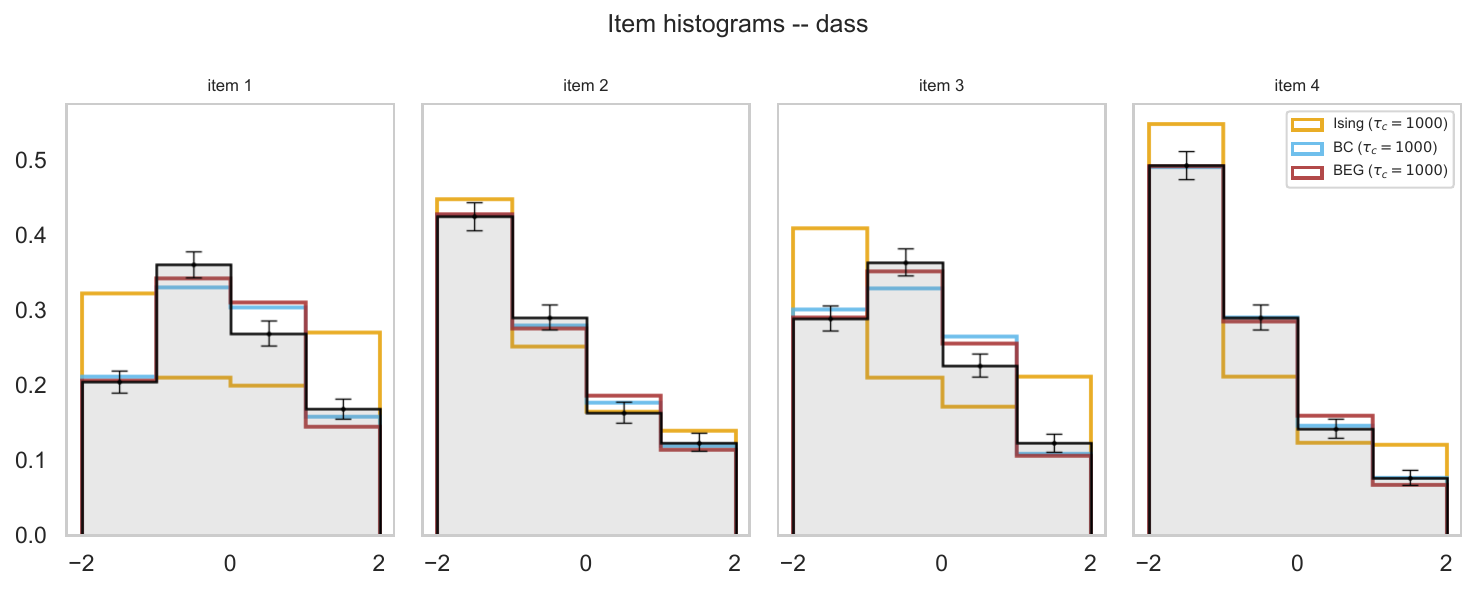} 
\includegraphics[width=0.45\columnwidth]{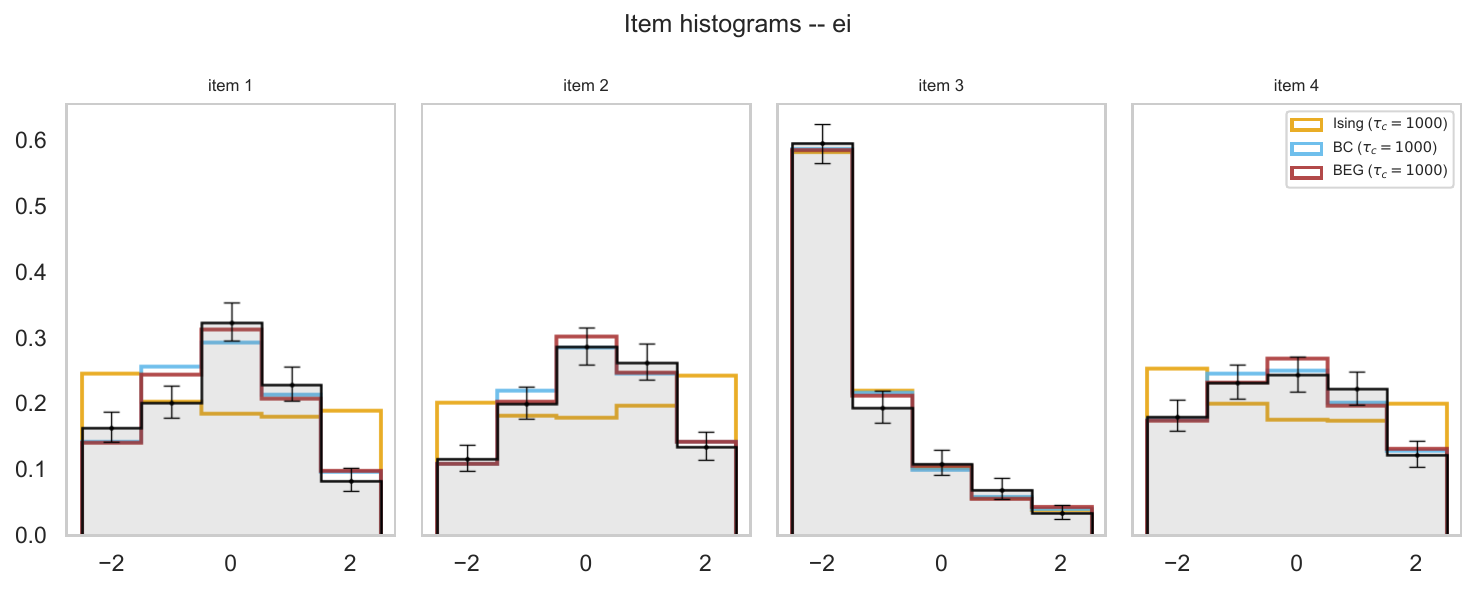} \\
\includegraphics[width=0.45\columnwidth]{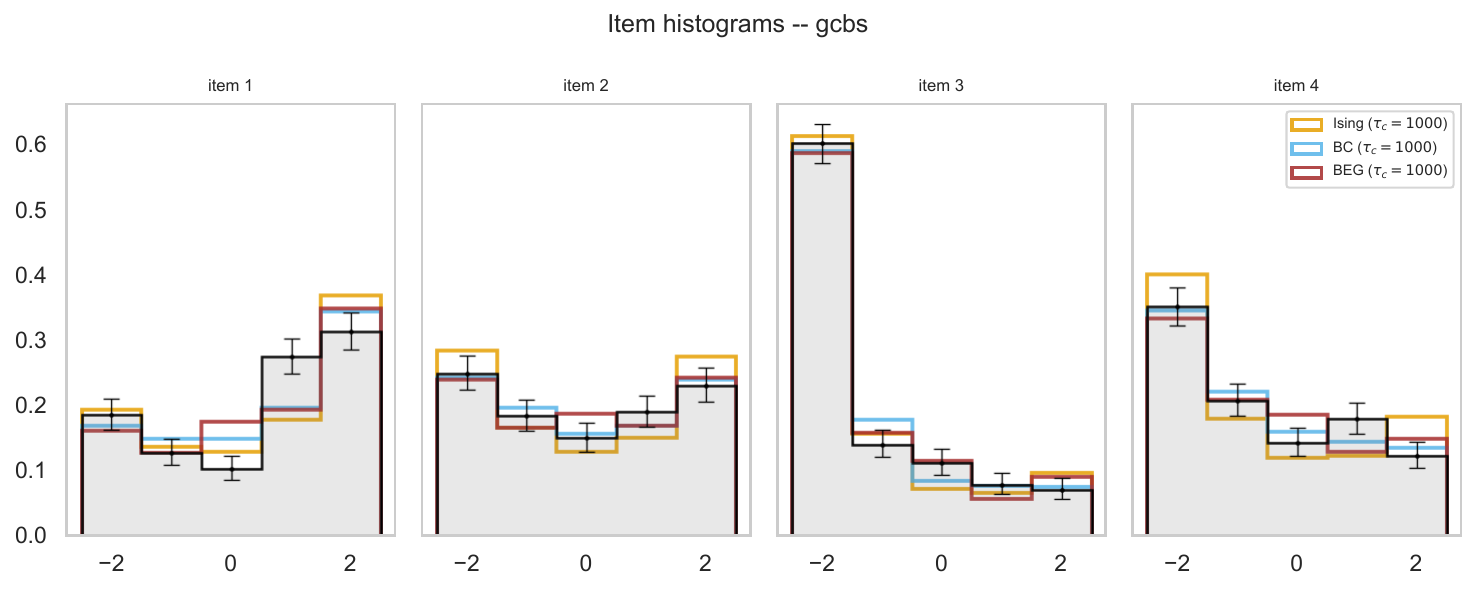}
\includegraphics[width=0.45\columnwidth]{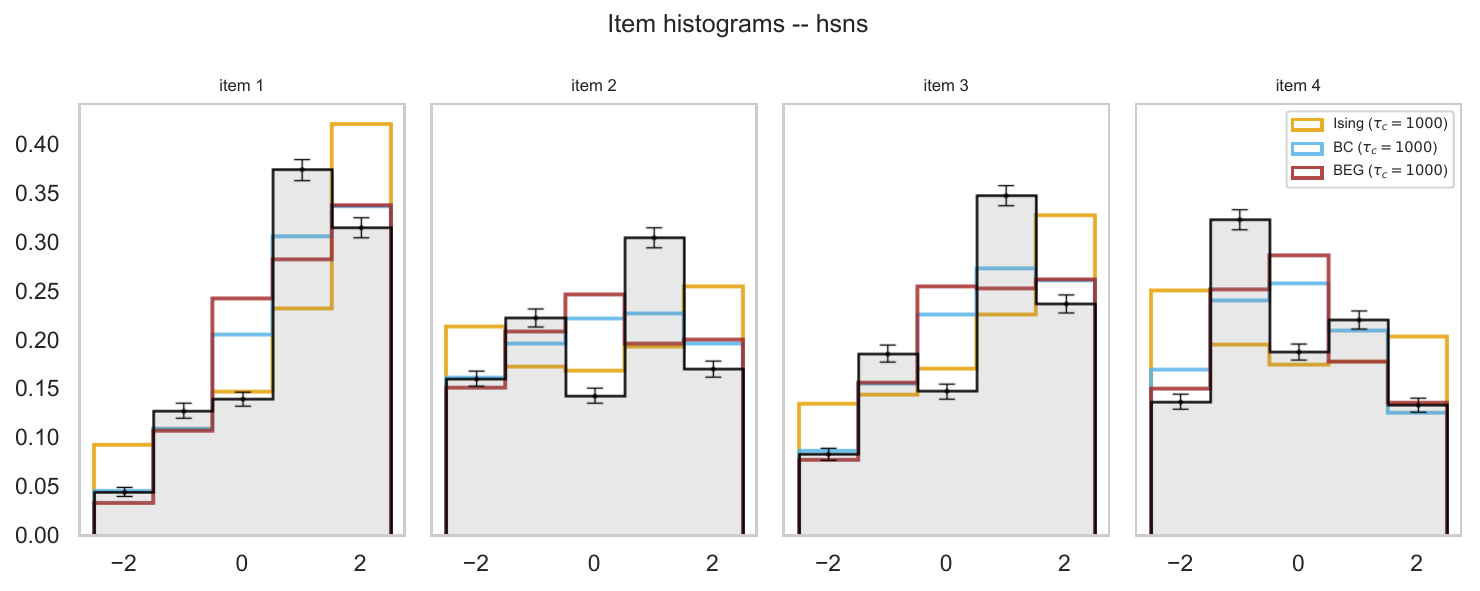} \\
\caption{Histograms of item responses ${\sf h}_{x_i}$ for the first 6 analyzed questionnaires: comparison between empirical data and the three spin models (Ising, BC, BEG). The spin models approximately reproduce the empirical item histograms despite these not being constrained by construction as sufficient statistics.}
\label{fig:itemhist_all}
\end{center}
\end{figure}

\begin{figure}[H]
\begin{center}
\includegraphics[width=0.45\columnwidth]{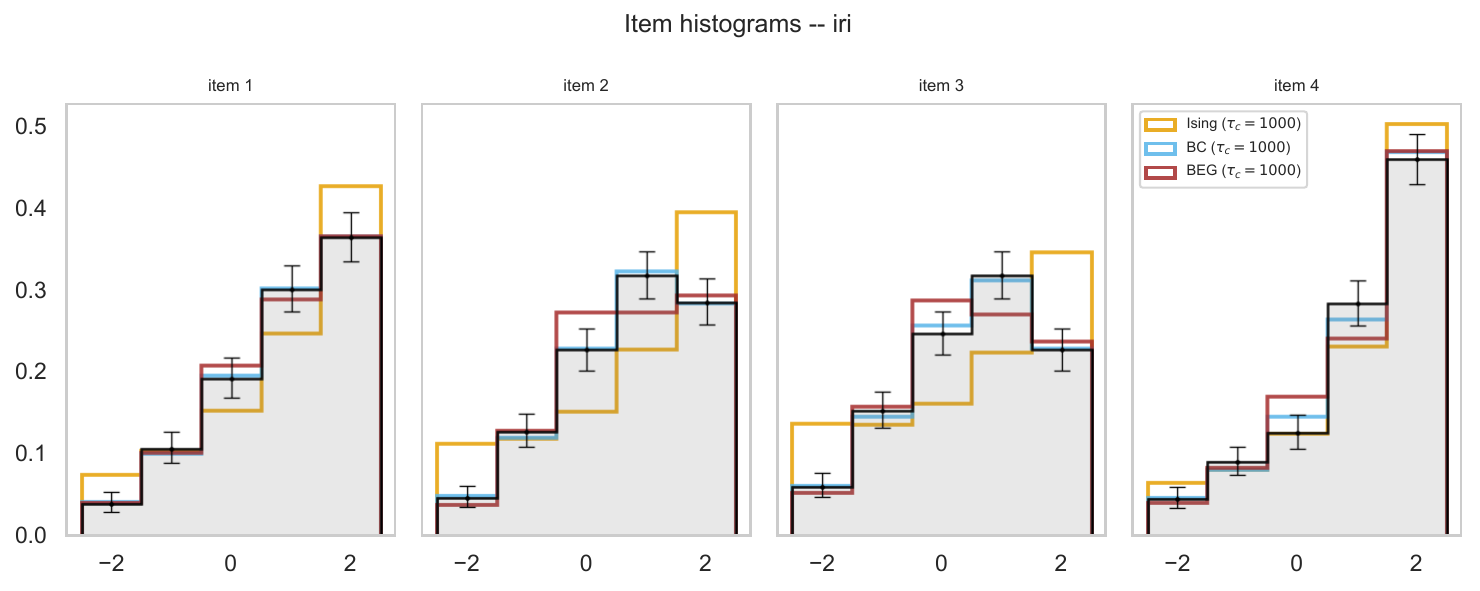}
\includegraphics[width=0.45\columnwidth]{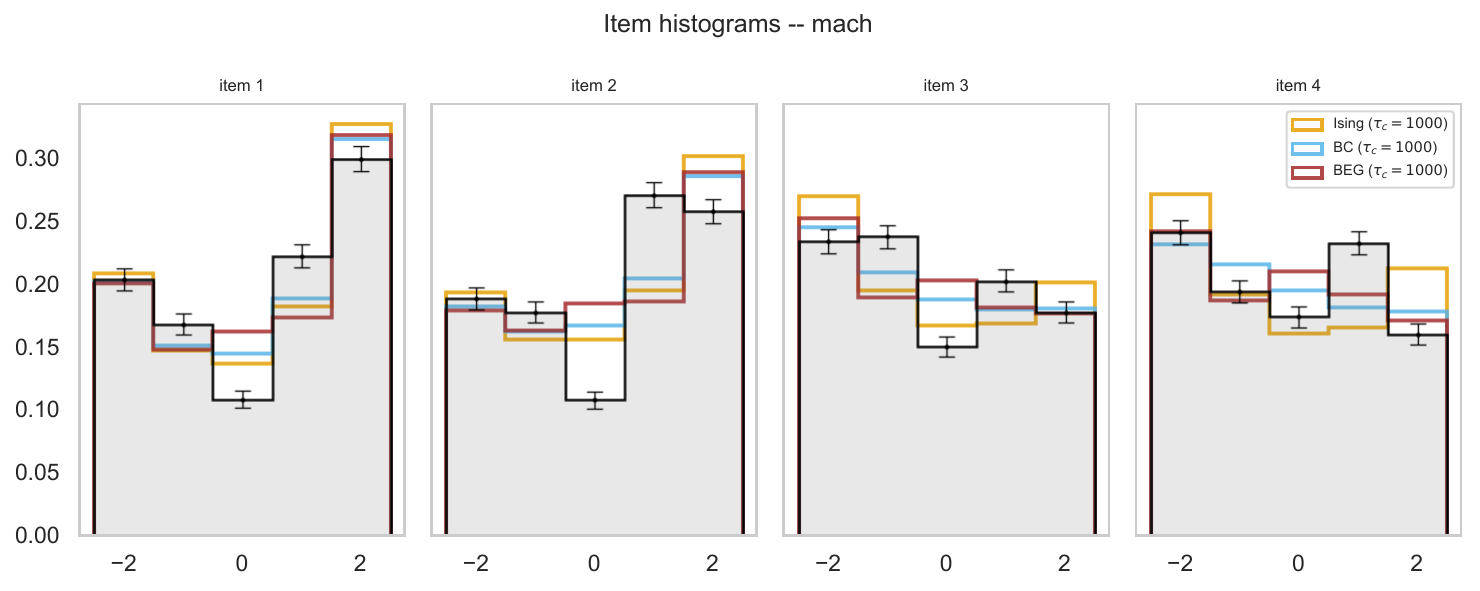}\\
\includegraphics[width=0.45\columnwidth]{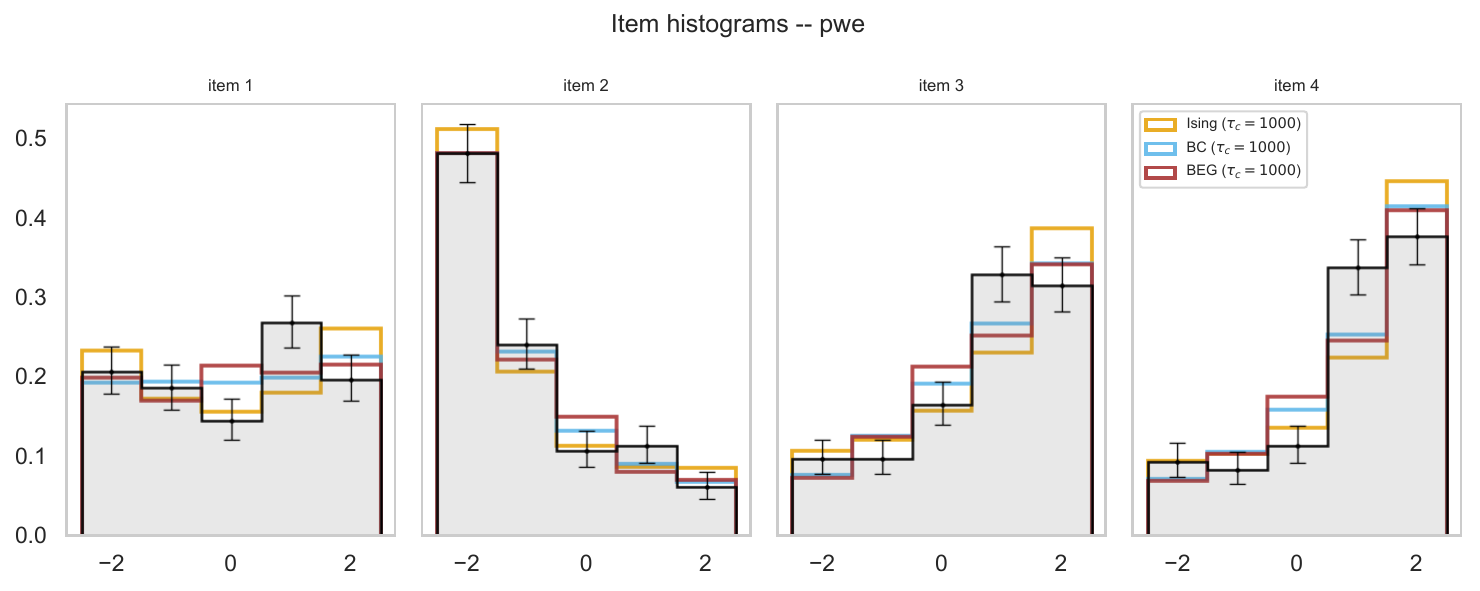} 
\includegraphics[width=0.45\columnwidth]{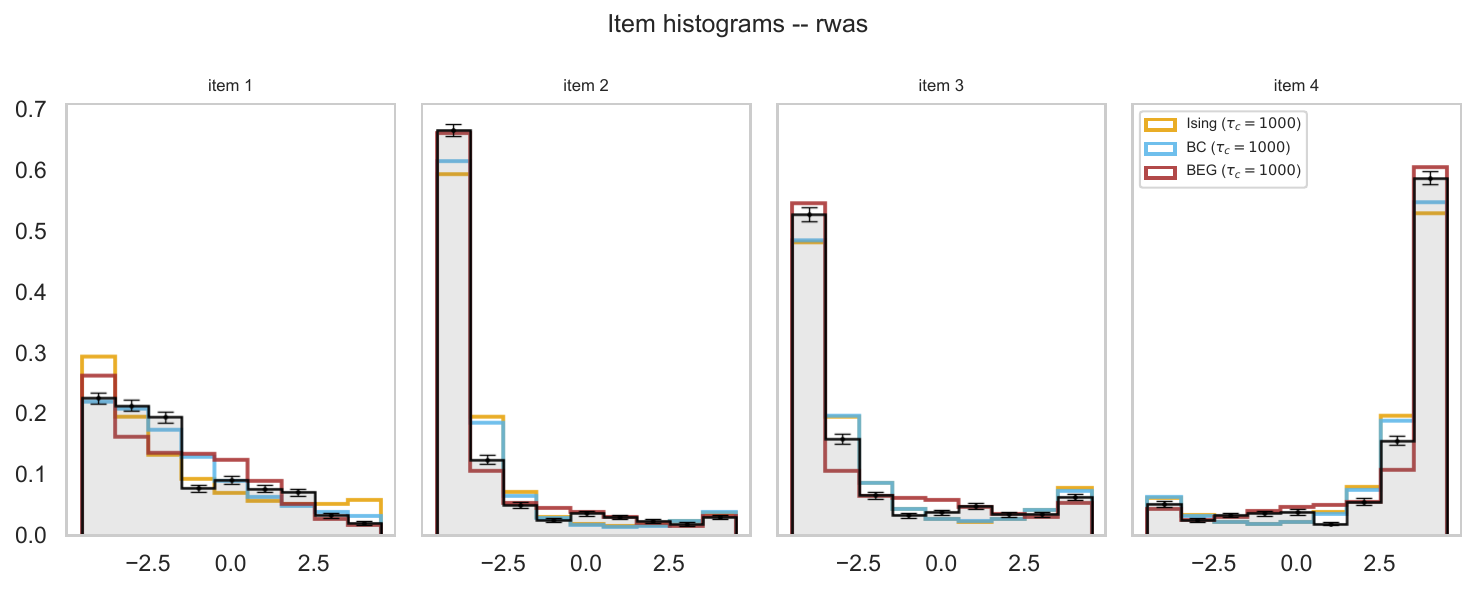}\\
\includegraphics[width=0.45\columnwidth]{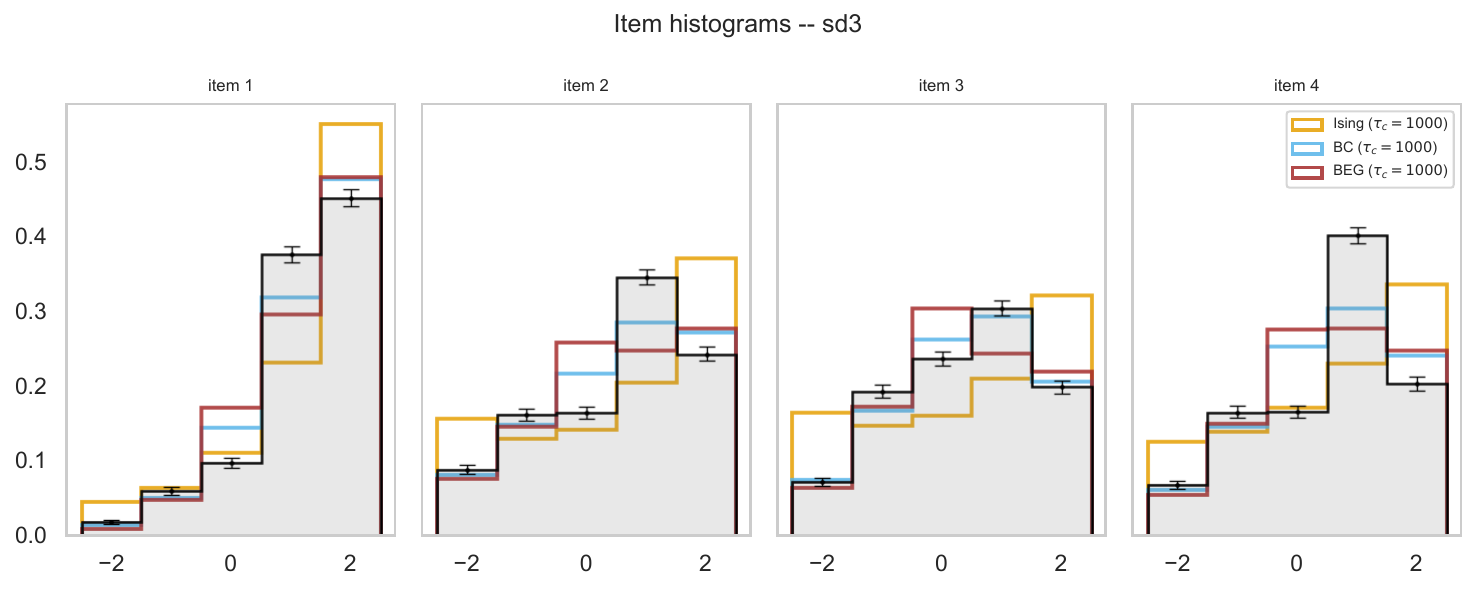}
\caption{Histograms of item responses ${\sf h}_{x_i}$ for the last 5 analyzed questionnaires: comparison between empirical data and the three spin models (Ising, BC, BEG). The spin models approximately reproduce the empirical item histograms despite these not being constrained by construction as sufficient statistics.}
\label{fig:itemhist_all2}
\end{center}
\end{figure}

\begin{figure}[H]
\begin{center}
\includegraphics[width=0.45\columnwidth]{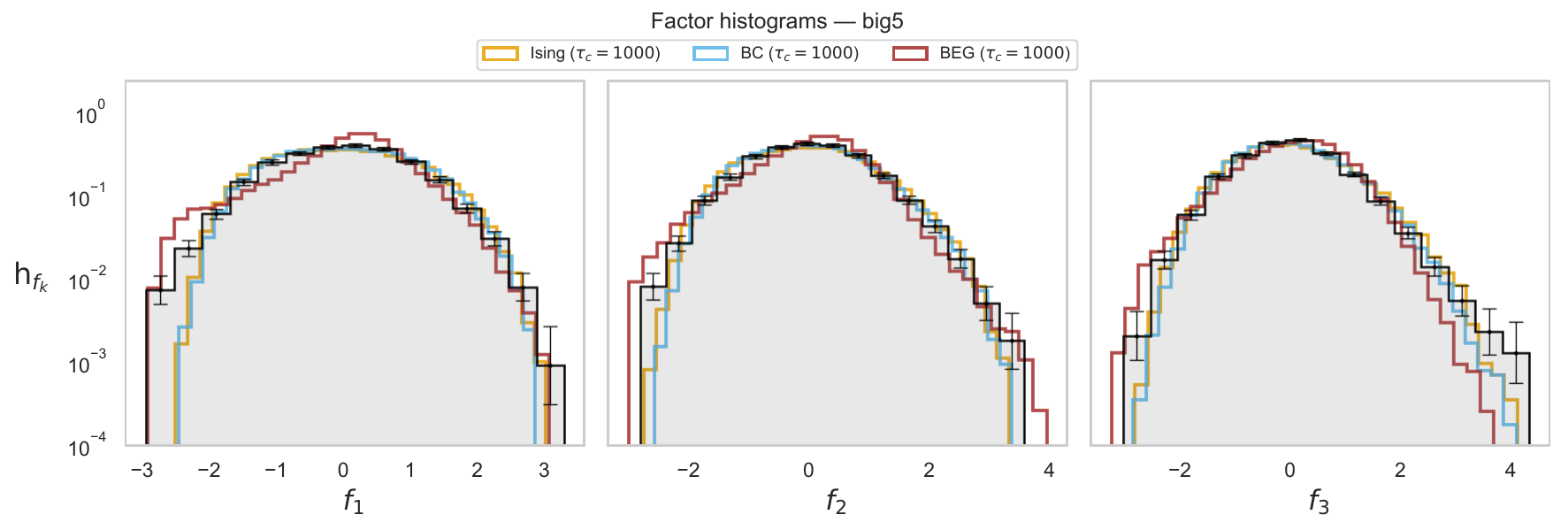}
\includegraphics[width=0.45\columnwidth]{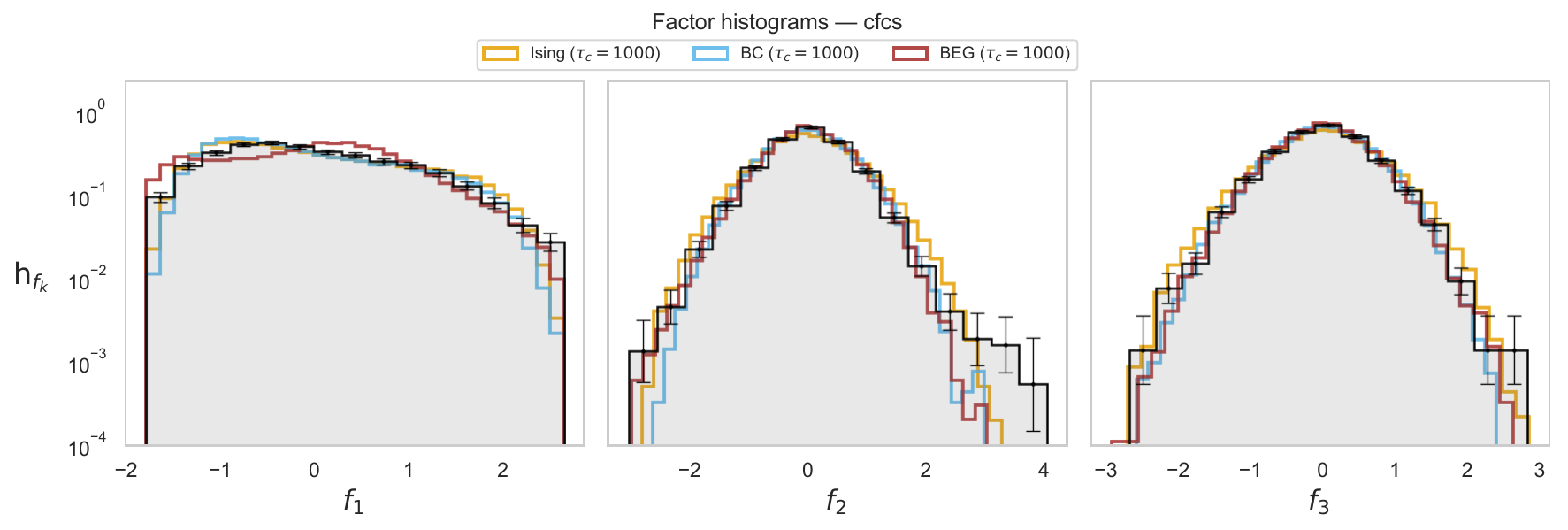}\\
\includegraphics[width=0.45\columnwidth]{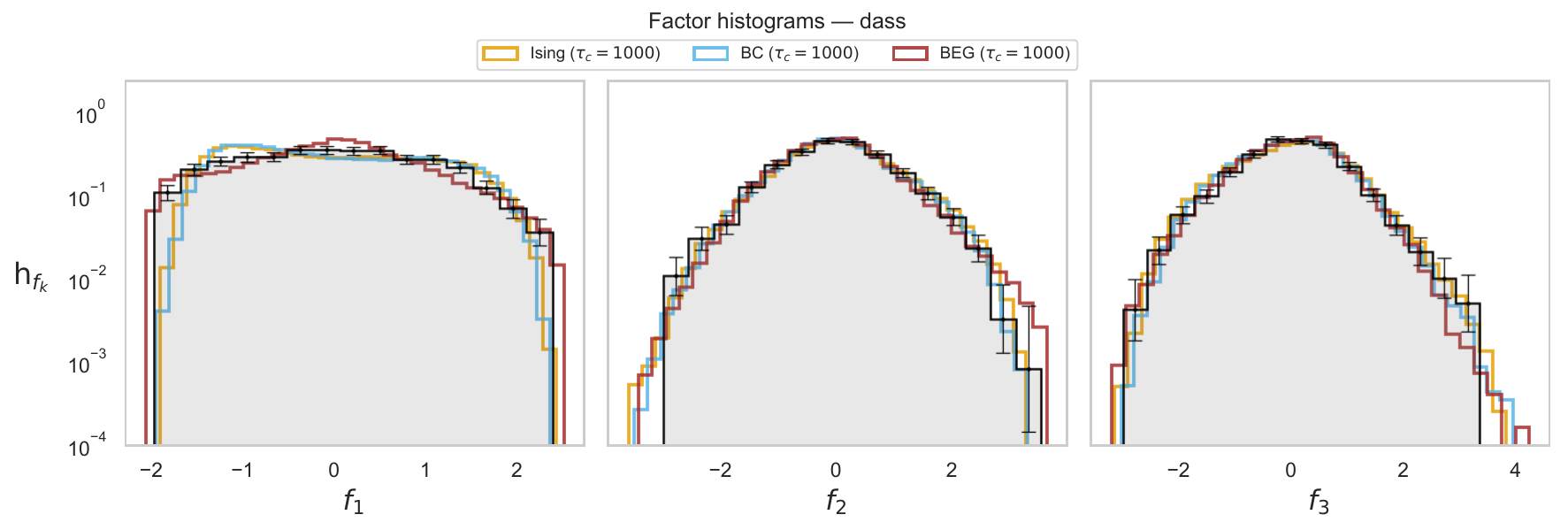} 
\includegraphics[width=0.45\columnwidth]{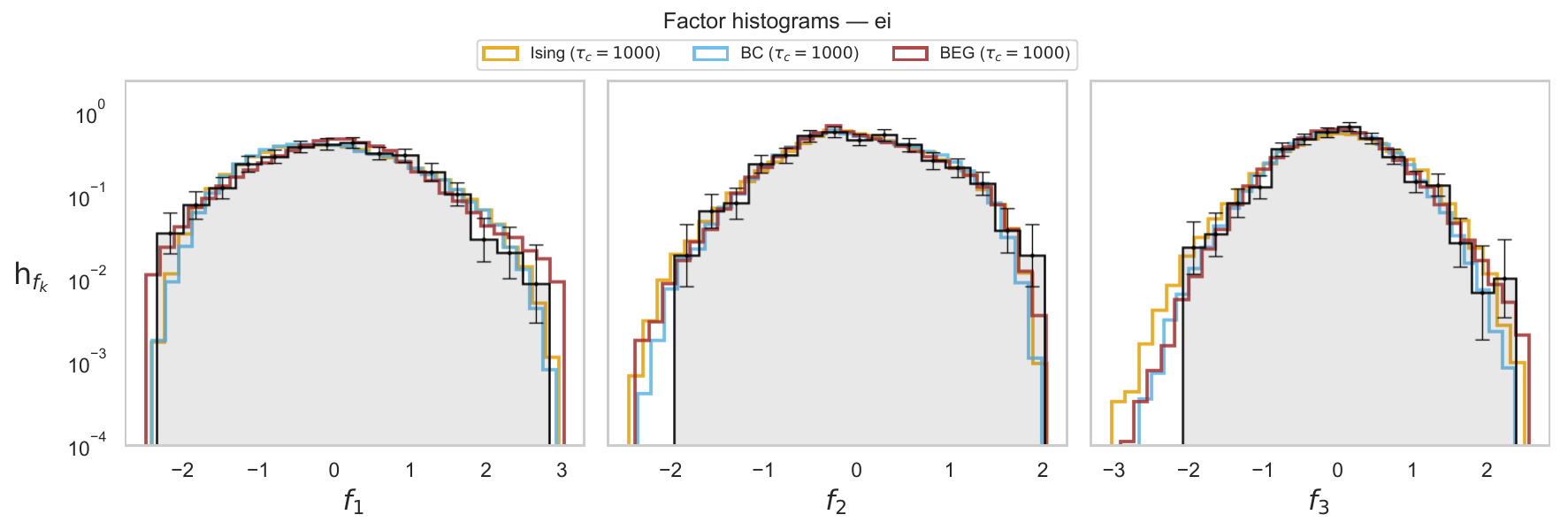}\\
\includegraphics[width=0.45\columnwidth]{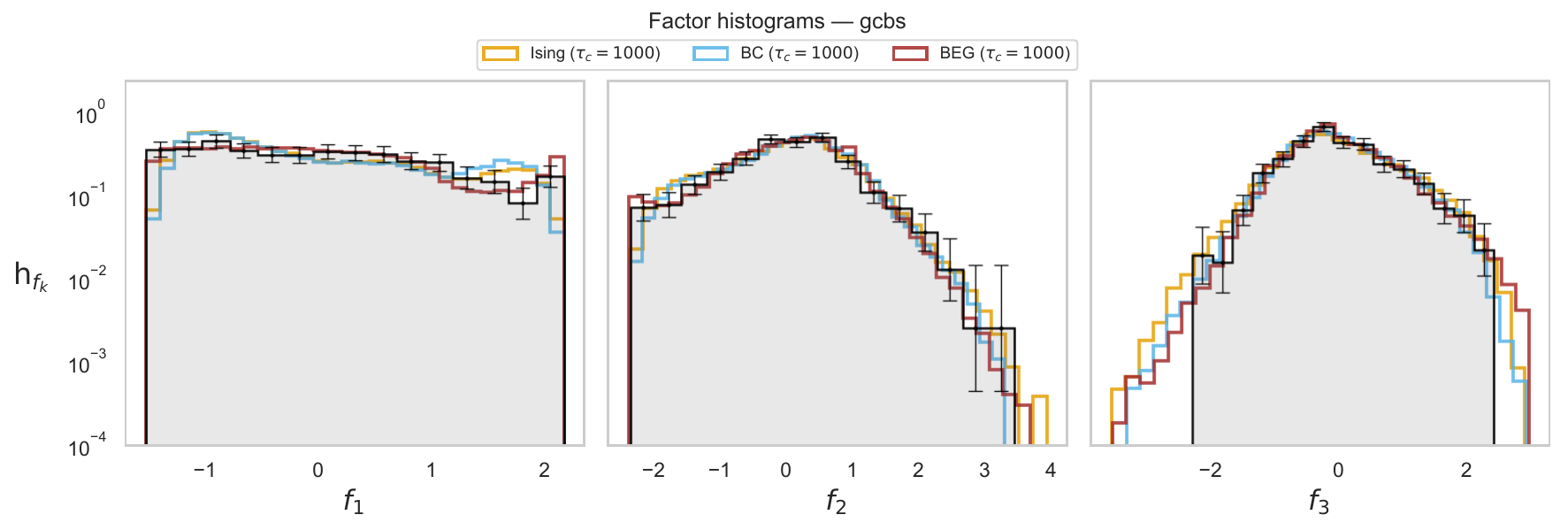}
\includegraphics[width=0.45\columnwidth]{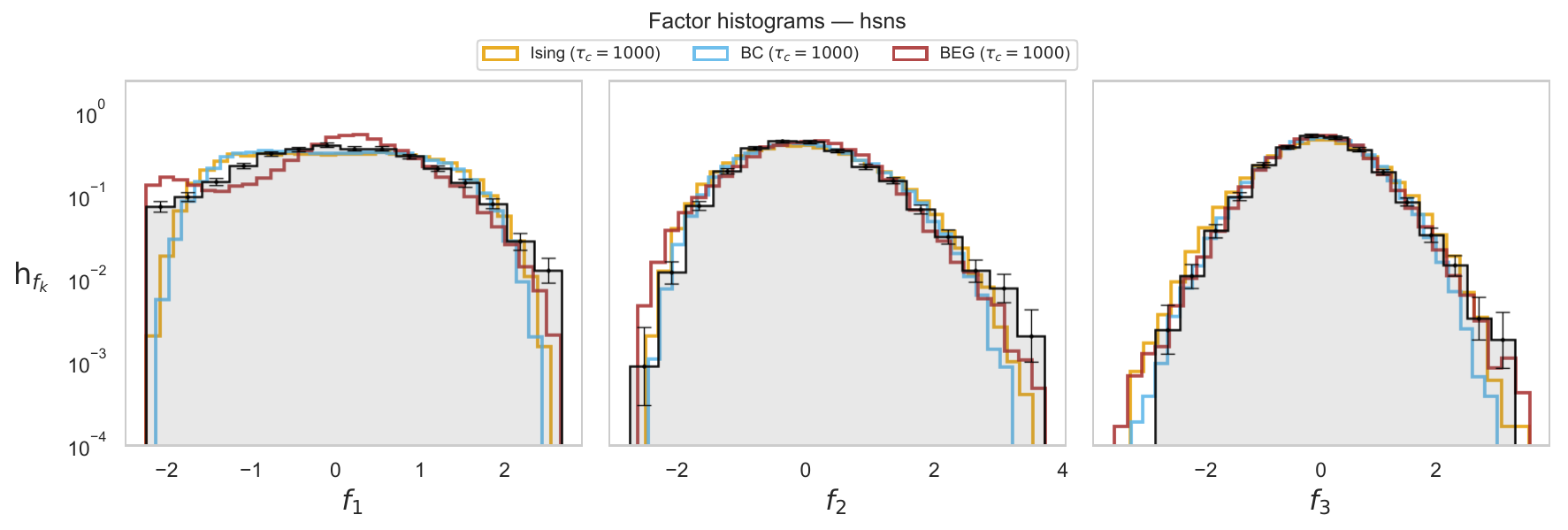} \\
\includegraphics[width=0.45\columnwidth]{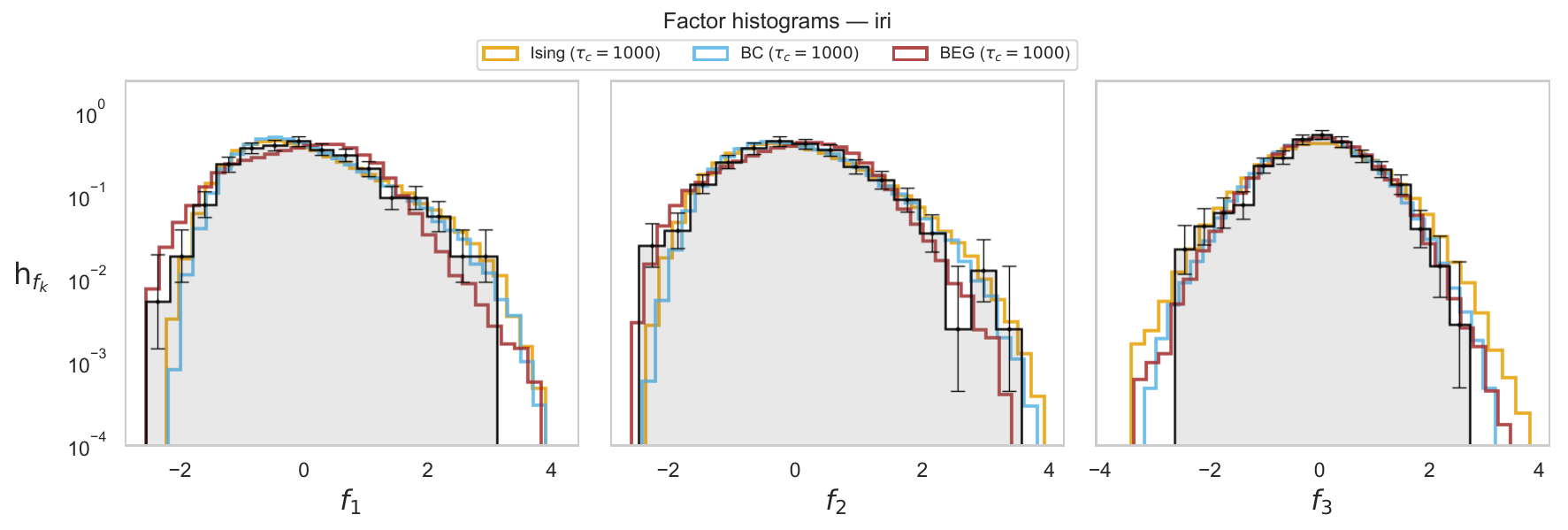}
\includegraphics[width=0.45\columnwidth]{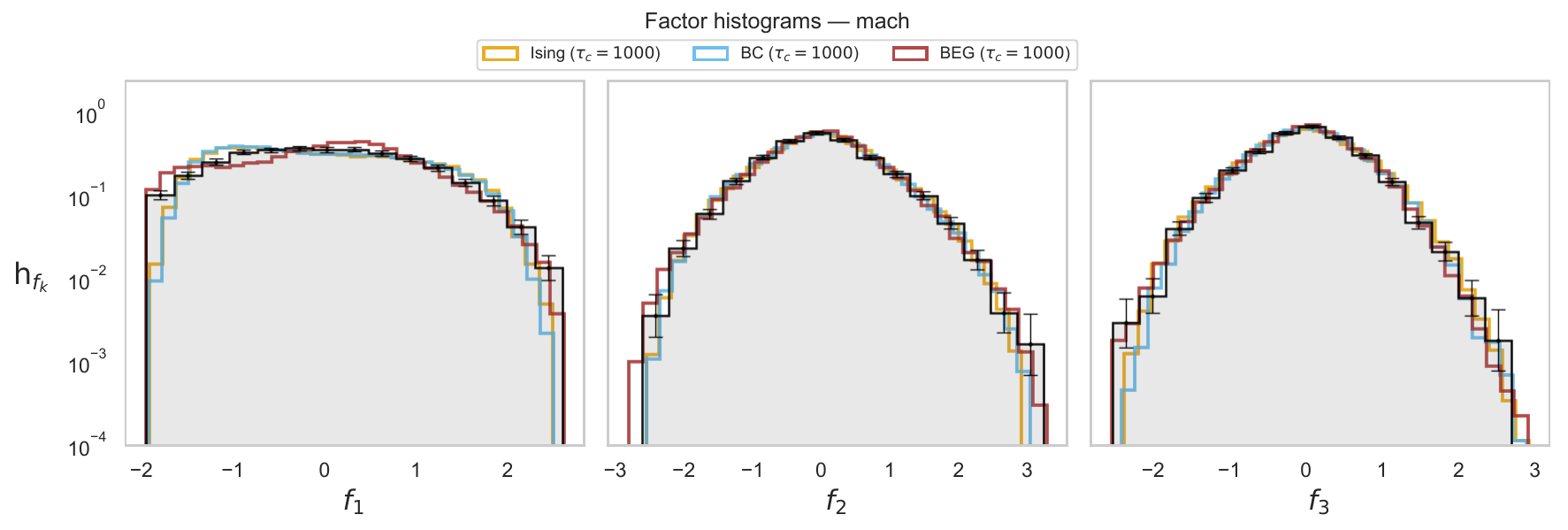}\\
\includegraphics[width=0.45\columnwidth]{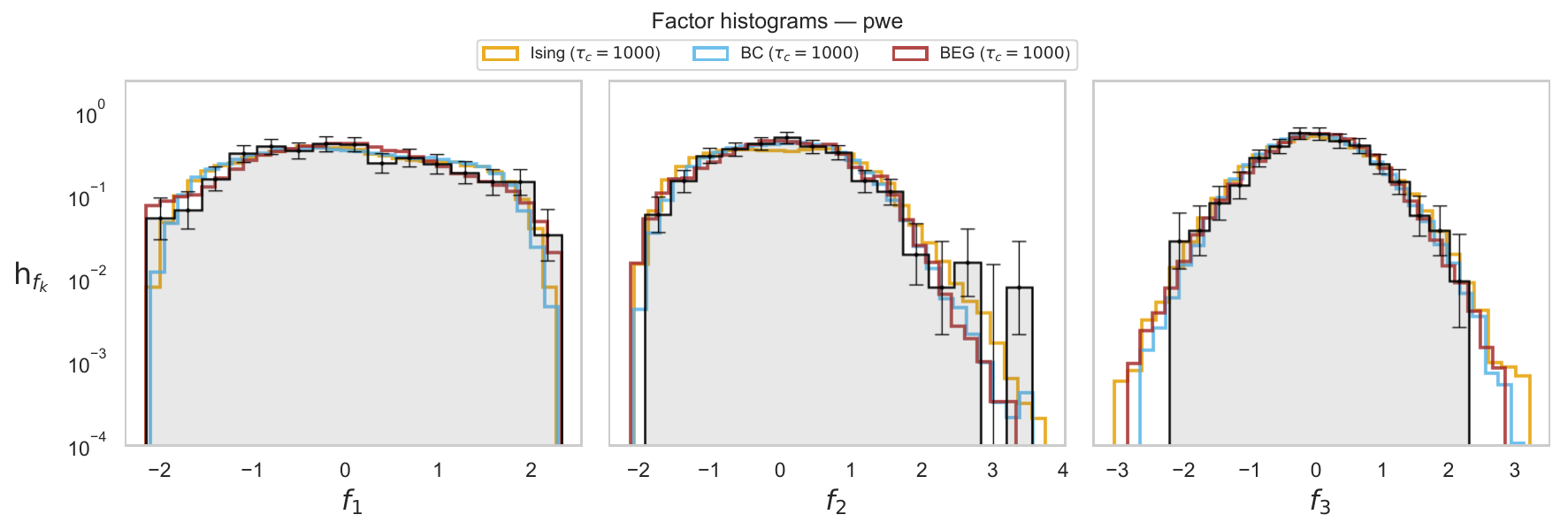} 
\includegraphics[width=0.45\columnwidth]{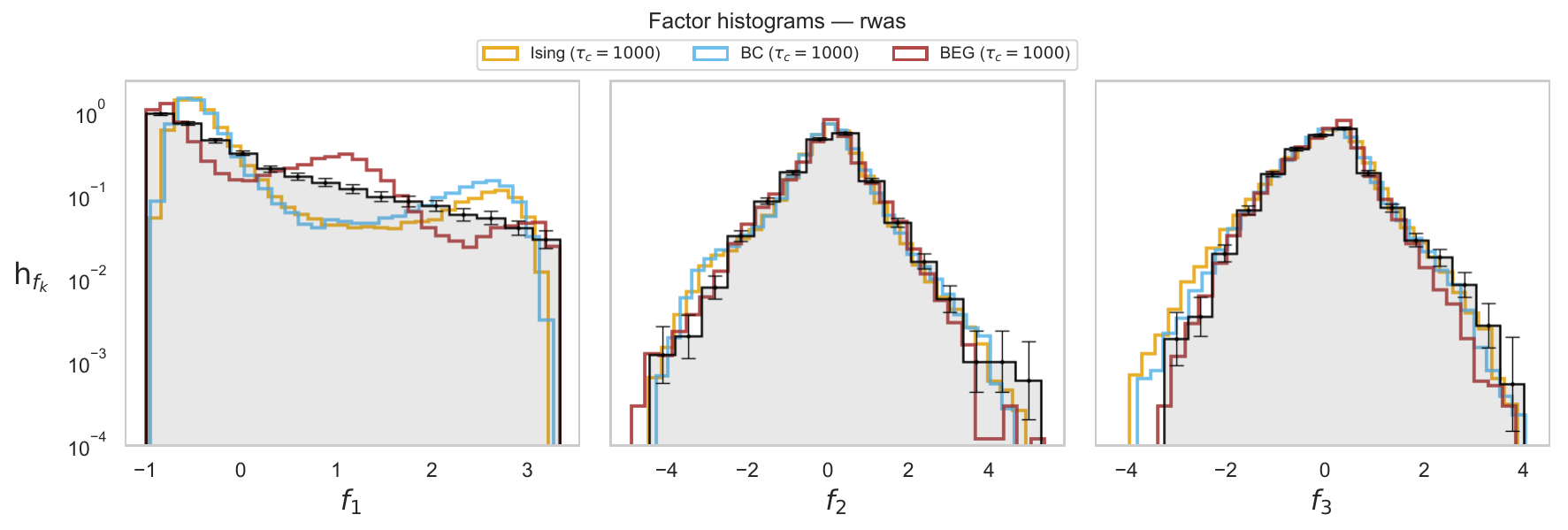}\\
\includegraphics[width=0.45\columnwidth]{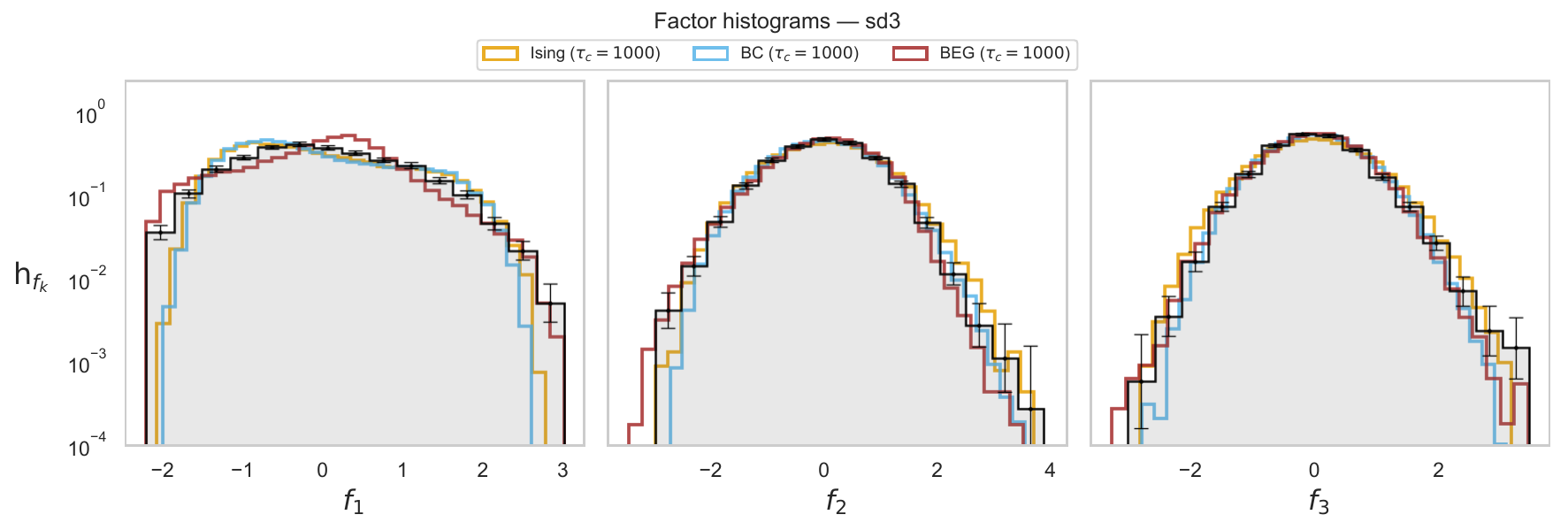}
\caption{Histograms of the first three factors ${\sf h}_{f_k}$ for all the analyzed questionnaires: comparison between empirical data and the three spin models (Ising, BC, BEG). Non-Gaussian and multi-modal structures are visible in several datasets. The most evident changes in the convexity occur for the datasets {\sf cfcs} ($f_1$), {\sf gcbs} ($f_1,f_2,f_3$), {\sf hsns} ($f_1$), {\sf pwe} ($f_2$), {\sf rwas} ($f_1,f_3$), {\sf sd3} ($f_3$). Error bars are Wilson score confidence intervals at $\alpha=0.05$.}
\label{fig:factorhist_all}
\end{center}
\end{figure}

\begin{figure}[H]
\begin{center}
\includegraphics[width=0.45\columnwidth]{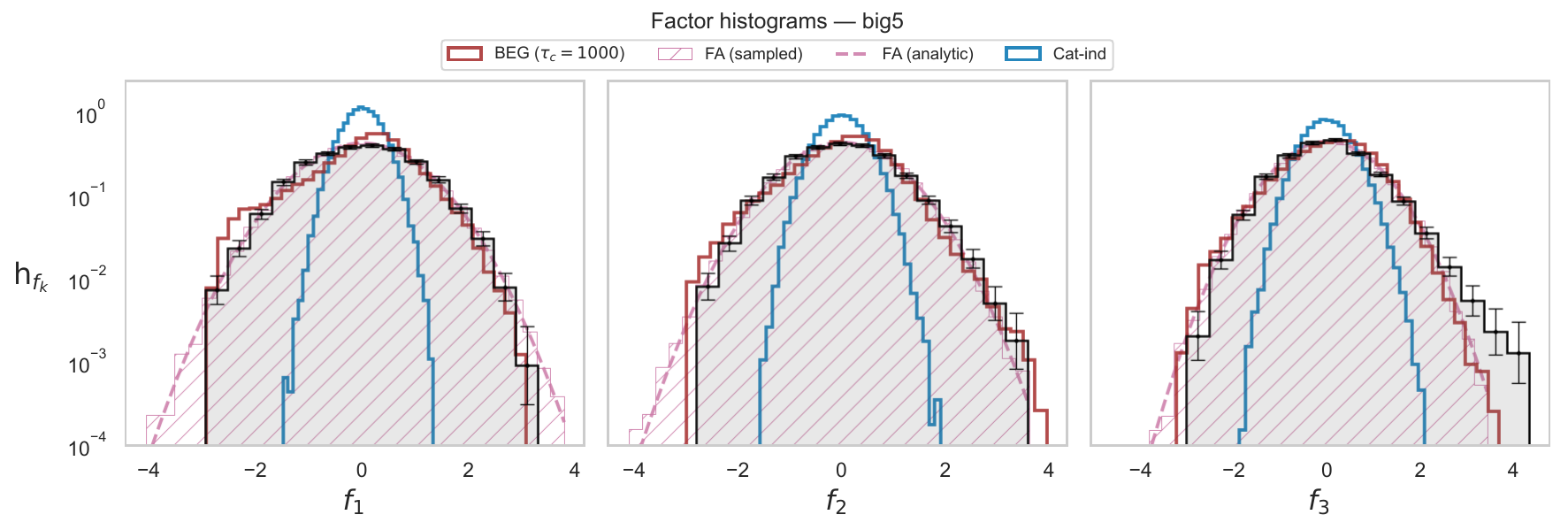}
\includegraphics[width=0.45\columnwidth]{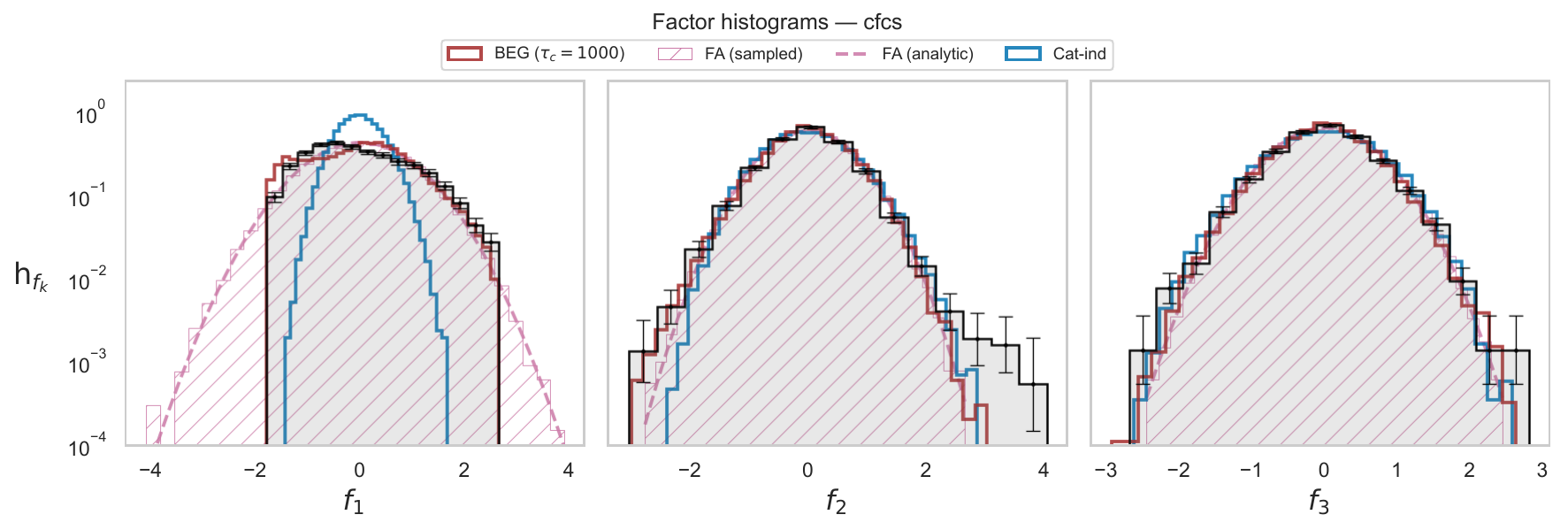}\\
\includegraphics[width=0.45\columnwidth]{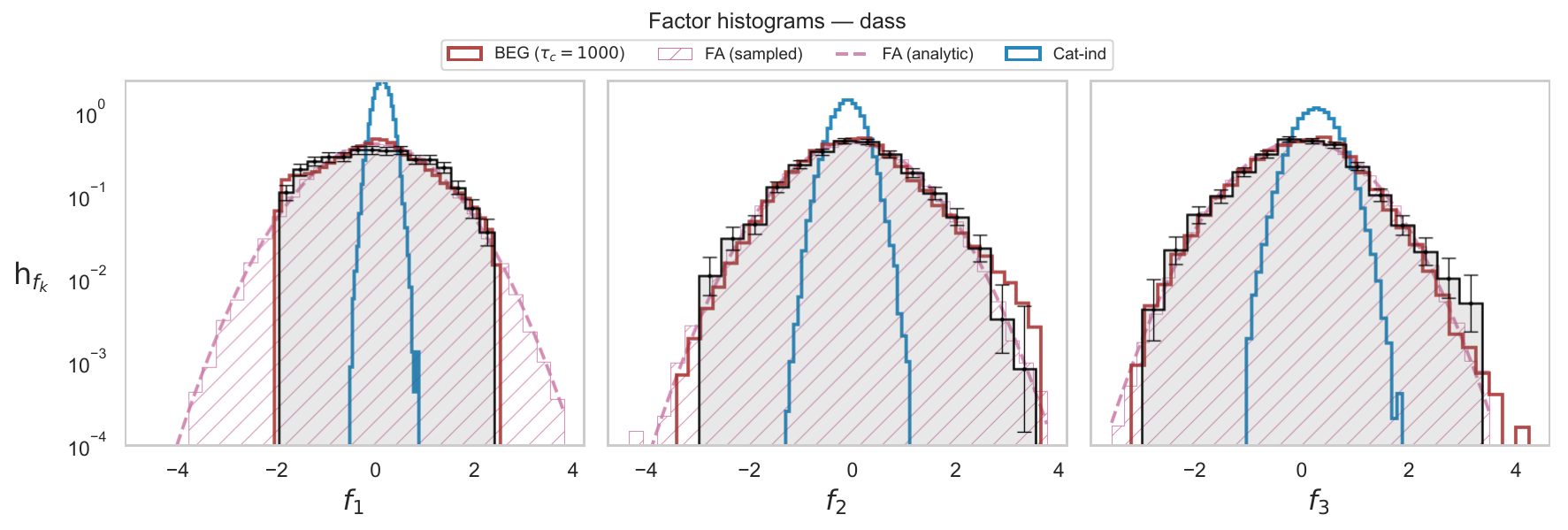} 
\includegraphics[width=0.45\columnwidth]{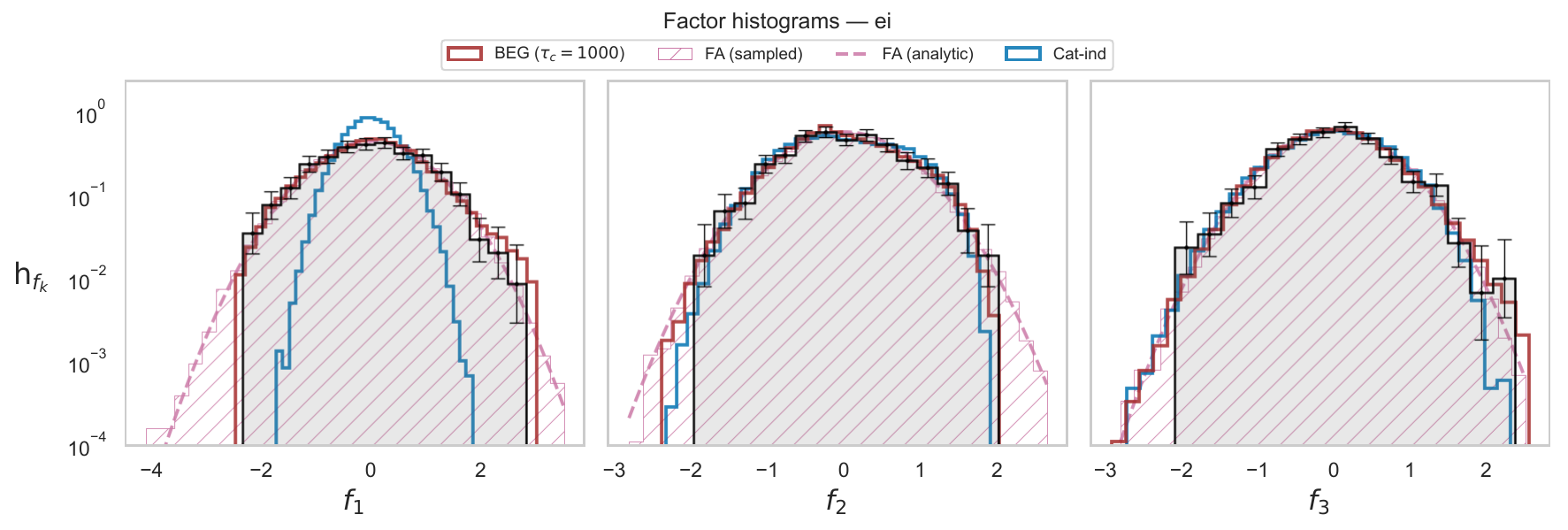}\\
\includegraphics[width=0.45\columnwidth]{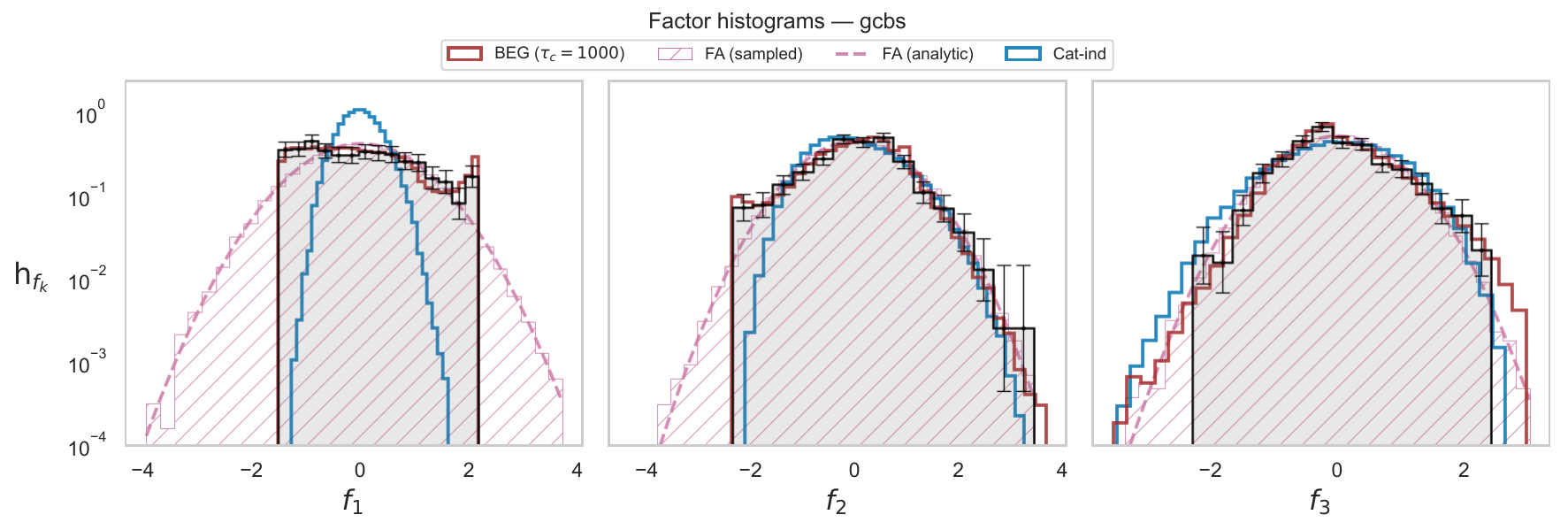}
\includegraphics[width=0.45\columnwidth]{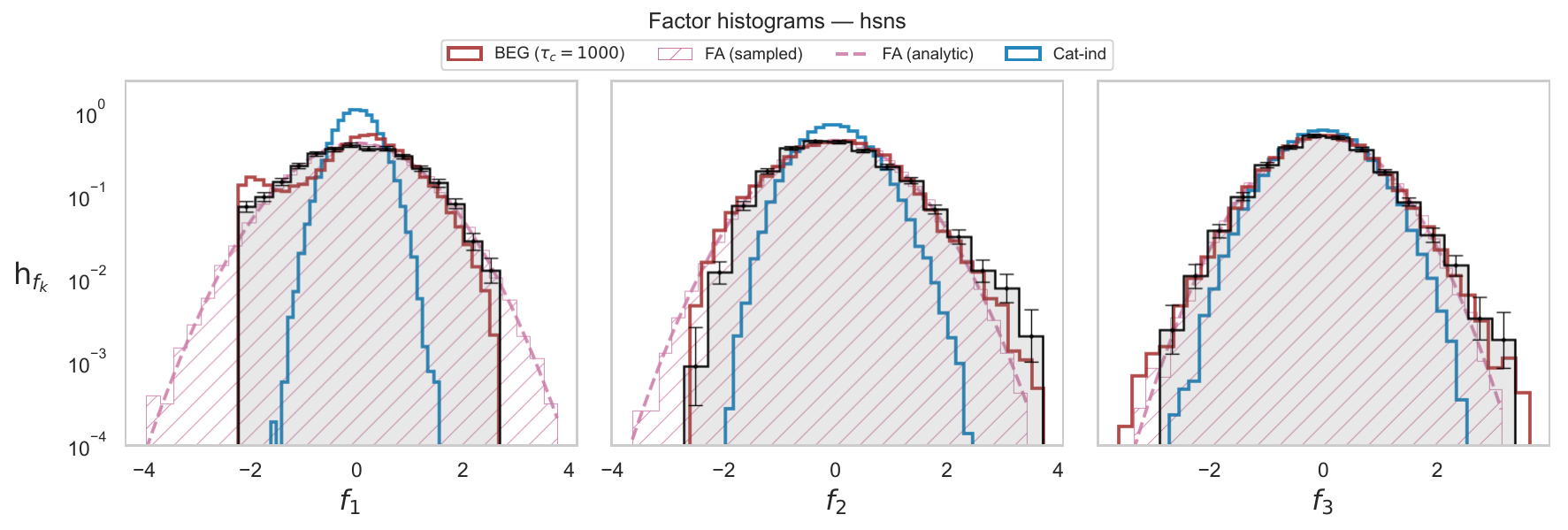} \\
\includegraphics[width=0.45\columnwidth]{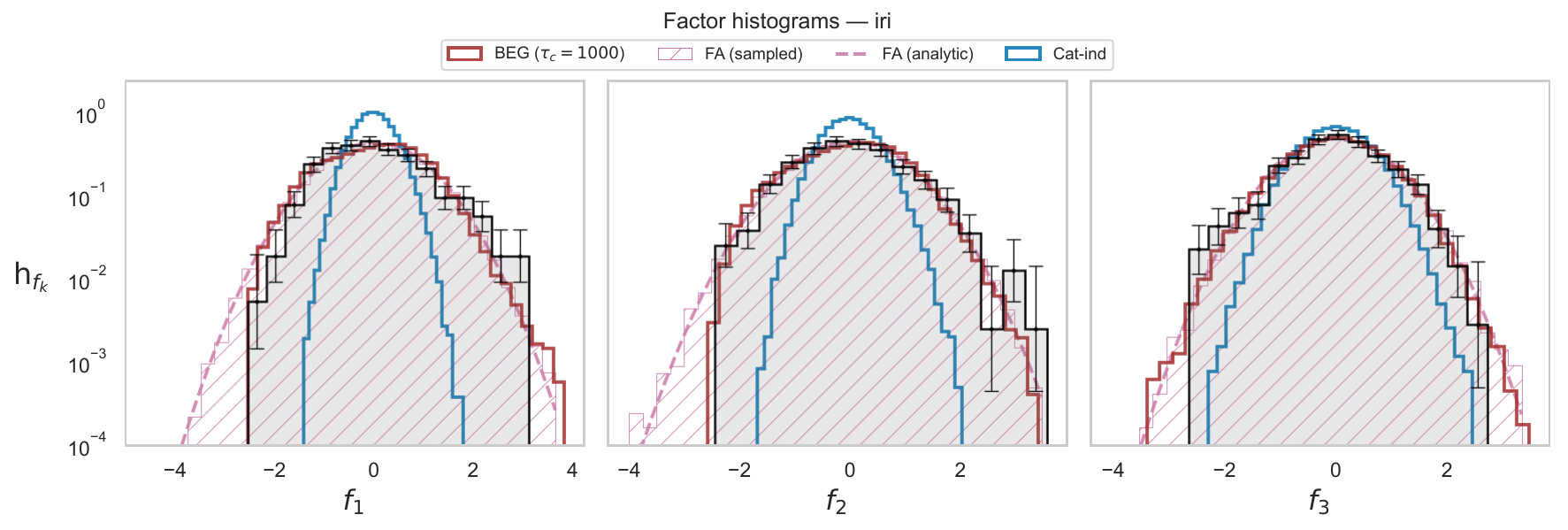}
\includegraphics[width=0.45\columnwidth]{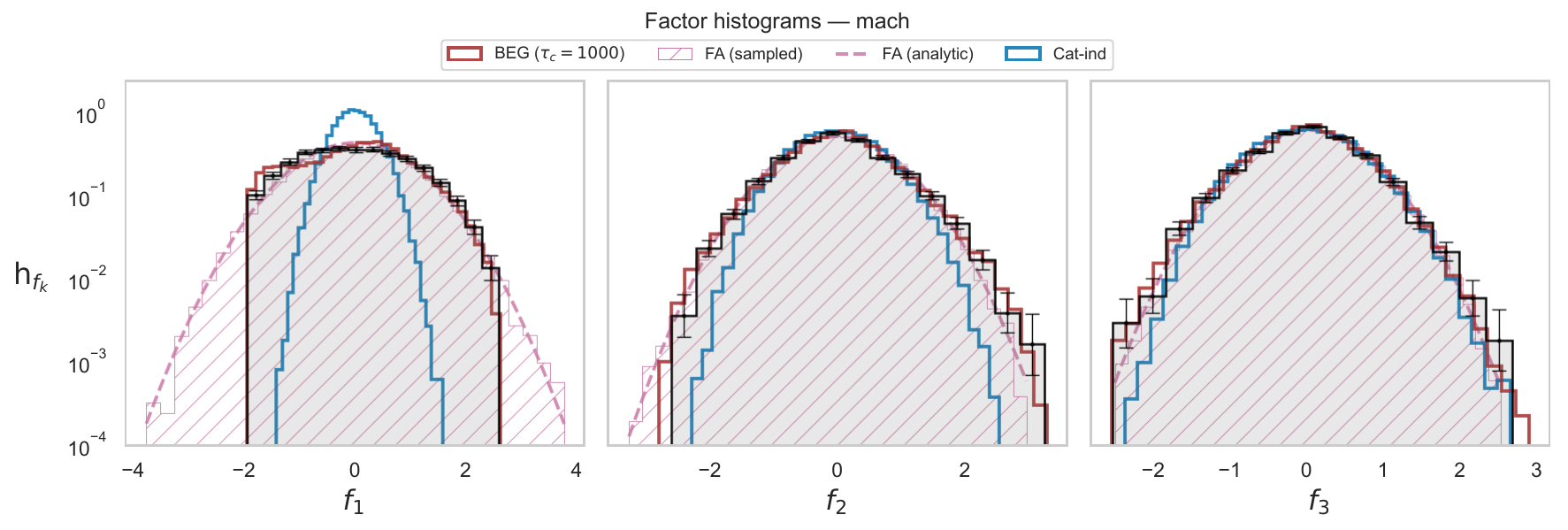}\\
\includegraphics[width=0.45\columnwidth]{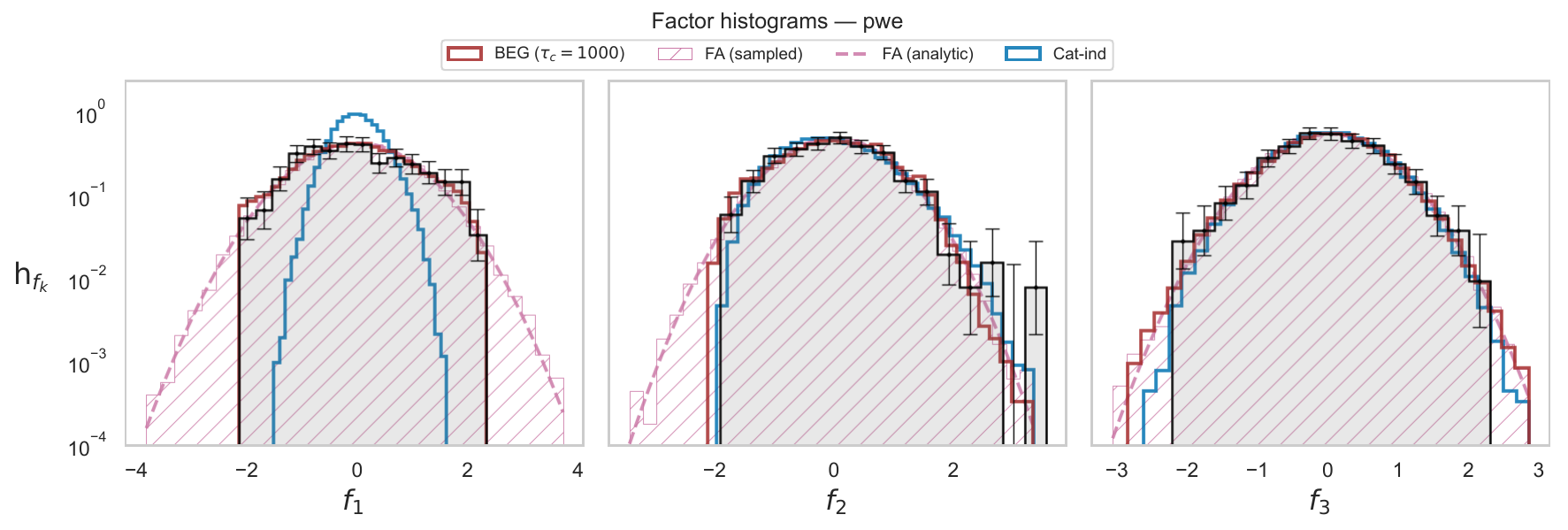} 
\includegraphics[width=0.45\columnwidth]{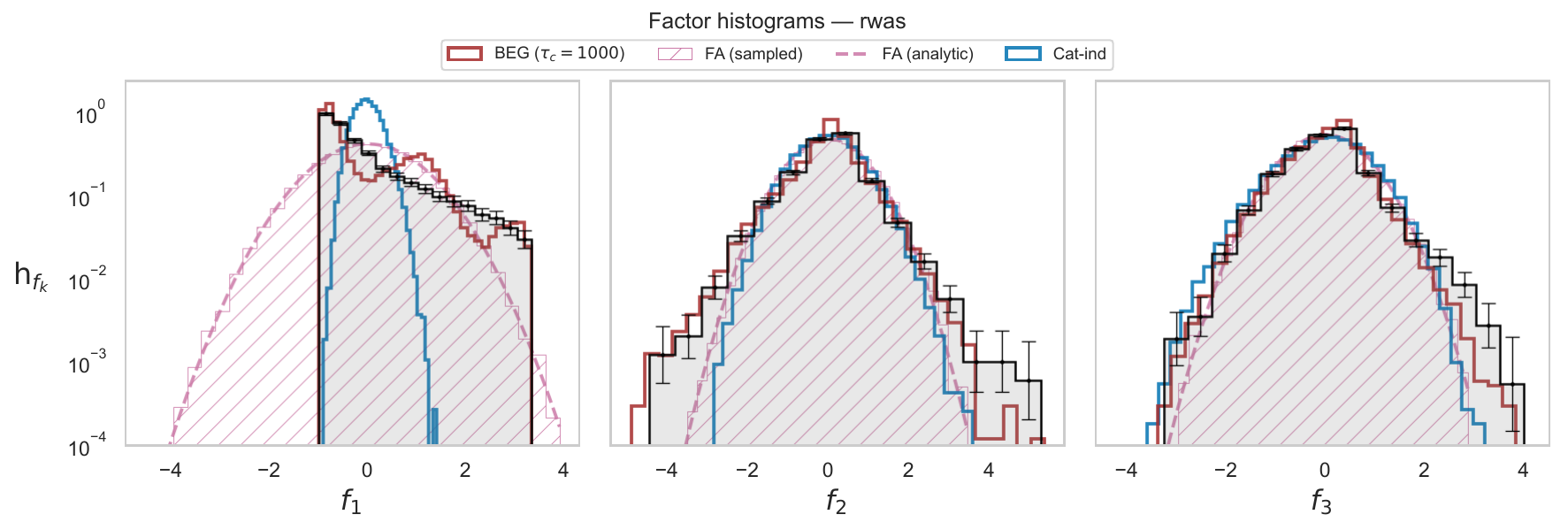}\\
\includegraphics[width=0.45\columnwidth]{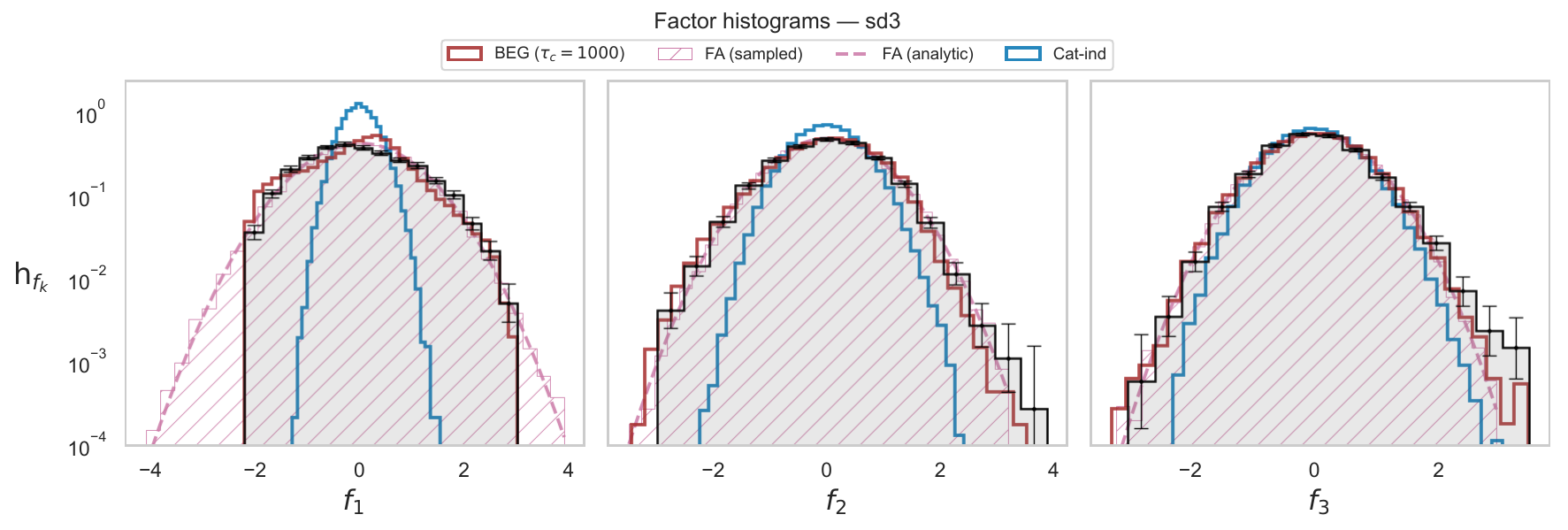}
\caption{Histograms of the first three factors ${\sf h}_{f_k}$ for all the analyzed questionnaires: comparison between empirical data and the simple models ({\sf gauss}, {\sf cat-ind}). Error bars are Wilson score confidence intervals at $\alpha=0.05$.}
\label{fig:factorhist_simple_all}
\end{center}
\end{figure}

\begin{figure}[H]
\begin{center}
\includegraphics[width=0.45\columnwidth]{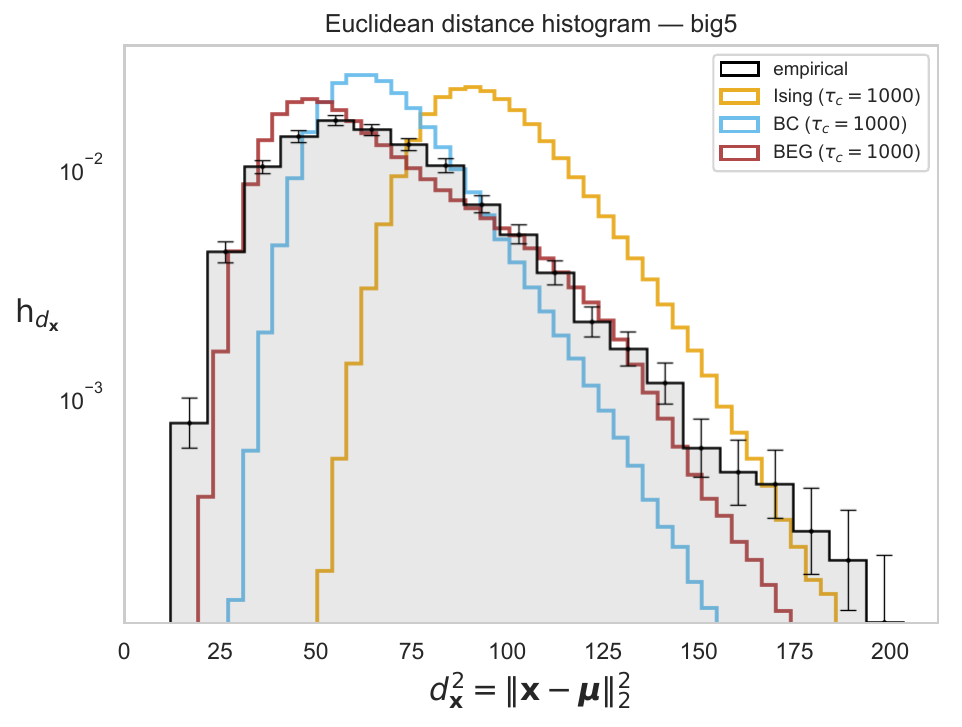}
\includegraphics[width=0.45\columnwidth]{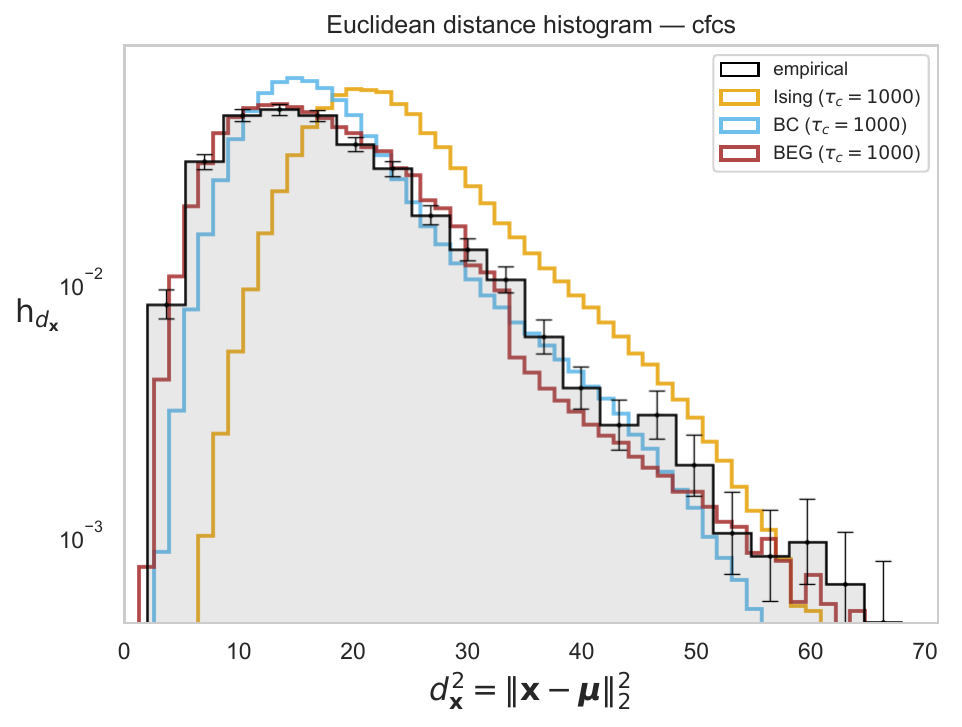}\\
\includegraphics[width=0.45\columnwidth]{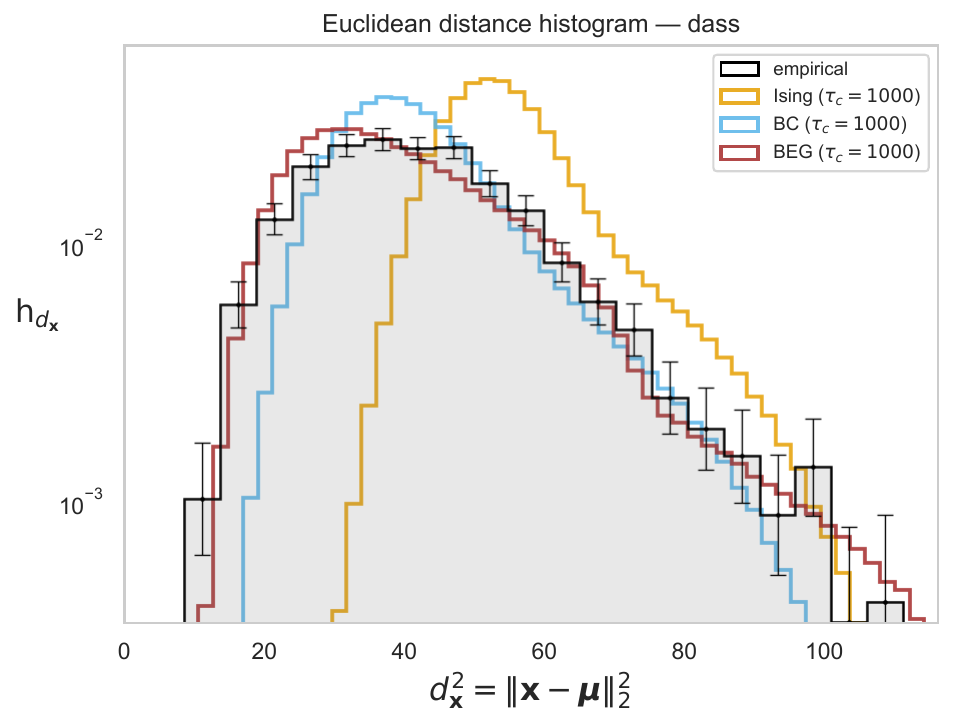} 
\includegraphics[width=0.45\columnwidth]{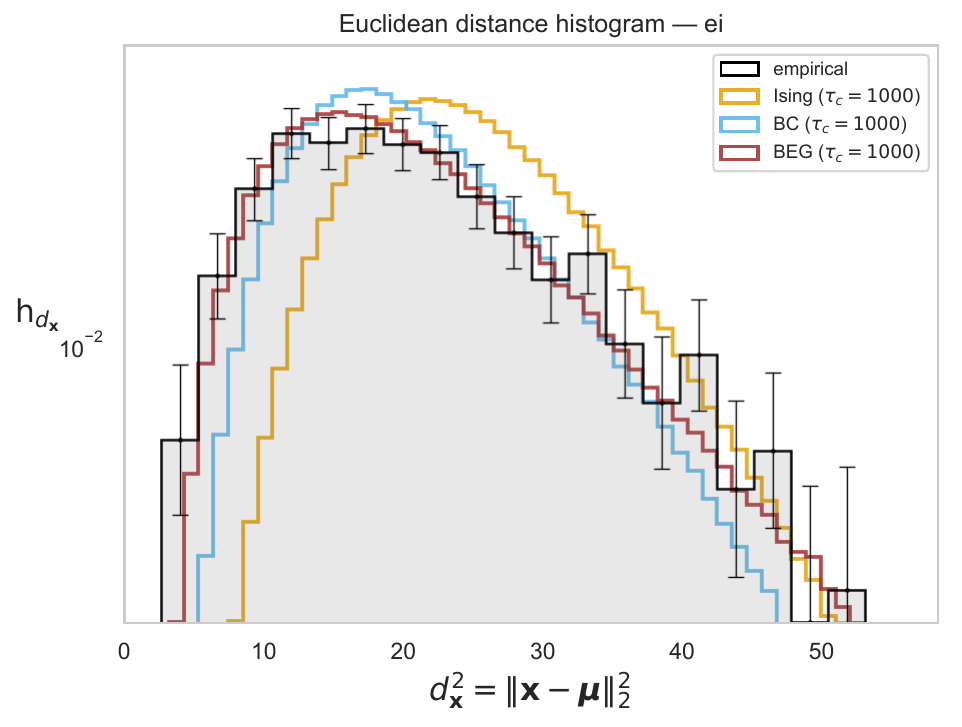}\\
\includegraphics[width=0.45\columnwidth]{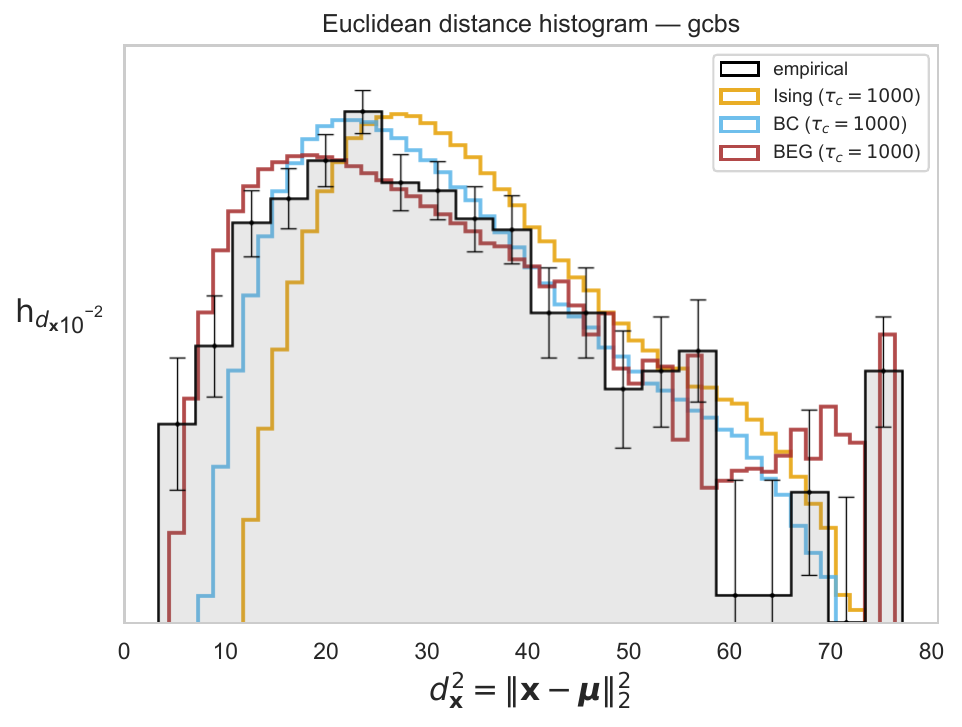}
\includegraphics[width=0.45\columnwidth]{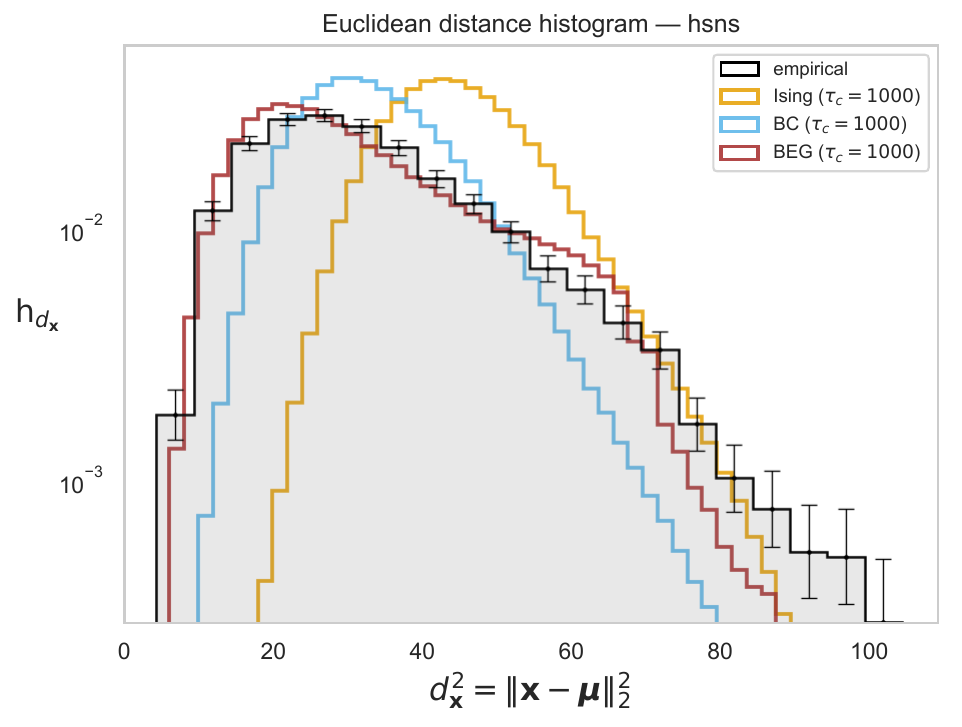} 
\caption{Histograms of the Euclidean distance to the mean, ${\sf h}_{d_\x}$, for all the analyzed questionnaires: comparison between empirical data and the three spin models (Ising, BC, BEG). Equivalent of left panel in Fig.~\ref{fig:E2dhist_gcbs} for all datasets.}
\label{fig:E2dhist_all}
\end{center}
\end{figure}

\begin{figure}[H]
\begin{center}
\includegraphics[width=0.45\columnwidth]{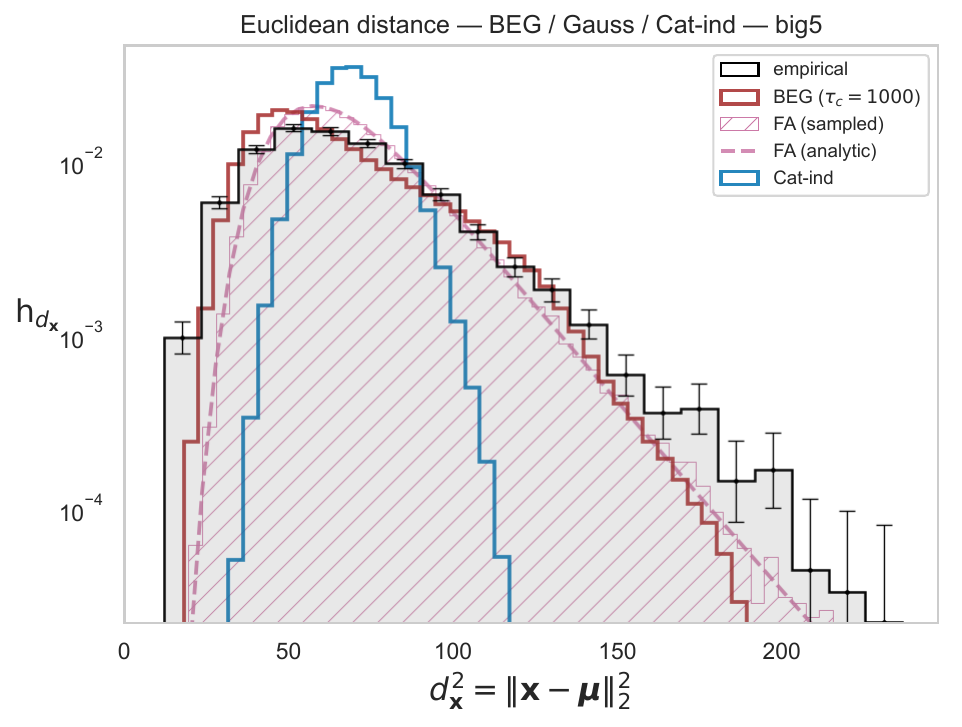}
\includegraphics[width=0.45\columnwidth]{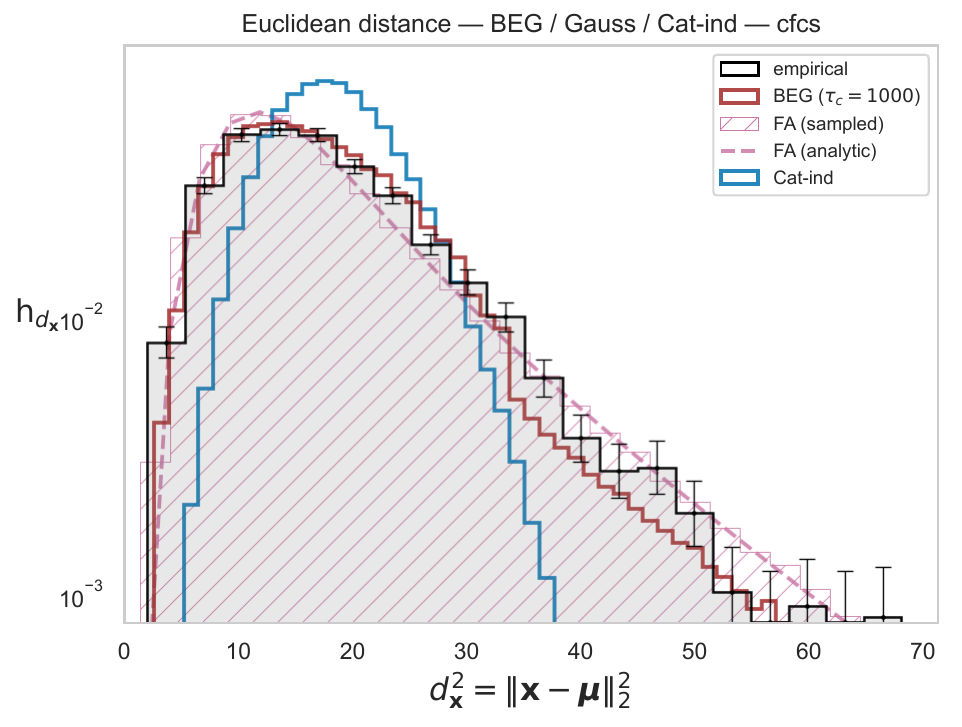}\\
\includegraphics[width=0.45\columnwidth]{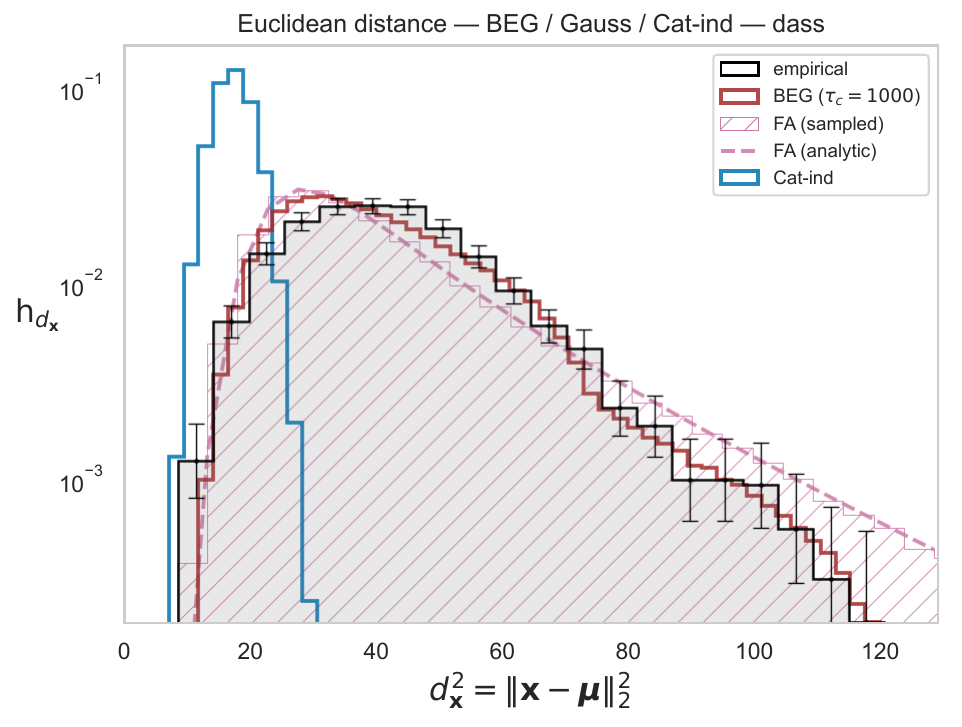}
\includegraphics[width=0.45\columnwidth]{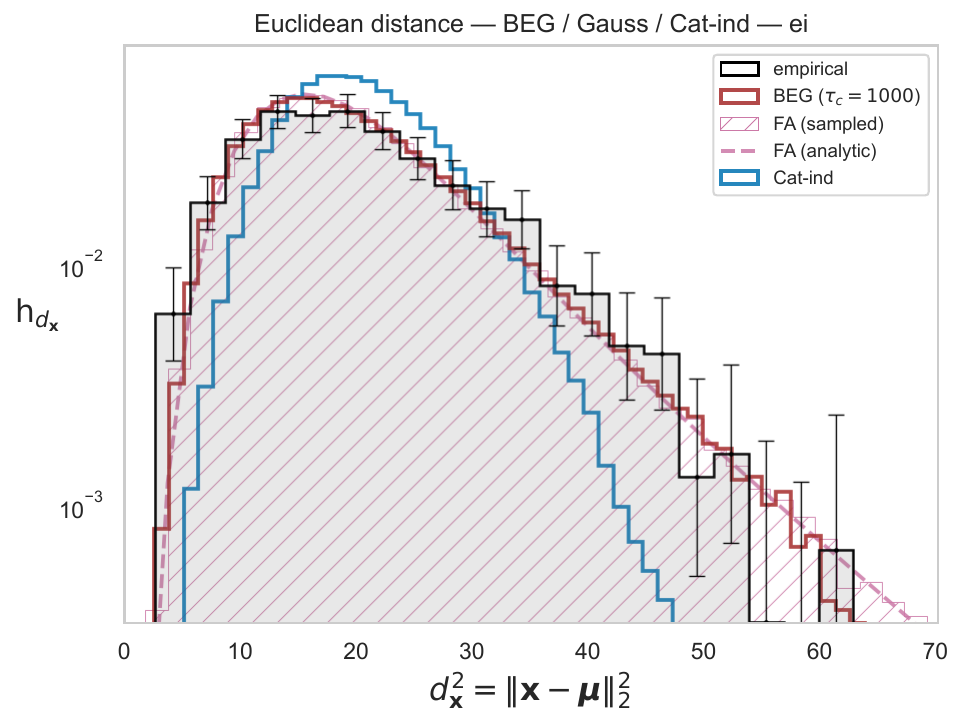}\\
\includegraphics[width=0.45\columnwidth]{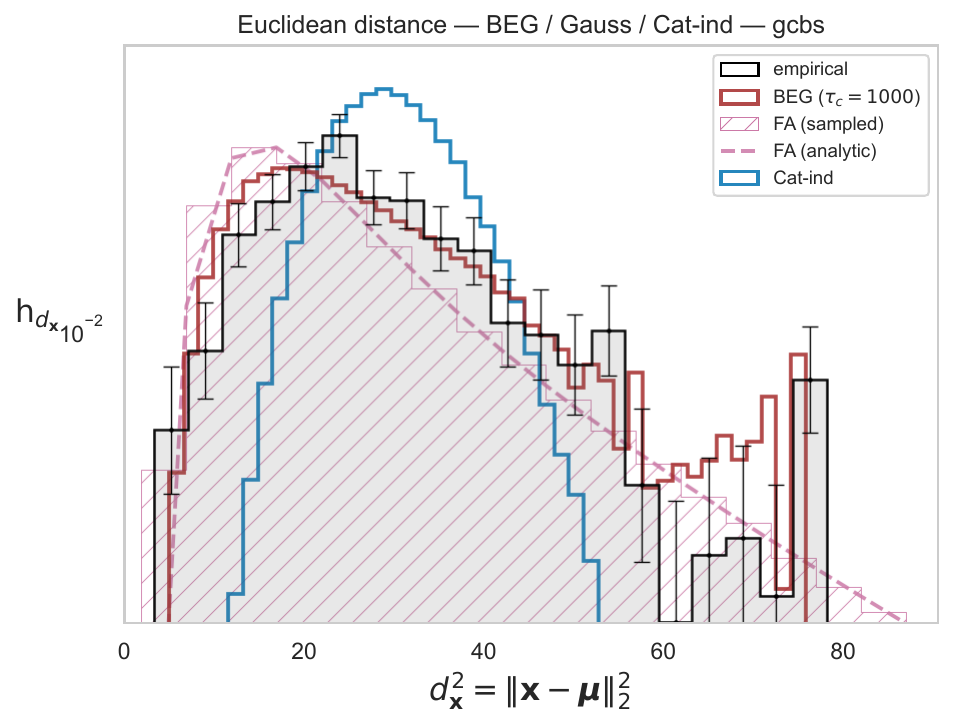}
\includegraphics[width=0.45\columnwidth]{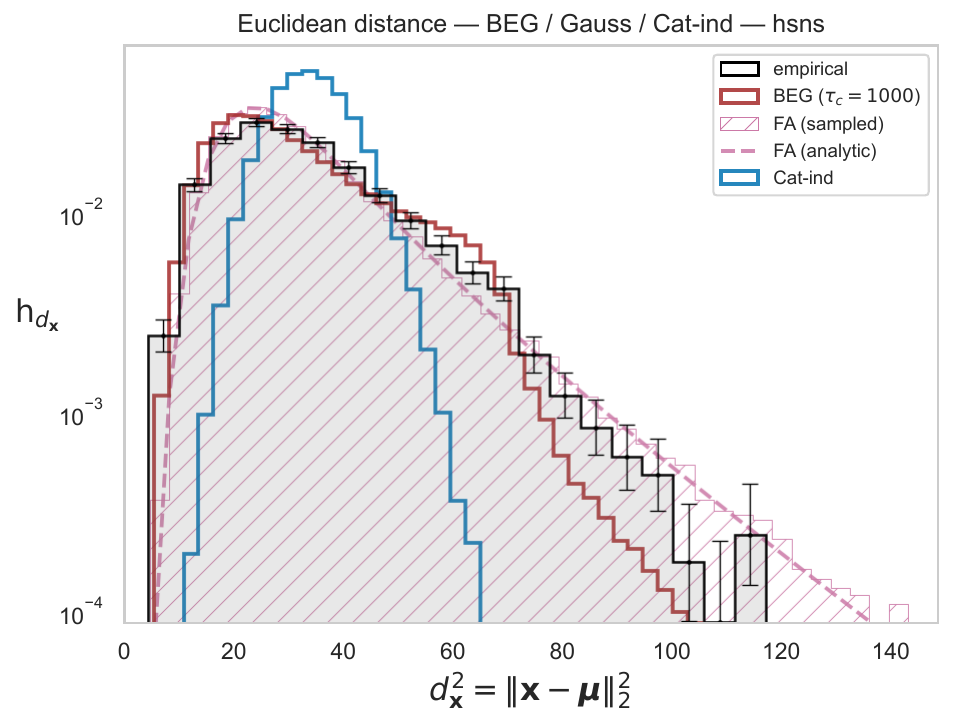}
\caption{Histograms of the Euclidean distance to the mean, ${\sf h}_{d_\x}$, for all the analyzed questionnaires: comparison between empirical data and the simple models ({\sf gauss}, {\sf cat-ind}). Equivalent of right panel in Fig.~\ref{fig:E2dhist_gcbs} for all datasets.Error bars are Wilson score confidence intervals at $\alpha=0.05$.}
\label{fig:E2dhist_simple_all}
\end{center}
\end{figure}

\begin{figure}[H]
\begin{center}
\includegraphics[width=0.45\columnwidth]{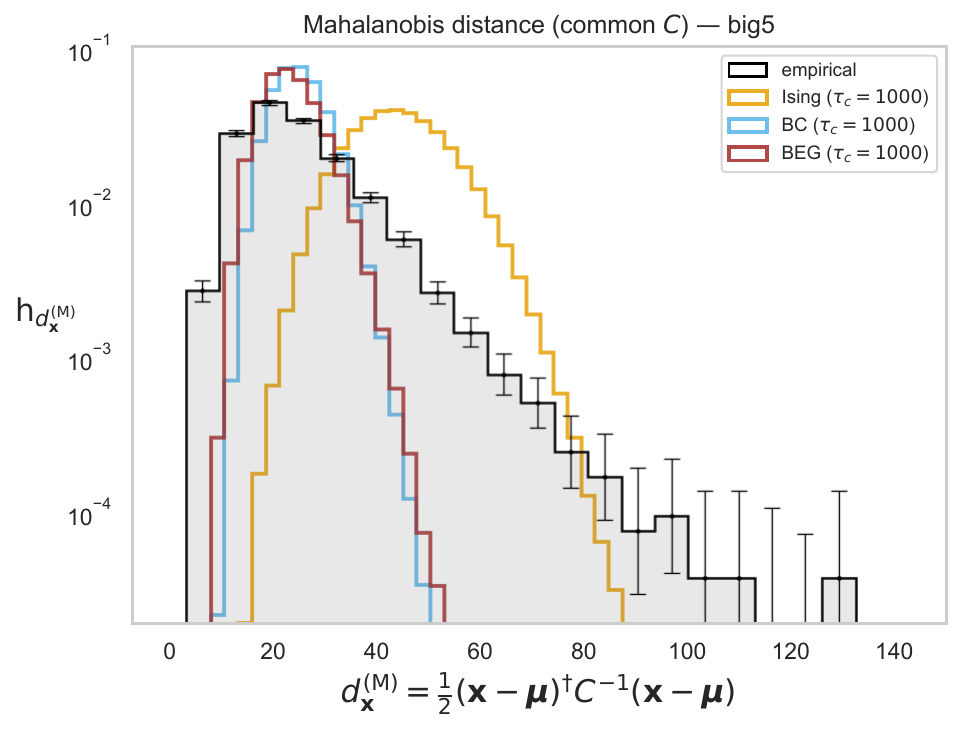}
\includegraphics[width=0.45\columnwidth]{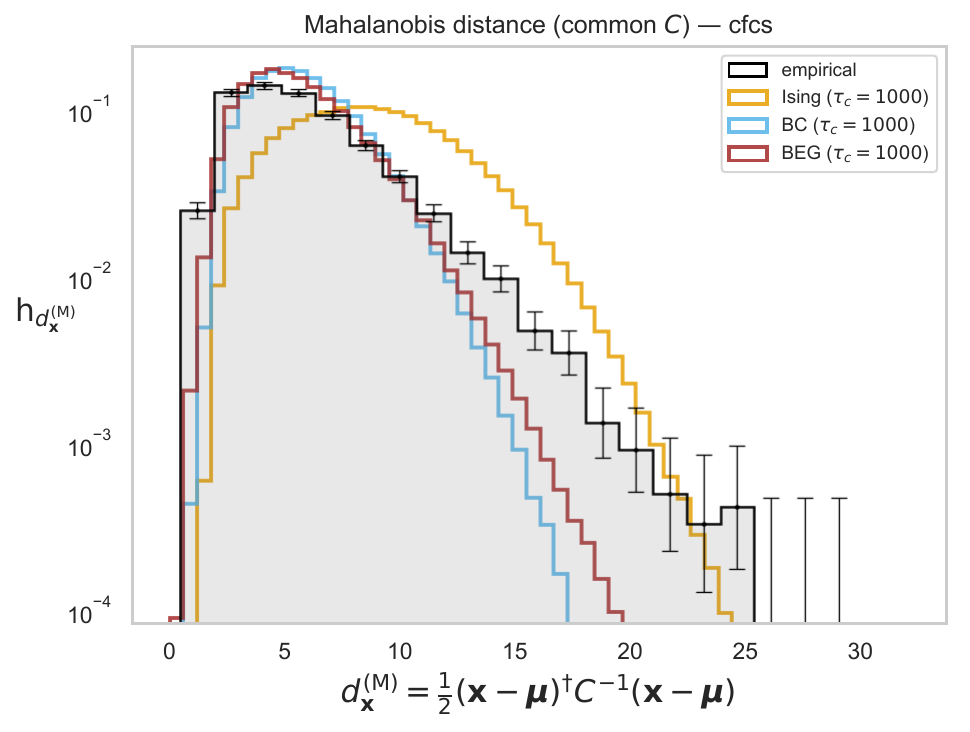}
\includegraphics[width=0.45\columnwidth]{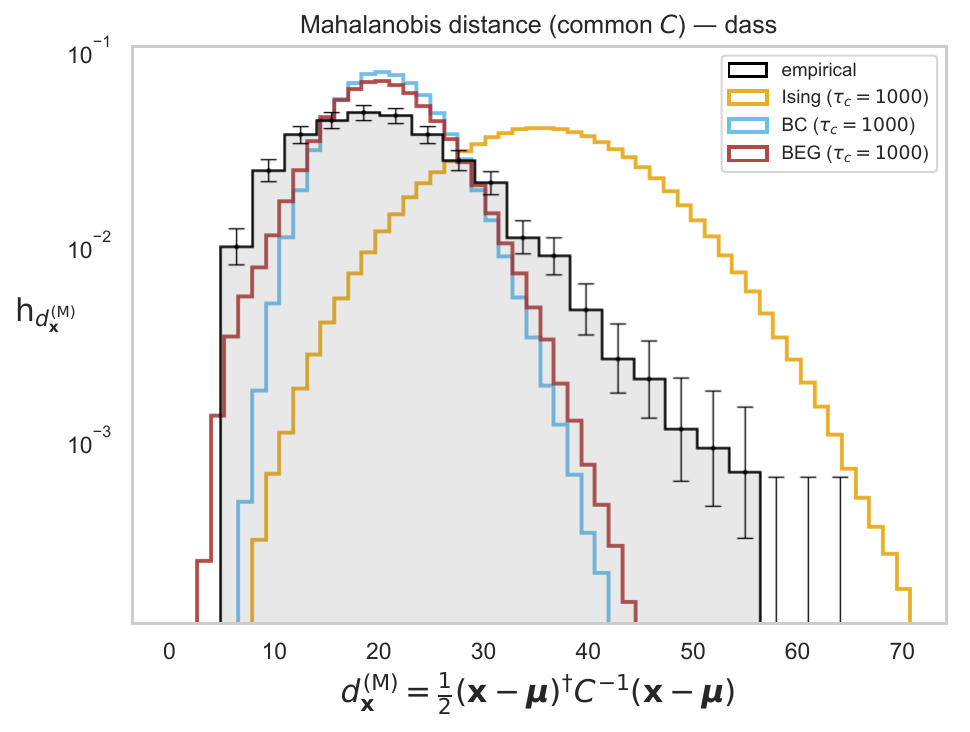}
\includegraphics[width=0.45\columnwidth]{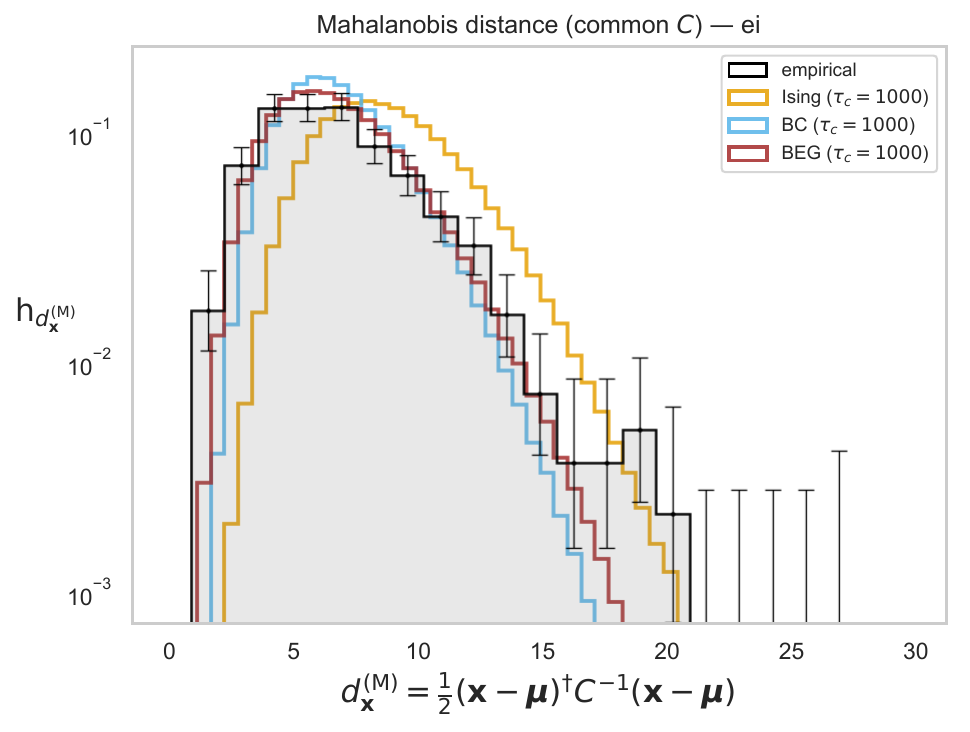}
\includegraphics[width=0.45\columnwidth]{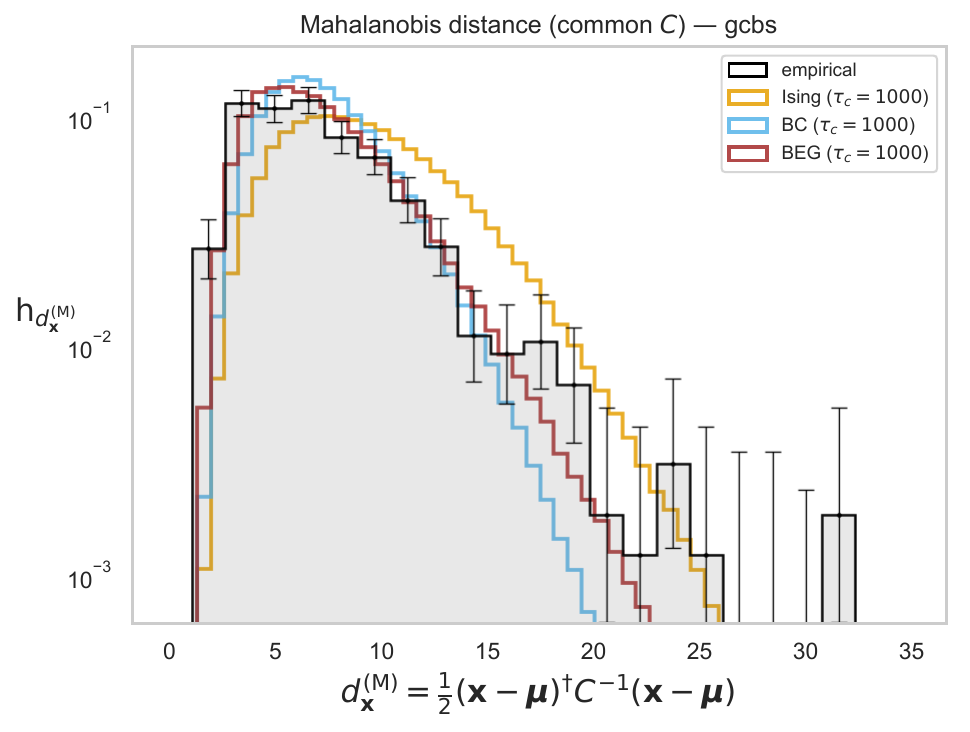}
\includegraphics[width=0.45\columnwidth]{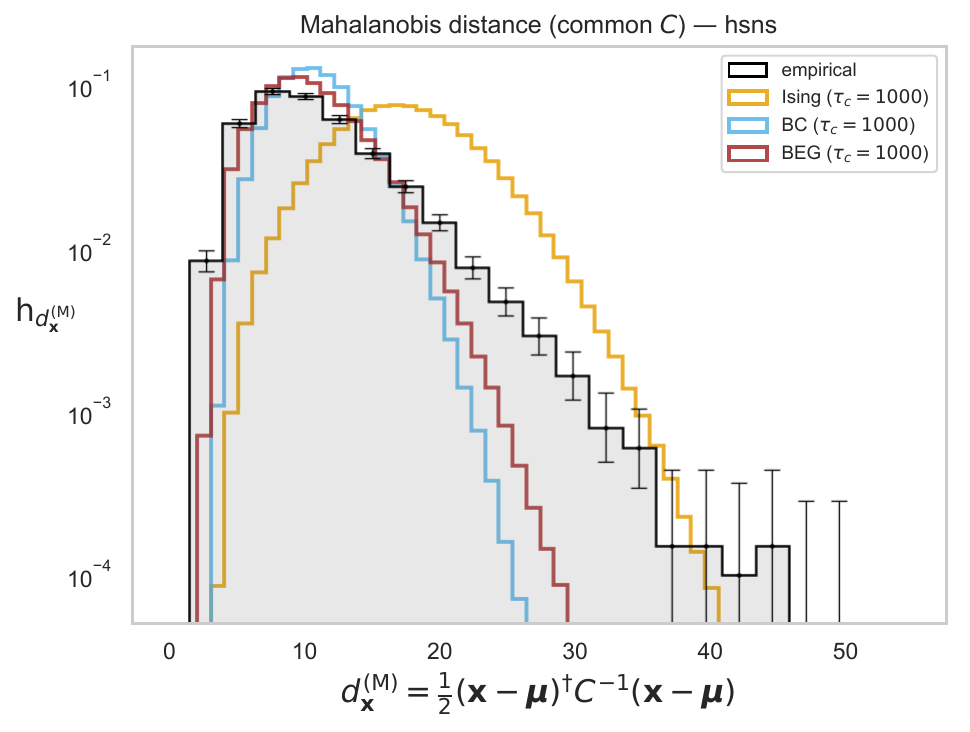}

\caption{Histograms of the Mahalanobis distance to the mean, ${\sf h}_{d_\x^{({\rm M})}}$, for all the analyzed questionnaires: comparison between empirical data and the three spin models (Ising, BC, BEG). The empirical covariance matrix $C$ is used in Eq.~(\ref{eq:maha}) for all models. All models systematically underestimate the tails. }
\label{fig:energy_all}
\end{center}
\end{figure}

\begin{figure}[H]
\begin{center}
\includegraphics[width=0.45\columnwidth]{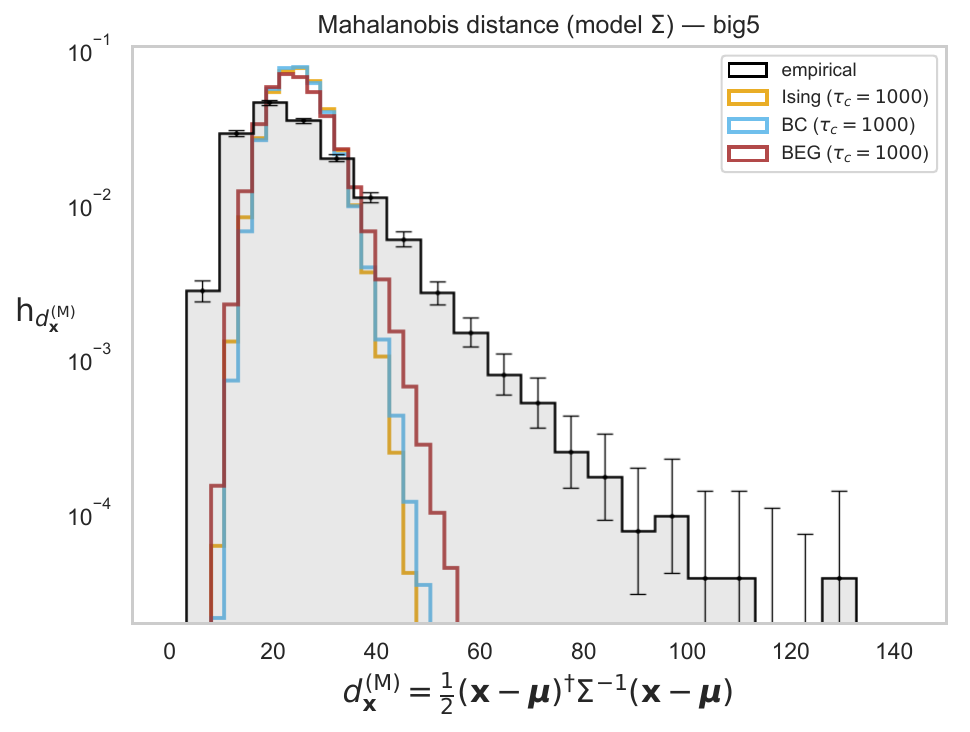}
\includegraphics[width=0.45\columnwidth]{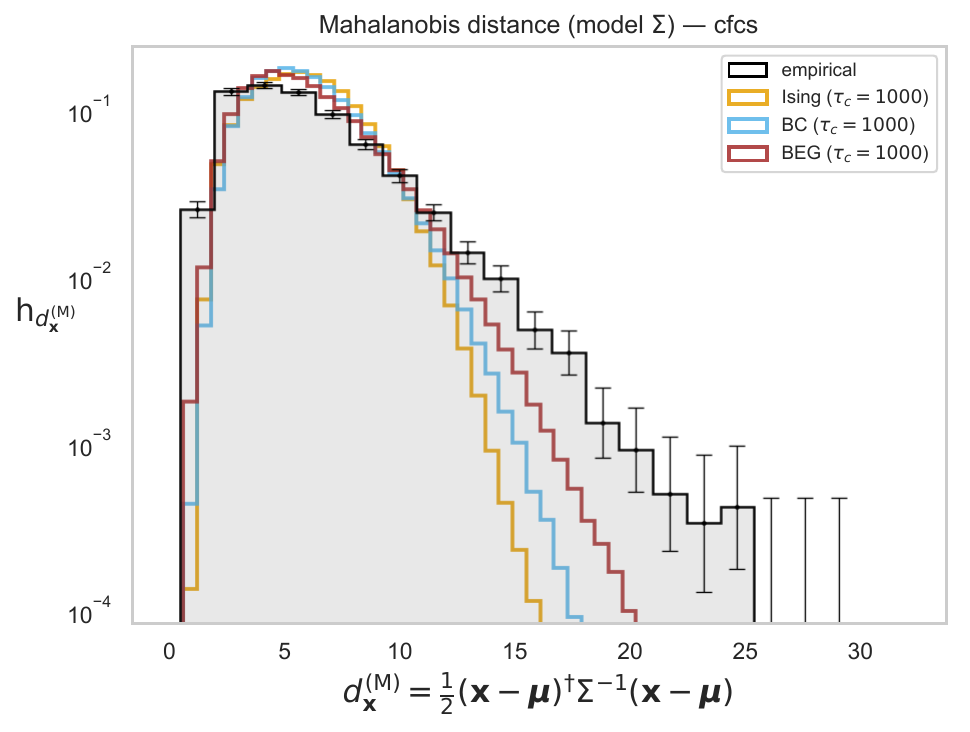}
\includegraphics[width=0.45\columnwidth]{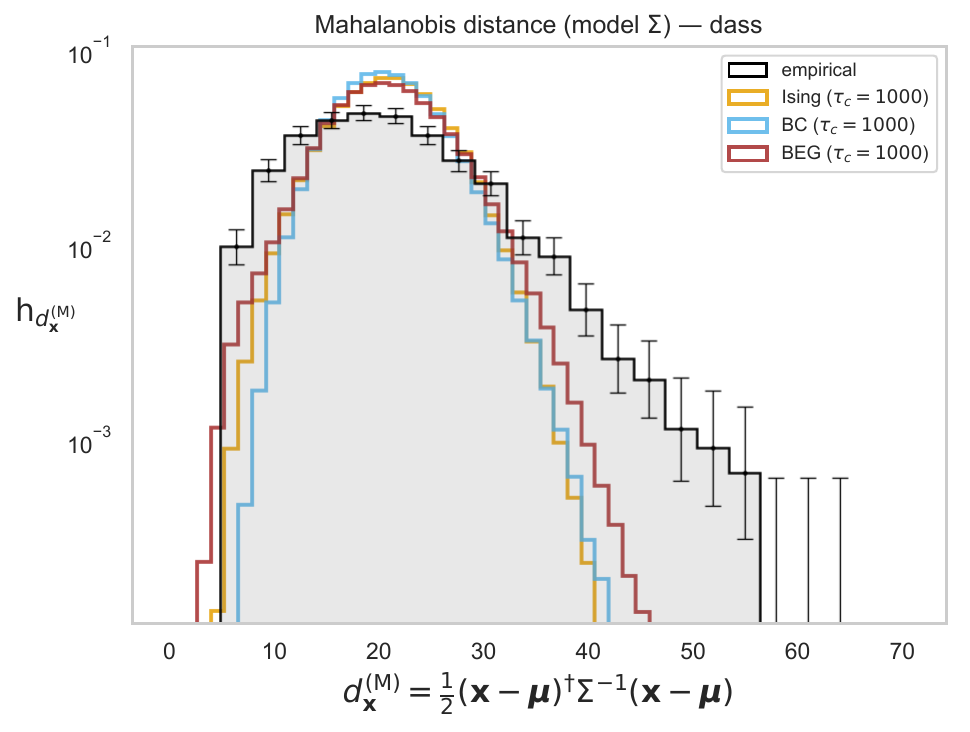}
\includegraphics[width=0.45\columnwidth]{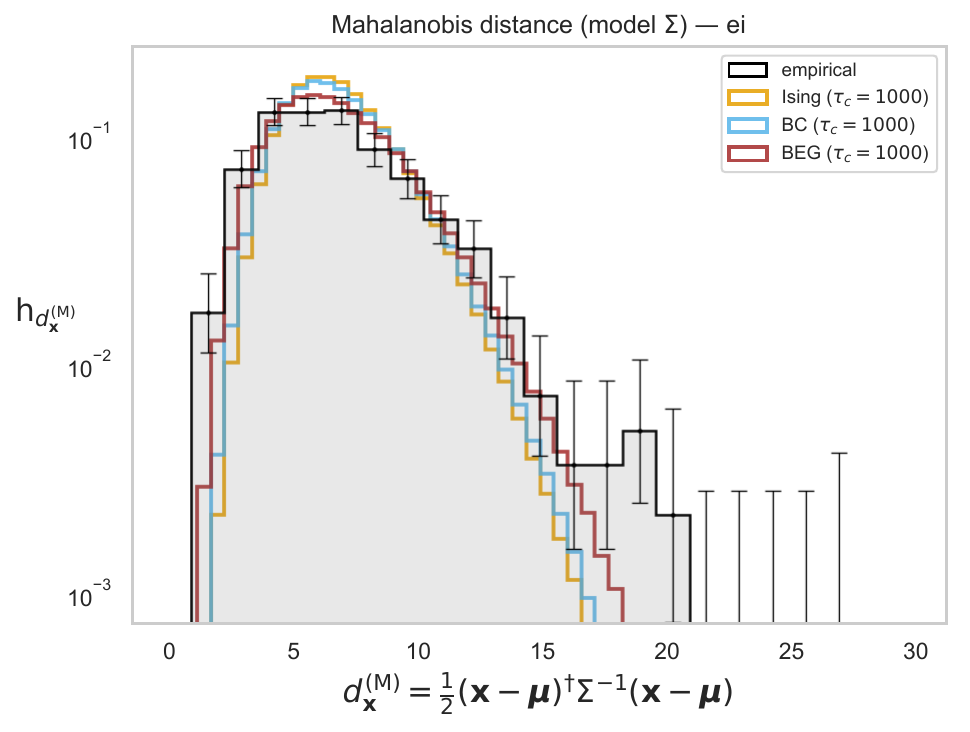}
\includegraphics[width=0.45\columnwidth]{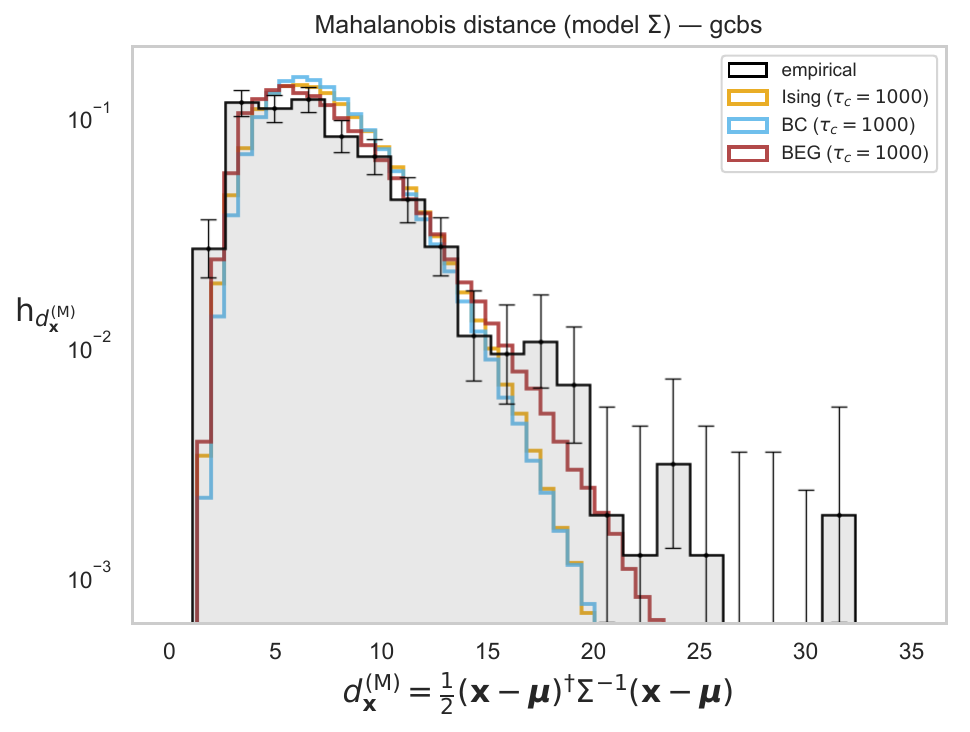}
\includegraphics[width=0.45\columnwidth]{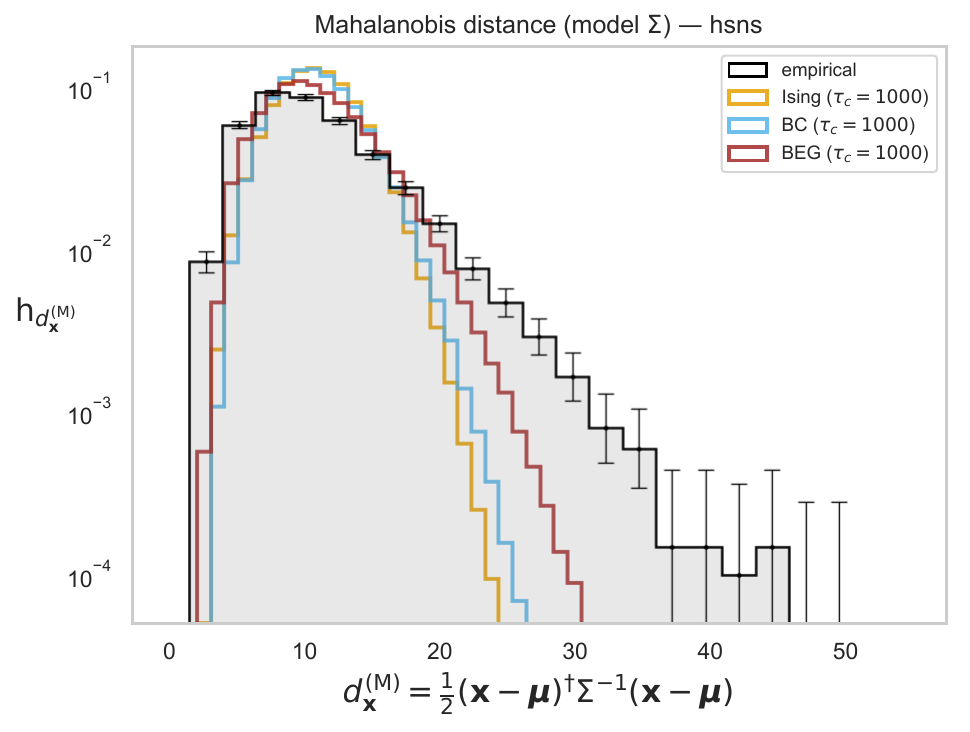}
\caption{Variant of Fig.~\ref{fig:energy_all}: histograms of the Mahalanobis distance $d_\x^{({\rm M})}=(1/2)\,\x^\dag\cdot\Sigma^{-1}\cdot\x$, where $\Sigma=\langle\x\x^\dag\rangle_P - \langle\x\rangle_P\langle\x^\dag\rangle_P$ is the model-dependent theoretical covariance matrix (instead of the empirical $C$). The discrepancy between empirical and theoretical histograms persists when model-dependent $\Sigma$ is used, ruling out finite-loss errors in the estimation of $C$ as its explanation. }
\label{fig:energyvar_all}
\end{center}
\end{figure}

\begin{figure}[H]
\begin{center}
\includegraphics[width=0.45\columnwidth]{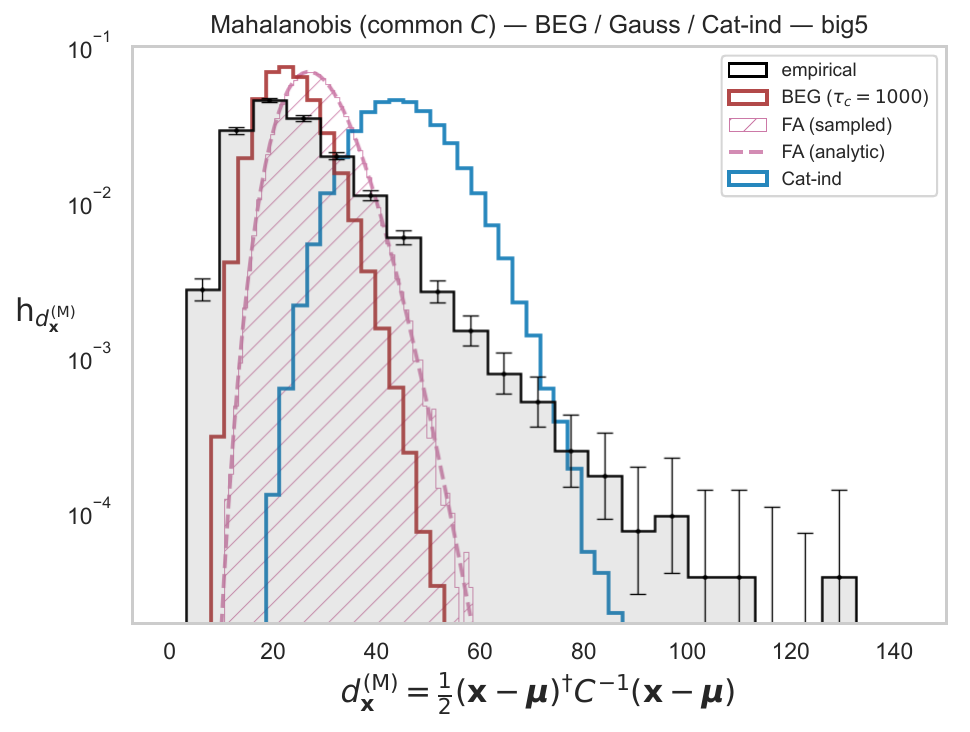}
\includegraphics[width=0.45\columnwidth]{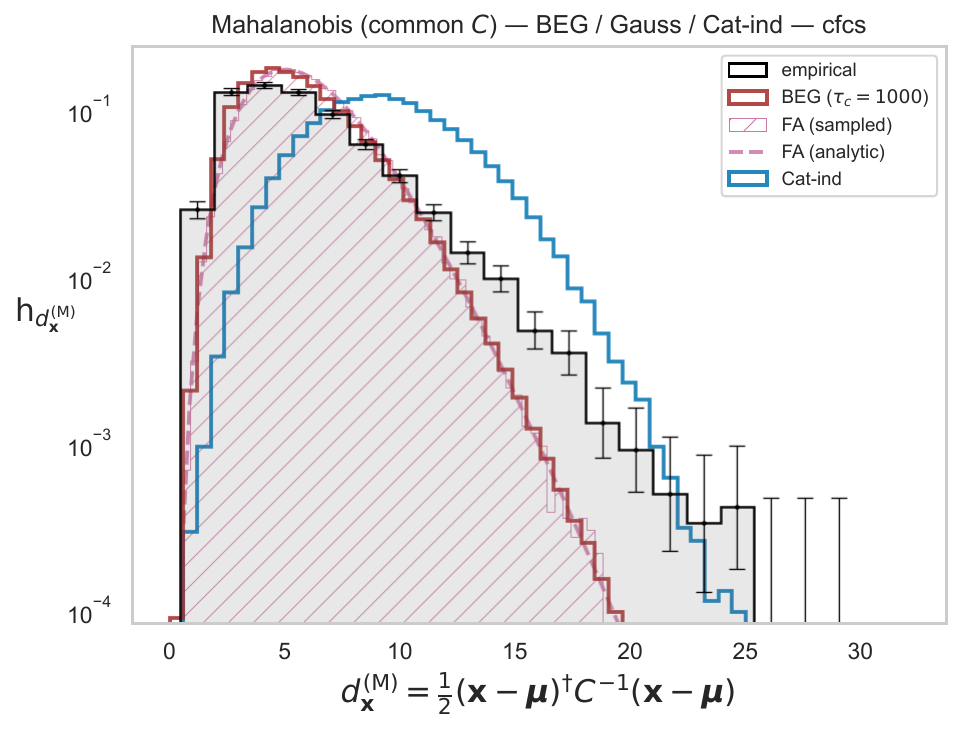}
\includegraphics[width=0.45\columnwidth]{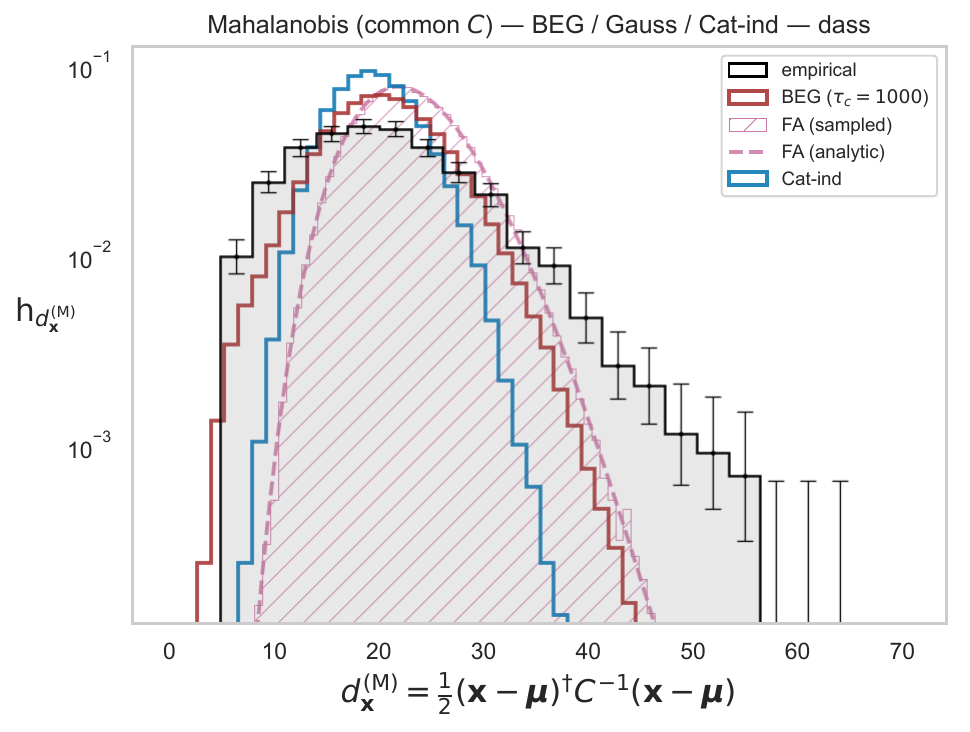}
\includegraphics[width=0.45\columnwidth]{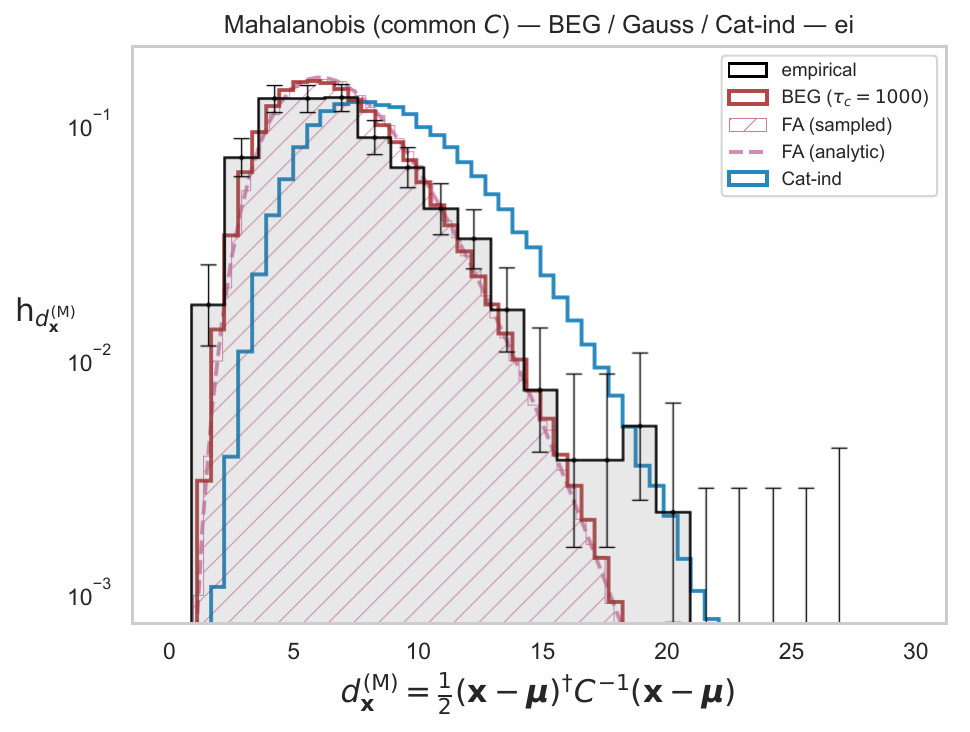}
\includegraphics[width=0.45\columnwidth]{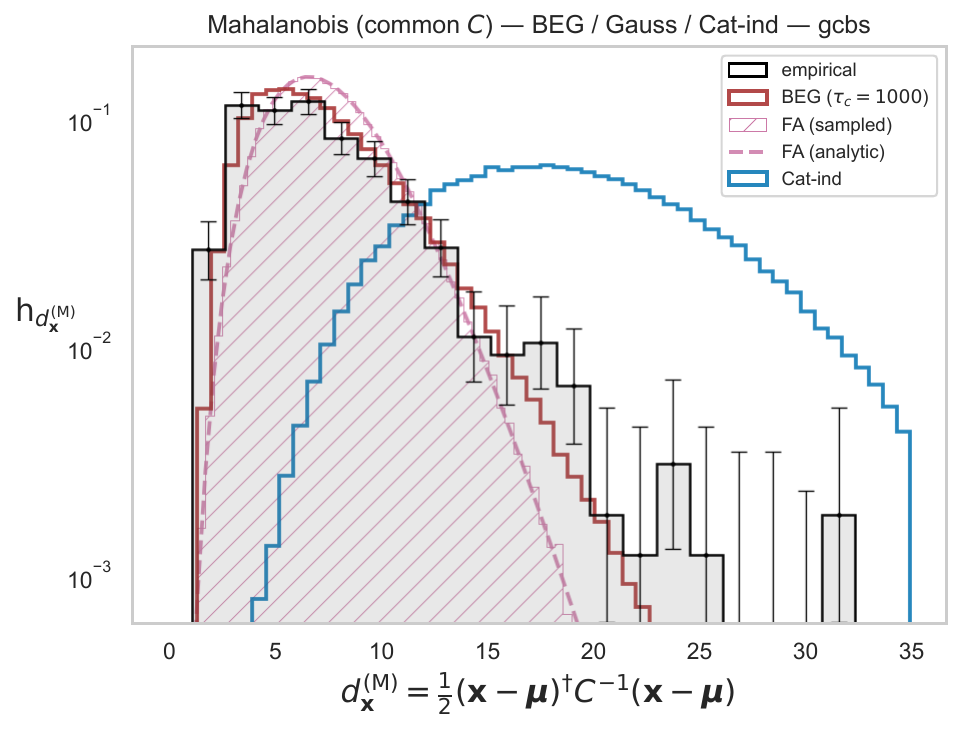}
\includegraphics[width=0.45\columnwidth]{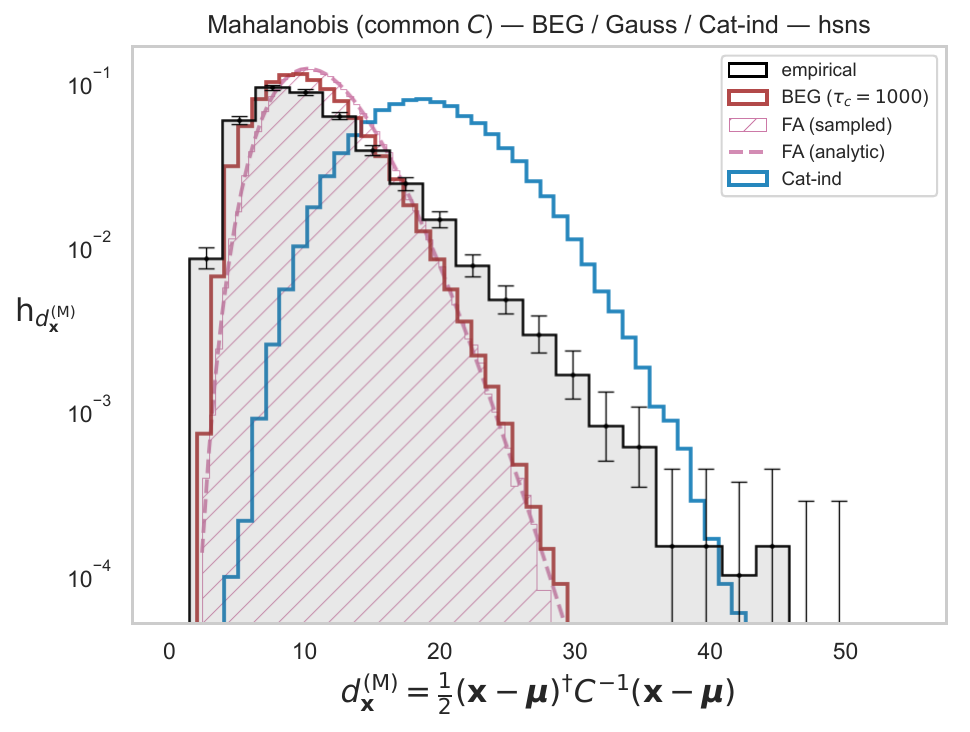}
\caption{Histograms of the Mahalanobis distance to the mean, ${\sf h}_{d_\x^{({\rm M})}}$, for all the analyzed questionnaires: comparison between empirical data and the simple models ({\sf gauss}, {\sf cat-ind}). The failure to reproduce the tails of the Mahalanobis distribution is not specific to the spin models but is shared by all simple models. Error bars are Wilson score confidence intervals at $\alpha=0.05$.}\label{fig:energy_simple_all}
\end{center}
\end{figure}

We report the variant of the figures in the main text for all the analyzed questionnaires: the losses as a function of the learning iterations in Fig. \ref{fig:losses_all}; the moment-matching during learning in the maximum likelihood algorithm in Fig. \ref{fig:momentmatching_sampling_all}; the empirical-spin comparison of item value histograms ${\sf h}_{x_i}$ in Figs. \ref{fig:itemhist_all},\ref{fig:itemhist_all2}; the empirical-spin comparison of histograms of factors in Fig. \ref{fig:factorhist_all}, and the comparison with simple models in Fig. \ref{fig:factorhist_simple_all}; the empirical-spin comparison of Euclidean distances to the mean in Fig. \ref{fig:E2dhist_all}, and the comparison with simple models in Fig. \ref{fig:E2dhist_simple_all}; the equivalent figures for the Mahalanobis distance are Figs. \ref{fig:energy_all},\ref{fig:energy_simple_all}; finally, Fig. \ref{fig:energyvar_all} is the variant of Fig. \ref{fig:energy_all} with model-dependent covariance matrix.

In all the figures of this section, error bars are Wilson score confidence intervals of the histogram heights at $\alpha=0.05$.

\begin{figure}[H]
\begin{center}
 \includegraphics[width=0.45\columnwidth]{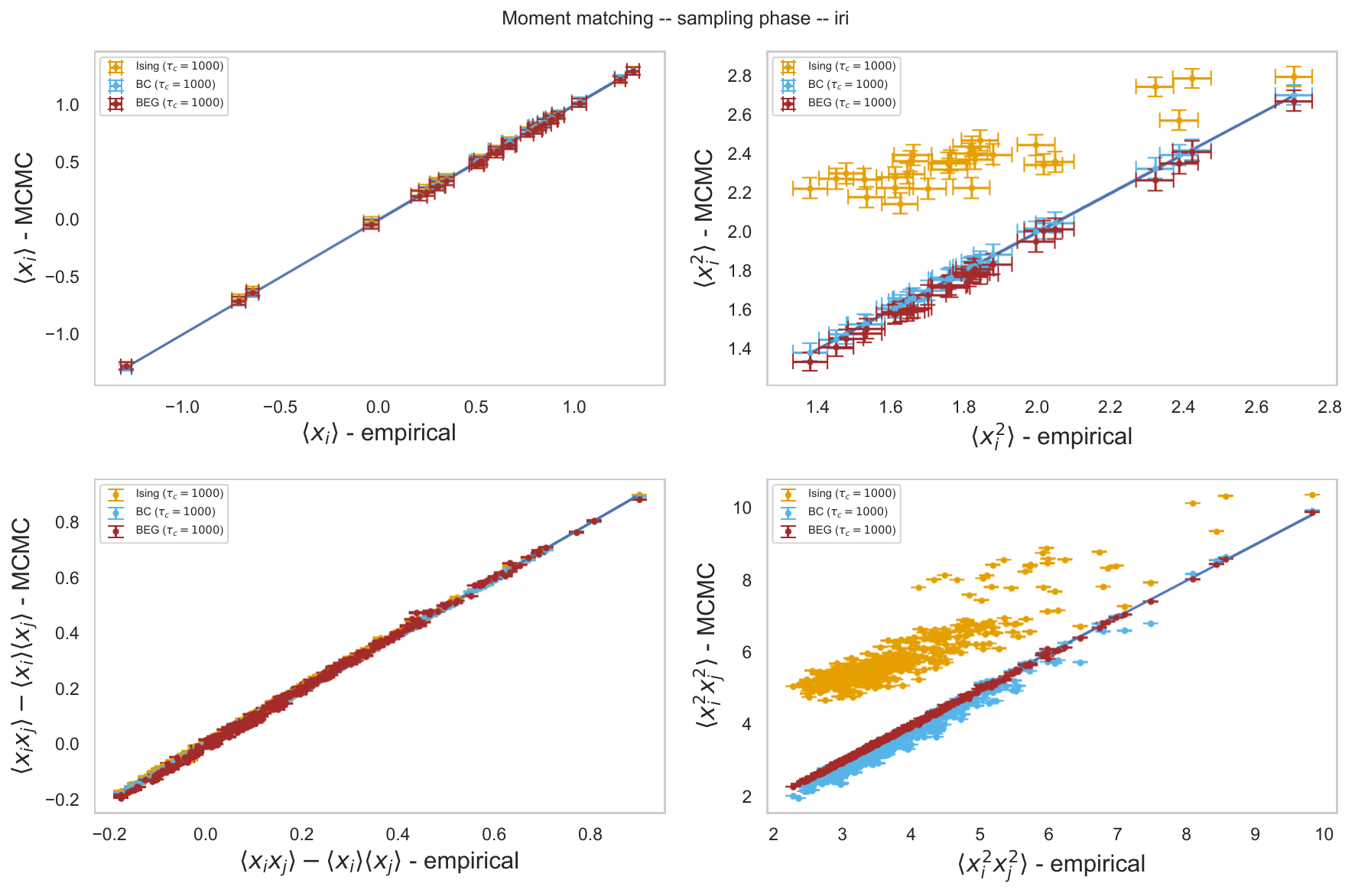}
\includegraphics[width=0.45\columnwidth]{figures/momentmatching_sampling/moment_matching_sampling_data=mach}\\
\includegraphics[width=0.45\columnwidth]{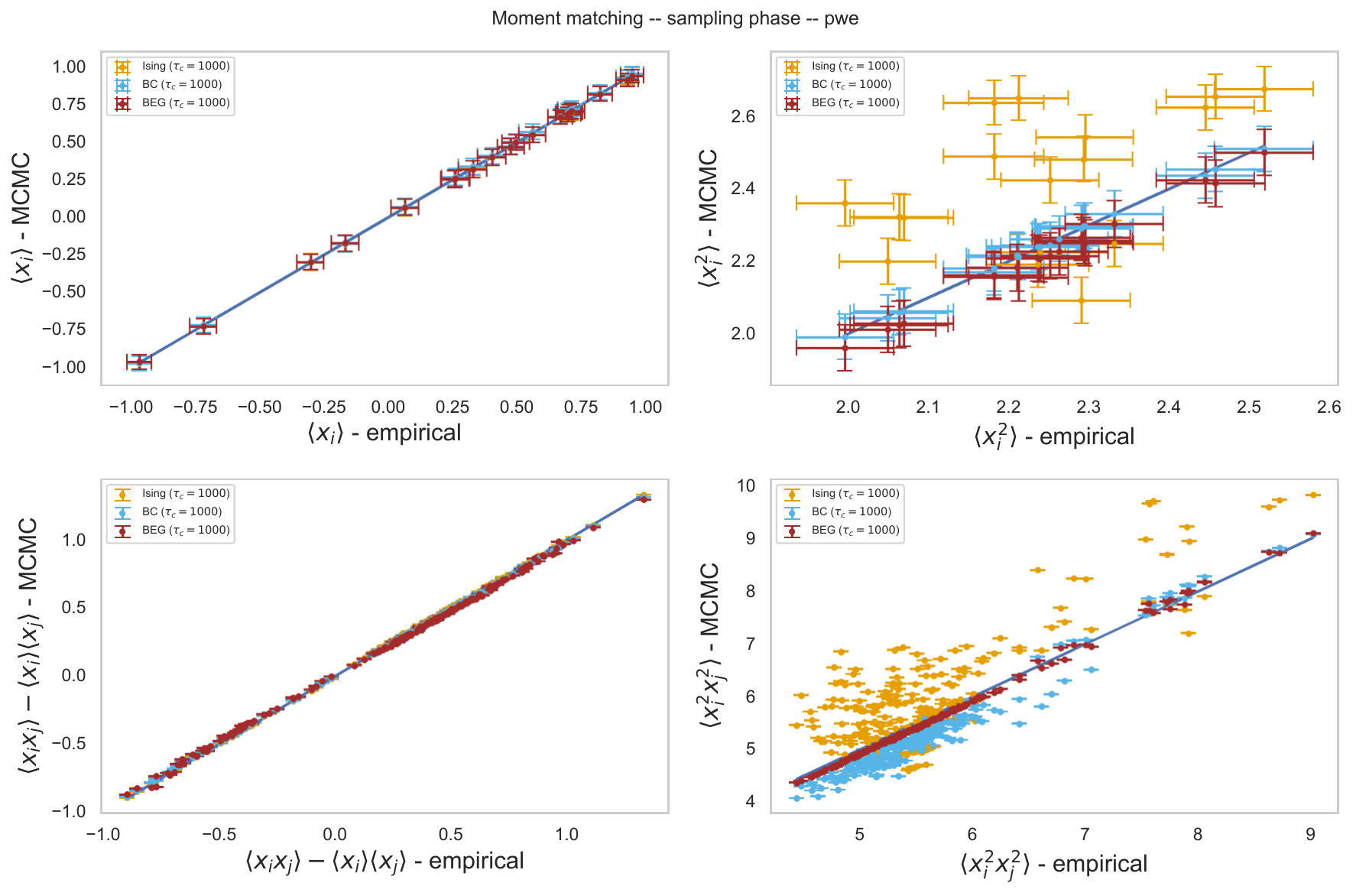} 
\includegraphics[width=0.45\columnwidth]{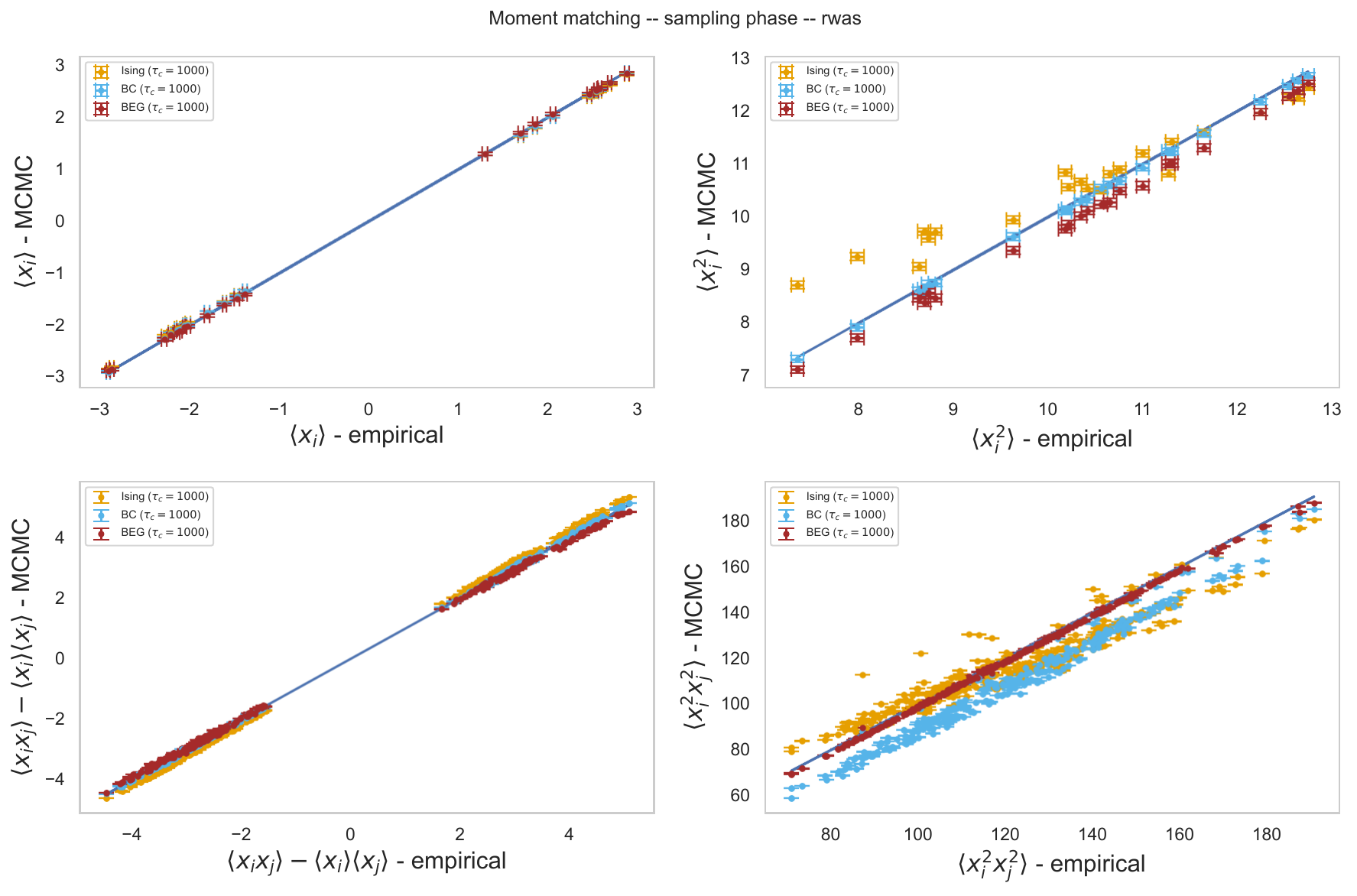}\\
\includegraphics[width=0.45\columnwidth]{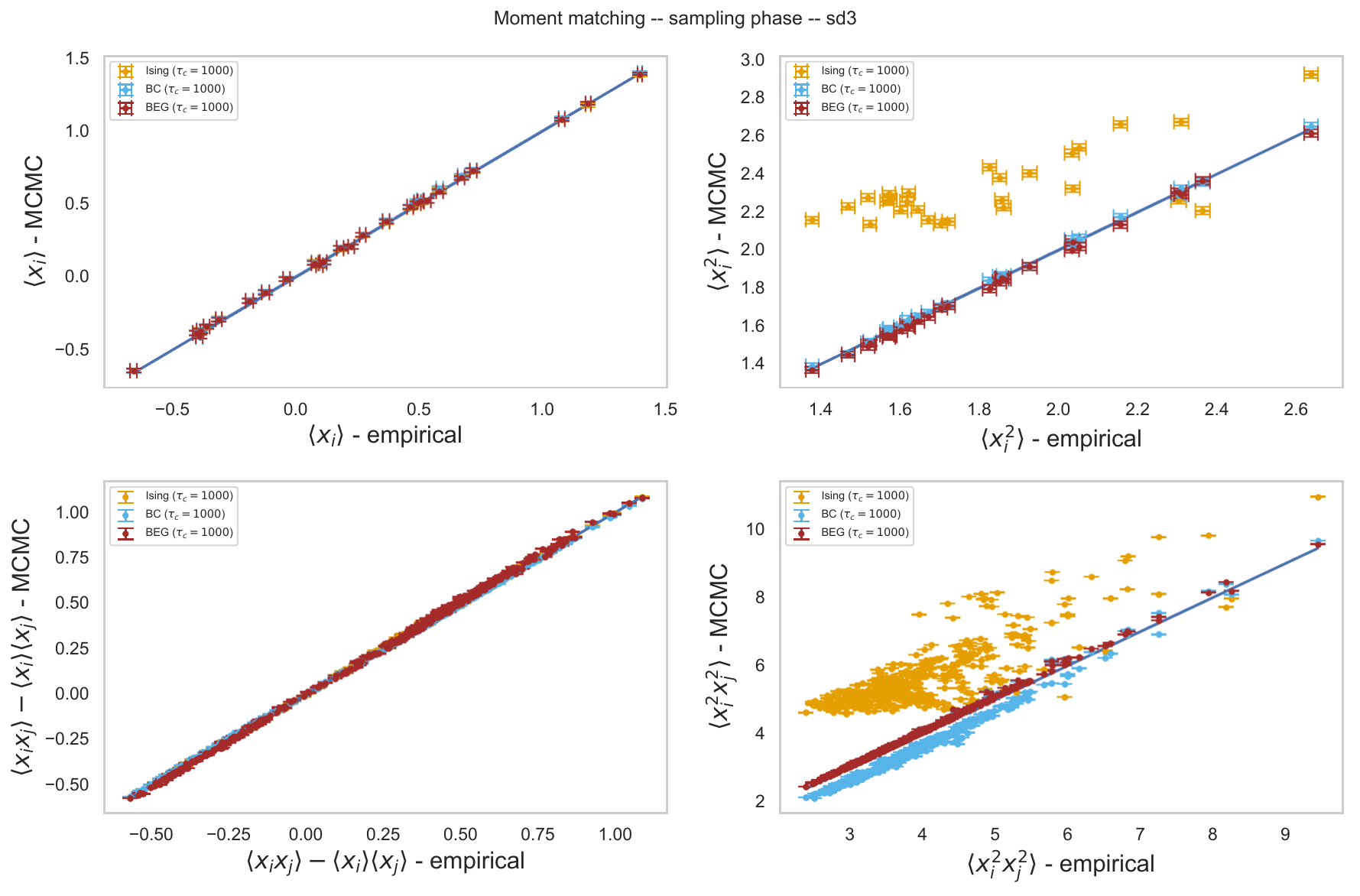}
\caption{Moment-matching consistency check in the sampling phase for all the analyzed questionnaires: comparison between empirical and theoretically sampled sufficient statistics. Equivalent of Fig.~\ref{fig:momentmatching_sampling_mach} for all datasets; see Sec.~\ref{sec:consistency}.}
\label{fig:momentmatching_sampling_all2}
\end{center}
\end{figure}

\begin{figure}[H]
\begin{center}
\includegraphics[width=0.45\columnwidth]{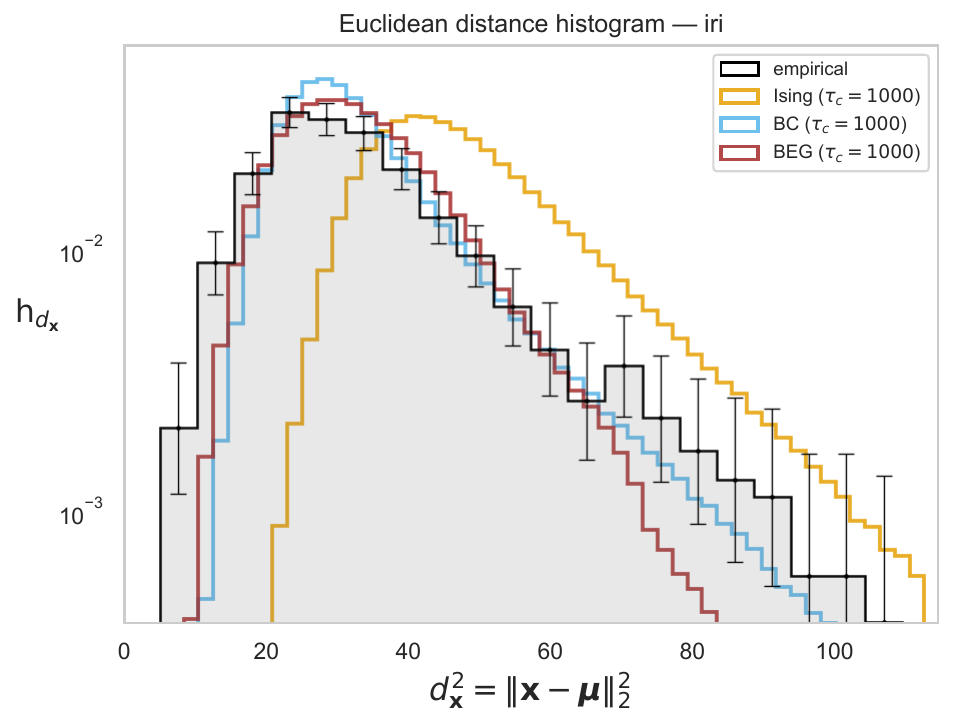}
\includegraphics[width=0.45\columnwidth]{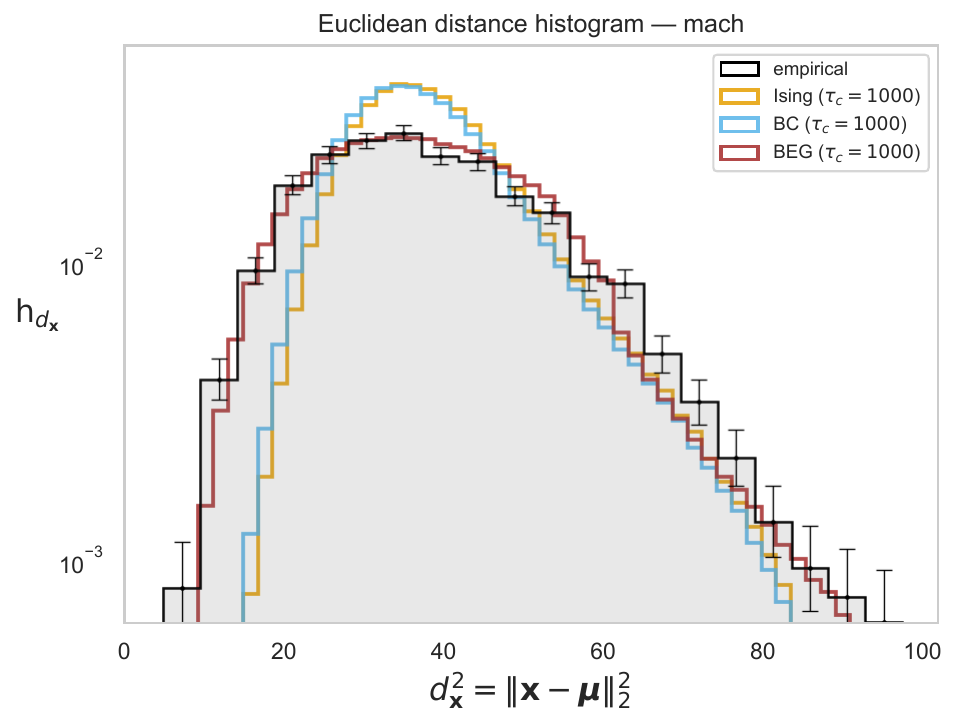}\\
\includegraphics[width=0.45\columnwidth]{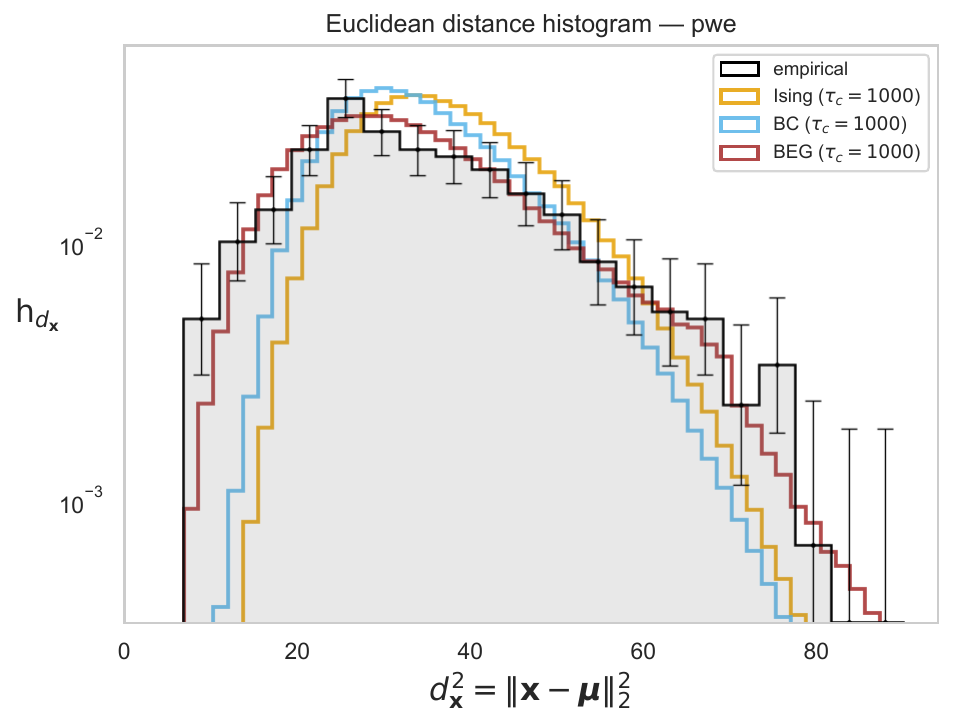} 
\includegraphics[width=0.45\columnwidth]{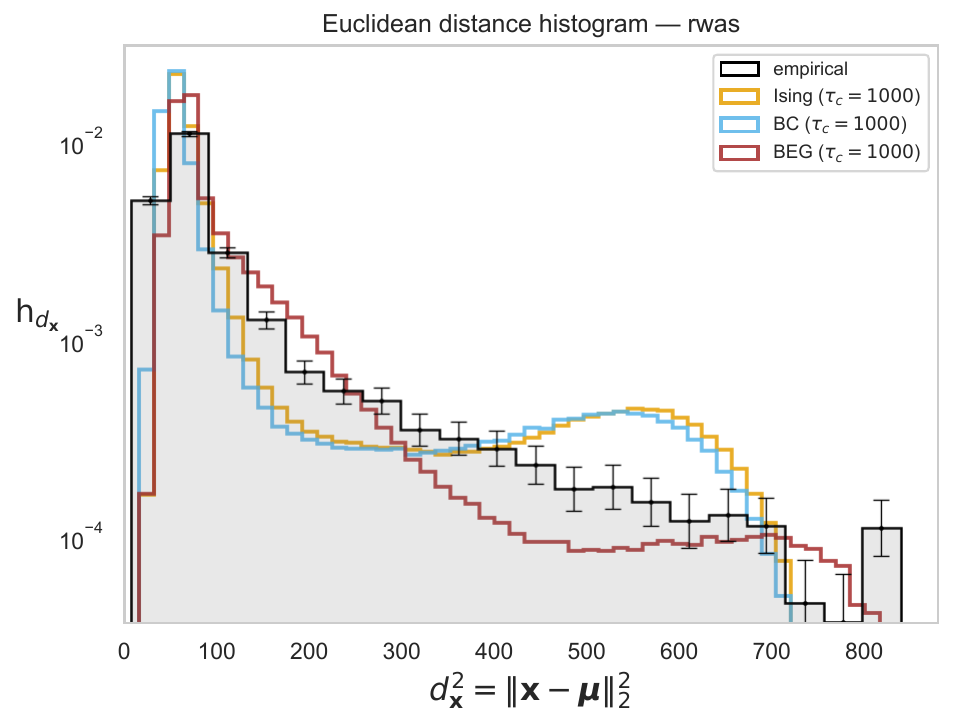}\\
\includegraphics[width=0.45\columnwidth]{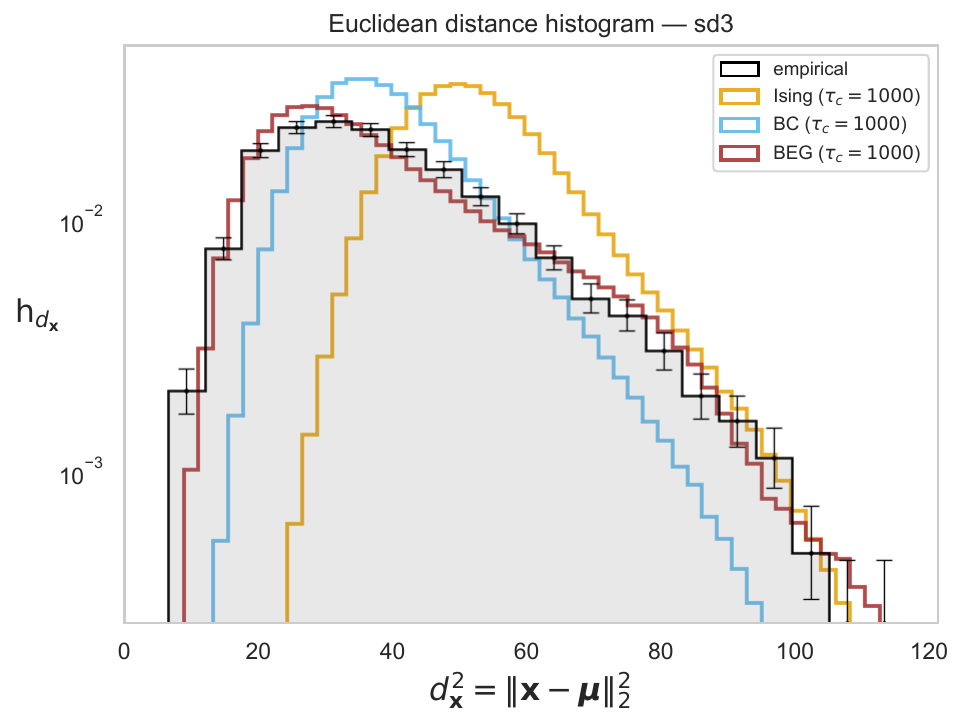}
\caption{Histograms of the Euclidean distance to the mean, ${\sf h}_{d_\x}$, for all the analyzed questionnaires: comparison between empirical data and the three spin models (Ising, BC, BEG). Equivalent of left panel in Fig.~\ref{fig:E2dhist_gcbs} for all datasets.}
\label{fig:E2dhist_all2}
\end{center}
\end{figure}

\begin{figure}[H]
\begin{center}
\includegraphics[width=0.45\columnwidth]{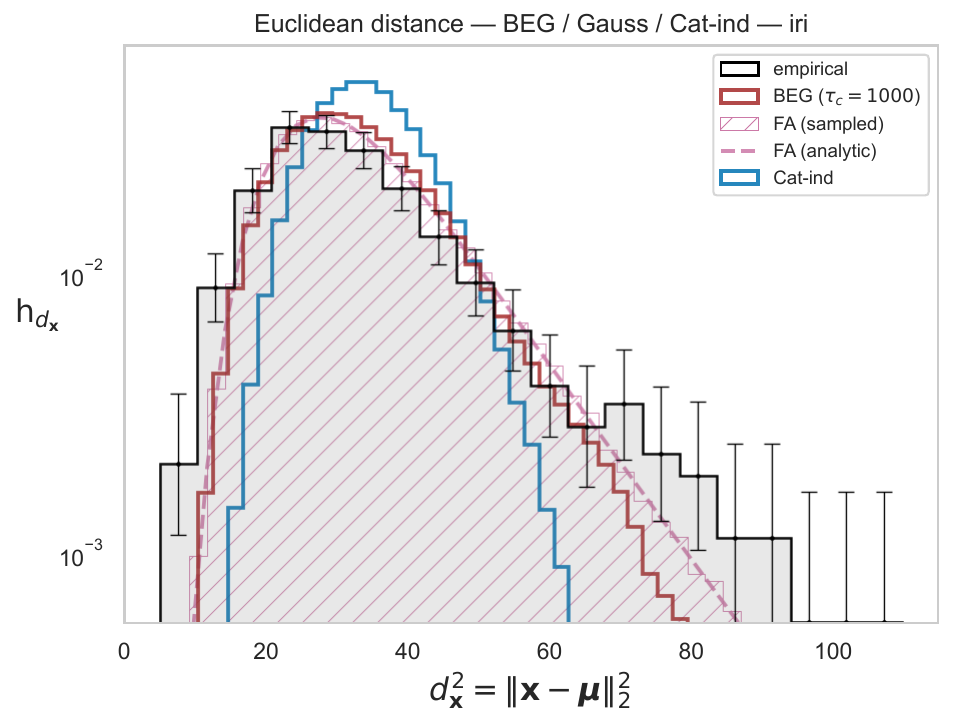}
\includegraphics[width=0.45\columnwidth]{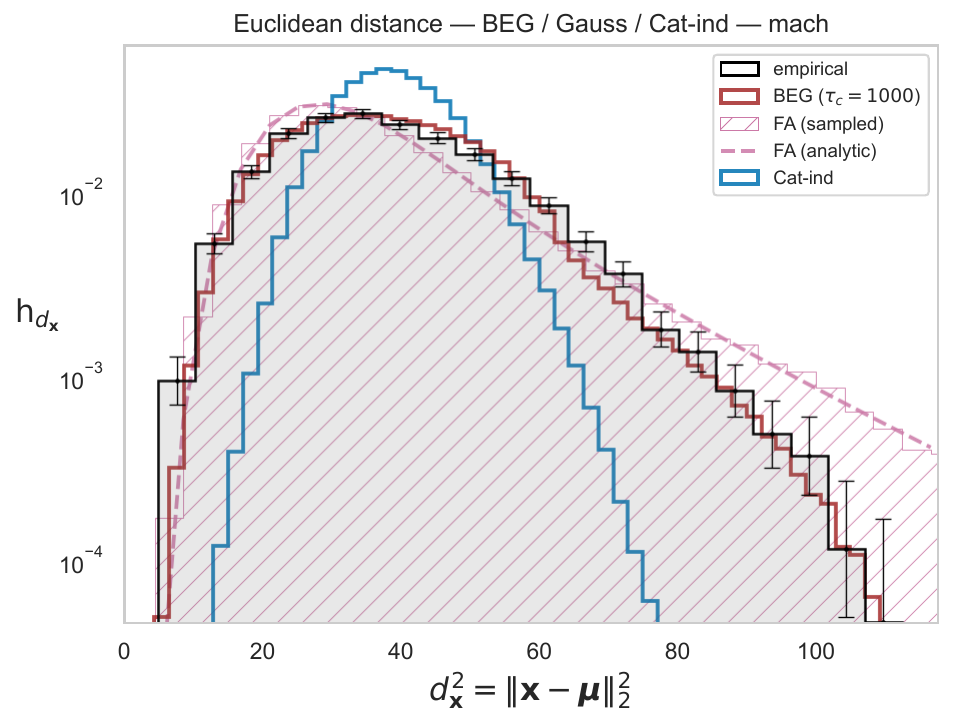}\\
\includegraphics[width=0.45\columnwidth]{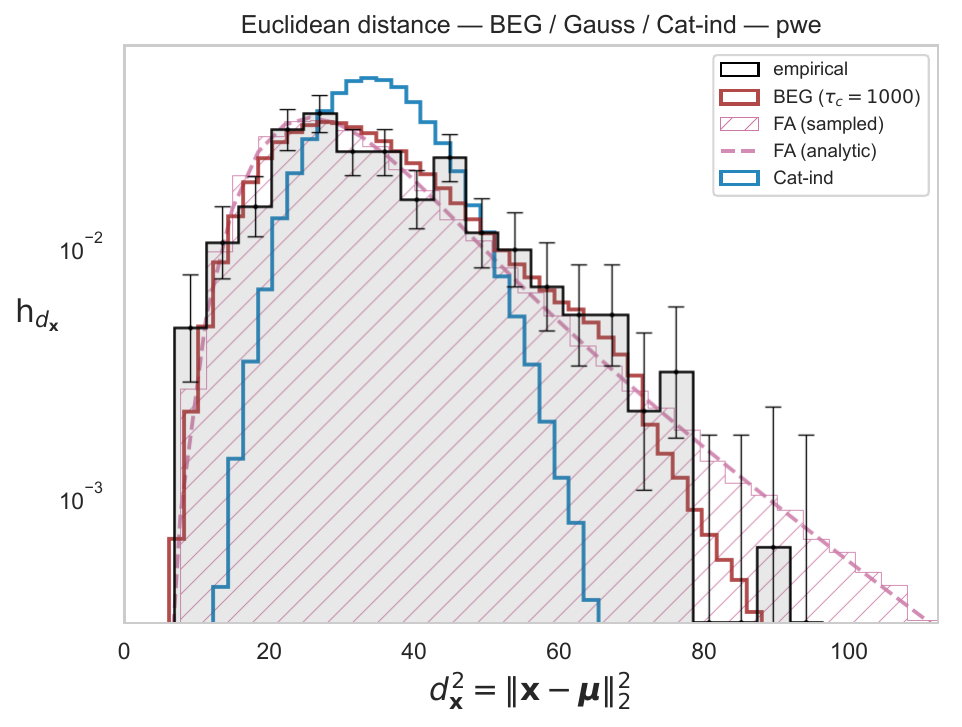}
\includegraphics[width=0.45\columnwidth]{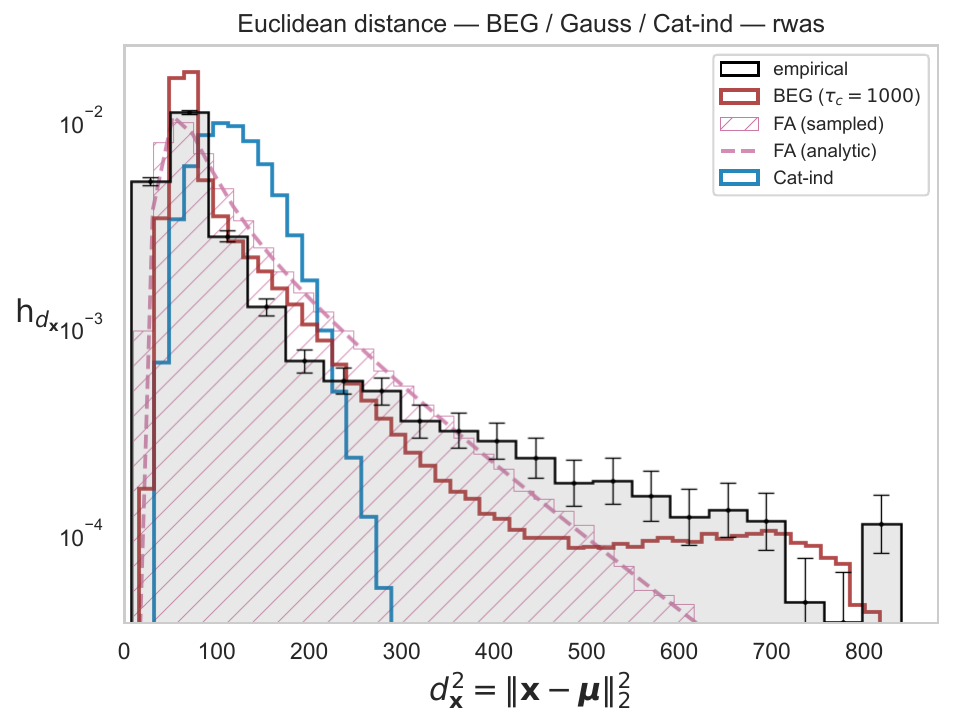}\\
\includegraphics[width=0.45\columnwidth]{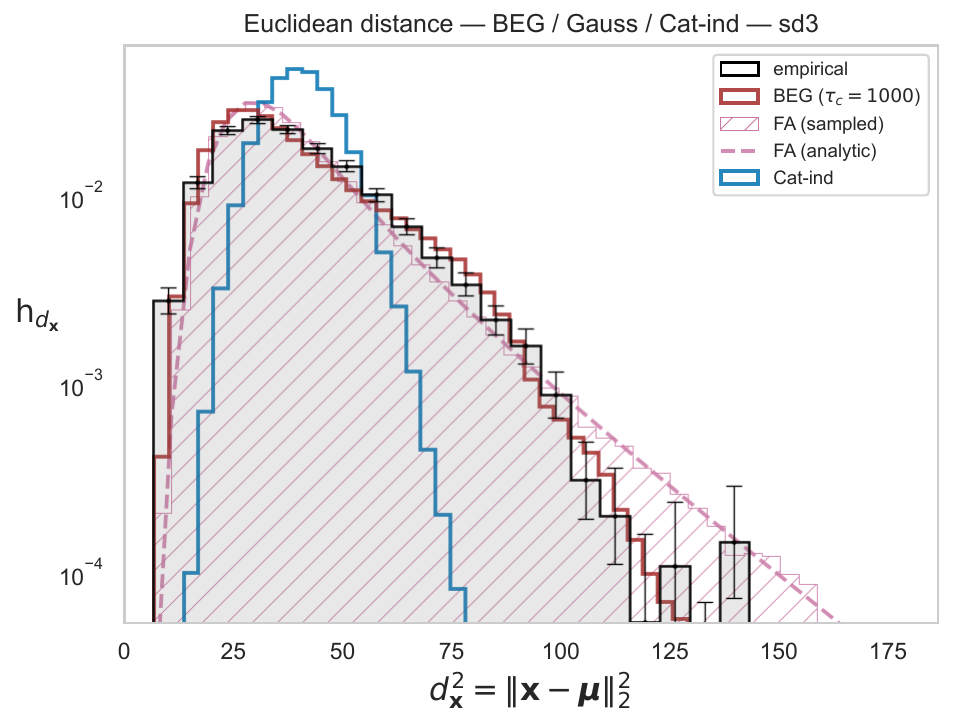}
\caption{Histograms of the Euclidean distance to the mean, ${\sf h}_{d_\x}$, for all the analyzed questionnaires: comparison between empirical data and the simple models ({\sf gauss}, {\sf cat-ind}). Equivalent of right panel in Fig.~\ref{fig:E2dhist_gcbs} for all datasets. Error bars are Wilson score confidence intervals at $\alpha=0.05$.}
\label{fig:E2dhist_simple_all2}
\end{center}
\end{figure}

\begin{figure}[H]
\begin{center}
\includegraphics[width=0.45\columnwidth]{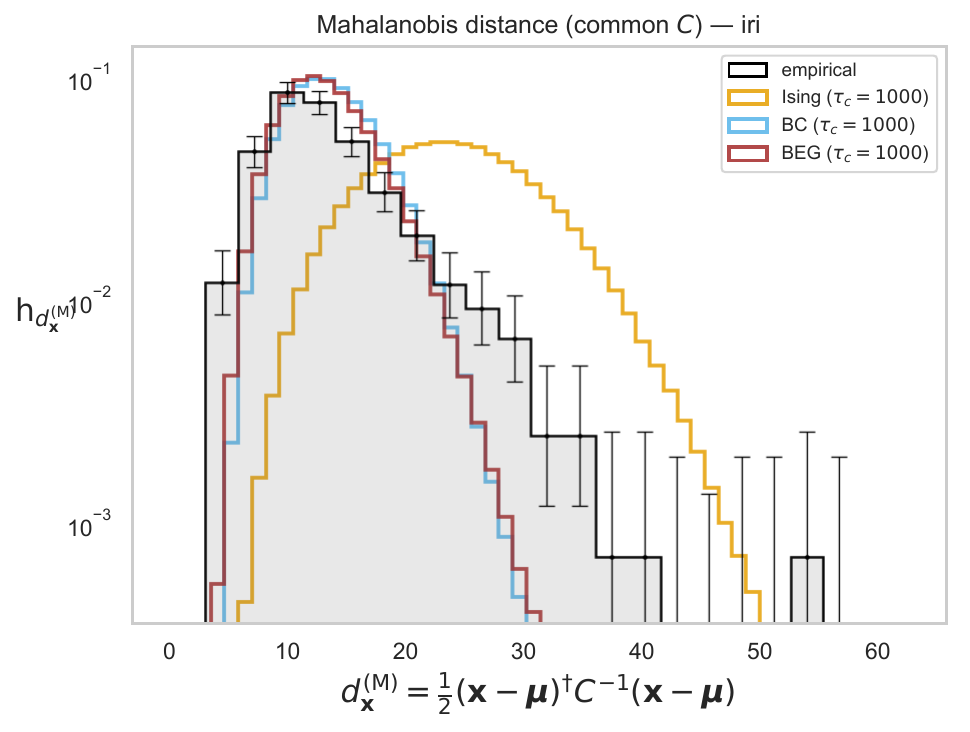}
\includegraphics[width=0.45\columnwidth]{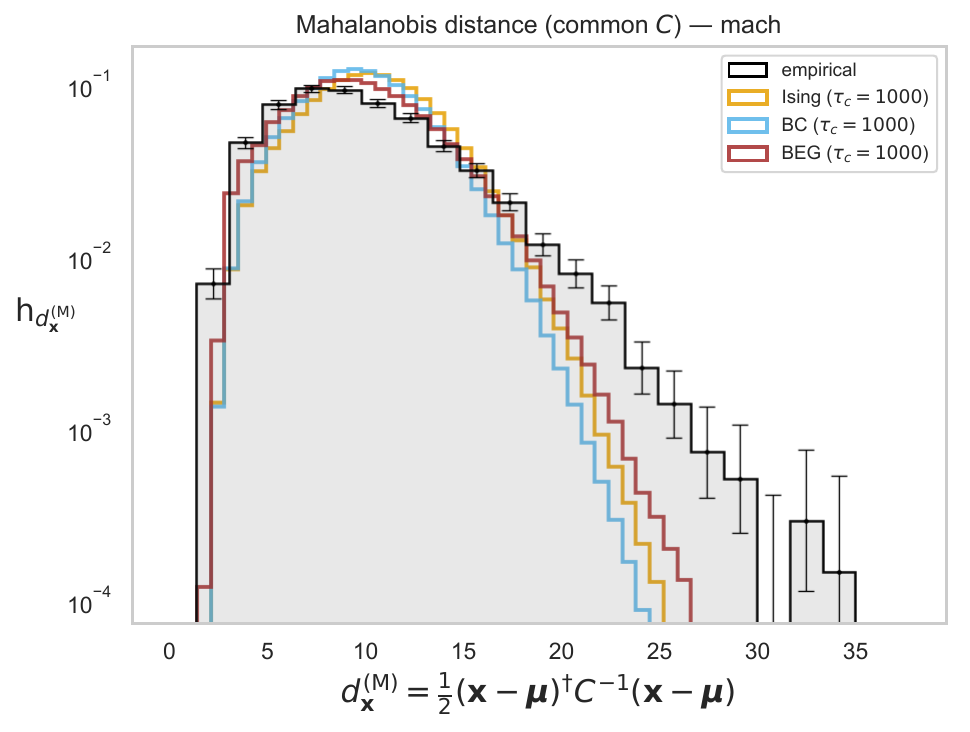}
\includegraphics[width=0.45\columnwidth]{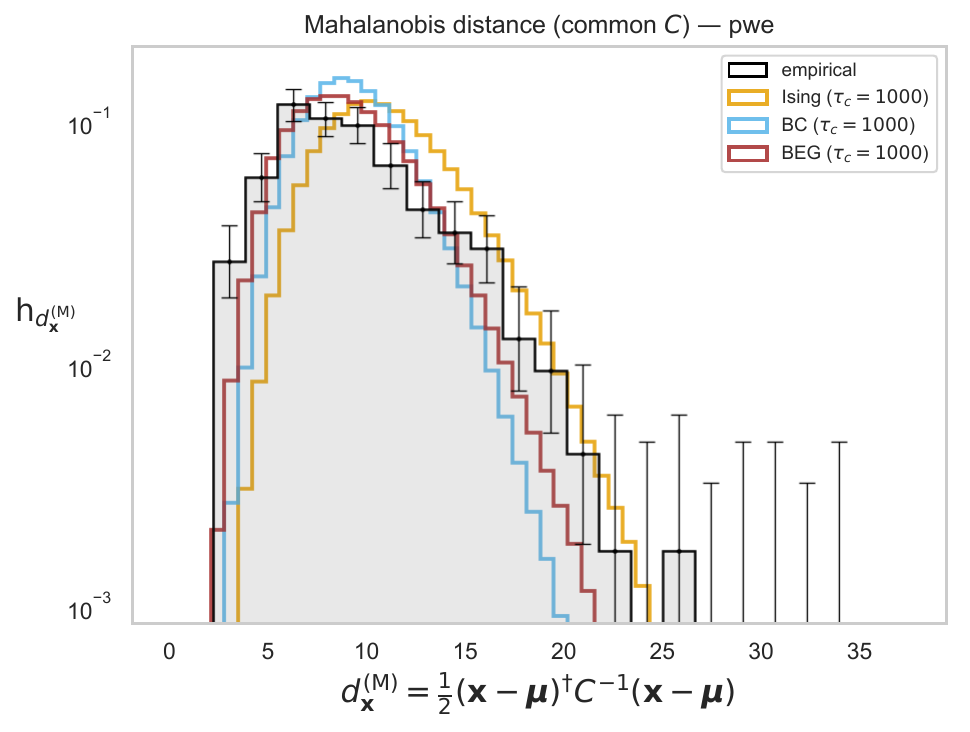}
\includegraphics[width=0.45\columnwidth]{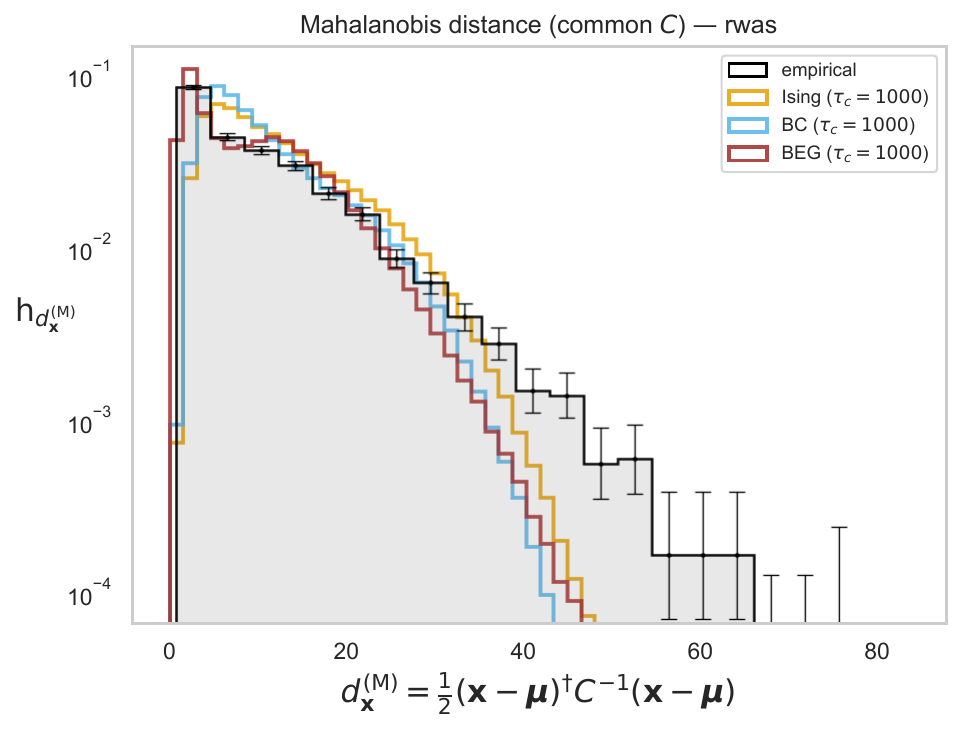}
\includegraphics[width=0.45\columnwidth]{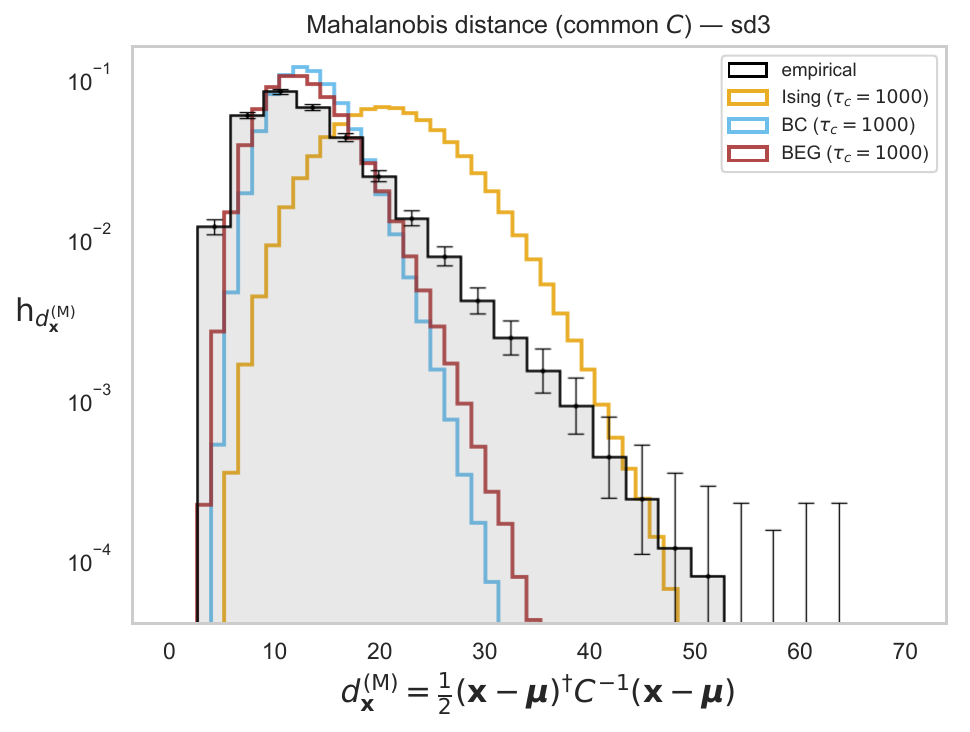}
\caption{Histograms of the Mahalanobis distance to the mean, ${\sf h}_{d_\x^{({\rm M})}}$, for all the analyzed questionnaires: comparison between empirical data and the three spin models (Ising, BC, BEG). The empirical covariance matrix $C$ is used in Eq.~(\ref{eq:maha}) for all models. All models systematically underestimate the tails. }
\label{fig:energy_all2}
\end{center}
\end{figure}

\begin{figure}[H]
\begin{center}
\includegraphics[width=0.45\columnwidth]{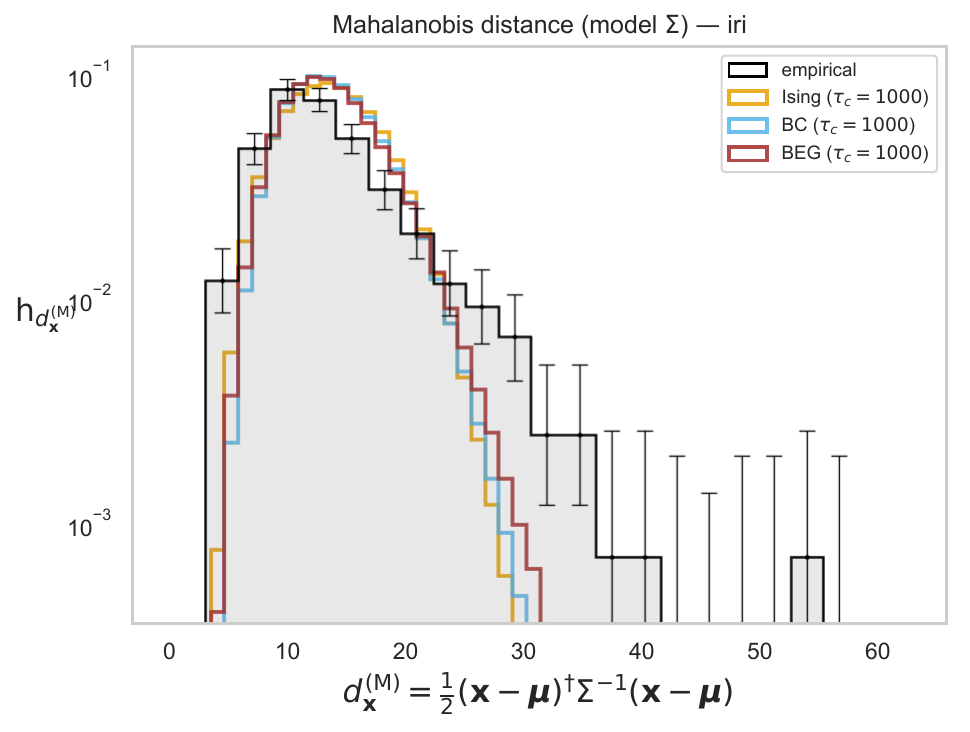}
\includegraphics[width=0.45\columnwidth]{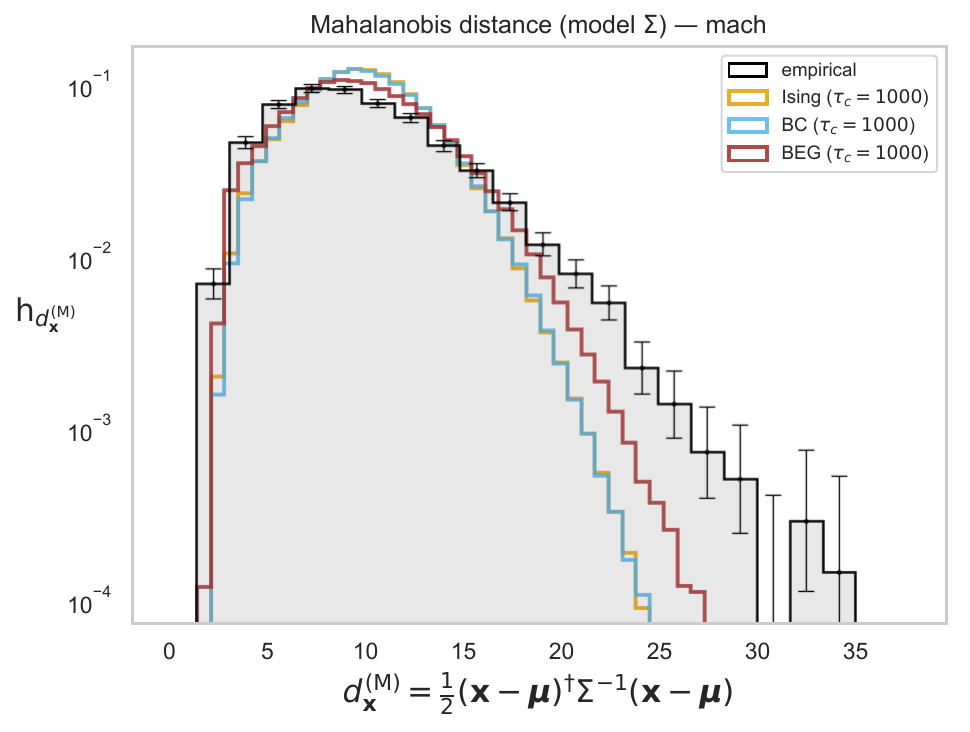}
\includegraphics[width=0.45\columnwidth]{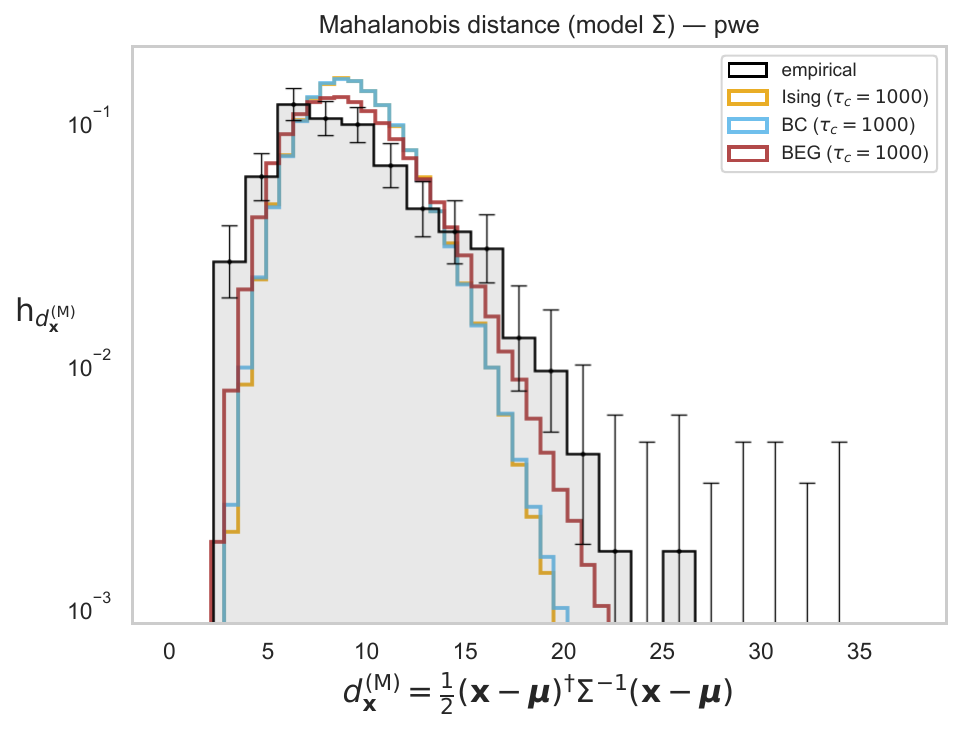}
\includegraphics[width=0.45\columnwidth]{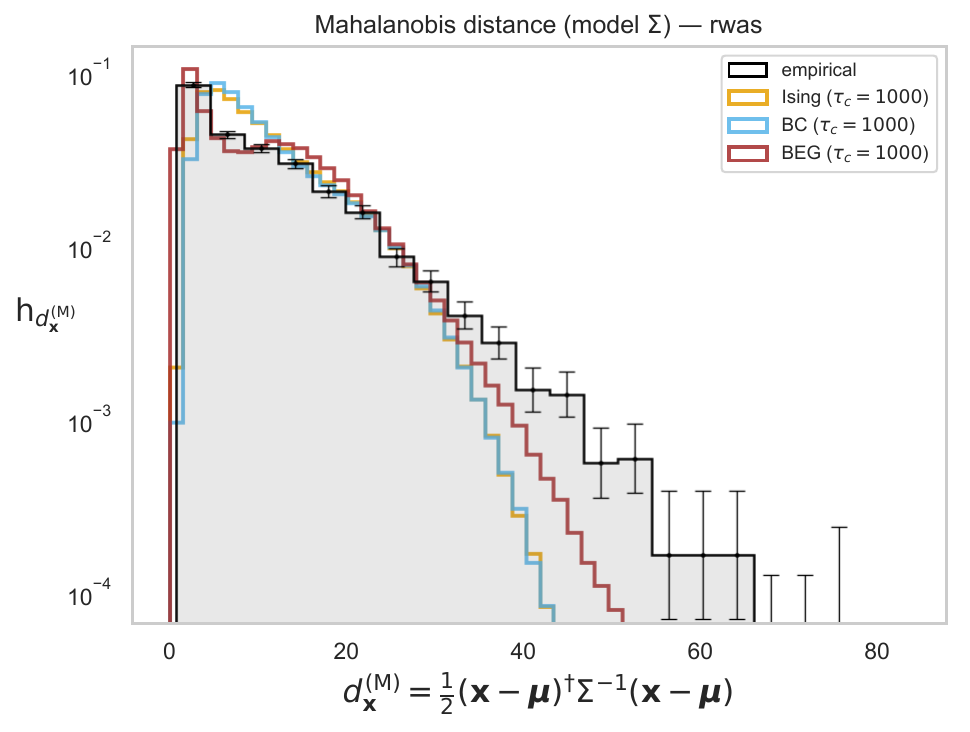}
\includegraphics[width=0.45\columnwidth]{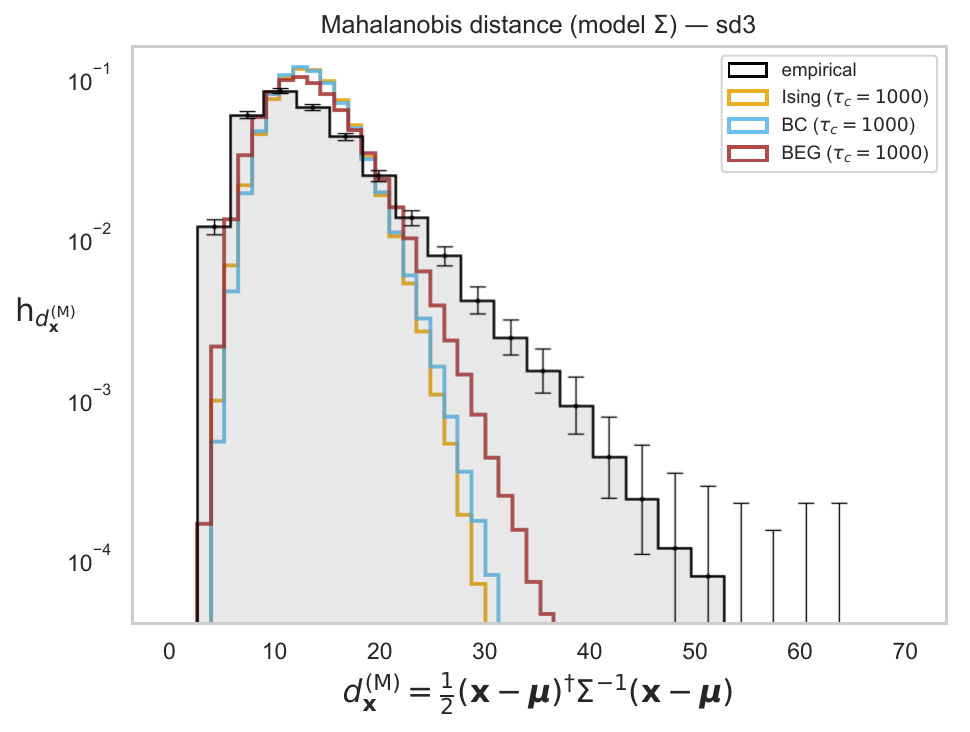}
\caption{Variant of Fig.~\ref{fig:energy_all}: histograms of the Mahalanobis distance $d_\x^{({\rm M})}=(1/2)\,\x^\dag\cdot\Sigma^{-1}\cdot\x$, where $\Sigma=\langle\x\x^\dag\rangle_P - \langle\x\rangle_P\langle\x^\dag\rangle_P$ is the model-dependent theoretical covariance matrix (instead of the empirical $C$). The discrepancy between empirical and theoretical histograms persists when model-dependent $\Sigma$ is used, ruling out finite-loss errors in the estimation of $C$ as its explanation. }
\label{fig:energyvar_all2}
\end{center}
\end{figure}

\begin{figure}[H]
\begin{center}
\includegraphics[width=0.45\columnwidth]{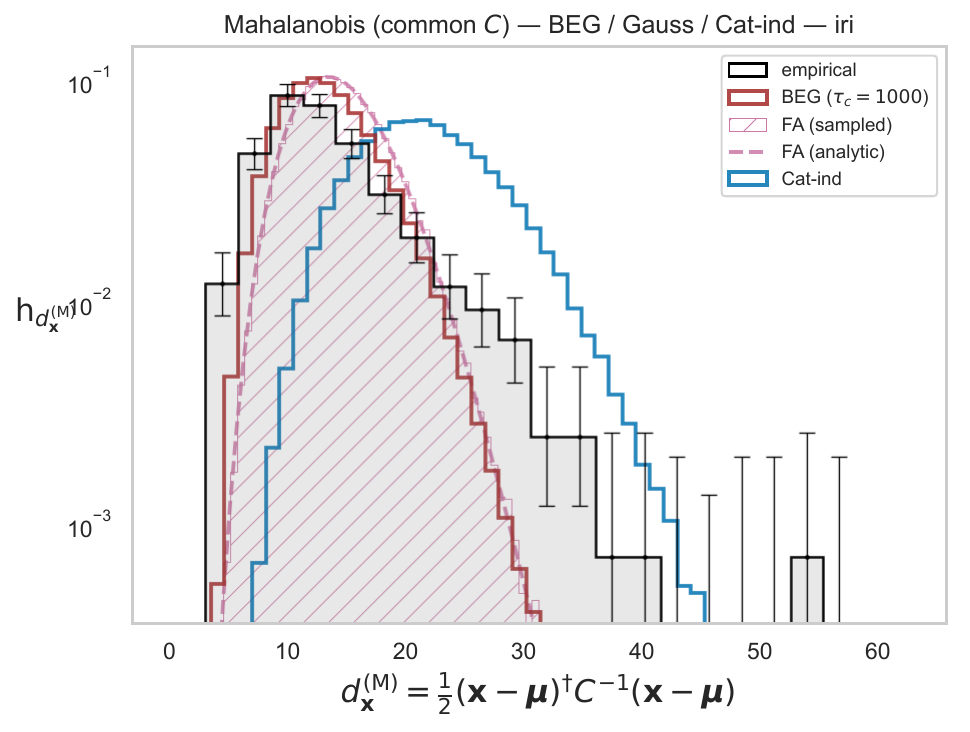}
\includegraphics[width=0.45\columnwidth]{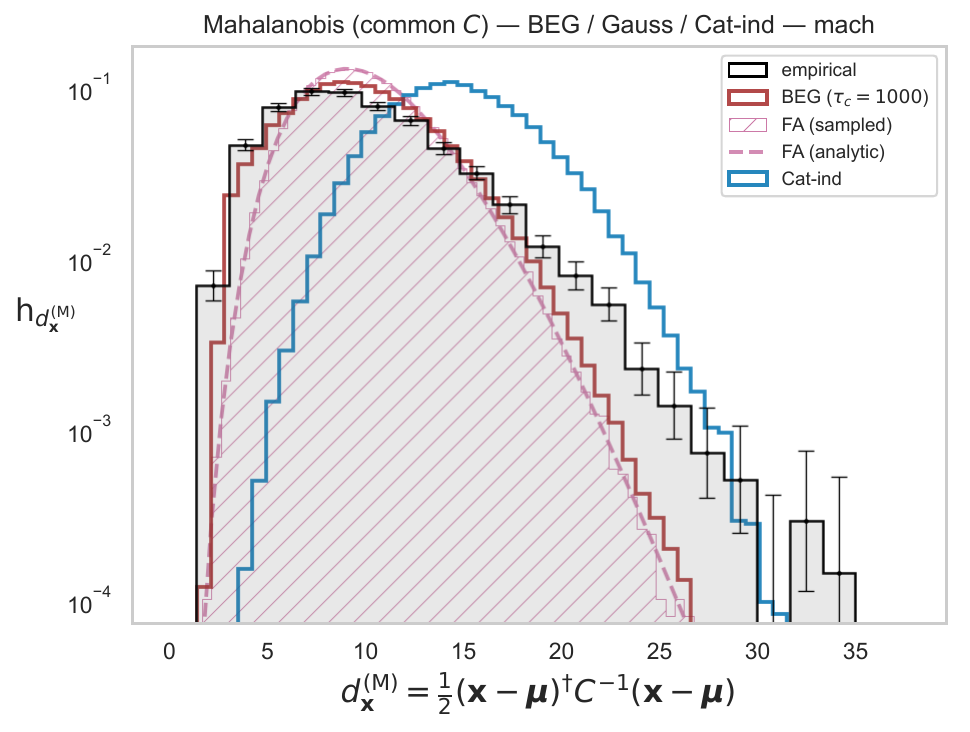}
\includegraphics[width=0.45\columnwidth]{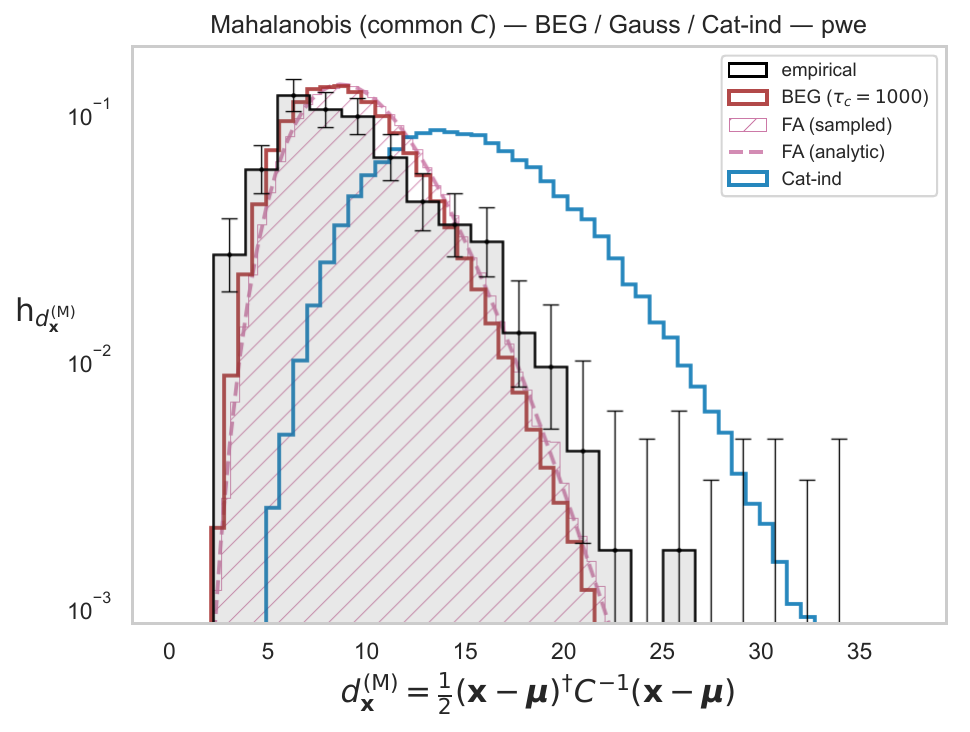}
\includegraphics[width=0.45\columnwidth]{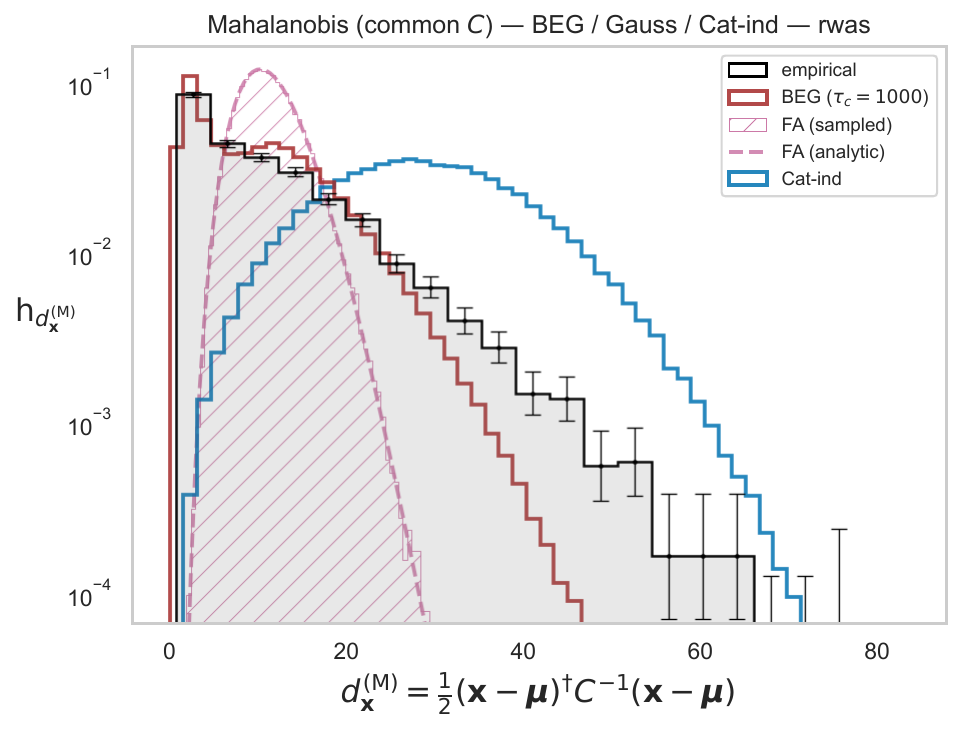}
\includegraphics[width=0.45\columnwidth]{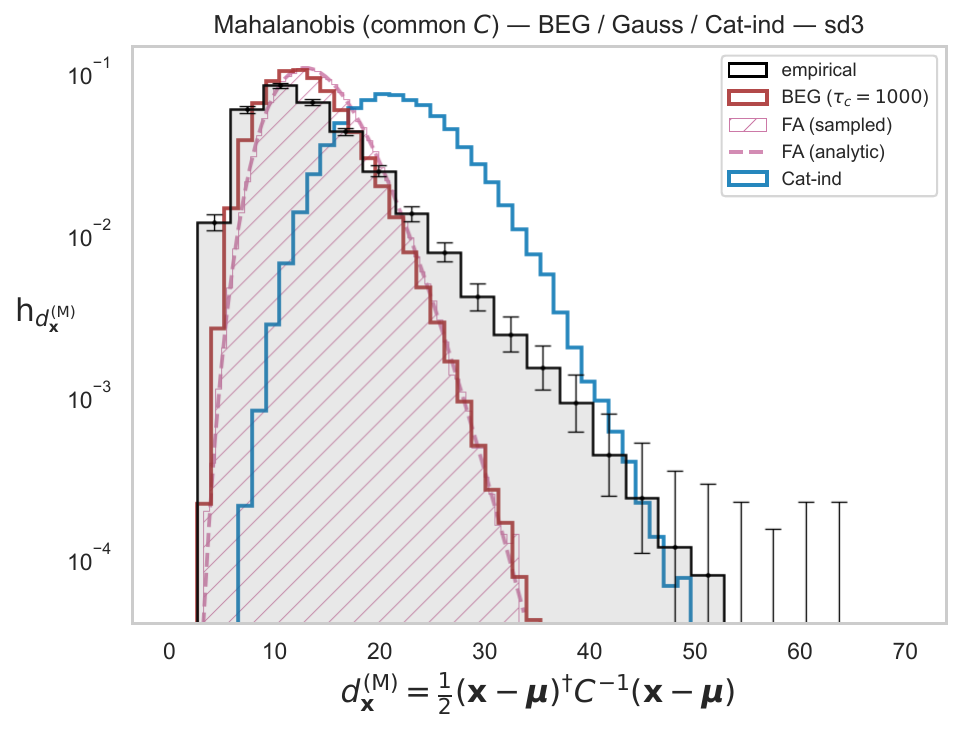}
\caption{Histograms of the Mahalanobis distance to the mean, ${\sf h}_{d_\x^{({\rm M})}}$, for all the analyzed questionnaires: comparison between empirical data and the simple models ({\sf gauss}, {\sf cat-ind}). The failure to reproduce the tails of the Mahalanobis distribution is not specific to the spin models but is shared by all simple models. Error bars are Wilson score confidence intervals at $\alpha=0.05$.}\label{fig:energy_simple_all2}
\end{center}
\end{figure}

\section{Significance analysis of the inferred couplings \label{sec:significance}}

We perform a simple assessment of significance of the inferred couplings of the spin models, for the {\sf big5} dataset. Such an assessment consists in the following procedure. We infer the maximum likelihood parameters $\theta^*$ maximizing the joint likelihood $P(X|\theta^*)$ of the training-set data $X$, with the PCD algorithm. Afterwards, we sample $N$ synthetic vectors $Y\sim P(\cdot|\theta^*)$, where $Y$ is the $N\times M$ matrix of $N$ synthetic subjects. We then infer the maximum likelihood parameters $\theta^{**}$ maximizing $P(X|\theta^{**})$. We compare the parameters $\theta^*$ and $\theta^{**}$ in terms of the element-wise relative error $|\theta^*_\mu-\theta^{**}_\mu|/|\theta_\mu^*|$. We retain as barely significant those elements for which the relative error is lower than $0.1$, i.e., those elements whose maximum likelihood value differs less than $10\%$ of their value from their sampling-plus-refitting variant. Such a bare significance analysis takes into account both bias and variance errors. If the training losses were zero, the difference between $\theta^*$ and $\theta^{**}$ would be an effect of the variance error only (the error induced by the fluctuations of the sufficient statistics $\<o_\mu\>_{\rho_X}$ for finite $N$, then propagated to $\theta_\mu$), which is here indirectly simulated through the sampling of $N$ synthetic subjects from $P(\cdot|\theta^*)$. If the variance error were zero (if $N$ were infinity), the difference between $\theta^*$ and $\theta^{**}$ would be an effect of the bias error only, induced by the stochastic character of the gradient ascent estimation in the PCD algorithm, by the finiteness of the number of gradient ascent iterations and of the learning rate. 

\begin{figure}[H]
\begin{center}
\includegraphics[width=\columnwidth]{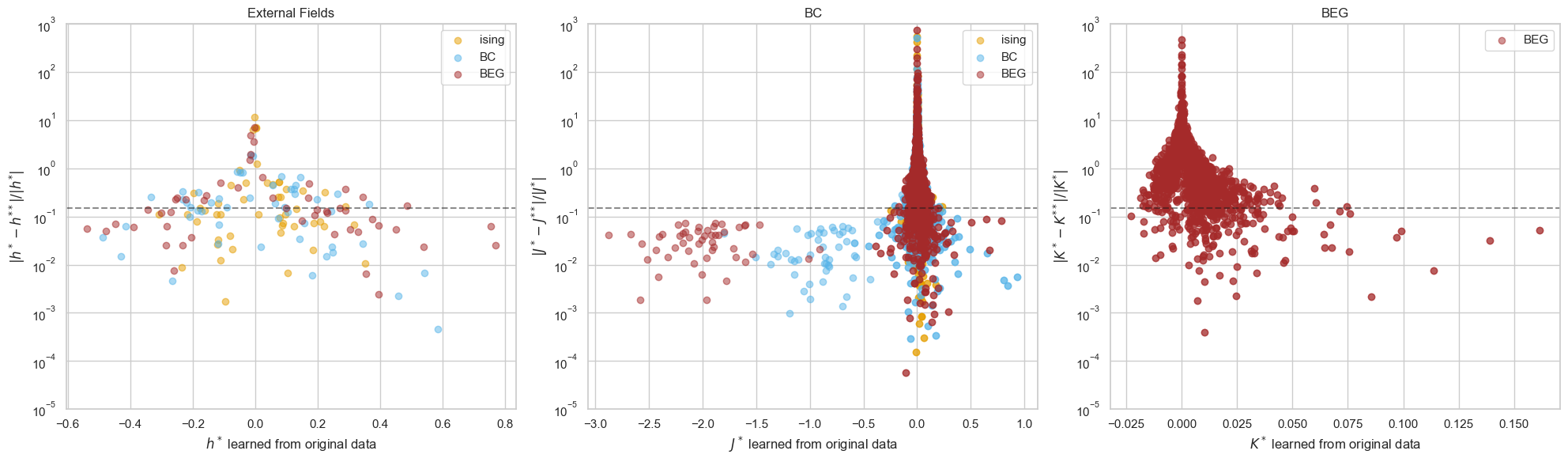}
\caption{Bare significance analysis of the couplings. We show the element-wise relative error $|\theta^*_\mu-\theta^{**}_\mu|/|\theta_\mu^*|$ of the maximum likelihood couplings inferred from the data ($\theta^*$) and from the synthetic data sampled from $P(\cdot|\theta^*)$ ($\theta^{**}$) (A,B,C) correspond, respectively, to $\theta={\bf h}$, $=J$ and $=K$ respectively. The horizontal dashed line indicates the relative error $0.1$ below which the couplings are considered to be barely significant.}
\label{fig:robustness}
\end{center}
\end{figure}

\begin{figure}[H]
\begin{center}
\includegraphics[width=0.49\columnwidth]{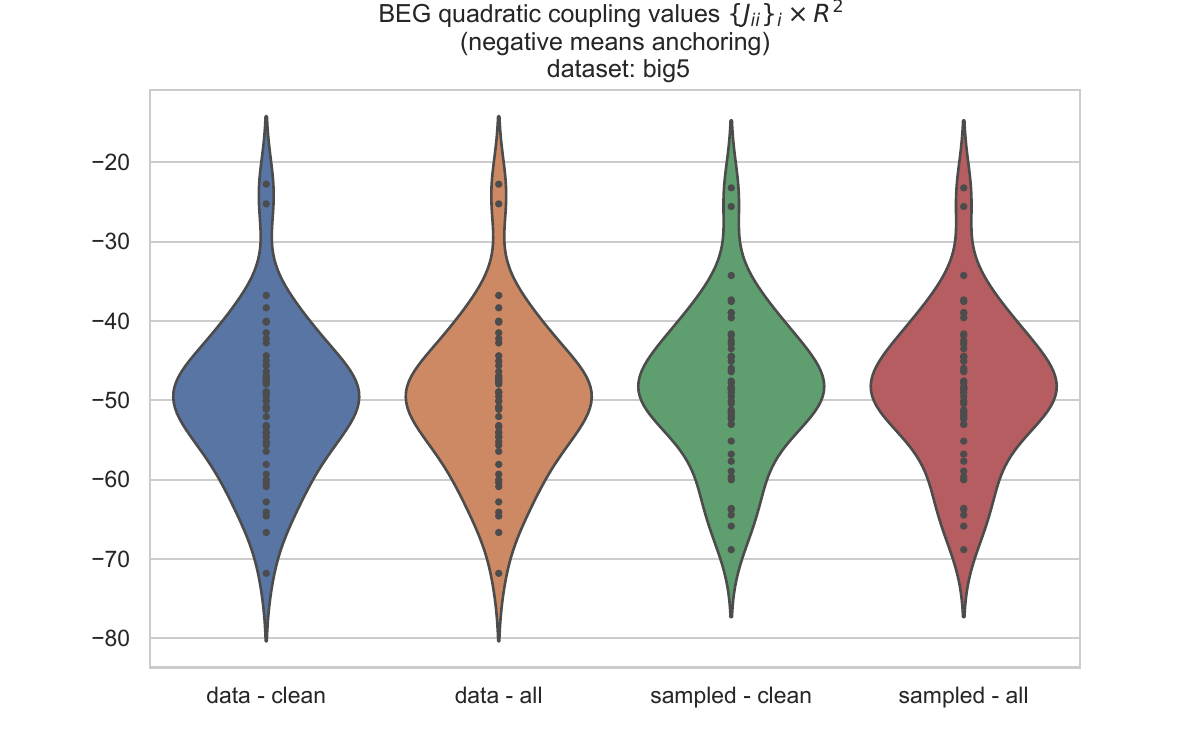}
\includegraphics[width=0.49\columnwidth]{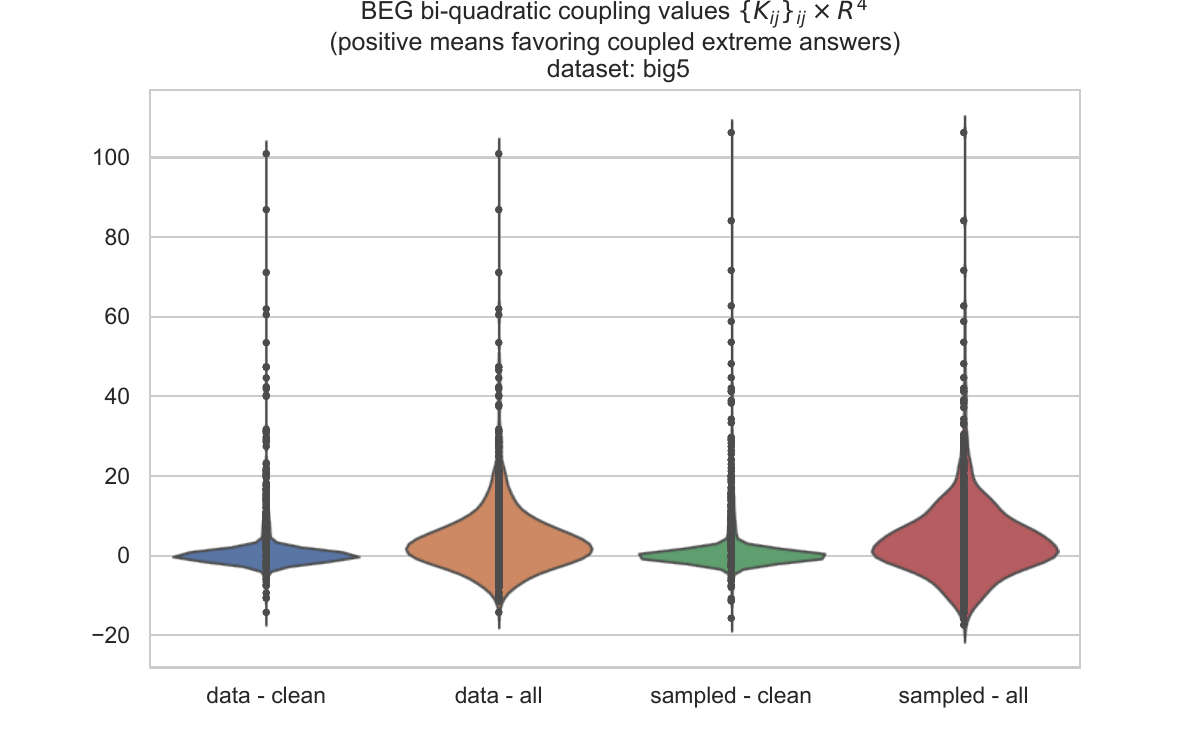}
\caption{Histograms of BEG model inferred quadratic $(J_{ii})_i$ (A) and bi-quadratic $(K_{ij})_{ij}$ (B) couplings for the {\sf big5} questionnaire. This is the robustness analysis of the results in Fig. \ref{fig:couplings} in the main test.}
\label{fig:couplings_robustness}
\end{center}
\end{figure}

\begin{figure}[H]
\begin{center}
\includegraphics[width=0.65\columnwidth]{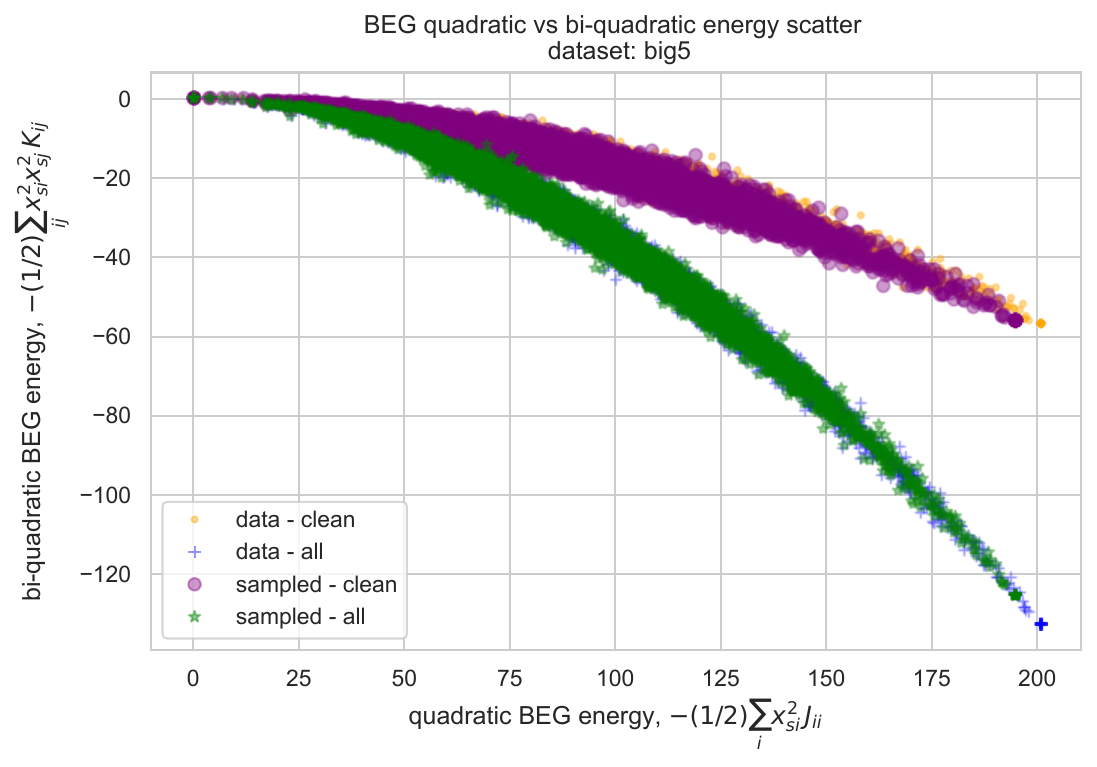}
\caption{Scatter plot of bi-quadratic vs quadratic energy terms, $H_{22}(\x_s)$ vs $H_{2}(\x_s)$, according to the inferred BEG model, for all the subjects in the training set (each point correspond to a subject $s$), for the {\sf big5} questionnaire. This is the robustness analysis of Fig. \ref{fig:energy_scatter} in the main test.}
\label{fig:energy_scatter_robustness}
\end{center}
\end{figure}

The results of the significance analysis are shown in Fig. \ref{fig:robustness}. The elements below the dashed line (marking relative error $=0.1$) are the tensor elements considered significant. As explained in the main article text, we do not perform a systematic significance assessment of the model parameters. We rather employ the above significance criterion to assess the robustness of our analysis regarding the interpretation of the couplings in Sec. \ref{sec:interpretation} of the main article. We therefore plot, as in Fig. \ref{fig:couplings}, the quadratic and bi-quadratic couplings, but for the refitted couplings $\theta^{**}$, and only for those maximum likelihood couplings elements $\theta^*_\mu$ that result significant in the above analysis. The results are in Fig. \ref{fig:couplings_robustness}. The equivalent robustness analysis of Fig. \ref{fig:energy_scatter} is Fig. \ref{fig:energy_scatter_robustness}. The conclusions of Sec. \ref{sec:interpretation} remain qualitatively equal and are not induced by non-significant values of the coupling constants.

\section{Out-of-sample analysis \label{sec:out}}

\begin{figure}[H]
    \centering
    \includegraphics[width=\linewidth]{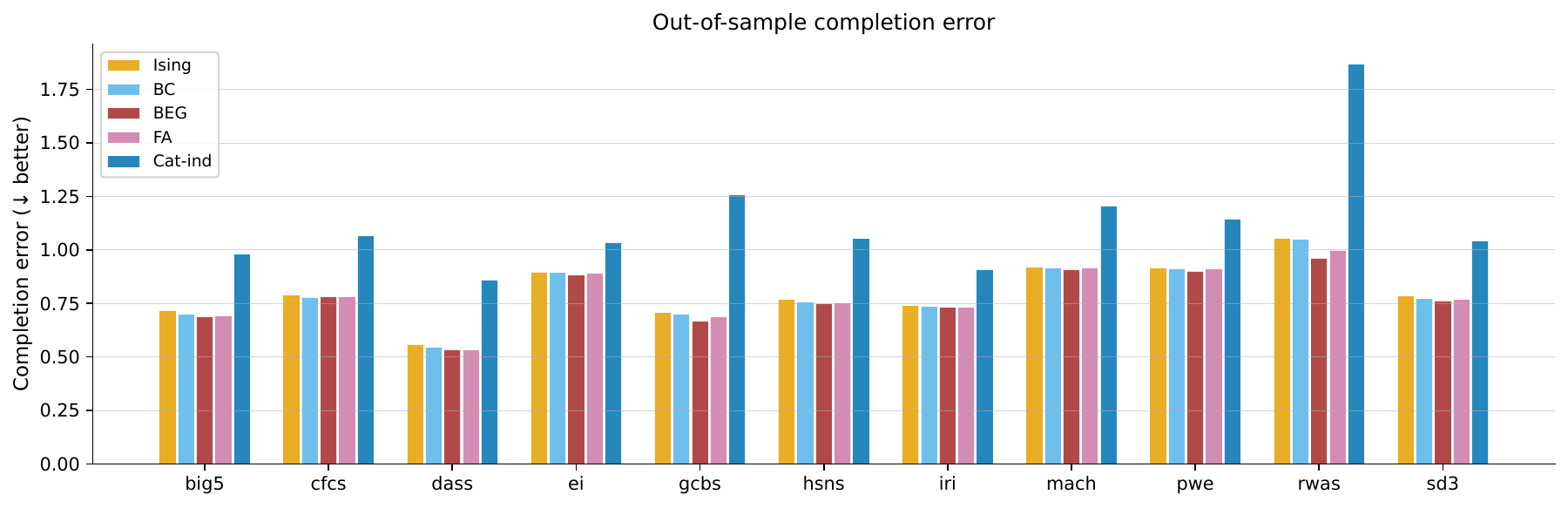}
    \caption{Comparison between out-of-sample completion error-1 of the spin models in Table \ref{table:testcomperr}, with the completion error of the cat-ind and the FA models.}
    \label{fig:compl-error}
\end{figure}

 We present the out-of-sample (test) pseudo-likelihoods of the maxent models for all the considered questionnaires in Table \ref{table:testpselik}, as well as the test completion errors-1 in Table \ref{table:testcomperr}. As said in the main text, the direct out-of-sample model comparison in terms of model likelihoods is intractable for the questionnaires at hand, as it requires a sum over $R^M$ terms. The completion error-1 of a model with inferred parameters $\theta^*$ on a test subject with response vector $\x$ is defined as the mean absolute deviation between each observed answer $x_i$ and its conditional expectation given all other answers according to the fitted model,
\begin{equation}
    \varepsilon_1(\x) = \frac{1}{M}\sum_{i=1}^M \left|x_i - \langle x_i\rangle_{P_i(x_i|\x_{\setminus i},\,\theta^*)}\right|,
\end{equation}
where $P_i(x_i|\x_{\setminus i},\theta^*)$ is the single-spin conditional distribution (Eq.~\ref{eq:condlik}) and $M$ is the number of items. The conditional mean $\langle x_i\rangle_{P_i}$ is thus the model's best prediction of question $i$ given the subject's answers to all other questions. The quantity reported in Table~\ref{table:testcomperr} is the average of $\varepsilon_1(\x)$ over all $N_\text{te}$ test subjects. The completion errors of the spin models are compared with those of the simple models in Fig. \ref{fig:compl-error}.

\begin{figure}
    \centering
    \includegraphics[width=0.9\linewidth]{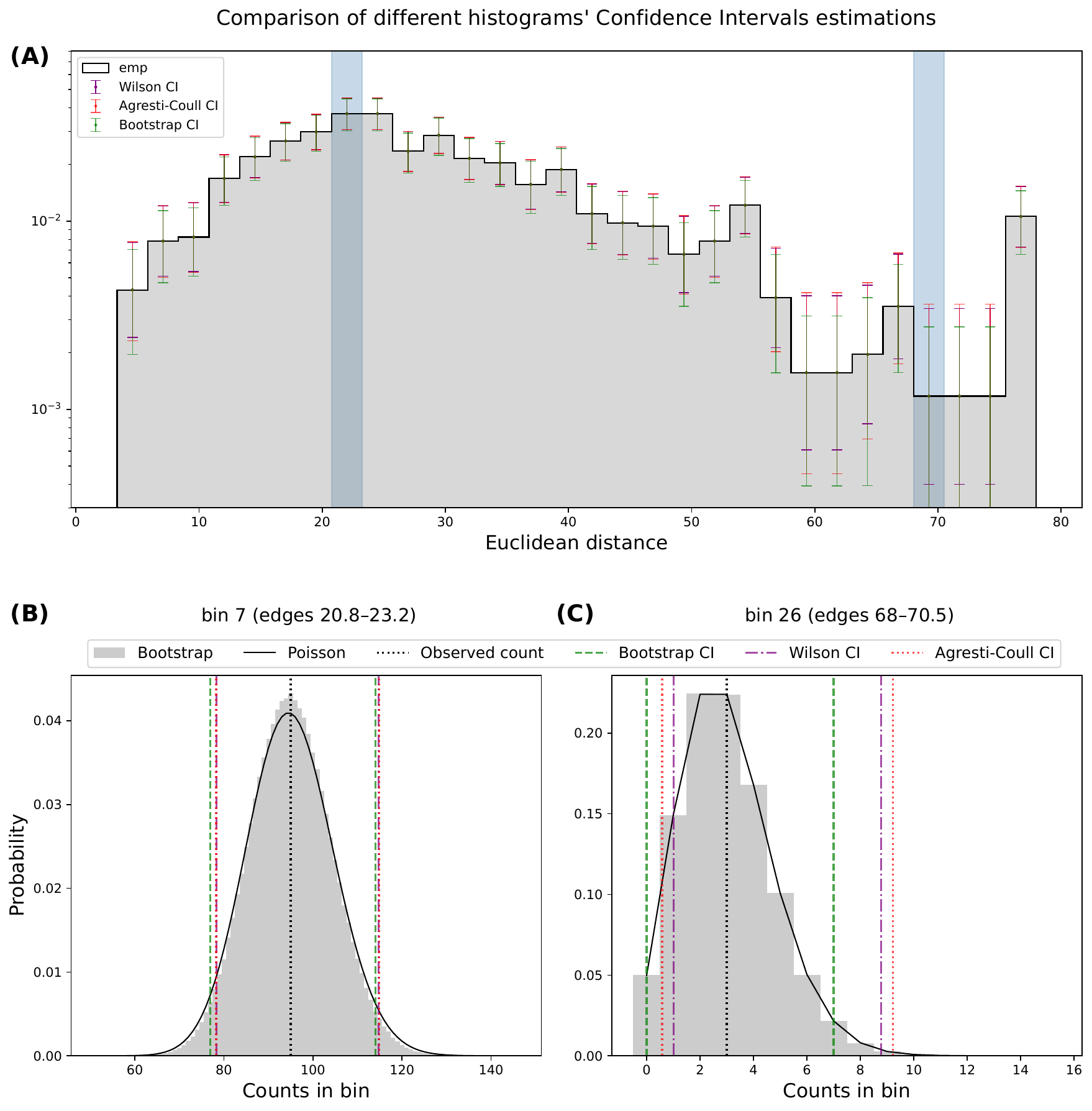}
    \caption{Comparison of confidence interval (CI) estimators for the empirical histogram of pairwise Euclidean distances. Top panel (A): Empirical density histogram (grey) with 95\% CIs computed via the Wilson, Agresti–Coull, and bootstrap methods overlaid on each bin (y-axis in log scale). Shaded blue bands mark a densely populated bin (bin 7) and a sparsely populated bin (bin 26), selected for detailed comparison below. Bottom panel (B-C): Bootstrap distributions (grey histograms, $n=10^6$ resamples) of the bin counts for bin 7 (B) and bin 26 (C), overlaid with the theoretical Poisson probability mass function evaluated at $\lambda$ equal to the observed count (black curve; dotted vertical line marks the observed count). Vertical dashed/dotted lines indicate the 95\% CI bounds from the bootstrap (green), Wilson (purple), and Agresti–Coull (red) methods.}
    \label{fig:ci-est}
\end{figure}

 The statistical significance of the empirical histograms shown in the results is assessed via confidence intervals on the bin frequencies. Each bin contains $k$ counts out of $n$ total observations, giving an empirical frequency $p = k/n$. We use the Wilson score interval \cite{wilson1927}, which provides a confidence interval $[p_-, p_+]$ at significance level $\alpha$ without assuming normality and without requiring $np \gg 1$:
\begin{equation}
    p_\pm = \frac{p + \dfrac{z^2}{2n} \pm \dfrac{z}{2n}\sqrt{4np(1-p)+z^2}}{1+\dfrac{z^2}{n}},
\end{equation}
where $z = \Phi^{-1}(1-\alpha/2)$ is the quantile of the standard normal distribution and $\alpha = 0.05$ throughout. Two alternative methods --the Agresti-Coull interval and non-parametric bootstrap re-sampling of the subject indices histograms-- were implemented, yielding completely consistent results (Fig. \ref{fig:ci-est}).

\begin{table}[H] 
\caption{Test pseudo-likelihood $\ln P_{\pl}(\Xte|\theta^*)$ for the three maxent models. \label{table:testpselik}}
\begin{ruledtabular}
\begin{tabular}{c|c|c|c}
questionnaire 		&  Ising	& BC 		& BEG 	\\
\hline
{\sf big5} 	& -0.0493(4)	& -0.0444(6)	& -0.0427(6)	\\	
{\sf cfcs} 	& -0.0508(4) 	& -0.0479(5)	& -0.0469(5)	\\	
{\sf dass} 	& -0.0387(8)	& -0.035(1)	& -0.035(1)	\\	
{\sf ei} 	& -0.050(1)	& -0.049(1)	& -0.049(1)	\\	
{\sf gcbs} 	& -0.041(1)	& -0.040(1)	& -0.038(2)	\\	
{\sf hsns} 	& -0.0504(4)	& -0.0466(5)	& -0.0449(5) 	\\	
{\sf iri} 	& -0.050(1)	& -0.046(1)	& -0.045(2)	\\	
{\sf mach} 	& -0.0489(5)	& -0.0485(5)	& -0.0470(6)	\\	
{\sf msscq} 	& -0.0440(5)	& -0.0395(7)	& -0.0370(7)	\\	
{\sf pwe} 	& -0.049(1)	& -0.048(2)	& -0.047(2)	\\	
{\sf rwas} 	& -0.0512(8)	& -0.0505(8)	& -0.0468(9)	\\	
{\sf sd3} 	& -0.0504(4)	& -0.0469(5)	& -0.0454(5)
\end{tabular}
\end{ruledtabular}
\end{table}

\begin{table}[H] 
\caption{Completion error-1 for the three maxent models. \label{table:testcomperr}}
\begin{ruledtabular}
\begin{tabular}{c|c|c|c}
questionnaire 		&  Ising	& BC 		& BEG 	\\
\hline
{\sf acme} &0.5543& 0.5418& 0.5165\\
{\sf big5} &0.7113& 0.6940& 0.6808\\
{\sf cfcs} &0.7840& 0.7741& 0.7747\\
{\sf dass} &1.124& 1.0932& 1.0692\\
{\sf ei} &0.9115& 0.9074& 0.8956\\
{\sf gcbs} &0.7207& 0.7124& 0.6753\\
{\sf hsns} &0.7617& 0.7480& 0.7385\\
{\sf iri} &0.7386& 0.7323& 0.7258\\
{\sf mach} &0.9232&  0.9218& 0.9135\\
{\sf msscq} &0.6369& 0.6162& 0.5897\\
{\sf pwe} &0.9030& 0.8951& 0.8850\\
{\sf rwas} &1.0455&  1.0372& 0.9484\\
{\sf sd3} &0.7885& 0.7766& 0.7665\\
\end{tabular}
\end{ruledtabular}
\end{table}


\section{Impact of the ADAM protocol.\label{sec:adam}}

As said in the main text, we perform maximum likelihood inference, without regularization, employing the ADAM algorithm. The last protocol turns out to be useful to infer some datasets. This is the case of the {\sf rwas} questionnaire, as we show in Fig. \ref{fig:losses_rwas} (compare with Fig. \ref{fig:losses_all} in Appendix \ref{sec:other}), in which we show the evolution of the losses during learning in absence of the ADAM protocol (using, instead, naif Euler iterations).

\begin{figure}[H]
\begin{center}
\includegraphics[width=0.99\columnwidth]{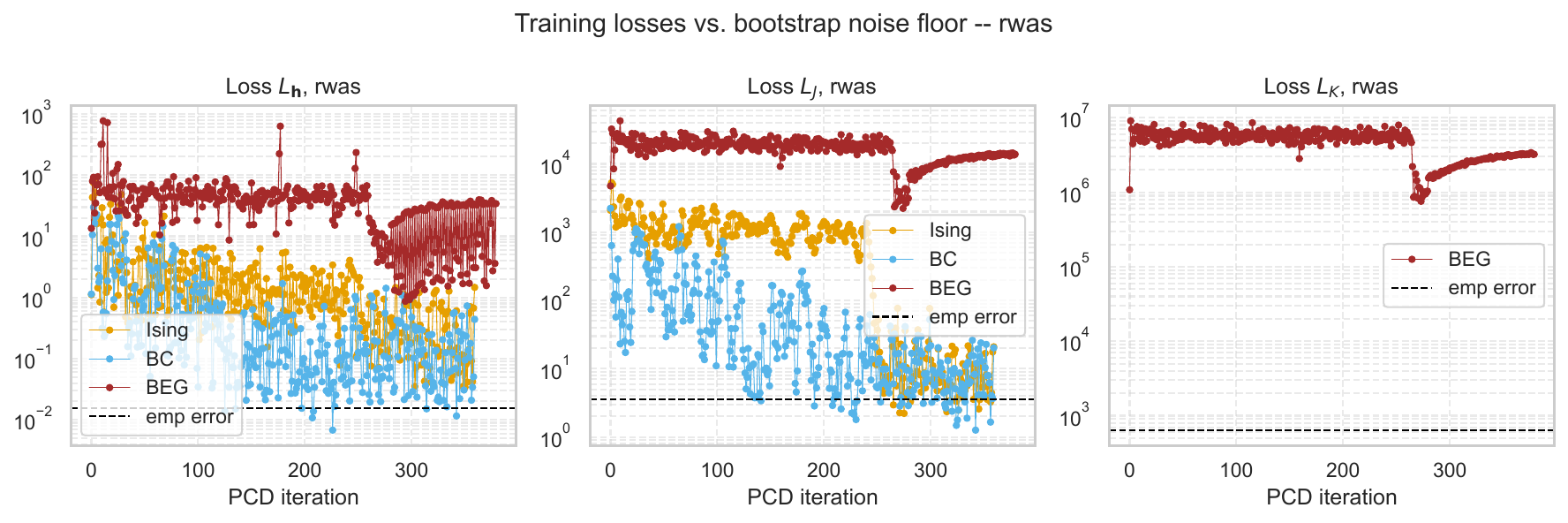}
\caption{Loss functions $L_{\bf h}$, $L_J$ (and $L_K$ for the BEG model) as a function of PCD iteration number, for the three spin models (Ising, BC, BEG) trained on the {\sf rwas} questionnaire without the ADAM protocol (compare with Fig. \ref{fig:losses_all} in Appendix \ref{sec:other}). \label{fig:losses_rwas}}
\end{center}
\end{figure}

As discussed in Sec. \ref{sec:validity} of the main text, for some datasets it could be worth to employ regularization in the learning procedure (at the expenses of moment matching). One should take into account, however, that the interaction between ADAM and regularization is not straightforward \cite{yun2021}. It is possible that the introduction of a certain amount of regularization makes ADAM less necessary. In any case, this question is out of the scope of the present work.

\section{Distances between empirical and theoretical histograms \label{sec:gof_hist}}
To provide a quantitative summary of the goodness of fit across all models and datasets, we compute two distribution-comparison metrics for each pair (model, questionnaire): the Jensen--Shannon Divergence (JSD) \cite{lin1991} and the Wasserstein Distance (WD, also called Earth Mover's Distance) \cite{peyre2019}.

The \textit{Jensen--Shannon Divergence} between two discrete distributions $P$ and $Q$ is defined as
\begin{equation}
    \mathrm{JSD}(P\|Q) = \frac{1}{2}\mathrm{KL}(P\|M) + \frac{1}{2}\mathrm{KL}(Q\|M),
    \qquad M = \frac{P+Q}{2},
\end{equation}
where $\mathrm{KL}$ denotes the Kullback--Leibler divergence. The JSD is symmetric, bounded in $[0,\ln 2]$, and equals zero if and only if $P=Q$.

The \textit{Wasserstein Distance} (of order 1) between two distributions on $\mathbb{R}$ is
\begin{equation}
    W_1(P,Q) = \int_{-\infty}^{+\infty} |F_P(x) - F_Q(x)|\,\d x,
\end{equation}
where $F_P, F_Q$ are the respective cumulative distribution functions. Unlike the JSD, the WD is sensitive to the \textit{geometry} of the support and penalizes mass displaced over large distances.

\begin{figure}[H]
    \centering
    \includegraphics[width=0.9\linewidth]{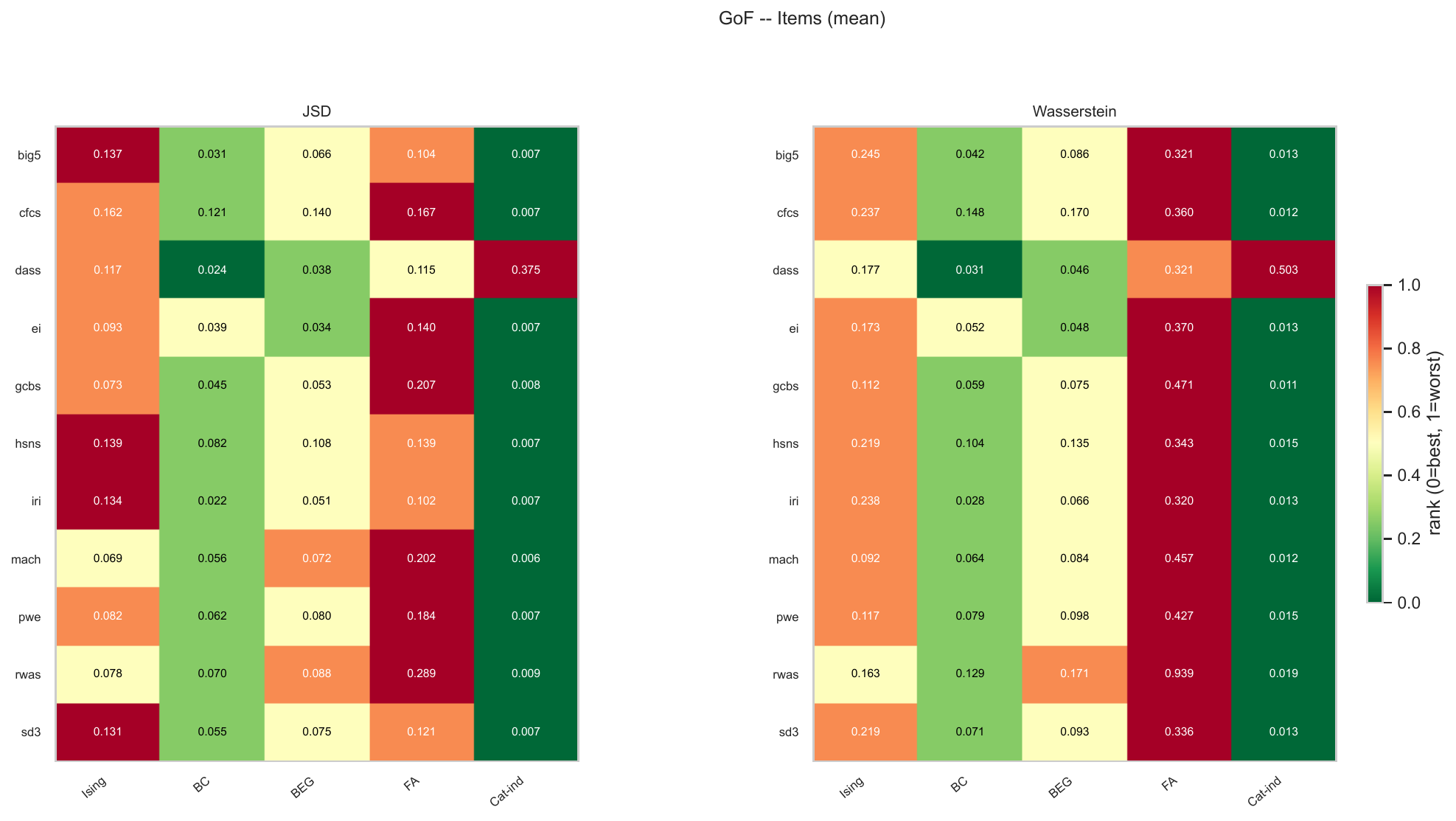}
    \caption{Goodness-of-fit heatmap for the histogram of the item value, ${\sf h}_{x_i}$, averaged over items $i$. Each cell reports the Jensen--Shannon Divergence (JSD, left panel) and Wasserstein Distance (WD, right) between the empirical histogram and the model histogram, for each questionnaire (rows) and each model (columns). Smaller values indicate better agreement. For each row, the cells are colored according to the ranking of the goodness of fit where dark green indicates the best fit and red the worst.}
    \label{fig:gof_item}
\end{figure}

\begin{figure}[H]
    \centering
    \includegraphics[width=0.9\linewidth]{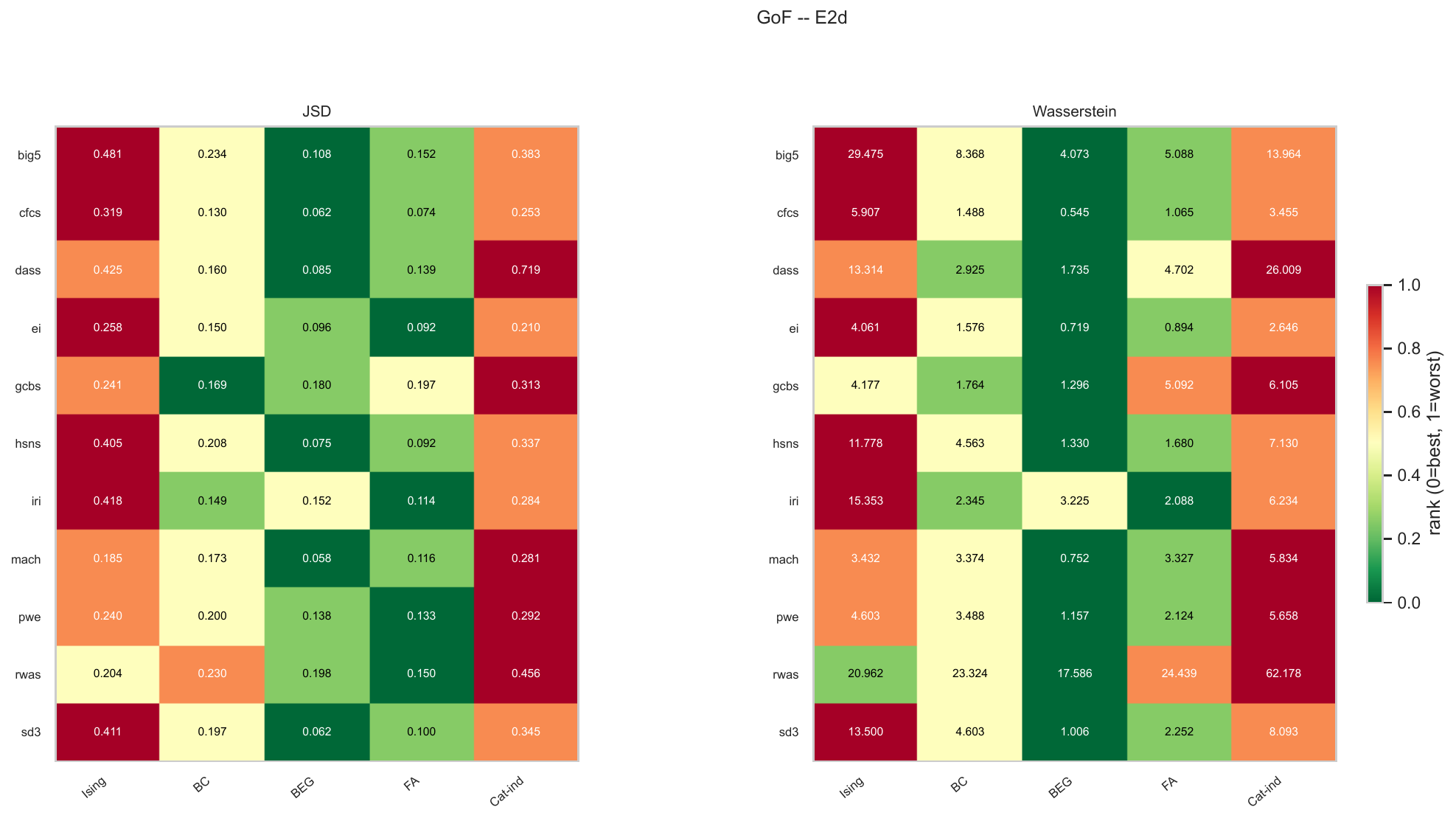}
    \caption{Goodness-of-fit heatmap for the histogram of the Euclidean distance to the mean, ${\sf h}_{d_\x}$. Each cell reports the Jensen--Shannon Divergence (JSD, left panel) and Wasserstein Distance (WD, right) between the empirical histogram and the model histogram, for each questionnaire (rows) and each model (columns). Smaller values indicate better agreement. For each row, the cells are colored according to the ranking of the goodness of fit where dark green indicates the best fit and red the worst.}
    \label{fig:gof_e2d}
\end{figure}

\begin{figure}[H]
    \centering
    \includegraphics[width=0.9\linewidth]{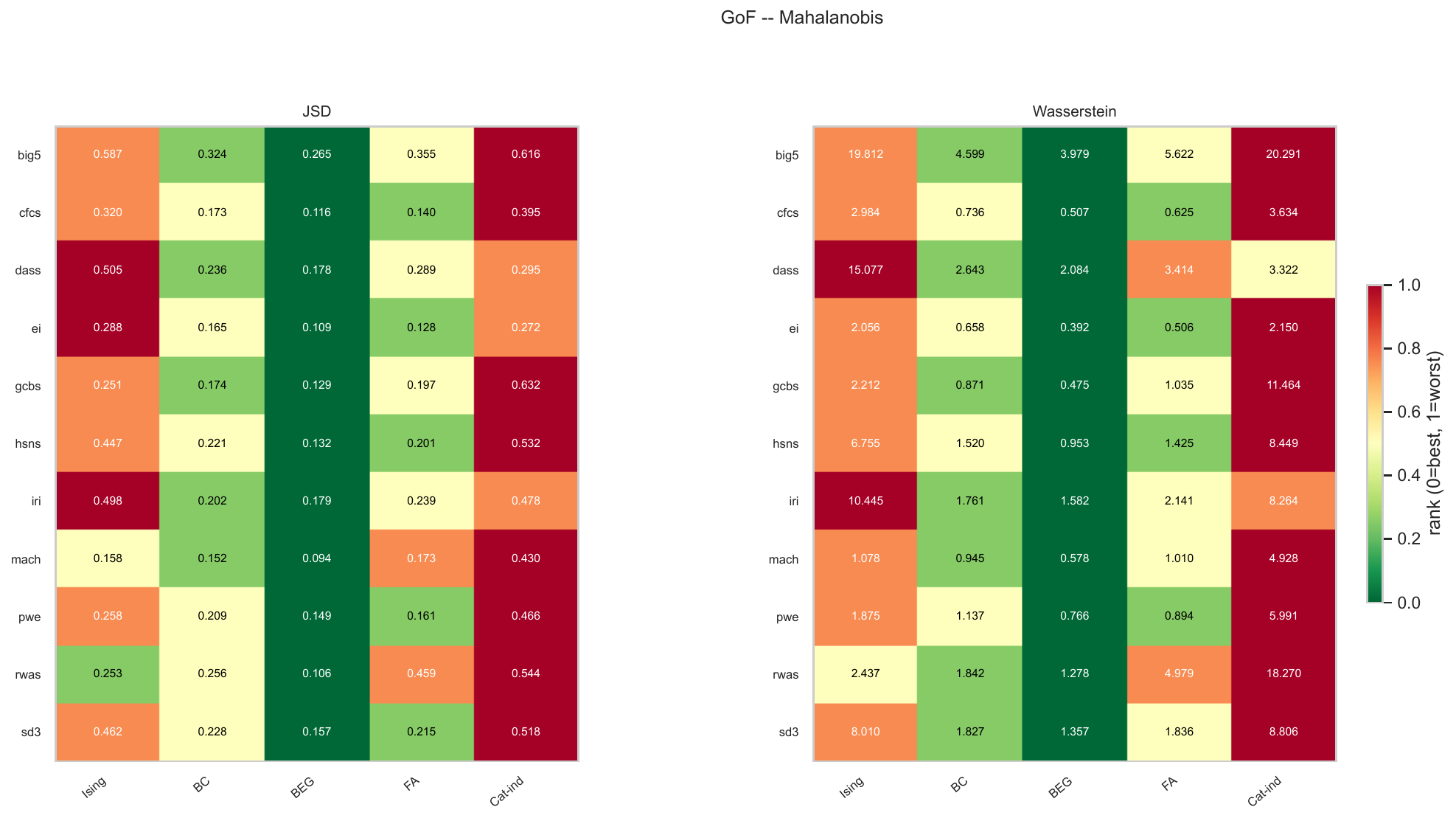}
    \caption{Goodness-of-fit heatmap for the histogram of the Mahalanobis distance to the mean, ${\sf h}_{d_\x^{({\rm M})}}$. Each cell reports the Jensen--Shannon Divergence (JSD, left panel) and Wasserstein Distance (WD, right) between the empirical histogram and the model histogram, for each questionnaire (rows) and each model (columns). Smaller values indicate better agreement. For each row, the cells are colored according to the ranking of the goodness of fit where dark green indicates the best fit and red the worst.}
    \label{fig:gof_mahala}
\end{figure}

\begin{figure}[H]
    \centering
    \includegraphics[width=0.9\linewidth]{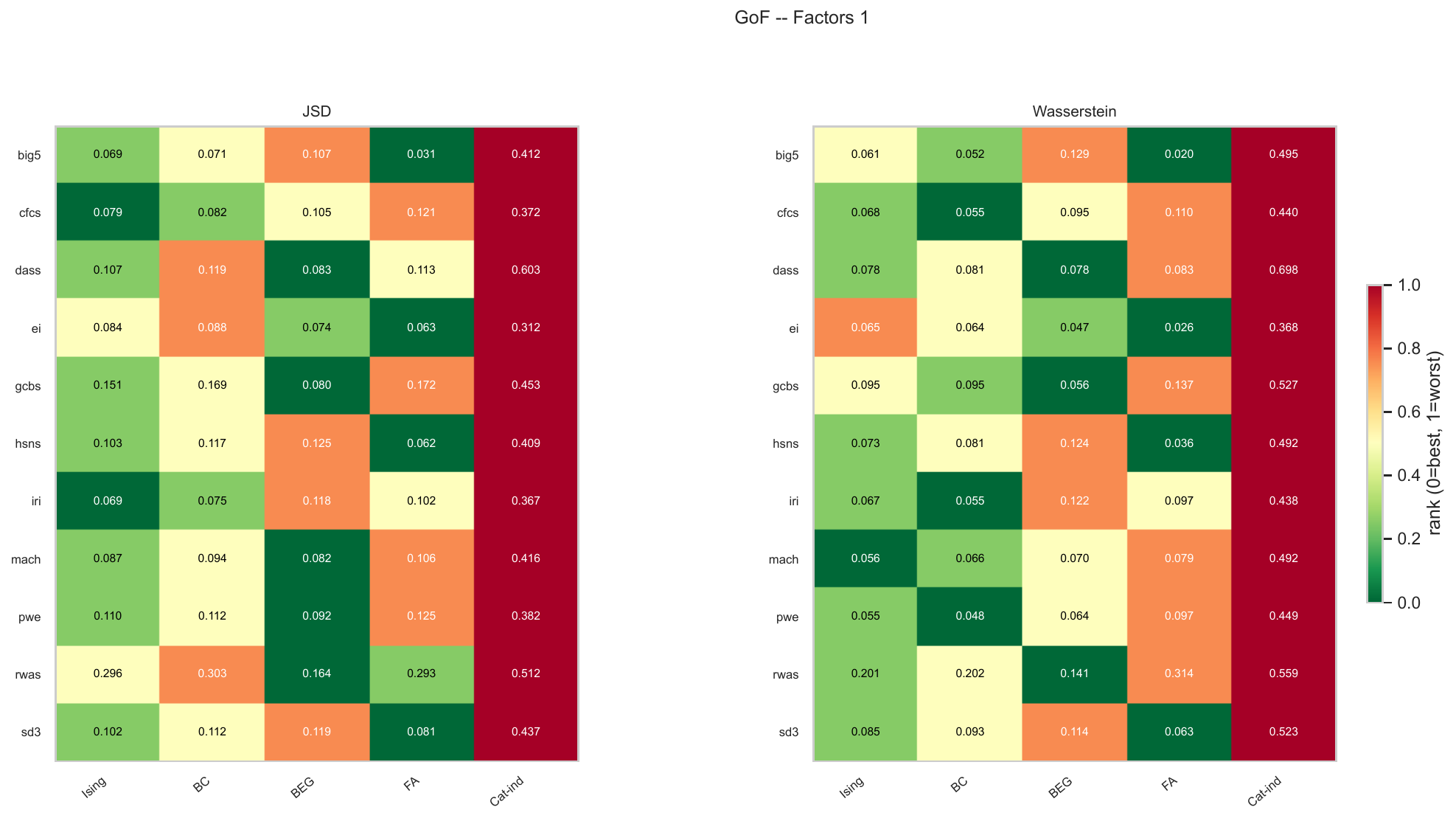}
    \caption{Goodness-of-fit heatmap for the histogram of the first factor, ${\sf h}_{f_1}$. Each cell reports the Jensen--Shannon Divergence (JSD, left panel) and Wasserstein Distance (WD, right) between the empirical histogram and the model histogram, for each questionnaire (rows) and each model (columns). Smaller values indicate better agreement. For each row, the cells are colored according to the ranking of the goodness of fit where dark green indicates the best fit and red the worst.}
    \label{fig:gof_f1}
\end{figure}

\begin{figure}[H]
    \centering
    \includegraphics[width=0.9\linewidth]{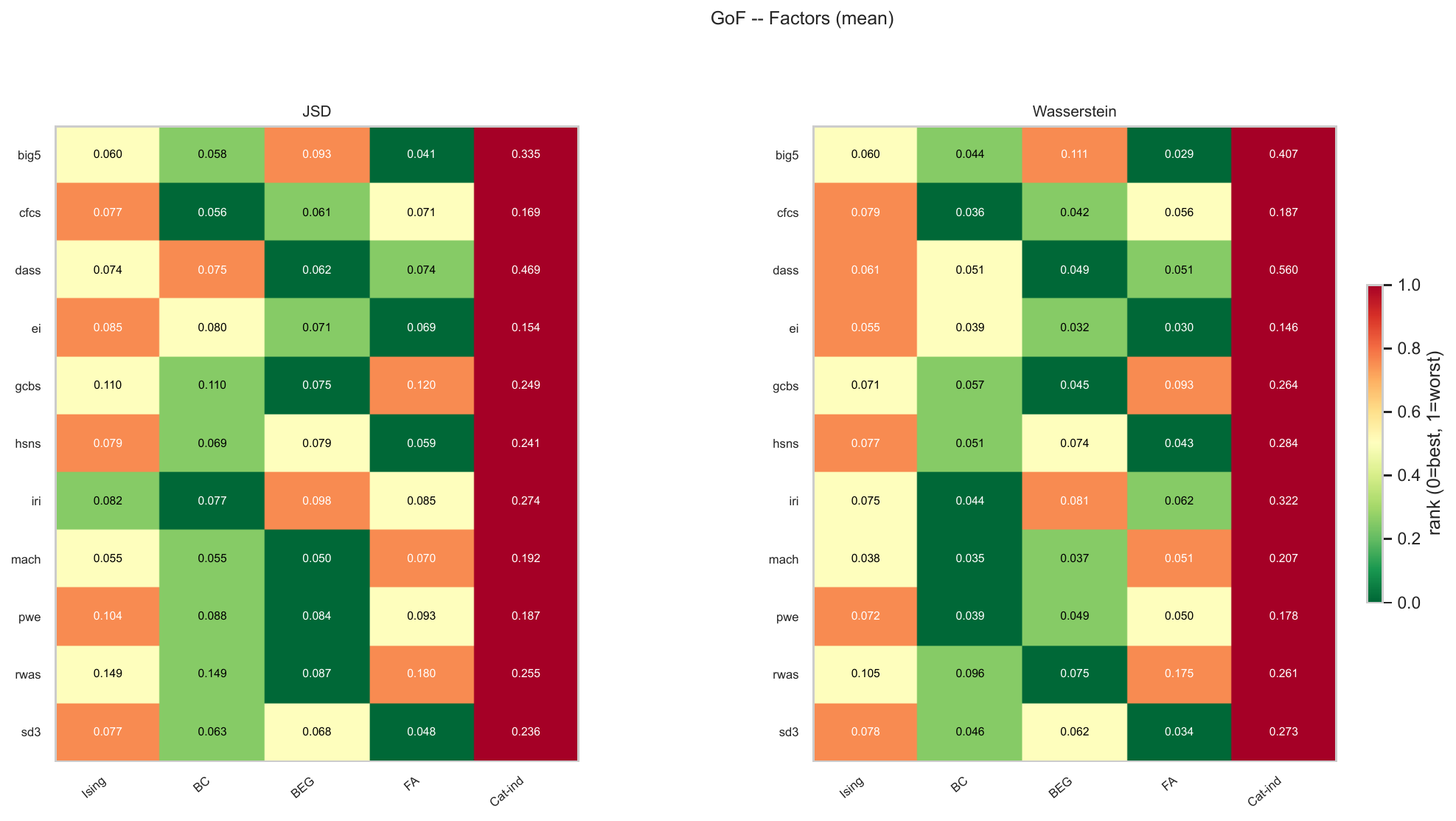}
    \caption{Goodness-of-fit heatmap for the histogram of the factors ${\sf h}_{f_j}$, averaged over factors $j$. Each cell reports the Jensen--Shannon Divergence (JSD, left panel) and Wasserstein Distance (WD, right) between the empirical histogram and the model histogram, for each questionnaire (rows) and each model (columns). Smaller values indicate better agreement. For each row, the cells are colored according to the ranking of the goodness of fit where dark green indicates the best fit and red the worst.}
    \label{fig:gof_fmean}
\end{figure}

Both metrics are computed between the empirical histogram ${\sf h}_{o,\rho_X}$ and the model histogram ${\sf h}_{o,P_{\cal M}}$ for each observable $o$ (Euclidean distance $d_\x$, Mahalanobis distance $d_\x^{({\rm M})}$, first factor $f_1$, all factors $f_j$ averaged over $j$) and each model $\mathcal{M}$. The results are displayed as heatmaps in Figs. \ref{fig:gof_item},\ref{fig:gof_e2d},\ref{fig:gof_mahala},\ref{fig:gof_f1},\ref{fig:gof_fmean} with rows corresponding to questionnaires and columns to models.

\section{Naif mean field theory, generalized magnetization and metastable phase onset \label{sec:meanfield}}

Consider the generalized Ising model, $x_j\in\{-r,-r+1,\ldots,r\}$ with $R=2r+1$ possible values of the spin. The variational free energy \cite{chaikin1995,mackay2003} associated to the probability distribution $Q(\x;{\bm a})=\prod_{j=1}^M q(x_j;a_j)$, with $q(x;a):= \exp(x a)/Z(a)$ reads, as a function of the mean fields $a_j$ and of the means ${\bm \mu}=\<\x\>_Q$:

\begin{align}
F &= -\frac{1}{2} {\bm\mu}^\dag\cdot J \cdot{\bm\mu}-{\bf h}^\dag\cdot{\bm\mu}-\beta^{-1}\sum_{j=1}^M \left( \ln Z(a_j)-a_j\mu_j \right) \label{eq:Fising} \\
Z(a) &: = \sum_{x=-r}^r e^{a x}
\end{align}

Minimizing the variational free energy leads to the mean field equations:

\begin{subequations}
\label{eq:meanfield}
\begin{align}
{\bm a} &= \beta\left( J\cdot {\bm\mu} + {\bf h} \right) \\
\mu_j &= \frac{R}{2}\coth\left(\frac{a_j R}{2}\right) - \frac{1}{2}\coth\left(\frac{a_j}{2}\right)
.
\end{align}
\end{subequations}

We are interested in the weak coupling regime, for low enough $\beta J$ and $\beta{\bf h}$. To order $O(a^4)$, the single-particle partition function reads (we expand the exponential with vanishing odd power terms, because of the symmetry of the sum in $x$):

\begin{align}
Z(a) &= R \left( 1 + \frac{a^2}{2} B_2 + \frac{a^4}{4!} B_4 \right) +  O(a^6)  \label{eq:Za} \\
B_2&:=\frac{1}{R}\sum_{x=-r}^r x^2 =\frac{R^2-1}{12} = \frac{r(r+1)}{3} \label{eq:B2} \\
B_4&:=\frac{1}{R}\sum_{x=-r}^r x^4 =\frac{(R^2-1)(3R^2-7)}{240} = \frac{r(r+1)(3r^2+3r-1)}{15} \label{eq:B4} 
\end{align}
where we have used the sum of squares and fourth powers identities: $\sum_{k=1}^n k^2 =n(n+1)(2n+1)/6$ and $\sum_{k=1}^n k^4 =n(n+1)(2n+1)(3n^2+3n-1)/30$. Therefore, using $\ln(1+y)=y-y^2/2+O(y^3)$: 

\begin{align}
\ln Z(a) & =  \ln R + \frac{B_2}{2}a^2  + \frac{B_{42}}{24} a^4 + O(a^6) \label{eq:logz} \\
B_{42} &:= B_4-3 B_2^2
\end{align}

The means $\bm \mu$ are related to the partition function as $\mu_j = \partial \ln Z(a_j)/\partial a_j$. Using Eq. \ref{eq:logz}, we get: 

\begin{align}\label{eq:muvsa}
\mu_j = B_2 a_j + (B_{42}/6) a^3 + O(a^5) 
.
\end{align}

Finally, the free energy as a function of the means takes the form, from Eqs. (\ref{eq:Fising},\ref{eq:Za},\ref{eq:muvsa}):

\begin{align}\label{eq:Fmus}
F &= -\beta^{-1}M\ln R -{\bf h}^\dag\cdot {\bm\mu} +\\
 &+\frac{1}{2}{\bm\mu}^\dag\cdot\left( (B_2\beta)^{-1} 1_M -  J \right)\cdot{\bm\mu}- \\
 &-\frac{1}{24\beta}\frac{B_{42}}{B_2^4}\sum_{j=1}^M\mu_j^4 + O({\mu}^5) 
\end{align}
the coefficient of the term proportional to the fourth power of the means can be seen to be actually positive. Indeed, from Eqs. (\ref{eq:B2},\ref{eq:B4}), we get an expression for $B_{42} = -r(r+1)(2r^2+2r+1)/15$, which is always negative. We will now analyze the behavior of the free energy in two representative cases. 

\subsection{Zero fields: continuous symmetry-breaking transition}

In the weak coupling approximation to $O(a^2)$, the solution of the mean field equations (\ref{eq:meanfield}) is ${\bm a}_0=B_2^{-1}{\bm\mu}_0$, with

\begin{align}\label{eq:mu0}
{\bm\mu}_0 = \left( (\beta B_2)^{-1} 1_M - J\right)^{-1}\cdot {\bf h}
.
\end{align}

Substituting in Eq. (\ref{eq:Fmus}), and defining $\z:={\bm\mu}-{\bm\mu}_0$, the deviations with respect to the minimum-free energy value ${\bm\mu}_0$ in Eq. (\ref{eq:mu0}), we get:

\begin{align}
 F &= F_0 + \frac{1}{2}{\z}^\dag\cdot\tilde J\cdot{\z}  -\frac{1}{24\beta}\frac{B_{42}}{B_2^4}\sum_{j=1}^M(z_j+\mu_{0j})^4 + O({z}^5) 
\end{align}
where $\tilde J :=((\beta B)^{-1} 1_M - J)$ and $F_0:=-\beta^{-1}M\ln R-\frac{1}{2}{\bm\mu_0^\dagger \tilde J\bm\mu_0}$. In this subsection we will first consider the vanishing field case ${\bm\mu}_0={\bf 0}$. In this situation, the free energy becomes an even function of $z_j$. Let now $J=\sum_{k=1}^M \epsilon_k {\u_k\u_k^\dag}$ be the spectral decomposition of $J$, and suppose $\beta<\epsilon_1^{-1}$ (where $\epsilon_1>\epsilon_2>\cdots$ is the largest $J$ eigenvalue). In this basis, it is: 

\begin{align}
F = F_0 + \frac{(\beta B_2)^{-1}}{2} \sum_{k=1}^M {z'_k}^2 ( 1 - \beta B_2 \epsilon_k ) + O(z^4)
,
\end{align}
where $z'_k=\u_k^\dag\cdot({\bm\mu}-{\bm\mu}_0)$ is the $k$-th principal component of $\bm\mu$. At $\beta = \beta^* = (B_2\epsilon_1)^{-1}$, the stability of the ${\z}={\bf 0}$ solution breaks down: this is the critical inverse temperature in the naif mean field approximation for weak couplings at $O({\bm a}^4)$ and zero external fields (assuming $\epsilon_1>0$). Immediately above this inverse temperature for low $\beta-\beta^*$, the free energy develops two symmetric local minima. The phase transition occurring at $\beta^*$ is the standard second-order paramagnetic-ferromagnetic transition of the mean field universality class, with broken symmetry $z'_1\to-z'_1$. Put it differently, using the above relation $\mu_j = B_2 a_j + O(a_j^2)$, and the mean field self-consistency equations (\ref{eq:meanfield}), we have ${\bm\mu} = \left( (B_2\beta)^{-1}1_M-J \right)^{-1}\cdot{\bf h}+O(a^3)$, from which we learn that the susceptibility matrix $\chi_{ij}=\partial\mu_i/\partial h_j$ takes the form: 

\begin{align}\label{eq:chi}
\chi = \left( (B_2\beta)^{-1}1_M-J \right)^{-1}
,
\end{align}
where $1_M$ is the identity matrix in $M$ dimensions. If we now rotate matrix $\chi$ in the basis of $J$ eigenvectors, we obtain $\chi'_{km}=\partial \mu'_k/\partial h'_m = \delta_{km} (B_2\beta)/(1-(B_2\beta \epsilon_k))$, which is regular for $\beta<\beta^*$, and diverges at $\beta^*=(B_2\epsilon_1)^{-1}$. From Eq. (\ref{eq:chi}), we also see that the eigenvectors of $\chi$ are in common with those of $J$ in this weak coupling scenario. If $\lambda_1,\epsilon_1$ are the largest eigenvalues of $\chi,J$ respectively, they are related as $\lambda_1=\beta B_2/(1-\beta B_2 \epsilon_1)$. Therefore, the axis $\u_1$ along which the generalized magnetization is defined, is actually the axis of largest variability: the first principal component of the data. 

\subsection{Non-zero fields and weak disorder: discontinuous onset of a metastable solution}

We will now adopt the {\it weak disorder approximation}, according to which the second eigenvalue of $J$ is much lower than the first, $\epsilon_1\gg\epsilon_2$. The free energy in Eq. (\ref{eq:Fmus}) as a function of the (non-centred) principal components $\mu_j':={\bf u}_j^\dag\cdot {\bm\mu}$ reads:

\begin{align}
F &= F_0  - {\bf h}'\cdot{\bm\mu}' + \frac{1}{2} \sum_{k=1}^M A_k(\beta) {\mu'_k}^2 -\frac{1}{24\beta}\frac{B_{42}}{B_2^4}\sum_{j=1}^M\left(\sum_{k=1}^M u_{kj}\mu'_k\right)^4 \label{eq:Fmuprime}  \\
A_k(\beta) &:= (B_2\beta)^{-1}   - \epsilon_k
\end{align}
where $F_0 = -\beta^{-1}M\ln\,R$ and ${\bf h}'=U\cdot {\bf h}$. In the weak disorder approximation, $\epsilon_2$ is enough smaller than $\epsilon_1$ to neglect the values of $\mu'_{k>1}$ in Eq.~(\ref{eq:Fmuprime}), which are exponentially suppressed by the quadratic term $A_{k>1}(\beta)$, in front of $\mu'_1$, so as to disentangle the different modes in the quartic term of Eq.~(\ref{eq:Fmuprime}). Consequently, the dependence of the variational free energy on the first principal component $\mu'_1$ is approximately equal to the {\it effective free energy for the mode $\mu'_1$}, $\tilde F_1$, depending on $\mu'_1$ only:

\begin{align} \label{eq:Fmuprime1}
\tilde F_1(\mu'_1) = -\beta^{-1}\ln R  - h'_1 \mu'_1 + \frac{1}{2} A_1(\beta) {\mu'_1}^2  +\frac{G_1(\beta)}{24}{\mu'_1}^4
\end{align}
where $G_k(\beta):=-(B_{42}/B_2^4)\beta^{-1}\sum_{j=1}^M u_{kj}^4$. The effective free energy $\tilde F_k$ develops two local minima whenever the discriminant of the cubic equation $\d\tilde F_k(z)/\d z=0$ becomes positive. This arises whenever:

\begin{align} \label{eq:discriminant}
A_1(\beta) \le -\left(\frac{9}{8}  G_1(\beta)\, {h'_1}^2\right)^{1/3}
.
\end{align}

For sufficiently low $h'_1$, there exists a critical value of inverse temperature $\beta_1\ge \beta^*$ (i.e., larger than the order-disorder inverse temperature transition $\beta^*$ defined in the above subsection, where we are still supposing $\epsilon_1>0$) such that the inequality (\ref{eq:discriminant}) holds for $\beta=\beta_1$ with an equal symbol and it holds with a $<$ symbol for any $\beta>\beta_1$. Immediately above this value $\beta\gtrsim\beta_1$, the free energy develops two separated local minima, corresponding, respectively, to the stable and to the metastable solution at a higher free energy. Therefore, in the presence of external fields, there is no second-order transition nor a low-temperature symmetry-broken phase. The mean field and weak disorder approximations predict instead a smooth crossover from low to large generalized magnetization ($\mu'_1$, which is, again, the first non-centered principal component of the data) and, for weak enough values of the (first principal component of the) external field $h'_1$, the onset of a second, metastable minimum of the free energy. The logarithm of the probability distribution of $x_1'$, ${\sf h}_{x'_1}(z_1)$ of the generalized Ising model can be approximated, in the weak disorder mean field approximation, and up to an additive constant, by $-\tilde F_1(z_1)$, since the probability density of $\mu'_1$ and of the other principal components factorize.

To sum up, the naif mean-field and weak-disorder approximations predict that, immediately above the critical temperature $\beta_1$, whenever it exists, the probability histogram of (only) the first principal component develops two local maxima for the inferred Ising model. This is what we actually observe (in the Ising model, not necessarily in the corresponding data, as discussed in Section~\ref{sec:results}) for the datasets {\sf gcbs} and {\sf rwas}.

\subsection{Validity of the mean-field approximation} So far, we have seen that the naif mean field and weak disorder approximations explain qualitatively the onset of a bi-modal histogram of the first principal component only, that we observe in some datasets. In order to address the validity of such approximations in these datasets, we have compared the (logarithm of the) histogram of the first principal components $x'_k$ of the Ising model inferred from these datasets, with the effective single-component free energies resulting from the mean field and weak disordered approximations:

\begin{align} \label{eq:Fmuprimek}
\tilde F_k(\mu'_k) = -\beta^{-1}\ln R  - h'_k \mu'_k + \frac{1}{2}A_k(\beta) {\mu'_k}^2  +\frac{1}{24}G_k(\beta){\mu'_k}^4
.
\end{align}

The comparison reveals that the picture explains the onset of the two modes of ${\sf h}_{x'_1}$ (and the absence of two modes of ${\sf h}_{x'_k}$ for $k>1$) only qualitatively. In fact, the mean field approximation is not accurate for the inferred Ising model presenting bi-modality, in the sense that, when substituting the inferred (maximum likelihood) couplings ${\bf h},J$ and the (maximum likelihood) $\beta=1$ in Eq.~(\ref{eq:Fmuprimek}), this function does not reproduce well the empirical histogram of the non-centred principal components. However, if we use, in the expression for $\tilde F_k$, not the maximum-likelihood value $\beta=1$ but a lower, fine-tuned value $\beta'<1$ (to correct for the inaccuracy of the mean-field approximation, which tends to underestimate the value of $\beta$~\cite{amit2005}), chosen so that the local maxima of $-\tilde F_1$ approximately reproduce those of ${\sf h}_{x'_1}$, we observe that in a wide interval around such $\beta'$ only $\tilde F_1(\mu'_1)$ presents two maxima, i.e.\ $\beta'>\beta_k$ only for $k=1$.

\begin{figure}[H]
\begin{center}
\includegraphics[width=0.7\columnwidth]{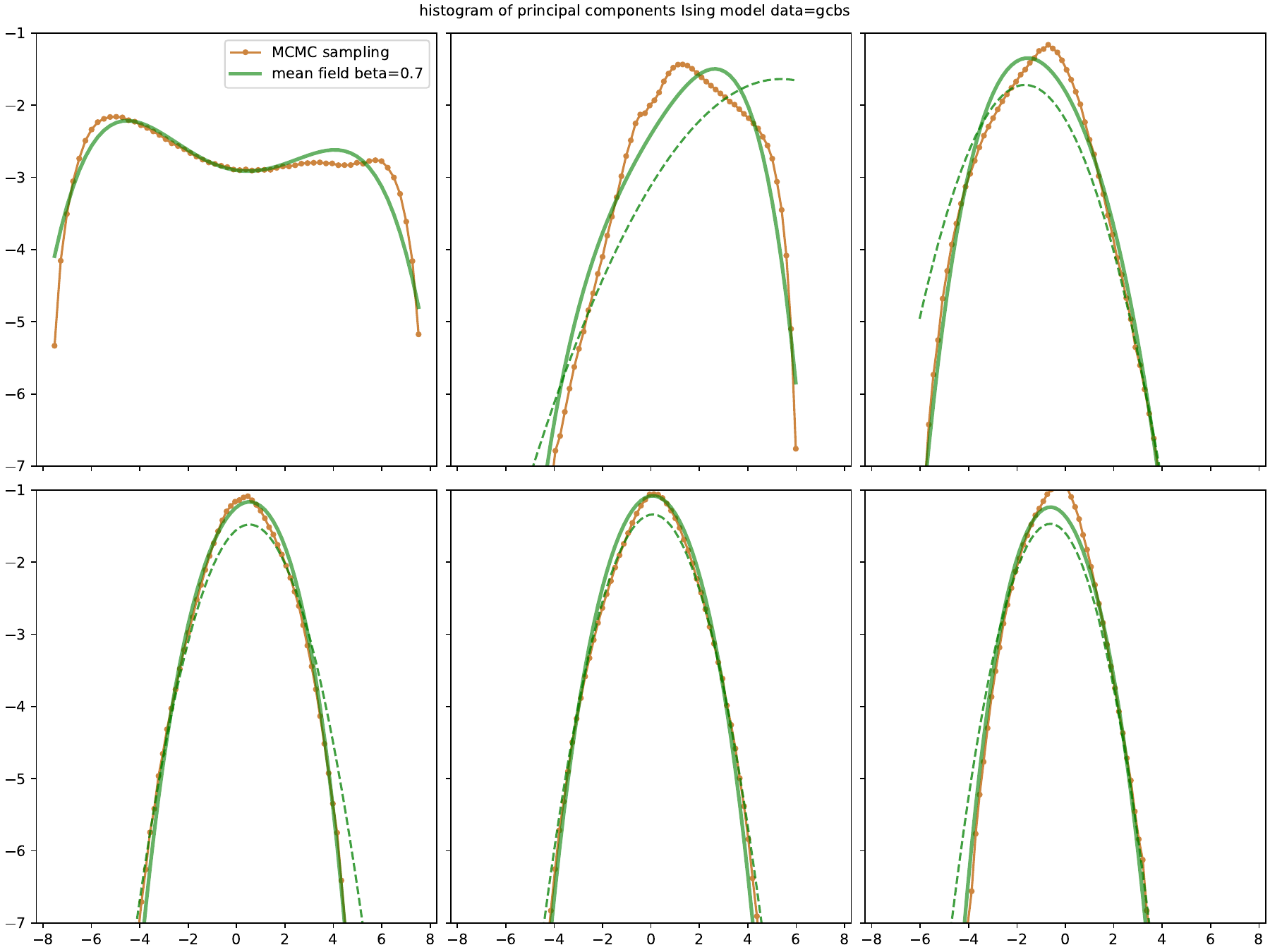}%
\caption{Histogram of the first 6 (non-centred) principal components ${\sf h}_{x'_i}$ ($i \in \{ 1, \,  \dots ,\, 6\}$) for the Ising model with parameters inferred from the {\sf gcbs} questionnaire: comparison between Ising data sampled from MCMC (points-lines) and the weak disorder mean field approximation $-\tilde F_k+c_k$ in Eq. (\ref{eq:Fmuprimek}) using an effective inverse temperature $\beta'=0.7$ (continuous green lines). The dashed line is Eq. (\ref{eq:Fmuprimek}) with zero quartic term. }\label{fig:naifmeanfield_gcbs}
\end{center}
\end{figure}

\begin{figure}[H]
\begin{center}
\includegraphics[width=0.7\columnwidth]{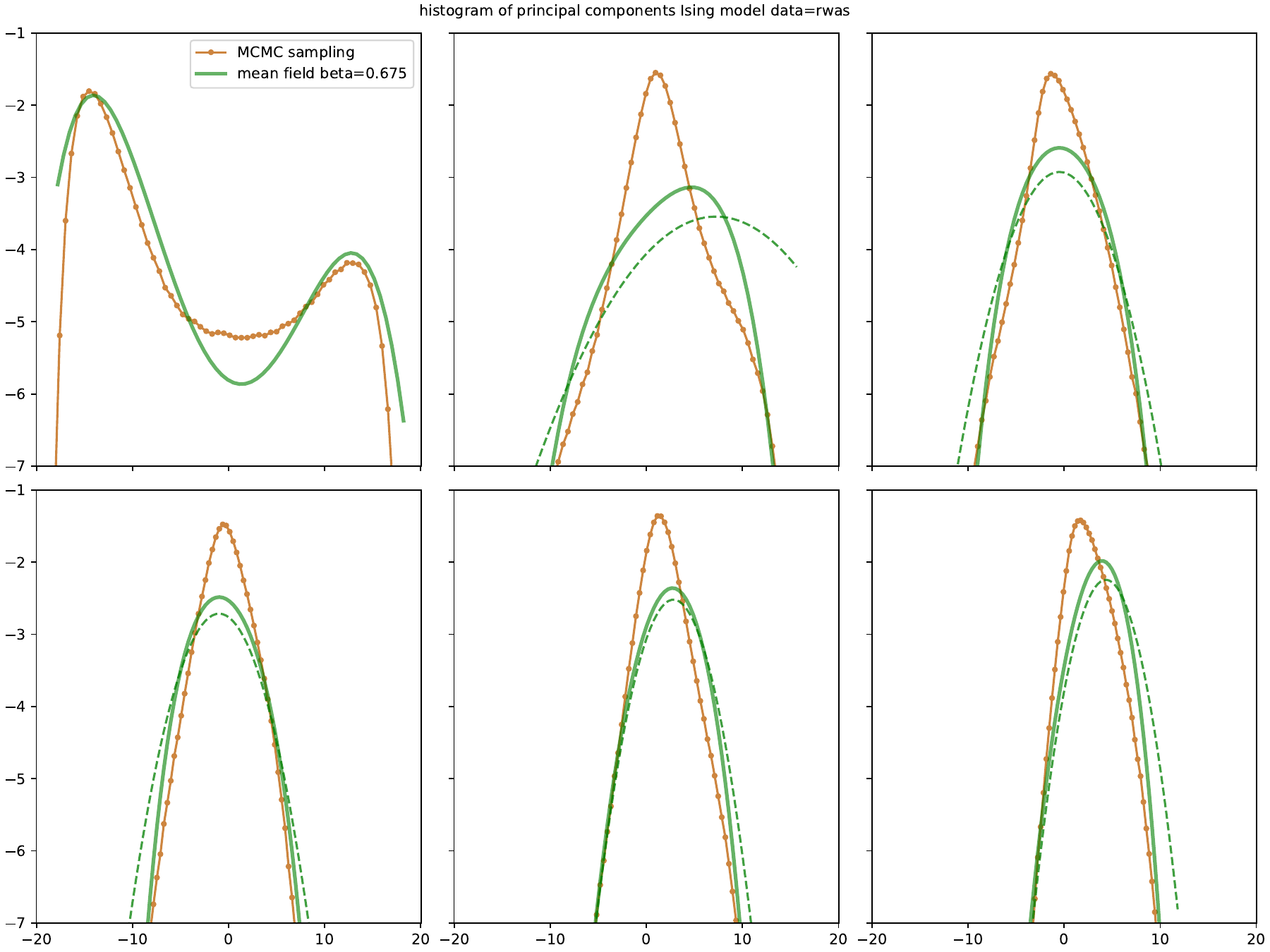}%
\caption{As in Fig. \ref{fig:naifmeanfield_gcbs} but for the {\sf rwas} data.}\label{fig:naifmeanfield_rwas}
\end{center}
\end{figure}

\begin{figure}[H]
\begin{center}
\includegraphics[width=0.7\columnwidth]{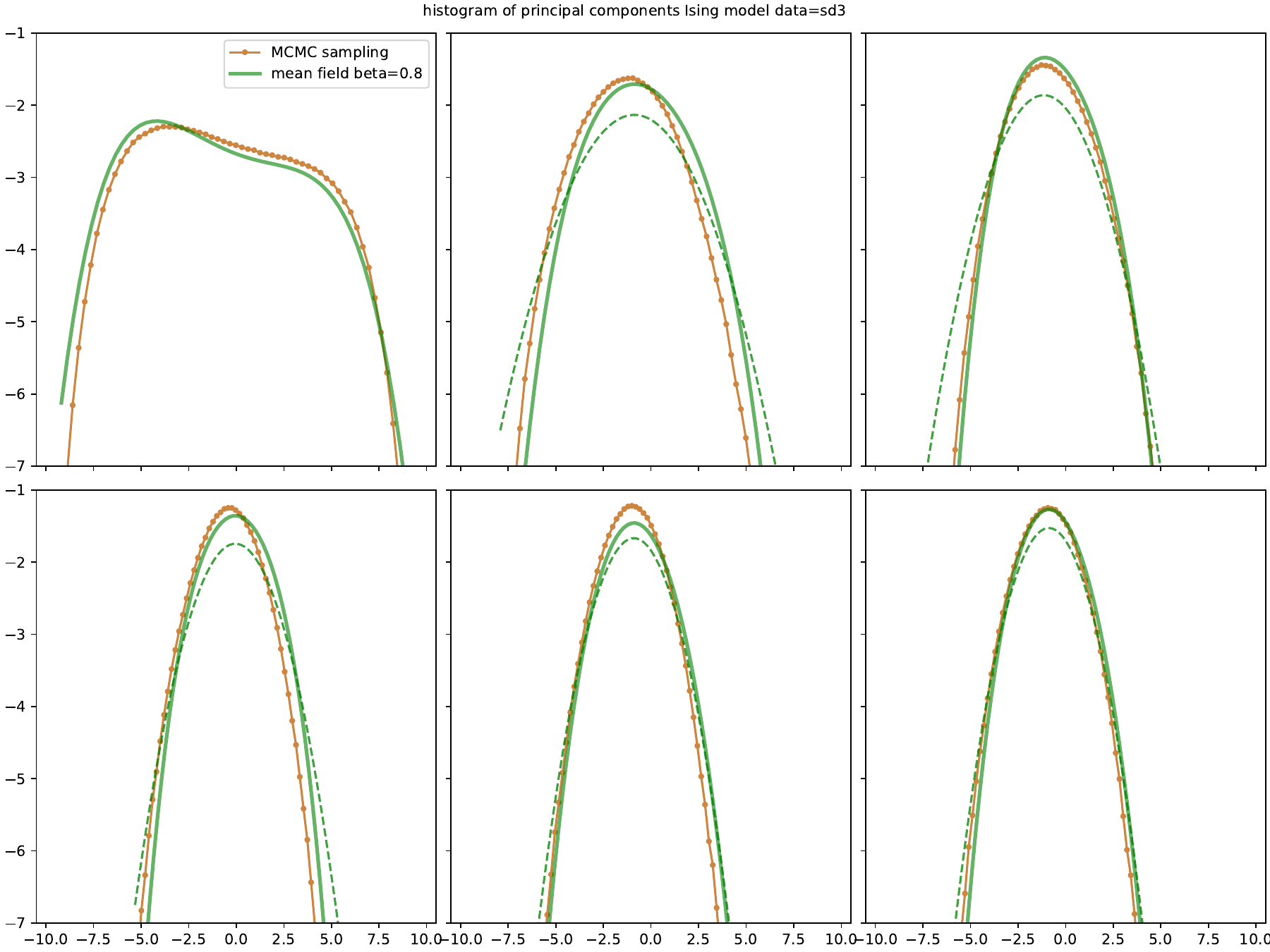}%
\caption{As in Fig. \ref{fig:naifmeanfield_gcbs} but for the {\sf sd3} data.}\label{fig:naifmeanfield_sd3}
\end{center}
\end{figure}

This is shown in Figs.~\ref{fig:naifmeanfield_gcbs},\ref{fig:naifmeanfield_rwas} for the {\sf gcbs}, {\sf rwas} datasets. In these figures, we show $\ln {\sf h}_{x'_k}$ for the Ising model versus $-\tilde F_k+c_k$, where $c_k$ is an offset minimizing the mean squared error with the Ising histogram in the given histogram points. The chosen values of $\beta'<1$ are indicated in the legend. We also show the case of the {\sf sd3} data in Fig.~\ref{fig:naifmeanfield_sd3}, to illustrate the fact that the above argument applies not only to the onset of a second maximum in ${\sf h}_{x'_k}$ (restricted to $k=1$), but also to the onset of a convex region (i.e., of two roots of $\d^2{\sf h}_{x'_k}(z)/\d z^2$) for $k=1$ only. A qualitatively identical picture holds for {\sf cfcs}, {\sf dass} and {\sf hsns}. 

We conclude that, although the naif mean-field approximation is not accurate enough to describe the Ising models that we infer from the data, the above explanation in the weak-disorder approximation is consistent: correcting the inaccuracy of the mean-field approximation with an effective inverse temperature $\beta'<1$, the effective free energies $\tilde F_k$ in~(\ref{eq:Fmuprimek}) correctly predict {\it which principal components of the inferred Ising model will develop bi-modality or convexity, and which will not, from the actual inferred values of the couplings} ${\bf h},J$.

A systematic assessment of the validity of mean-field theory for the models inferred in this article, as well as the cases of the BC and BEG models, will be examined elsewhere.

\subsection{Approximating the histogram of $x'_k$ with the free variational energy as a function of $\mu'_k$} 
In the above subsection, we have considered the variational free energy $-\tilde F_1$, evaluated in $z_1$, as the mean-field, weak-disorder approximation of the empirical histogram $\ln{\sf h}_{x'_1}(z_1)$. This comparison between $\ln{\sf h}_{x'_1}(z_1)$ and $-\tilde F_1(z_1)$ can be shown to be licit (despite $-\tilde F_1$ is a function of the principal component of the mean $\mu'_1$, and not of the variable $x'_1$) through a simple large-deviation like argument as the following. Up to an additive constant, the log-histogram $\ln{\sf h}_{x'_1}(z_1)$ is exactly given by what we can dub the constrained free energy, or $F_{z_1}=-\ln\sum_{\x}\delta(x'_1-z_1)\exp[-H(\x)]$. For large $M$, this constrained free energy can be seen to be bounded from above by the variational free energy $F[Q_{z_1}]$, $F_{z_1}\le F[Q_{z_1}]$, where $Q_{z_1}(\x)$ is the auxiliary probability distribution in which the first principal component $x'_1$ is constrained {\it in average}, as:

\begin{align}
Q_{z_1}(\x;\lambda,{\bm a})=:e^{\lambda(x'_1-z_1)}\prod_j q(x_j;a_j)
,
\end{align}
where $q(x;a):=\exp(a x)/Z(a)$ as above, and where the variational parameters are now $(\lambda,{\bm a})$. The hard constraint in $F_{z_1}$ corresponds, in other words, to a soft constraint in $Q_{z_1}$, with $\lambda$ being understood as a linear Lagrange multiplier shifting the mean of $Q_{z_1}$ so that $\mu'_1=\<x'_1\>_{Q_{z_1}} = z_1$. The fact that, in the large-$M$ limit, it is $F_{z_1}\le F[Q_{z_1}]$, can be seen by provisionally substituting the $\delta$ function in $F_{z_1}$ by an harmonic potential with a positive elastic constant $v_M$, of which we will take the $v_M\to\infty$ limit:

\begin{align} 
F_{z_1;v_M}&=\ln\sum_{\x}\exp[-H(\x;v_M)] \\
H(\x;v_M) &:= H(\x)+\frac{v_M}{2}\,(x'_1-z_1)^2
.
\end{align}
The variational free energy corresponding to the Hamiltonian $H(\x;v_M)$ reads:

\begin{align} 
F[Q_{z_1};v_M]&= \< H \>_{Q_{z_1}} + \frac{v_M}{2} \< (x_1' - z_1)^2 \>_{Q_{z_1}} -  S[Q_{z_1}] = \\
&= F[Q_{z_1}]  + \frac{v_M}{2}  \left( \sigma_1^2 + (\mu_1' - z_1)^2 \right) \label{eq:constrainedF}
,
\end{align}
where $S[\cdot]$ is the entropy of a distribution, and where $\sigma_1^2 = \text{Var}_{Q_{z_1}} (x'_1)=\< (x_1' - \mu_1')^2 \>_{Q_{z_1}}$. If the eigenvector ${\bf u}_1$ is de-localized in such a way that its components scale as $M^{-1/2}$, $x_1' = \sum_j u_{1j} x_j$ is a sum of independent random variables, whose variance scales as $O(M^0)$. For this same reason, $\mu'_1$ should scale as $O(M^{1/2})$, as well as the reference value $z'_1$. For large enough $M$, and under these assumptions, $\sigma_1^2\sim O(M^0)$ can be considered to be negligible in front of $(x'_1-z_1)^2\sim O(M)$. Therefore, in the large-$v_M$ limit, $F[Q_{z_1};v_M]$ converges to $F[Q_{z_1}]$, while $F_{z_1,v_M}$ converges to $F_{z_1}$. The larger is the value of $v_M$, the lower will be the deviations of the optimal parameter $\lambda$ minimizing the free energy with respect to the value $\lambda^*({\bm a})$ for which $\<x'_1\>_{Q(\x;{\lambda^*({\bm a}),{\bm a}})}=z_1$. 
This means that the best mean-field approximation for $F_{z_1}$ is $\min_{{\bm a}} F[Q_{z_1}(\cdot;\lambda^*({\bm a}),{\bm a})]$, or the minimum variational free energy among those whose mean first principal component $\mu'_1$ is $=z_1$. Hence, one can approximate $F_{z_1}$ by our free energy $\tilde F_1(\mu'_1)$, expressed as a function of the means only (where the ${\bm a}^*$ have been substituted by their naif mean field expression in terms of the $\bm\mu$'s), point by point in $z_1$, with $\mu'_1=z_1$.

\section{Alternative Mahalanobis distance with model-dependent covariance matrix \label{sec:mahalanobis}}

In order to assess to what extent the differences between the empirical and inferred Mahalanobis distances are due to the fact that the $L_J$ losses are not negligible and, therefore, the theoretical covariance matrix may be different from $C$ (something that can have an impact especially in the inverse of $C$ in Eq. \ref{eq:maha} \cite{ibanez2023}), we report as well the comparison between the empirical and theoretical histograms of a (model-dependent) variant of the Mahalanobis distance such that the covariance matrix in Eq. \ref{eq:maha}) is replaced by the covariance matrix according to each model. 
In this variant, ${\sf h}_{d_{\x}^{({\rm M})},P}$ is the histogram of the quantity $d_{\x}^{({\rm M})}=(1/2)\x^\dag\cdot\Sigma^{-1}\cdot\x$, where $\Sigma = \<\x\x^\dag\>_P-\<\x\>_P\<\x^\dag\>_P$. The results of such analysis (See Fig. \ref{fig:energyvar_all} in Appendix \ref{sec:other}) reveal that the differences between theoretical and empirical Mahalanobis distances cannot be attributed to the finite loss error in the estimation of $C$. 
%

\section{Closer to the mean outliers in {\sf rwas}. \label{sec:rwasenergy}} 

\begin{figure}[H]
\begin{center}
\includegraphics[width=0.49\columnwidth]{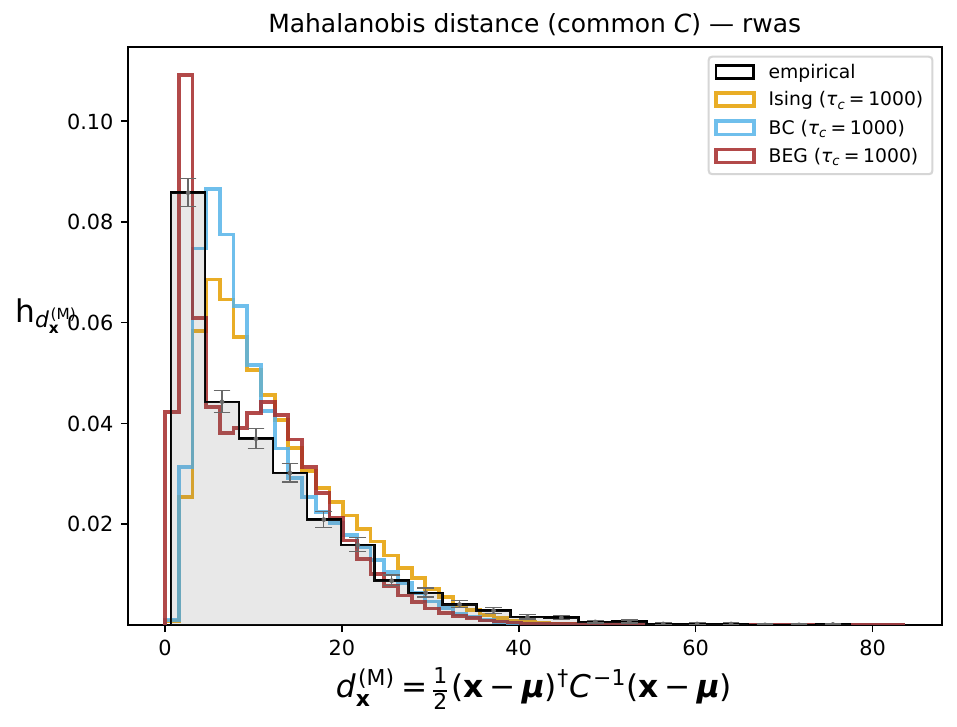}
\includegraphics[width=0.49\columnwidth]{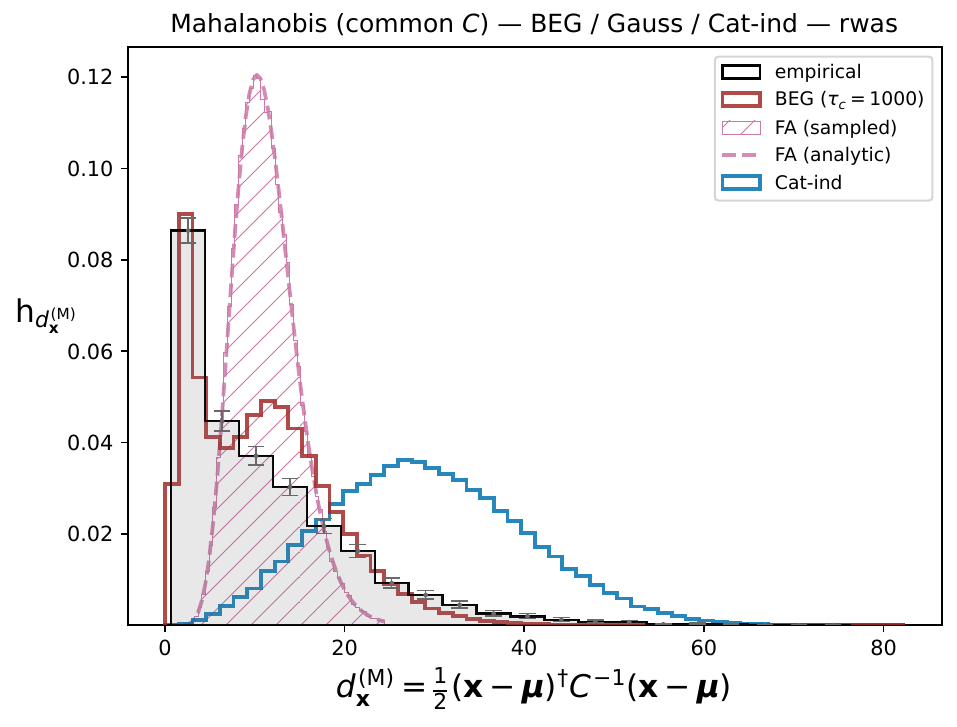}
\caption{Histogram of the Mahalanobis distance to the mean, ${\sf h}_{d_\x^{({\rm M})}}$ (linear scale), for the {\sf rwas} questionnaire. Left: comparison between the three spin models (Ising, BC, BEG) and the empirical distribution; only the BEG model accounts for the empirical peak of subjects near the origin (small $d_\x^{({\rm M})}$). Right: same comparison for the simple models (FA, cat-ind) and BEG model); both simple models predict a maximum at a significantly higher Mahalanobis distance. Error bars are Wilson score confidence intervals at $\alpha=0.05$.}
\label{fig:energy_rwas}
\end{center}
\end{figure}

The {\sf rwas} questionnaire presents a proliferation of subjects close to the mean in terms of the Mahalanobis distance, for which, among the spin models, only the BEG model accounts. This effect is shown in Fig. \ref{fig:energy_rwas}. The simple models -- {\sf gauss} and {\sf cat-ind} models -- predict a maximum at a much higher value of the Mahalanobis distance from the centre.

\section{Correlation time analysis \label{sec:correlation}}

We have estimated the correlation time of the MCMC induced by the Gibbs sampling algorithm on the maxent models with the values of the parameters $\theta^*$ inferred with the maximum likelihood prescription. This is done to exclude the possibility that the correlation time is of the same order of the number of MCMC sweeps from which we sample the theoretical PC and distance histograms presented in this article. In order for our estimation to be reliable, the number of MCMC sweeps $\tau_{\rm s}$ should be much larger than such a correlation time. Indeed, one should not talk about about a single correlation time, but about a correlation time for each observable $o$ of interest. When sampling from $\tau_{\rm s}$ MCMC sweeps, the standard deviation of the average of $o$ over the $\tau_{\rm s}$ samples decreases, when averaged over many realizations of the Markov Chain, as one over the square root of the {\it effective number of MCMC sweeps} $\tau_{\rm s}/\tau_{o}$, where $\tau_{o}$ is the correlation time associated to the observable $o$ \cite{pelissetto1993,sokal1997}. We have estimated, with two different methods, the correlation time of the first principal components of the data $o(\x)=x'_j$, with $j=1,\ldots,7$. The first method is the integral of the autocorrelation function $g_{o}(t)$:

\begin{align} 
g_{o}(t) := \frac{\<o(t')o(t'+t)\>-\<o(t')\>\<o(t')\>}{\<o(t')^2\>-\<o(t')\>^2}
\end{align} 
while the second is the quotient of the naif and Jack-Knife estimators of the variance of $o$ when the data is coarse-grained in blocks of increasingly large block size \cite{amit2005}. Both methods provide a consistent answer: the first principal components $x'_
1$ exhibit the largest correlation times for all the models; while for any given $x'_j$, the BEG model exhibits a larger correlation time than the BC model, in its turn larger than for the Ising model; the largest correlation time, $\tau_{x'_1}$ for the BEG model, is of the order of dozens (for instance, in the {\sf big5} data, the confidence interval for $\tau_{x'_1}$ at $\alpha=0.05$, assuming a $\chi^2$ distribution, is $[48.6,57.9]$). This implies the correlation time of the slowest quantity that we have considered is at least $10^{4}$ times smaller than the total number of MCMC sweeps, and at least two orders of magnitude smaller than the number of MCMC sweeps between resetting the initial configuration of the Markov chain to an empirical configuration, equal to ${\tt t}_{\rm r}\times \tau_{\rm PCD}=1.2\cdot 10^{4}$ sweeps in the PCD learning algorithm (see Sec. \ref{sec:PCD}).

\begin{figure}
\includegraphics[width=0.33\columnwidth]{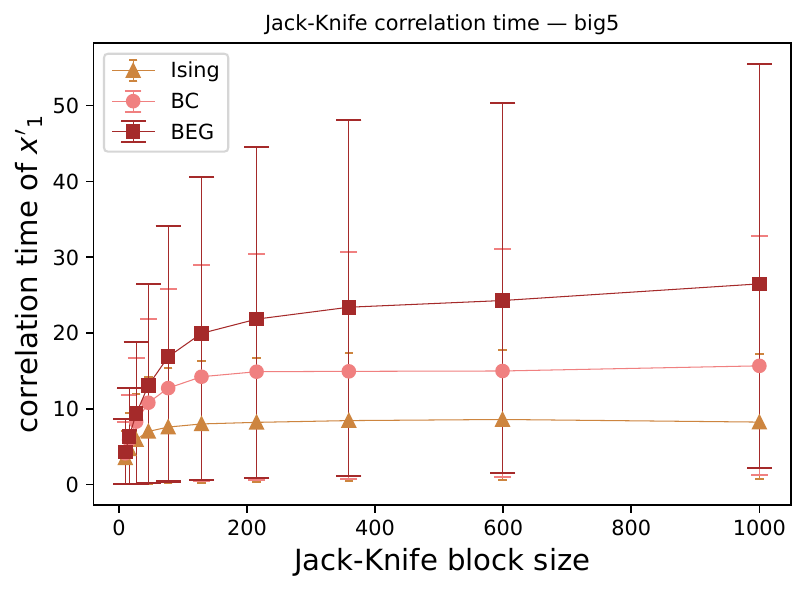}%
\includegraphics[width=0.33\columnwidth]{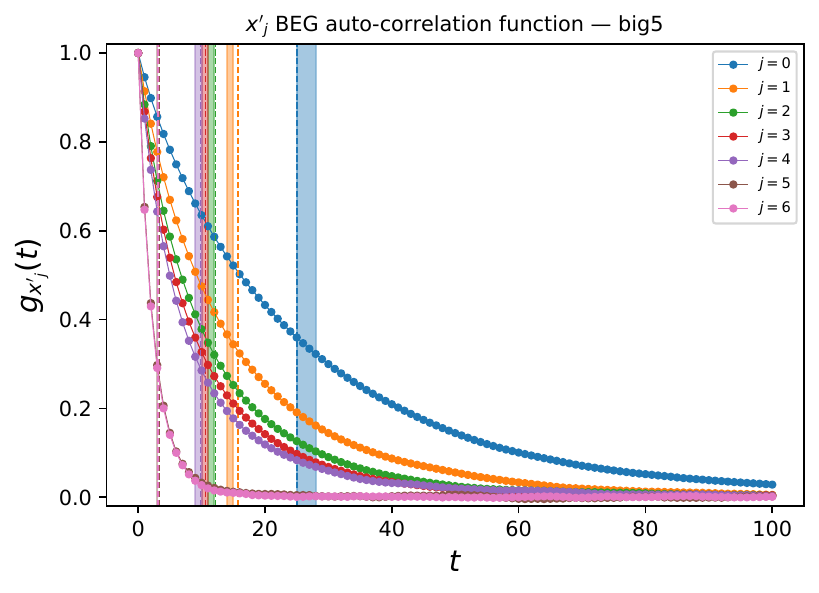}%
\includegraphics[width=0.33\columnwidth]{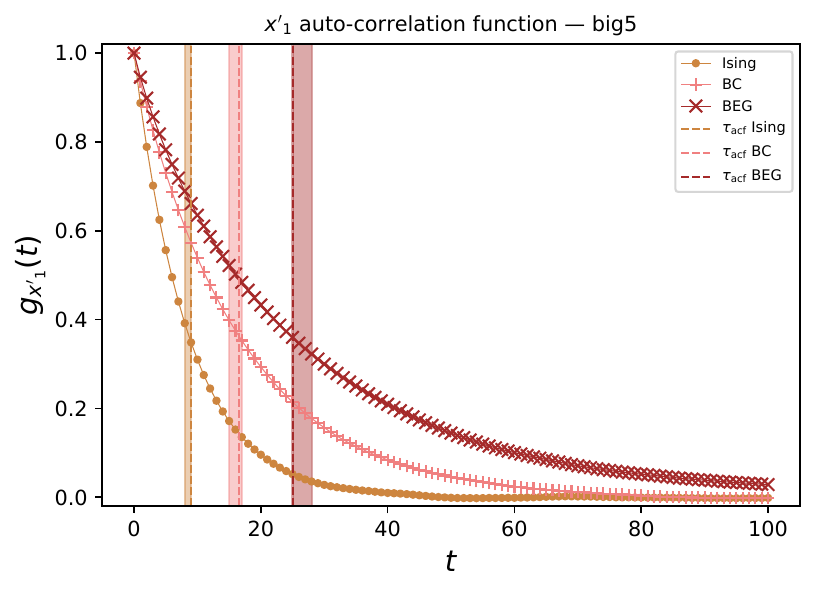}%
\caption{Correlation time analysis for the {\sf big5} questionnaire. Left: estimated correlation times $\tau_{x'_1}$ for the first principal components $x'_1$ of the three spin models, obtained via Jack-Knife method. Center: normalized autocorrelation function $g_{x'_j}(t)$ ($j=1,\ldots,7$) for the BEG model across the first 7 principal components. Right: comparison of the normalized autocorrelation function for the first principal component $g_{x'_1}(t)$ across the three spin models. Error bars in the left figure show error estimates of $\tau_{x'_1}$ for the three models computed via Jack-Knife, while the width of the vertical strips in the centre figure represents the error estimate of $\tau_{x'_j}$ for the BEG model, computed via Jack-Knife method, and the dotted vertical lines are the estimates of $\tau_{x'j}$ via the integrated autocorrelation function method; the right figure is similar, but it regards all the three models just for $\tau_{x_1'}$. The largest estimated correlation time (BEG model, $x'_1$) lies in the interval $[24.3, \,28.9]$ at $\alpha=0.05$ (assuming a $\chi^2$ distribution), which is at least $10^4$ times smaller than the total number of MCMC sweeps $\tau_{\rm s}=10^7$ used for sampling; see Sec.~\ref{sec:correlation}.}
\label{fig:correlation_big5}
\end{figure}

We illustrate these results for the {\sf big5} questionnaire in Fig. \ref{fig:correlation_big5}. The analysis of the rest of the questionnaires leads to the same conclusions. 

\end{document}